\documentclass[aps,prb,twocolumn,superscriptaddress,longbibliography,floatfix]{revtex4-2}
\usepackage{amsmath,amssymb,graphicx,bm,xcolor,slashed,dsfont}
\usepackage[colorlinks=true,linkcolor=blue,citecolor=blue,urlcolor=blue]{hyperref}
\newcommand{\Tr}{\mathrm{Tr}}
\newcommand{\one}{\mathds{1}}
\newcommand{\X}{\mathcal{X}}
\newcommand{\Xb}{\bar{\mathcal{X}}}
\newcommand{\ii}{\mathrm{i}}

\begin{document}

\title{Dzyaloshinskii--Moriya interactions in the SU(2) gauge theory of the Shastry--Sutherland Dirac spin liquids}

\author{Yasir Iqbal}
\email{yiqbal@physics.iitm.ac.in}
\affiliation{Department of Physics, Indian Institute of Technology Madras, Chennai 600036, India}

\begin{abstract}
We classify the fermion bilinears of the SU(2) gauge theory of the
Shastry--Sutherland Dirac spin liquids under the spin--orbit-locked space
group and treat the Dzyaloshinskii--Moriya (DM) interactions of
SrCu$_2$(BO$_3$)$_2$ at leading order in $1/N$ and at one loop. At linear
order DM can neither gap the spinons nor destabilize the SO(5) point above
a scale $T^*$. Symmetry forbids every local gauge-singlet fermion mass at any
order. Among the singlet bilinears, the in-plane couplings induce only a
conserved spin current of dimension exactly 2, which sets
$T^*=4\lambda\simeq0.03$--$0.17$\,K, and the out-of-plane
coupling only a flavored spin-2 current, irrelevant at a unitary critical
point. DM-induced Yukawa couplings start below $2\%$ of the one that
bounds the pseudocritical window. In zero field, measured samples order above $1$\,K,
at or above $T^*$. Below $T^*$, in a two-dimensional sample at the critical
pressure, the Dirac cones reconstruct into a chirally protected $d$-wave
quadratic band touching whose only gauge-singlet gap channels are the two
valence-bond components and the scalar chirality. A one-loop flow, with
the gauge field screened by the band touching, ends at
$T^{**}\sim(0.03$--$0.1)\lambda$ on a ray containing valence-bond order
for the material's exchange and for most random couplings,
and otherwise on a gapped chiral $\mathbb{Z}_2$ spin liquid. The DM bias
between the two plaquette patterns is at most of order $10^{-4}$, with a direction not fixed at $O(1/N)$, whereas non-magnetic quenched disorder couples a
relevant random field to the plaquette doublet and none to the chirality.
A field along $[100]$ induces a $(\pi,\pi)$ N\'eel moment linear in the
field; a field along $[001]$ does not. This selection
rule also holds in the dimer phase and is testable by $^{11}$B NMR under
pressure.
\end{abstract}

\maketitle

\section{Introduction}
\label{sec:intro}

Quantum phase transitions of two-dimensional antiferromagnets can fall
outside the Landau--Ginzburg--Wilson paradigm when the critical degrees of
freedom are fractionalized spinons coupled to emergent gauge fields~\cite{Senthil2004a,Senthil2004b}. The Shastry--Sutherland (SS) lattice~\cite{ShastrySutherland} is realized by SrCu$_2$(BO$_3$)$_2$~\cite{Kageyama1999,MiyaharaUeda1999,MiyaharaUeda2003}, the
most studied material candidate for such a transition~\cite{LYSV2019}. Its
crystal structure was determined by Smith and Keszler~\cite{SmithKeszler1991}.
Under pressure its plaquette-singlet phase gives way to an
antiferromagnet~\cite{Zayed2017,Guo2020,LarreaJimenez2021}, also in
applied fields~\cite{LarreaJimenez2021,Shi2022}, and field-induced
magnetization plateaus persist into the plaquette phase~\cite{Haravifard2016}.
Field-tuned $^{11}$B nuclear magnetic resonance (NMR) reports scaling near a proximate deconfined
critical point~\cite{Cui2023}, zero-field calorimetry finds a first-order
transition with a bicritical point~\cite{Guo2025}, and calorimetry in
applied fields finds a specific heat linear in $T$ whose origin is open,
fractionalized spin excitations being one possibility~\cite{Guo2026}. Between the plaquette and
N\'eel phases of the SS model, numerical studies find a narrow gapless
spin-liquid phase~\cite{Yang2022,Viteritti2025,Corboz2026} (identified as a $\mathbb Z_2$ Dirac
spin liquid~\cite{Maity2025}), a weakly first-order transition~\cite{Xi2023}, or a continuous one with emergent symmetry~\cite{Wang2024}. An infinite projected entangled-pair state (iPEPS) study extrapolated to infinite bond dimension places this phase at $0.785(5)\le J'/J\le0.82(1)$, compatible with a gapless spin liquid, entered from the plaquette phase by a weakly first-order transition and left continuously into the N\'eel phase, and attributes the differing conclusions of Refs.~\cite{Xi2023,Wang2024} to limited bond dimension and, for Ref.~\cite{Wang2024}, open boundaries~\cite{Corboz2026}. A fermionic gauge theory unifies the SU(2) $\pi$-flux
state, the U(1) staggered-flux state U800 and the gapless $\mathbb Z_2$
Dirac spin liquid Z3000~\cite{Maity2025} within a single Higgs framework~\cite{Feuerpfeil2026,Maity2026}, which carries over to the SS lattice the
SU(2) gauge theory with adjoint Higgs fields constructed for the
square-lattice antiferromagnet~\cite{ShackletonThomsonSachdev2021}.

These field theories retain the SU(2) spin symmetry of the Heisenberg
limit. The material has spin--orbit-induced Dzyaloshinskii--Moriya (DM)
interactions as its leading symmetry-lowering perturbation. The
out-of-plane inter-dimer component $D'_\perp\simeq0.09$--$0.18$\,meV~\cite{Cepas2001,Romhanyi2015,Mazurenko2008} is the largest inter-dimer
component~\cite{Cepas2001,Mazurenko2008} and, by the symmetry analysis of
Sec.~\ref{sec:model} in agreement with Ref.~\cite{Zorko2004}, the only
DM coupling that the structure allows above the buckling transition at
$T_s\simeq395$\,K~\cite{Sparta2001}. Below $T_s$ in-plane components
appear on the intra-dimer bonds, where they are the largest DM couplings~\cite{Zorko2004,Kodama2005,Mazurenko2008}, and on the inter-dimer bonds~\cite{Cheng2007,Romhanyi2015,Mazurenko2008}. Together with symmetric
anisotropies~\cite{McClarty2017} they control the triplon splittings of
the dimer phase~\cite{Cepas2001,Gaulin2004,Romhanyi2015,McClarty2017}.
DM interactions have so far been included only in the dimer phase, at
the dimer--plaquette transition by iPEPS~\cite{LarreaJimenez2021}, and in
spin-wave analyses of the N\'eel phase~\cite{Fogh2024PRL,Cederholm2026}, and their
absence from the theory of the deconfined scenario has been noted~\cite{Cui2024two}.

On the kagome lattice the out-of-plane DM component projects at linear
order onto a time-reversal-even spin-triplet mass of the Dirac theory, the
hexagon vector chirality, and was argued to induce $\bm q=0$ order~\cite{Hermele2008}, as found by exact diagonalization above a finite $D_c\simeq0.1J$~\cite{Cepas2008}
and variational Monte Carlo~\cite{Kagome2407}. The DM-driven transition
has been analyzed as one out of a $\mathbb Z_2$ spin liquid of bosonic
spinons~\cite{Huh2010}. The singlet projective symmetry groups carry over to $z$-component triplet ans\"atze
when each lattice operation also acts on spin~\cite{Dodds2013}, DM
together with a spin polarization generates a net gauge flux through the unit cell~\cite{KangLee2024}, and DM drives a node-creating band inversion in a
$1/9$-plateau spinon state~\cite{Kagome2503}. At a Higgs transition of the
square-lattice Dirac spin liquid the critical Higgs fields break the
emergent SO(5) with spin rotations intact~\cite{ShackletonThomsonSachdev2021,ShackletonSachdev2022}.
The effect of couplings that reduce the spin-rotation symmetry itself on
the SS Dirac spin liquid has not been addressed.

The question matters for the SO(5) regime. In the theory of
Ref.~\cite{Feuerpfeil2026}, with $N_f=2$ Dirac fermions and $N_b=2$
adjoint Higgs fields, the SO(5)-symmetric point at vanishing Yukawa
coupling $y$ is left along the relevant $y$. Its renormalization-group
(RG) eigenvalue at large $N$ is $\simeq1.4$, so the flow stays near that point over
a pseudocritical window only for a small bare $y$. The large-$y$ limit, a
critical point without Lorentz invariance and with weakly broken SO(5), is the square-lattice
theory of Ref.~\cite{ShackletonThomsonSachdev2021}. In the scenario of
Refs.~\cite{WangNahum2017,Nahum2015} the flow instead passes slowly near a
pair of complex fixed points. Emergent SO(5) then requires the rank-4
symmetric tensor to be effectively irrelevant, whereas the rank-2 tensor
contains the tuning parameter and is relevant, as on the fuzzy sphere, where the rank-4 tensor has $\Delta=3.89(3)$ and the rank-2 tensor $1.456(12)$~\cite{Zhou2024fuzzy}. In the leading-order large-$N$ theory used here,
whose tuning parameter is the Higgs mass, the symmetry-breaking part of the rank-2
tensor is irrelevant ($\Delta=3.46$, Sec.~\ref{sec:selection}). Which operators DM induces, and with which
dimensions, therefore decides whether spin--orbit coupling can act on the
deconfined regime.

The companion Letter~\cite{companionLetter} states the symmetry results:
time reversal and the spin--orbit-locked space group forbid every gauge-singlet fermion
mass at any order in DM. At linear order the in-plane couplings enter the gauge-singlet bilinears only
as the combination $\mathcal O_{\rm DM}=J^x_x-J^y_y$ ($x\parallel[100]$)
of conserved spin-current components, of dimension exactly 2 at the
spin-rotation-invariant fixed point, and $D'_\perp$ only through a
flavored spin-2 current, irrelevant at any unitary critical point; and a
magnetic field activates the N\'eel moment selectively by orientation.
The present paper derives these statements and follows their
consequences, above the crossover scale $T^*$ set by $\mathcal O_{\rm DM}$
and below it. The derivations rest on a complete classification of the
fermion bilinears of the SS gauge theory, gauge-singlet and Yukawa, at
zeroth and first gradient order, under the spin--orbit-locked space
group, in which every lattice operation acts on spin through its
axial-vector rotation. The Higgs fields are spin singlets, so the locked
group acts on them as the unlocked one and their potential is unchanged.
The classification agrees with the bilinear transformation tables of
Ref.~\cite{Feuerpfeil2026}, and with those of Ref.~\cite{ThomsonSachdev} except for three time-reversal
signs of the N\'eel-flavored gauge-adjoint currents, which we correct in
Appendix~\ref{app:protocol}; our signs are used throughout, in particular
for the Yukawa couplings of Sec.~\ref{sec:yukawa}. The continuum symmetry
action follows from projecting the lattice projective symmetry group (PSG)
onto the Dirac subspace, the DM textures agree with
Refs.~\cite{Cepas2001,Romhanyi2015}, and a lattice mean-field calculation
confirms the operator selection. Appendix~\ref{app:protocol} states the
conventions.

The main results of the paper are summarized below.
\begin{enumerate}
\item \emph{Crossover scale.} The coefficient $\lambda$ of
$\mathcal O_{\rm DM}$ is $\lambda\simeq0.085\,D'_\Sigma$ at mean field,
with $D'_\Sigma$ equal to $\sqrt2$ times the in-plane inter-dimer DM component along
the tetragonal axis of each bond, so that
$T^*=4\lambda\simeq0.34\,|D'_\Sigma|$, i.e.\ $0.03$--$0.17$\,K from the only density-functional-theory (DFT) calculation of the DM vectors, including the
tabulated precision and a factor-$1.7$ prefactor uncertainty (Secs.~\ref{sec:model} and~\ref{sec:crossover}).
\item \emph{Band touching below $T^*$.} Near the critical point the Dirac
cones reconstruct into a chirally protected $d$-wave quadratic band
touching whose only gauge-singlet gap channels are the two valence-bond-solid (VBS)
components and the scalar chirality (Sec.~\ref{sec:qbt}).
\item \emph{SO(5)-breaking perturbations above $T^*$.} At second order DM
seeds the rank-2 SO(5) tensor, easy-axis from $D'_\perp$ and easy-plane
from $\lambda$ at mean field (Sec.~\ref{sec:selection}). The DM-induced
Yukawa couplings to the critical Higgs fields come only from the
intra-dimer $D$, start at $6\times10^{-4}$--$2\times10^{-2}$ of the
Yukawa coupling that bounds the pseudocritical window and, barring a
cancellation among the Z3000 amplitudes, stay below it by at least that
factor (Sec.~\ref{sec:yukawa}).
\item \emph{Instabilities of the band touching.} A Fierz-complete one-loop
flow ends on a ray containing valence-bond order for most initial couplings, including the
gauge-singlet ones and the exchange of the material, and otherwise on a
gapped chiral $\mathbb{Z}_2$ spin liquid with Bogoliubov--de Gennes (BdG) Chern number $\mp4$, at
$T^{**}\simeq(0.03$--$0.1)\lambda$ (Secs.~\ref{sec:qbtir}
and~\ref{sec:chiral}).
\item \emph{Plaquette selection.} The DM correction to the plaquette
pattern is a bias of at most of order $10^{-4}$ whose direction is not determined
at $O(1/N)$, where the mixing of the SO(5)-singlet channel is solved but
its Higgs sector is not controlled (Sec.~\ref{sec:whichvbs}).
\item \emph{Magnetic fields.} A field along $[110]$ adds a
valence-bond-imbalance mass at order $B^2D^2$, and the in-plane component
of a tilted field rounds a critical point on a scale that grows steeply as
the scaling dimension of the N\'eel order parameter decreases (Sec.~\ref{sec:field}).
\item \emph{Disorder.} Non-magnetic, time-reversal-invariant quenched
disorder couples a relevant random field to the plaquette doublet and,
to all orders in the disorder and in $1/N$, none to the chirality or the
N\'eel channels. Random $T_c$ is relevant at $O(1/N)$, a result that is
not controlled at $N=12$ (Sec.~\ref{sec:disorder}).
\end{enumerate}
The U(1) descendant U800 carries exactly one symmetry-allowed monopole,
the SO(5) singlet, with or without DM, so for U800 the symmetry argument
excludes local masses but not confinement (Appendix~\ref{app:monopole}).
The measurements that test these statements are an $^{11}$B NMR
comparison of $\bm B\parallel[100]$ with $\bm B\parallel[001]$ at
$2.4$--$2.8$\,GPa, calorimetry below $T^*$ near the critical pressure, a
zero-field thermal Hall search below $T^{**}$ (millikelvin scale or below
under pressure) after field cooling, and an
independent determination of the sign of $D'_\perp$, which decides the
direction of the DM plaquette bias only together with the SO(5)-singlet
spectrum (Sec.~\ref{sec:convergences}).

The paper is organized as follows. Section~\ref{sec:model} sets up the model, the symmetry-allowed DM
textures and the material parameters, and Sec.~\ref{sec:continuum}
constructs the spin--orbit-locked symmetry group of the continuum gauge
theory. Sections~\ref{sec:inventory}--\ref{sec:spin2} derive the results
of the Letter: the operator content and the absence of gauge-singlet masses, the induced
couplings and $\lambda$, the Higgs--Yukawa sector and node stability, and
the dimension of the spin-2 current. Section~\ref{sec:selection} treats
the second-order SO(5) anisotropies, Sec.~\ref{sec:rg} the crossover
scale and the band touching, Sec.~\ref{sec:instab} the instabilities
of the band touching and the plaquette selection, Sec.~\ref{sec:field}
the magnetic field, and Sec.~\ref{sec:disorder} quenched disorder.
Section~\ref{sec:experiment} compares the theory with experiment probe by
probe, Sec.~\ref{sec:scope} lists the scope and limitations, and Sec.~\ref{sec:conclusions} concludes. The appendices
collect conventions, proofs and computational details. ``Exact''
denotes a statement fixed by symmetry, current conservation or unitarity.
``Leading order'' denotes $O(1/N)$ or one-loop results, and ``estimate''
denotes mean-field prefactors.

\section{Model and symmetry-allowed DM couplings}
\label{sec:model}

The Shastry--Sutherland model with DM interactions is
\begin{align}
\hat H&=J_s\!\!\sum_{\langle ij\rangle_\square}\!\!\hat{\bm S}_i\!\cdot\!
\hat{\bm S}_j
+J_d\!\!\sum_{\langle ij\rangle_{\rm dim}}\!\!\hat{\bm S}_i\!\cdot\!
\hat{\bm S}_j
+\!\!\sum_{\rm bonds}\!\bm D_{ij}\!\cdot\!(\hat{\bm S}_i\times\hat{\bm S}_j),
\label{eq:H}
\end{align}
with $J_d=J$ on the dimer bonds and $J_s=J'$ on the square-lattice bonds.
Throughout this paper $J$ denotes the intra-dimer coupling. Several
pressure studies label the two couplings the other way round~\cite{Guo2020,Guo2025,Guo2026,Cui2023,Cui2024two,Wu2026}, and their numbers are quoted below in the
notation of this paper. We use the conventions of
Ref.~\cite{Feuerpfeil2026}, with square-lattice sites $\bm i=(i_x,i_y)$
and a $2\times2$ unit cell with sublattices $(m_x,m_y)\in\{A,B\}^2$. The SS
dimers lie along $(1,1)$, centered at $(\tfrac12,\tfrac12)+2\bm r$
(material dimer $A$, along the crystallographic $[110]$ direction), and
along $(-1,1)$, centered at $(\tfrac32,\tfrac32)+2\bm r$ (dimer $B$). The
square-lattice axes $\hat x,\hat y$ correspond to the crystallographic
$[100],[010]$ axes of SrCu$_2$(BO$_3$)$_2$.

The material crystallizes in the tetragonal space group $I\bar42m$ with
CuBO$_3$ layers stacked along $c$~\cite{SmithKeszler1991}. Above
$T_s=395$\,K the structure is $I4/mcm$ and the CuO$_4$ plaquettes are
planar. At $T_s$ a second-order displacive transition buckles the layers~\cite{Sparta2001}. The bending angle between neighboring CuO$_4$
plaquettes, measured by single-crystal x-ray diffraction at ambient
pressure, grows on cooling from zero at $T_s$ to about $11^\circ$ at
$100$\,K, the lowest temperature of that study~\cite{Sparta2001}.
High-resolution neutron powder diffraction from room temperature down to
$2$\,K refines every data set in $I\bar42m$~\cite{Vecchini2009}. Below
$34$\,K, where the dimer singlet forms, the lattice shows a small negative
thermal expansion and the bending angle decreases slightly, to about
$10.9^\circ$ at $2$\,K. The layers stay buckled, the CuBO$_3$ layer is
not a mirror plane for the dimers, and every DM component allowed by
$I\bar42m$ remains allowed~\cite{Vecchini2009}. We take the in-plane DM
couplings introduced below to be those of the buckled structure and
neglect this change of a few percent in the bending angle. Under pressure the tetragonal structure persists up to a continuous transition to a monoclinic phase, at $4.7$\,GPa at room temperature~\cite{Loa2005}. Between $25$ and $300$\,K the onset of the monoclinic distortion stays between $4.0$ and $5.0$\,GPa~\cite{Zayed2014}. The plaquette phase and the antiferromagnet adjacent to it, studied at $1.8$--$3$\,GPa~\cite{Cui2023,Cui2024two,Guo2025,LarreaJimenez2021} and up to about $4$\,GPa in Ref.~\cite{Guo2020}, lie in the tetragonal phase, where the locked group $I\bar42m$ applies; the antiferromagnetic order found above $4$\,GPa~\cite{Zayed2017,Fogh2024PRL,Wu2026} lies at or above the monoclinic onset, which $^{11}$B NMR places near $4$\,GPa at low temperature~\cite{Wu2026}. Whether the
plaquette phase itself lowers the tetragonal symmetry is not settled.
Reference~\cite{Waki2007} finds at $2.4$\,GPa, in a $c$-axis field of $7$\,T,
a gradual loss of fourfold symmetry of the $^{11}$B magnetic shifts below
about $30$\,K, above its $3.6$\,K transition, and assigns it to an
orthorhombic $Fmm2$ symmetry of primarily magnetic origin, since the
quadrupole couplings show no such loss. Reference~\cite{Shi2022} describes a
tendency toward an orthorhombic distortion at intermediate pressure, and
Ref.~\cite{Boos2019} expects no clear orthorhombic distortion of the
crystal if the orientation of the weak intra-dimer bonds alternates
between layers. Reference~\cite{Loa2005} attributes part of the anomalous in-plane compressibility to a reduction of the layer corrugation, which suggests that the in-plane DM couplings weaken under pressure. No low-temperature structure refinement under pressure quantifies this. The crystals
contain inversion twins with volume fraction $0.50(2)$~\cite{Sparta2001},
and the two twin domains carry opposite signs of every in-plane DM
component~\cite{Suetsugu2022}. Inversion, which relates the two domains~\cite{Sparta2001}, leaves the magnetic field, the
magnetization, the thermal Hall conductivity $\kappa_{xy}$ and the NMR
spectra unchanged, so these observables are the same in both domains.
For the triplon thermal Hall effect, Ref.~\cite{Suetsugu2022} notes that
the two domains give $\kappa_{xy}$ of the same sign. In that treatment the triplon
dispersion and Berry curvature depend only on the square of the
effective in-plane inter-dimer DM coupling.

\begin{figure}[t]
\includegraphics[width=\linewidth]{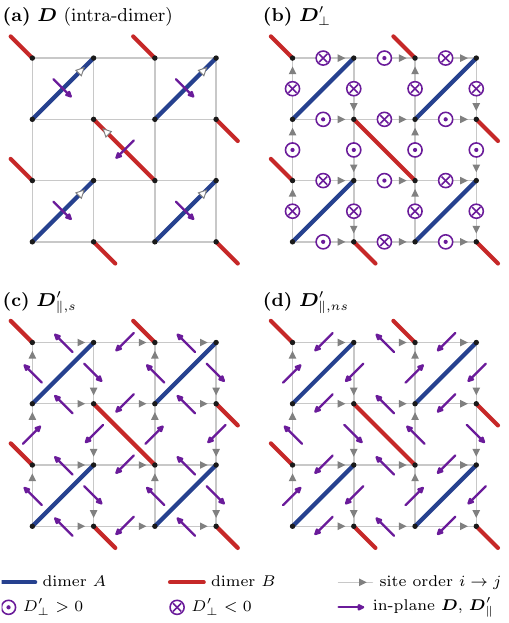}
\caption{Symmetry-allowed DM textures on the SS lattice, generated as the
null space of the spin--orbit-locked space-group action on bond-vector
space. They agree bond by bond with Ref.~\cite{Romhanyi2015} (all panels, panel (d) through its
Supplementary Note~1) and Ref.~\cite{Cepas2001} (panel (b)):
(a) the intra-dimer $\bm D$, (b) the out-of-plane inter-dimer $D'_\perp$,
(c),(d) the staggered and non-staggered in-plane inter-dimer components
$\bm D'_{\parallel,s}$, $\bm D'_{\parallel,ns}$. Blue/red: dimer
orientations $A$/$B$; small arrowheads (white-filled on the dimers, gray
on the inter-dimer bonds): site ordering $i\to j$ in
$\bm S_i\times\bm S_j$; circled dot/cross: the sign of the $D'_\perp$
mode on that bond, $+1$ (out of the plane) or $-1$ (into the plane);
violet arrows: in-plane components. The intra-dimer $D$ and the
in-plane inter-dimer components exist only below $T_s\simeq395$\,K.}
\label{fig:dm}
\end{figure}

\begin{table*}[t!]
\caption{Bond-resolved DM textures for one unit cell, with the bond
orientations $i\to j$ listed and the site labels $A_1{=}(A,A)$,
$A_2{=}(B,B)$, $B_1{=}(B,A)$, $B_2{=}(A,B)$ of Ref.~\cite{Romhanyi2015}.
Four of the eight orientations are reversed relative to that reference. Primed
labels denote the same sublattice site in a neighboring cell,
$A_1'=(2,0)$, $A_1''=(0,2)$, $B_1'=(1,2)$, $B_2'=(2,1)$. Each column is one
symmetry mode normalized to its coupling constant, and the three
inter-dimer modes span the full three-dimensional invariant space. The
high-temperature structure admits only $\bm D^{(\perp)}$.}
\label{tab:dm}
\begin{ruledtabular}
\begin{tabular}{llccc}
bond $i\to j$ & sites & $\bm D^{(\perp)}/D'_\perp$ & $\bm D^{(s)}/D'_{\parallel,s}$ & $\bm D^{(ns)}/D'_{\parallel,ns}$ \\
\colrule
$(0,0)\to(1,0)$ & $A_1\to B_1$ & $(0,0,1)$ & $(-\tfrac{1}{\sqrt2},\tfrac{1}{\sqrt2},0)$ & $(-\tfrac{1}{\sqrt2},-\tfrac{1}{\sqrt2},0)$ \\
$(0,0)\to(0,1)$ & $A_1\to B_2$ & $(0,0,-1)$ & $(-\tfrac{1}{\sqrt2},\tfrac{1}{\sqrt2},0)$ & $(\tfrac{1}{\sqrt2},\tfrac{1}{\sqrt2},0)$ \\
$(0,1)\to(1,1)$ & $B_2\to A_2$ & $(0,0,-1)$ & $(-\tfrac{1}{\sqrt2},\tfrac{1}{\sqrt2},0)$ & $(-\tfrac{1}{\sqrt2},-\tfrac{1}{\sqrt2},0)$ \\
$(1,1)\to(1,0)$ & $A_2\to B_1$ & $(0,0,-1)$ & $(\tfrac{1}{\sqrt2},-\tfrac{1}{\sqrt2},0)$ & $(-\tfrac{1}{\sqrt2},-\tfrac{1}{\sqrt2},0)$ \\
$(1,0)\to(2,0)$ & $B_1\to A_1'$ & $(0,0,-1)$ & $(-\tfrac{1}{\sqrt2},-\tfrac{1}{\sqrt2},0)$ & $(-\tfrac{1}{\sqrt2},\tfrac{1}{\sqrt2},0)$ \\
$(0,1)\to(0,2)$ & $B_2\to A_1''$ & $(0,0,1)$ & $(\tfrac{1}{\sqrt2},\tfrac{1}{\sqrt2},0)$ & $(-\tfrac{1}{\sqrt2},\tfrac{1}{\sqrt2},0)$ \\
$(1,2)\to(1,1)$ & $B_1'\to A_2$ & $(0,0,1)$ & $(-\tfrac{1}{\sqrt2},-\tfrac{1}{\sqrt2},0)$ & $(\tfrac{1}{\sqrt2},-\tfrac{1}{\sqrt2},0)$ \\
$(1,1)\to(2,1)$ & $A_2\to B_2'$ & $(0,0,1)$ & $(-\tfrac{1}{\sqrt2},-\tfrac{1}{\sqrt2},0)$ & $(-\tfrac{1}{\sqrt2},\tfrac{1}{\sqrt2},0)$ \\
\colrule
\multicolumn{2}{l}{intra-dimer: $(0,0)\to(1,1)$ [$A_1\to A_2$]} & \multicolumn{3}{l}{$\bm D^{(D)}/D = (\tfrac{1}{\sqrt2},-\tfrac{1}{\sqrt2},0)$}\\
\multicolumn{2}{l}{\phantom{intra-dimer:} $(2,1)\to(1,2)$ [$B_2'\to B_1'$]} & \multicolumn{3}{l}{$\bm D^{(D)}/D = (-\tfrac{1}{\sqrt2},-\tfrac{1}{\sqrt2},0)$}\\
\end{tabular}
\end{ruledtabular}
\end{table*}

\emph{DM textures from symmetry.} With spin--orbit coupling, a lattice
operation $O$ acts on spins through the axial-vector rule
$\bm S\to\det(O)\,O\,\bm S$, and a DM pattern $\{\bm D_{ij}\}$ is allowed
if and only if $\bm D_{O(i)O(j)}=\det(O)\,O\,\bm D_{ij}$ for all locked
operations. The allowed patterns form the null space of this condition
on the space of bond vectors. For the high-temperature structure the mirror $\sigma_h$
is itself a symmetry, so both $O$ and $O\sigma_h$ act. In the buckled
structure of space group $I\bar42m$ the glide $G_x$ and the fourfold
rotation survive only combined with $\sigma_h$, while the diagonal mirrors
$\sigma_{xy},\sigma_{x\bar y}$ act purely~\cite{Sparta2001,Romhanyi2015}.
The result is shown in Fig.~\ref{fig:dm} and Table~\ref{tab:dm}. At high
temperature a single coupling is allowed, the alternating out-of-plane
inter-dimer $D'_\perp$. Reference~\cite{Cepas2001} proposed this
pattern. It excluded the intra-dimer coupling by the approximate
inversion center of each dimer, took the Cu plane as an approximate
mirror plane, and neglected the in-plane inter-dimer components as
smaller. Below $T_s$ three further
couplings appear: the staggered in-plane intra-dimer $D$, perpendicular to
each dimer, and the staggered and non-staggered in-plane inter-dimer
components $D'_{\parallel,s}$, $D'_{\parallel,ns}$. On the reference bond
$A_1\to B_1$ we take $\bm D'=(-D'_{\parallel,ns},D'_{\parallel,s},D'_\perp)$
in the frame $(\hat x_R,\hat y_R,\hat z)$ of Ref.~\cite{Romhanyi2015},
related to ours in Appendix~\ref{app:dm}. Supplementary Note~1 of
Ref.~\cite{Romhanyi2015} writes this vector as
$(-D'_{\parallel,x},D'_{\parallel,y},D'_\perp)$, and the triangle rule
stated there makes the $y$ component the staggered one. In our frame the
same vector is $\bigl(-(D'_{\parallel,s}+D'_{\parallel,ns})/\sqrt2,\,
(D'_{\parallel,s}-D'_{\parallel,ns})/\sqrt2,\,D'_\perp\bigr)$, as in
Table~\ref{tab:dm}. The patterns agree with the
published figures arrow by arrow, up to the overall sign convention for
$D$ (Appendix~\ref{app:dm}).

\emph{Material parameters.} The standard exchange set is $J\simeq85$\,K
and $J'\simeq54$\,K, $J'/J=0.635$~\cite{MiyaharaUeda2003}.
Reference~\cite{Cepas2001} quotes the same $J$ and $J'$. Its 20-site
exact diagonalization, compared with the measured ratio of the
Raman-active singlet energy to the triplet gap, gives $J'/J\simeq0.62$,
and it extracts $D'_\perp$ at that ratio. Another fit gives $J=71.5$\,K
with $J'/J=0.603$~\cite{Knetter2000}. The iPEPS study of
Ref.~\cite{Shi2022} uses $J=81.5$\,K at ambient pressure, a value
between earlier estimates that agrees well with the observed onsets of
the 1/4 and 1/3 plateaus. It takes $J'/J=0.63$ from high-field
magnetization data~\cite{Matsuda2013}. Those pulsed-field data, taken up
to $118$\,T with $\bm B\parallel c$ at $2.1$--$2.2$\,K, are best
described by numerical results for the pure Shastry--Sutherland model at
$J'/J\simeq0.63$, and Ref.~\cite{Matsuda2013} states no value of $J$. First-principles
values are $J=7.3$\,meV and $J'=4.1$\,meV~\cite{Mazurenko2008}.

The DM couplings are known with different degrees of certainty. The
out-of-plane inter-dimer coupling $D'_\perp$ is the only DM coupling
present above $T_s$ and the largest inter-dimer anisotropy. Its raw
observable is the zero-field splitting of the one-triplet electron spin resonance (ESR) lines for
$B\parallel c$, $85$\,GHz $=0.35$\,meV~\cite{Nojiri2003,Cepas2001}. The
conversion to $D'_\perp$ depends on how $J'$ is treated. To lowest order
in $J'/J$ the splitting is $\delta(q{=}0)=4D'_\perp$~\cite{Cepas2001,Gaulin2004}.
The exact diagonalization of Ref.~\cite{Cepas2001} finds that finite $J'$ reduces it to $\delta\simeq2.0D'_\perp$ at
$J'/J=0.62$, which gives $D'_\perp\simeq0.18$\,meV, the value used here.
Reference~\cite{Gaulin2004} notes that Ref.~\cite{Cepas2001} estimates a reduction of the
bandwidth by about a factor of two from quantum fluctuations. The
triplon-hopping model of Ref.~\cite{Romhanyi2015}, which keeps terms linear
in the DM couplings and takes $J$ as the measured spin gap, has
$\delta(q{=}0)=4|D'_\perp|$ and uses $D'_\perp=-21$\,GHz, i.e.\
$|D'_\perp|\simeq0.087$\,meV, the magnitude also found by far-infrared
spectroscopy~\cite{Room2004}. The two estimates of $D'_\perp$ differ by a
factor of two. The DFT value $0.10$\,meV~\cite{Mazurenko2008} lies between
them. The intra-dimer coupling is larger, with $D/J=0.030$--$0.038$
from the field-induced staggered magnetization seen by $^{11}$B NMR~\cite{Kodama2005}, i.e.\ $D\simeq0.22$--$0.28$\,meV, $0.22$\,meV from
far-infrared spectroscopy~\cite{Room2004}, $0.31(4)$\,meV from the high-temperature ESR
linewidth, assuming an in-plane inter-dimer component of $30\%$ of
$D'_\perp$~\cite{Zorko2004}, and $0.35$\,meV in DFT~\cite{Mazurenko2008}. The in-plane inter-dimer components are the
least constrained. Triplon spectroscopy fixes only the combination
$\tilde D_\parallel=D'_{\parallel,s}\mp(J'/2J)D\simeq0.07$--$0.08$\,meV,
with the sign set by the convention for $D$~\cite{Cheng2007,Romhanyi2015},
and $D'_{\parallel,ns}$ does not enter the zero-field triplon energies at
linear order~\cite{Cheng2007}. The separate components come from a single
DFT calculation~\cite{Mazurenko2008}, which quotes the magnitudes
$0.06$ and $0.04$\,meV and lists the four inter-dimer vectors from
one site in its Table~II, for example $\bm D_{14}=(-0.07,-0.01,0.1)$\,meV. The bond vectors of that calculation~\cite{MazurenkoPC} place each vector on its bond, in agreement with the site
labels of its Figs.~6 and 8 and with the crystal coordinates of
Ref.~\cite{Sparta2001}, so that bond $1\to4$ is $(1,1)\to(1,2)$ here, and
Table~\ref{tab:dm} matches the tabulated vectors only pair by pair:
$\bm D_{14}$ and $\bm D_{15}$ give
$(D'_{\parallel,s},D'_{\parallel,ns},D'_\perp)\simeq(-0.057,+0.042,-0.10)$\,meV,
whereas $\bm D_{13}$ and $\bm D_{16}$ give $(+0.042,-0.057,-0.10)$\,meV.
The two pairs differ in the sign of the large ($0.07$\,meV) component
perpendicular to the tetragonal axis of the bond, which $I\bar42m$
requires to alternate between the two pairs of bonds and which the table
has uniform, so the tabulated vectors are not covariant under the crystal
symmetry (least-squares residual $0.14$\,meV). We adopt the second pair,
$D'_{\parallel,s}\simeq+0.042$ and $D'_{\parallel,ns}\simeq-0.057$\,meV,
because with the DFT intra-dimer coupling $D\simeq-0.35$\,meV in our
convention it gives $|\tilde D_\parallel|=|D'_{\parallel,s}+(J'/2J)D|
\simeq0.07$\,meV, the spectroscopic value, whereas the first pair gives
$0.17$\,meV. Both pairs give the same $D'_\perp\simeq-0.10$\,meV and the
same sum $D'_\Sigma=D'_{\parallel,s}+D'_{\parallel,ns}\simeq-0.014$\,meV,
the combination that controls the spin-liquid physics below, with
$|D'_\Sigma|=\sqrt2\,|D'_{\rm bond}|$, where $D'_{\rm bond}$ is the in-plane
component along the tetragonal axis of the inter-dimer bond (the Cu--Cu
bond itself is rotated by $28^\circ$ from that axis). That component has
magnitude $0.01$\,meV on all four DFT bonds. Hence
$D'_\Sigma\simeq-0.014$\,meV $\simeq-0.16$\,K. The tabulated precision of
one significant figure allows $|D'_\Sigma|\simeq0.007$--$0.021$\,meV~\footnote{The
arXiv version of the same calculation gives $D'_{\rm bond}=0$, and if the
in-plane components scale like the ESR-to-DFT ratio of $D'_\perp$,
$|D'_\Sigma|$ rises to about $0.025$\,meV $\simeq0.29$\,K}. A
Keffer-type estimate from the bridging BO$_3$ groups of the real structure
gives an in-plane inter-dimer component dominantly perpendicular to the
bond axis, with the relative sign of the adopted pair, in agreement with
these values. Throughout Secs.~\ref{sec:crossover} and~\ref{sec:experiment} we use
$|D'_\Sigma|\simeq0.014$\,meV, or $0.007$--$0.025$\,meV with the tabulated precision and the ESR rescaling. The sign of $D'_\perp$ enters the DM-induced plaquette selection of
Sec.~\ref{sec:whichvbs}. ESR and neutron spectra fix only $|D'_\perp|$.
The $^{11}$B NMR study of Ref.~\cite{Kodama2005} determines the sign from
the level repulsion at the critical field: of the two $D'$-split triplet
branches only one mixes with the singlet through the intra-dimer DM
coupling and the staggered $g$ tensor, and it must be the lower one.
Figure~1(a) of Ref.~\cite{Kodama2005} and Fig.~1 of Ref.~\cite{Cepas2001} use the convention of
Table~\ref{tab:dm}: a positive coupling points toward the viewer for bonds
ordered counterclockwise around a dimer-containing plaquette. The $D'<0$ of Ref.~\cite{Kodama2005} is
therefore $D'_\perp<0$ here, the sign of Table~II of Ref.~\cite{Mazurenko2008}. That determination uses an exact-diagonalization model
with $D'_\perp$ as the only inter-dimer DM component and takes
$|D'|/J=0.02$ from Ref.~\cite{Cepas2001}; Ref.~\cite{Cheng2007} notes that
the level repulsion also depends on the in-plane components. The
far-infrared analysis of Ref.~\cite{CepasZiman2004} concludes $D>0$ in the
notation of Ref.~\cite{Cepas2001}, but only given an assumed relative sign
of the phonon-induced DM vectors on the two dimers, and
Ref.~\cite{Cheng2007} describes the two determinations as opposite. The
published version of Ref.~\cite{Mazurenko2008}, 
states that its signs agree with both Refs.~\cite{Kodama2005} and~\cite{CepasZiman2004} and
attributes the apparent discrepancy to different conventions. We use
$D'_\perp<0$ below and state where the conclusion would reverse with the
sign.

\section{Continuum theory and spin--orbit-locked symmetries}
\label{sec:continuum}

We work in the Majorana-matrix formulation of
Refs.~\cite{ThomsonSachdev,Feuerpfeil2026}. The spinons
$f_{i\alpha}$ are grouped into
$\mathcal F_i=\bigl(\begin{smallmatrix} f_{i\uparrow}&-f_{i\downarrow}\\
f^\dagger_{i\downarrow}&f^\dagger_{i\uparrow}\end{smallmatrix}\bigr)$ with
the reality condition $\mathcal F^\dagger_i=\sigma^y\mathcal F^T_i\sigma^y$
(Ref.~\cite{Feuerpfeil2026} denotes all Pauli matrices acting on $\mathcal F$
by $\tau$).
On the lattice, matrix SU(2) gauge transformations act by left
multiplication, and spin rotations $\Omega$ by right multiplication,
$\mathcal F\to\mathcal F\,\sigma^z\Omega^T\sigma^z$. Around the
$\pi$-flux point the low-energy field $\X$ carries a Dirac index, on which
$\gamma^0=\rho^y$, $\gamma^x=\ii\rho^z$, $\gamma^y=\ii\rho^x$ act, a valley
index with Pauli matrices $\mu^a$, a spin index and a gauge index, with
$\Xb=\X^\dagger\gamma^0$. We fix the continuum notation by the index on
which each matrix acts rather than by a side of multiplication. The Dirac
and valley matrices and the global SO(5) generators
$T^j=\{\mu^y,\sigma^a,\mu^x\sigma^a,\mu^z\sigma^a\}$ and vector
$\Gamma^j=\{\mu^x,\mu^z,\mu^y\sigma^a\}$ act on the Dirac, valley and spin
indices and commute with the gauge SU(2). A Pauli matrix written outside
the pair $\Xb\cdots\X$, as in $\Tr[\sigma^a\Xb\cdots\X]$, acts on the gauge
index and marks a gauge-adjoint bilinear. In the gauge-singlet rows
$\Tr[\Xb M\X\sigma^b]$ of Tables~\ref{tab:unlocked}--\ref{tab:hightemp}
and in Eq.~\eqref{eq:ODM}, $\sigma^b$ is the global spin matrix of the row
label $M\sigma^b$. The Lagrangian, including the
three real gauge-adjoint Higgs fields $\Phi^a_{1,2,3}$ of
Ref.~\cite{Feuerpfeil2026}, is
\begin{align}
\mathcal L&=\ii\Tr[\Xb\gamma^\mu\partial_\mu\X]+V(\Phi)\nonumber\\
&+\Phi_1^a\Tr[\sigma^a\Xb(\gamma^x\mu^z
-\delta(\gamma^0\mu^y\ii\partial_x-\ii\partial_y))\X]\nonumber\\
&+\Phi_2^a\Tr[\sigma^a\Xb(\gamma^y\mu^x
+\delta(\gamma^0\mu^y\ii\partial_y-\ii\partial_x))\X]\nonumber\\
&+\Phi_3^a\Tr[\sigma^a\Xb\mu^y(\gamma^y\ii\partial_x+\gamma^x\ii\partial_y)\X].
\label{eq:Lmajorana}
\end{align}
Here, $\sigma^a$ to the left of $\Xb$ is the gauge adjoint, and the SU(2)
gauge field enters by minimal coupling. $V(\Phi)$ is the symmetry-allowed
potential up to quartic order, identical to the square-lattice case~\cite{Feuerpfeil2026}. The terms proportional to $\delta$ are of gradient
order and drop out of the critical theory. The mean-field phases of
$V(\Phi)$ are the $\pi$-flux Dirac spin liquid with all
$\langle\Phi\rangle=0$, the staggered-flux state U800 with
$\langle\Phi_3\rangle\neq0$, and the gapless $\mathbb Z_2$ state Z3000 with
$\langle\Phi_1\rangle,\langle\Phi_2\rangle\neq0$. The critical sector of
Ref.~\cite{Feuerpfeil2026} retains $\Phi_1$ and $\Phi_2$. Under the SS
generators $(G_x,\mathcal T,\sigma_{xy})$ the Higgs fields transform as
$\Phi_1\to(+\Phi_1,-\Phi_1,\Phi_2)$, $\Phi_2\to(-\Phi_2,-\Phi_2,\Phi_1)$
and $\Phi_3\to(-\Phi_3,+\Phi_3,-\Phi_3)$, and the same generators act on
$\X$ as
\begin{align}
G_x&:\ \X\to\rho^x\,\X(x,-y),\qquad
\mathcal T:\ \X\to\rho^y\mu^y\,\X,\ \ \ii\to-\ii,\nonumber\\
\sigma_{xy}&:\ \X\to-\rho^z\mu^z\,
e^{-\ii\pi\rho^y/4}e^{\ii\pi\mu^y/4}\,\X(y,x).
\label{eq:latact}
\end{align}
Projecting the lattice PSG action of Ref.~\cite{Feuerpfeil2026} onto the
Dirac zero-mode subspace gives the matrices of Eq.~\eqref{eq:latact} with
unit phase. In Eq.~\eqref{eq:latact} the antiunitary $\mathcal T$ acts on
the global spin index through the additional factor $\ii\sigma^y$, which
is implicit here and in the tables below; the full time reversal
reverses the spin-carrying bilinears.

\emph{Dirac form and large-$N$ count.} The large-$N$ computations of
Secs.~\ref{sec:spin2} and~\ref{sec:qbtir} and of
Appendices~\ref{app:polarization} and~\ref{app:integrals} use the Dirac
fermions $\psi_{a,v}$ of Ref.~\cite{Feuerpfeil2026}, Eq.~(53), which
collect the eight real components of valley $v$ into one complex gauge
doublet of two-component Dirac fermions, $a$ being the gauge index. In
this form the critical theory reads
\begin{align}
\mathcal L_\psi&=\ii\bar\psi_v\gamma^\mu D_\mu\psi_v,\qquad
\mathcal L_\Phi=\frac{1}{2g}\sum_{s=1,2}\bigl[(D_\mu\Phi_s)_a\bigr]^2,\nonumber\\
\mathcal L_{\Phi\psi}&=y\,\bigl(\Phi_1^a\,\bar\psi\mu^z\gamma^x\sigma^a\psi
+\Phi_2^a\,\bar\psi\mu^x\gamma^y\sigma^a\psi\bigr),\nonumber\\
D_\mu&=\partial_\mu-\ii A^a_\mu\sigma^a,\qquad
(D_\mu\Phi)_a=\partial_\mu\Phi_a+2\varepsilon_{abc}A^b_\mu\Phi_c,
\label{eq:Ldirac}
\end{align}
with $\bar\psi=\psi^\dagger\gamma^0$ and $\sigma^a$ acting on the gauge
index of $\psi_v$. The sign of the Higgs covariant derivative is the one
required by gauge covariance with $D_\mu\psi=(\partial_\mu-\ii
A^a_\mu\sigma^a)\psi$, as in Eqs.~(54)--(56) of Ref.~\cite{Feuerpfeil2026}. Spin is not manifest in this form. The spin
component $S^z$ acts as the $\psi$ number and $S^\pm$ as pairing terms, so
every flavored gauge-singlet bilinear carries only valley matrices in
$\psi$ form. For example, the flux mass is $2\bar\psi\psi$, the VBS masses
are $2\bar\psi\mu^{x,z}\psi$, the N\'eel$_z$ mass is $2\bar\psi\mu^y\psi$,
and the $S^z$ current is $2\bar\psi\gamma^\mu\psi$. The theory contains $N_f=2$ valleys, each a Dirac
doublet with gauge vertex $\gamma_\mu\sigma^a$, and $N_b=2$ real adjoint
triplets $\Phi_1,\Phi_2$. The gauge propagator is generated by matter loops
alone. A Dirac doublet contributes to the one-loop polarization with weight
$\operatorname{tr}(\sigma^a\sigma^b)=2\delta^{ab}$, and a real adjoint
triplet with weight $\tfrac12\operatorname{tr}(t^at^b)=4\delta^{ab}$ for
$(t^b)_{ac}=-2\ii\varepsilon_{bac}$, so the polarization is proportional to
$2N_f+4N_b$ (Appendix~\ref{app:polarization}). We define
\begin{equation}
N\equiv2N_f+4N_b=12
\label{eq:Ndef}
\end{equation}
and use $N$ as the single large-$N$ parameter throughout. The U(1)
sublimit, in which only $A^3_\mu$ is kept and $\sigma^3=\pm1$, contains
four unit-charge two-component fermions, the $N=4$ QED$_3$ of the
$\pi$-flux Dirac spin liquid, which fixes the fermion weight. It also
contains two charge-2 complex scalars of weight 4 each, so the total is
again $N=12$. The count $N=12$ applies at the multicritical point, where
the Higgs fields are critical. In the $\pi$-flux Dirac spin liquid with
gapped Higgs fields only the fermions screen, $N=2,~N_f=4$, and
$64/(\pi^2N)\simeq1.6$ makes the expansion unusable there. All large-$N$
numbers in this paper are evaluated at vanishing Yukawa coupling, $y=0$.

\emph{Locked actions.} With DM present, each generator carries its
axial-vector spin rotation as a right factor
$W_O=\sigma^z\Omega_O^T\sigma^z$:
\begin{align}
\hat G_x&=G_x\otimes C_{2x},&
\hat\sigma_{xy}&=\sigma_{xy}\otimes C_2^{[1\bar10]},\nonumber\\
\hat\sigma_{x\bar y}&=\sigma_{x\bar y}\otimes C_2^{[110]},&
\hat S_4&=C_4\otimes R_z(-\tfrac\pi2),
\label{eq:locked}
\end{align}
for the buckled structure, together with $\mathcal T$. The glide and the
fourfold rotation appear only composed with $\sigma_h$, whence the $C_{2x}$
and $R_z(-\tfrac\pi2)$ spin parts. Every axial spin part $\det(O)\,O$ is a
proper rotation. The high-temperature structure additionally
admits $\hat\sigma_h=\one \otimes C_{2z}$, the pure glide
$G_x\otimes C_{2y}$, the pure rotation $\hat C_4=\hat\sigma_h\hat
S_4=C_4\otimes R_z(+\pi/2)$ and, when only $D'_\perp$ is present, the full
spin U(1)$_z$.

\section{Complete operator content}
\label{sec:inventory}

\begin{table}[t!]
\caption{Transformation of all 36 gauge-singlet fermion bilinears
$\Tr[\Xb\,M\,\X\,\sigma^b]$ (rows labeled by $M\sigma^b$) under the
unlocked SS generators, with the global spin matrices set to unity. First
block: the six masses (SO(5) vector and flux singlet). Second block: the 30
conserved SO(5)-adjoint current components. References~\cite{ThomsonSachdev,Feuerpfeil2026} tabulate the gauge-adjoint
bilinears $\Tr[\sigma^a\Xb\cdots\X]$ (their Figs.~3--5 and Tables~2--4,
respectively). The present table covers the complementary gauge-singlet
sector at zeroth order in gradients, including its spin-carrying rows.}
\label{tab:unlocked}
\begin{ruledtabular}
\begin{tabular}{lccccc}
$\mathrm{Tr}[\bar{\mathcal X}\,\cdot\,\mathcal X]$ & $G_x$ & $\sigma_{xy}$ & $\sigma_{x\bar y}$ & $C_4$ & $\mathcal T$ \\
\colrule
$\one$ & $-$ & $-$ & $-$ & $+$ & $-$ \\
$\mu^x$ & $-$ & $-\mu^z$ & $-\mu^z$ & $+\mu^z$ & $+$ \\
$\mu^y\sigma^x$ & $-$ & $+$ & $+$ & $-$ & $-$ \\
$\mu^y\sigma^y$ & $-$ & $+$ & $+$ & $-$ & $-$ \\
$\mu^y\sigma^z$ & $-$ & $+$ & $+$ & $-$ & $-$ \\
$\mu^z$ & $-$ & $-\mu^x$ & $-\mu^x$ & $+\mu^x$ & $+$ \\
\colrule
$\gamma^0\sigma^x$ & $+$ & $+$ & $+$ & $+$ & $-$ \\
$\gamma^x\sigma^x$ & $+$ & $+\gamma^y\sigma^x$ & $-\gamma^y\sigma^x$ & $-\gamma^y\sigma^x$ & $+$ \\
$\gamma^y\sigma^x$ & $-$ & $+\gamma^x\sigma^x$ & $-\gamma^x\sigma^x$ & $+\gamma^x\sigma^x$ & $+$ \\
$\gamma^0\sigma^y$ & $+$ & $+$ & $+$ & $+$ & $-$ \\
$\gamma^x\sigma^y$ & $+$ & $+\gamma^y\sigma^y$ & $-\gamma^y\sigma^y$ & $-\gamma^y\sigma^y$ & $+$ \\
$\gamma^y\sigma^y$ & $-$ & $+\gamma^x\sigma^y$ & $-\gamma^x\sigma^y$ & $+\gamma^x\sigma^y$ & $+$ \\
$\gamma^0\sigma^z$ & $+$ & $+$ & $+$ & $+$ & $-$ \\
$\gamma^x\sigma^z$ & $+$ & $+\gamma^y\sigma^z$ & $-\gamma^y\sigma^z$ & $-\gamma^y\sigma^z$ & $+$ \\
$\gamma^y\sigma^z$ & $-$ & $+\gamma^x\sigma^z$ & $-\gamma^x\sigma^z$ & $+\gamma^x\sigma^z$ & $+$ \\
$\gamma^0\mu^x\sigma^x$ & $+$ & $+\gamma^0\mu^z\sigma^x$ & $+\gamma^0\mu^z\sigma^x$ & $+\gamma^0\mu^z\sigma^x$ & $+$ \\
$\gamma^x\mu^x\sigma^x$ & $+$ & $+\gamma^y\mu^z\sigma^x$ & $-\gamma^y\mu^z\sigma^x$ & $-\gamma^y\mu^z\sigma^x$ & $-$ \\
$\gamma^y\mu^x\sigma^x$ & $-$ & $+\gamma^x\mu^z\sigma^x$ & $-\gamma^x\mu^z\sigma^x$ & $+\gamma^x\mu^z\sigma^x$ & $-$ \\
$\gamma^0\mu^x\sigma^y$ & $+$ & $+\gamma^0\mu^z\sigma^y$ & $+\gamma^0\mu^z\sigma^y$ & $+\gamma^0\mu^z\sigma^y$ & $+$ \\
$\gamma^x\mu^x\sigma^y$ & $+$ & $+\gamma^y\mu^z\sigma^y$ & $-\gamma^y\mu^z\sigma^y$ & $-\gamma^y\mu^z\sigma^y$ & $-$ \\
$\gamma^y\mu^x\sigma^y$ & $-$ & $+\gamma^x\mu^z\sigma^y$ & $-\gamma^x\mu^z\sigma^y$ & $+\gamma^x\mu^z\sigma^y$ & $-$ \\
$\gamma^0\mu^x\sigma^z$ & $+$ & $+\gamma^0\mu^z\sigma^z$ & $+\gamma^0\mu^z\sigma^z$ & $+\gamma^0\mu^z\sigma^z$ & $+$ \\
$\gamma^x\mu^x\sigma^z$ & $+$ & $+\gamma^y\mu^z\sigma^z$ & $-\gamma^y\mu^z\sigma^z$ & $-\gamma^y\mu^z\sigma^z$ & $-$ \\
$\gamma^y\mu^x\sigma^z$ & $-$ & $+\gamma^x\mu^z\sigma^z$ & $-\gamma^x\mu^z\sigma^z$ & $+\gamma^x\mu^z\sigma^z$ & $-$ \\
$\gamma^0\mu^y$ & $+$ & $-$ & $-$ & $-$ & $-$ \\
$\gamma^x\mu^y$ & $+$ & $-\gamma^y\mu^y$ & $+\gamma^y\mu^y$ & $+\gamma^y\mu^y$ & $+$ \\
$\gamma^y\mu^y$ & $-$ & $-\gamma^x\mu^y$ & $+\gamma^x\mu^y$ & $-\gamma^x\mu^y$ & $+$ \\
$\gamma^0\mu^z\sigma^x$ & $+$ & $+\gamma^0\mu^x\sigma^x$ & $+\gamma^0\mu^x\sigma^x$ & $+\gamma^0\mu^x\sigma^x$ & $+$ \\
$\gamma^x\mu^z\sigma^x$ & $+$ & $+\gamma^y\mu^x\sigma^x$ & $-\gamma^y\mu^x\sigma^x$ & $-\gamma^y\mu^x\sigma^x$ & $-$ \\
$\gamma^y\mu^z\sigma^x$ & $-$ & $+\gamma^x\mu^x\sigma^x$ & $-\gamma^x\mu^x\sigma^x$ & $+\gamma^x\mu^x\sigma^x$ & $-$ \\
$\gamma^0\mu^z\sigma^y$ & $+$ & $+\gamma^0\mu^x\sigma^y$ & $+\gamma^0\mu^x\sigma^y$ & $+\gamma^0\mu^x\sigma^y$ & $+$ \\
$\gamma^x\mu^z\sigma^y$ & $+$ & $+\gamma^y\mu^x\sigma^y$ & $-\gamma^y\mu^x\sigma^y$ & $-\gamma^y\mu^x\sigma^y$ & $-$ \\
$\gamma^y\mu^z\sigma^y$ & $-$ & $+\gamma^x\mu^x\sigma^y$ & $-\gamma^x\mu^x\sigma^y$ & $+\gamma^x\mu^x\sigma^y$ & $-$ \\
$\gamma^0\mu^z\sigma^z$ & $+$ & $+\gamma^0\mu^x\sigma^z$ & $+\gamma^0\mu^x\sigma^z$ & $+\gamma^0\mu^x\sigma^z$ & $+$ \\
$\gamma^x\mu^z\sigma^z$ & $+$ & $+\gamma^y\mu^x\sigma^z$ & $-\gamma^y\mu^x\sigma^z$ & $-\gamma^y\mu^x\sigma^z$ & $-$ \\
$\gamma^y\mu^z\sigma^z$ & $-$ & $+\gamma^x\mu^x\sigma^z$ & $-\gamma^x\mu^x\sigma^z$ & $+\gamma^x\mu^x\sigma^z$ & $-$ \\
\end{tabular}
\end{ruledtabular}
\end{table}

\begin{table}[t!]
\caption{Same as Table~\ref{tab:unlocked} under the spin--orbit-locked
generators of Eq.~\eqref{eq:locked}, relevant in the presence of in-plane
DM. No mass row is invariant under all columns. The current rows
$\gamma^x\sigma^x$ and $\gamma^y\sigma^y$ map into each other with unit
signs under the swap-type generators, so that their sum is the unique
invariant, the operator $\mathcal O_{\rm DM}$ of
Eq.~\eqref{eq:ODM}.}
\label{tab:locked}
\begin{ruledtabular}
\begin{tabular}{lccccc}
$\mathrm{Tr}[\bar{\mathcal X}\,\cdot\,\mathcal X]$ & $\hat G_x$ & $\hat\sigma_{xy}$ & $\hat\sigma_{x\bar y}$ & $\hat S_4$ & $\mathcal T$ \\
\colrule
$\one$ & $-$ & $-$ & $-$ & $+$ & $-$ \\
$\mu^x$ & $-$ & $-\mu^z$ & $-\mu^z$ & $+\mu^z$ & $+$ \\
$\mu^y\sigma^x$ & $-$ & $+\mu^y\sigma^y$ & $-\mu^y\sigma^y$ & $+\mu^y\sigma^y$ & $-$ \\
$\mu^y\sigma^y$ & $+$ & $+\mu^y\sigma^x$ & $-\mu^y\sigma^x$ & $-\mu^y\sigma^x$ & $-$ \\
$\mu^y\sigma^z$ & $+$ & $-$ & $-$ & $-$ & $-$ \\
$\mu^z$ & $-$ & $-\mu^x$ & $-\mu^x$ & $+\mu^x$ & $+$ \\
\colrule
$\gamma^0\sigma^x$ & $+$ & $+\gamma^0\sigma^y$ & $-\gamma^0\sigma^y$ & $-\gamma^0\sigma^y$ & $-$ \\
$\gamma^x\sigma^x$ & $+$ & $+\gamma^y\sigma^y$ & $+\gamma^y\sigma^y$ & $+\gamma^y\sigma^y$ & $+$ \\
$\gamma^y\sigma^x$ & $-$ & $+\gamma^x\sigma^y$ & $+\gamma^x\sigma^y$ & $-\gamma^x\sigma^y$ & $+$ \\
$\gamma^0\sigma^y$ & $-$ & $+\gamma^0\sigma^x$ & $-\gamma^0\sigma^x$ & $+\gamma^0\sigma^x$ & $-$ \\
$\gamma^x\sigma^y$ & $-$ & $+\gamma^y\sigma^x$ & $+\gamma^y\sigma^x$ & $-\gamma^y\sigma^x$ & $+$ \\
$\gamma^y\sigma^y$ & $+$ & $+\gamma^x\sigma^x$ & $+\gamma^x\sigma^x$ & $+\gamma^x\sigma^x$ & $+$ \\
$\gamma^0\sigma^z$ & $-$ & $-$ & $-$ & $+$ & $-$ \\
$\gamma^x\sigma^z$ & $-$ & $-\gamma^y\sigma^z$ & $+\gamma^y\sigma^z$ & $-\gamma^y\sigma^z$ & $+$ \\
$\gamma^y\sigma^z$ & $+$ & $-\gamma^x\sigma^z$ & $+\gamma^x\sigma^z$ & $+\gamma^x\sigma^z$ & $+$ \\
$\gamma^0\mu^x\sigma^x$ & $+$ & $+\gamma^0\mu^z\sigma^y$ & $-\gamma^0\mu^z\sigma^y$ & $-\gamma^0\mu^z\sigma^y$ & $+$ \\
$\gamma^x\mu^x\sigma^x$ & $+$ & $+\gamma^y\mu^z\sigma^y$ & $+\gamma^y\mu^z\sigma^y$ & $+\gamma^y\mu^z\sigma^y$ & $-$ \\
$\gamma^y\mu^x\sigma^x$ & $-$ & $+\gamma^x\mu^z\sigma^y$ & $+\gamma^x\mu^z\sigma^y$ & $-\gamma^x\mu^z\sigma^y$ & $-$ \\
$\gamma^0\mu^x\sigma^y$ & $-$ & $+\gamma^0\mu^z\sigma^x$ & $-\gamma^0\mu^z\sigma^x$ & $+\gamma^0\mu^z\sigma^x$ & $+$ \\
$\gamma^x\mu^x\sigma^y$ & $-$ & $+\gamma^y\mu^z\sigma^x$ & $+\gamma^y\mu^z\sigma^x$ & $-\gamma^y\mu^z\sigma^x$ & $-$ \\
$\gamma^y\mu^x\sigma^y$ & $+$ & $+\gamma^x\mu^z\sigma^x$ & $+\gamma^x\mu^z\sigma^x$ & $+\gamma^x\mu^z\sigma^x$ & $-$ \\
$\gamma^0\mu^x\sigma^z$ & $-$ & $-\gamma^0\mu^z\sigma^z$ & $-\gamma^0\mu^z\sigma^z$ & $+\gamma^0\mu^z\sigma^z$ & $+$ \\
$\gamma^x\mu^x\sigma^z$ & $-$ & $-\gamma^y\mu^z\sigma^z$ & $+\gamma^y\mu^z\sigma^z$ & $-\gamma^y\mu^z\sigma^z$ & $-$ \\
$\gamma^y\mu^x\sigma^z$ & $+$ & $-\gamma^x\mu^z\sigma^z$ & $+\gamma^x\mu^z\sigma^z$ & $+\gamma^x\mu^z\sigma^z$ & $-$ \\
$\gamma^0\mu^y$ & $+$ & $-$ & $-$ & $-$ & $-$ \\
$\gamma^x\mu^y$ & $+$ & $-\gamma^y\mu^y$ & $+\gamma^y\mu^y$ & $+\gamma^y\mu^y$ & $+$ \\
$\gamma^y\mu^y$ & $-$ & $-\gamma^x\mu^y$ & $+\gamma^x\mu^y$ & $-\gamma^x\mu^y$ & $+$ \\
$\gamma^0\mu^z\sigma^x$ & $+$ & $+\gamma^0\mu^x\sigma^y$ & $-\gamma^0\mu^x\sigma^y$ & $-\gamma^0\mu^x\sigma^y$ & $+$ \\
$\gamma^x\mu^z\sigma^x$ & $+$ & $+\gamma^y\mu^x\sigma^y$ & $+\gamma^y\mu^x\sigma^y$ & $+\gamma^y\mu^x\sigma^y$ & $-$ \\
$\gamma^y\mu^z\sigma^x$ & $-$ & $+\gamma^x\mu^x\sigma^y$ & $+\gamma^x\mu^x\sigma^y$ & $-\gamma^x\mu^x\sigma^y$ & $-$ \\
$\gamma^0\mu^z\sigma^y$ & $-$ & $+\gamma^0\mu^x\sigma^x$ & $-\gamma^0\mu^x\sigma^x$ & $+\gamma^0\mu^x\sigma^x$ & $+$ \\
$\gamma^x\mu^z\sigma^y$ & $-$ & $+\gamma^y\mu^x\sigma^x$ & $+\gamma^y\mu^x\sigma^x$ & $-\gamma^y\mu^x\sigma^x$ & $-$ \\
$\gamma^y\mu^z\sigma^y$ & $+$ & $+\gamma^x\mu^x\sigma^x$ & $+\gamma^x\mu^x\sigma^x$ & $+\gamma^x\mu^x\sigma^x$ & $-$ \\
$\gamma^0\mu^z\sigma^z$ & $-$ & $-\gamma^0\mu^x\sigma^z$ & $-\gamma^0\mu^x\sigma^z$ & $+\gamma^0\mu^x\sigma^z$ & $+$ \\
$\gamma^x\mu^z\sigma^z$ & $-$ & $-\gamma^y\mu^x\sigma^z$ & $+\gamma^y\mu^x\sigma^z$ & $-\gamma^y\mu^x\sigma^z$ & $-$ \\
$\gamma^y\mu^z\sigma^z$ & $+$ & $-\gamma^x\mu^x\sigma^z$ & $+\gamma^x\mu^x\sigma^z$ & $+\gamma^x\mu^x\sigma^z$ & $-$ \\
\end{tabular}
\end{ruledtabular}
\end{table}

\begin{table}[t!]
\caption{Same as Table~\ref{tab:locked} for the two generators that the
high-temperature structure adds to the locked group, the mirror
$\hat\sigma_h=\one \otimes C_{2z}$ and the pure fourfold rotation
$\hat C_4=\hat\sigma_h\hat S_4=C_4\otimes R_z(+\pi/2)$
(Sec.~\ref{sec:continuum}). Together with Table~\ref{tab:locked} and the
spin U(1)$_z$ present for $D'_\perp$ alone, these columns give the
$D'_\perp$-only result of Sec.~\ref{sec:zeroth}: no row is invariant.}
\label{tab:hightemp}
\begin{ruledtabular}
\begin{tabular}{lcc}
$\mathrm{Tr}[\bar{\mathcal X}\,\cdot\,\mathcal X]$ & $\hat\sigma_{h}$ & $\hat C_4$ \\
\colrule
$\one$ & $+$ & $+$ \\
$\mu^x$ & $+$ & $+\mu^z$ \\
$\mu^y\sigma^x$ & $-$ & $-\mu^y\sigma^y$ \\
$\mu^y\sigma^y$ & $-$ & $+\mu^y\sigma^x$ \\
$\mu^y\sigma^z$ & $+$ & $-$ \\
$\mu^z$ & $+$ & $+\mu^x$ \\
\colrule
$\gamma^0\sigma^x$ & $-$ & $+\gamma^0\sigma^y$ \\
$\gamma^x\sigma^x$ & $-$ & $-\gamma^y\sigma^y$ \\
$\gamma^y\sigma^x$ & $-$ & $+\gamma^x\sigma^y$ \\
$\gamma^0\sigma^y$ & $-$ & $-\gamma^0\sigma^x$ \\
$\gamma^x\sigma^y$ & $-$ & $+\gamma^y\sigma^x$ \\
$\gamma^y\sigma^y$ & $-$ & $-\gamma^x\sigma^x$ \\
$\gamma^0\sigma^z$ & $+$ & $+$ \\
$\gamma^x\sigma^z$ & $+$ & $-\gamma^y\sigma^z$ \\
$\gamma^y\sigma^z$ & $+$ & $+\gamma^x\sigma^z$ \\
$\gamma^0\mu^x\sigma^x$ & $-$ & $+\gamma^0\mu^z\sigma^y$ \\
$\gamma^x\mu^x\sigma^x$ & $-$ & $-\gamma^y\mu^z\sigma^y$ \\
$\gamma^y\mu^x\sigma^x$ & $-$ & $+\gamma^x\mu^z\sigma^y$ \\
$\gamma^0\mu^x\sigma^y$ & $-$ & $-\gamma^0\mu^z\sigma^x$ \\
$\gamma^x\mu^x\sigma^y$ & $-$ & $+\gamma^y\mu^z\sigma^x$ \\
$\gamma^y\mu^x\sigma^y$ & $-$ & $-\gamma^x\mu^z\sigma^x$ \\
$\gamma^0\mu^x\sigma^z$ & $+$ & $+\gamma^0\mu^z\sigma^z$ \\
$\gamma^x\mu^x\sigma^z$ & $+$ & $-\gamma^y\mu^z\sigma^z$ \\
$\gamma^y\mu^x\sigma^z$ & $+$ & $+\gamma^x\mu^z\sigma^z$ \\
$\gamma^0\mu^y$ & $+$ & $-$ \\
$\gamma^x\mu^y$ & $+$ & $+\gamma^y\mu^y$ \\
$\gamma^y\mu^y$ & $+$ & $-\gamma^x\mu^y$ \\
$\gamma^0\mu^z\sigma^x$ & $-$ & $+\gamma^0\mu^x\sigma^y$ \\
$\gamma^x\mu^z\sigma^x$ & $-$ & $-\gamma^y\mu^x\sigma^y$ \\
$\gamma^y\mu^z\sigma^x$ & $-$ & $+\gamma^x\mu^x\sigma^y$ \\
$\gamma^0\mu^z\sigma^y$ & $-$ & $-\gamma^0\mu^x\sigma^x$ \\
$\gamma^x\mu^z\sigma^y$ & $-$ & $+\gamma^y\mu^x\sigma^x$ \\
$\gamma^y\mu^z\sigma^y$ & $-$ & $-\gamma^x\mu^x\sigma^x$ \\
$\gamma^0\mu^z\sigma^z$ & $+$ & $+\gamma^0\mu^x\sigma^z$ \\
$\gamma^x\mu^z\sigma^z$ & $+$ & $-\gamma^y\mu^x\sigma^z$ \\
$\gamma^y\mu^z\sigma^z$ & $+$ & $+\gamma^x\mu^x\sigma^z$ \\
\end{tabular}
\end{ruledtabular}
\end{table}

\emph{Operator content.} The space of Hermitian gauge-singlet fermion
bilinears at zeroth gradient order is 36-dimensional, the gauge-invariant
subspace of the 120 Hermitian quadratic Majorana forms
(Appendix~\ref{app:proof}). It consists of the six masses and the 30
components of the conserved SO(5)-adjoint currents. The masses are the
SO(5) vector $n_j=\Tr[\Xb\Gamma^j\X]$, two VBS and three N\'eel
components, and the $\mathcal T$-odd flux singlet $\Tr[\Xb\X]$. With
$\Xb=\X^\dagger\gamma^0$ these traces are Hermitian operators as written. The
currents are $\Tr[\Xb\gamma^\mu T^j\X]$ with the ten generators $T^j$. All
ten SO(5)-adjoint scalars vanish identically by the Majorana reality
condition (Appendix~\ref{app:proof}). There is thus no
quantum-spin-Hall-type mass $\Tr[\Xb\X\sigma^a]$, no
$\Tr[\Xb\mu^{x,z}\X\sigma^a]$ and no $\Tr[\Xb\mu^y\X]$. The complete
transformation laws are given in Tables~\ref{tab:unlocked},~\ref{tab:locked}
and~\ref{tab:hightemp}.

\emph{Absence of gauge-singlet masses.} Every existing mass is either $\mathcal T$-odd
(the N\'eel components and the flux singlet) or odd under locked
point-group elements (the VBS components, Table~\ref{tab:locked}). Since DM
preserves $\mathcal T$ and the locked space group, no explicit,
symmetry-allowed local gauge-singlet fermion mass is generated at any order in the DM
couplings. The locked space group alone already suffices: the locked
glide together with $\hat\sigma_{xy}$ leaves no mass invariant, and time
reversal is an additional reason for the N\'eel and flux-singlet masses.
This result concerns explicit local masses in the symmetric state and
requires the full layer group. A shear strain along $[110]$ lowers it and
allows a VBS mass, and a magnetic field allows the masses of
Sec.~\ref{sec:field}. Without these perturbations the Dirac nodes of the $\pi$-flux state cannot be
gapped by a gauge-singlet mass term. For U800 the same conclusion follows
from the $\Phi_3$-Yukawa classification of Sec.~\ref{sec:yukawa}, since the
locked group admits a zeroth-order $\Phi_3$ coupling only to a current and
none to a mass. U800 is, however, a compact U(1) theory whose SO(5)-singlet
monopole is invariant under both the unlocked and the locked group
(Appendix~\ref{app:monopole}), so there the symmetry argument excludes local
masses but not confinement by monopole proliferation. In the Higgs-condensed
Z3000 phase, where the Yukawa sector adds gauge-adjoint bilinears, the nodes
carry a chiral winding and can still disappear by pairwise annihilation if
DM displaces them far enough, which this argument does not exclude;
Sec.~\ref{sec:yukawa} finds no such annihilation over the material-parameter range.

\section{DM-induced couplings}
\label{sec:classify}

\subsection{Zeroth gradient order}
\label{sec:zeroth}
Of the 36 bilinears, DM can induce only those invariant under the locked
group that its couplings preserve, and Tables~\ref{tab:locked}
and~\ref{tab:hightemp} identify them. Without DM no operator is
induced, as required. With $D'_\perp$ alone, under the high-temperature
locked group and U(1)$_z$, no operator is induced either, even without
imposing $\mathcal T$. With in-plane DM, under the buckled locked group, a
single operator is generated,
\begin{equation}
\mathcal O_{\rm DM}
=\Tr[\Xb\gamma^x\X\,\sigma^x]+\Tr[\Xb\gamma^y\X\,\sigma^y].
\label{eq:ODM}
\end{equation}
Here, and in the tables, the spin labels are related to physical axes by
$x\to-x$ ($\sigma^x_{\rm label}=-S^x$), which
leaves the diagonal signs of the tables unchanged. In physical axes, with
$x\parallel[100]$ and $y\parallel[010]$, Eq.~\eqref{eq:ODM} reads
$\mathcal O_{\rm DM}=J^x_x-J^y_y$ up to an overall sign absorbed in $\lambda$.
Two further current combinations are allowed without $\mathcal T$ and
removed by it. $\mathcal O_{\rm DM}$ is the spatial-diagonal,
spin--orbit-locked component of the conserved spin currents. It acts as a
constant non-abelian spin vector potential,
$A^{\rm spin}_x\propto S^x$ and $A^{\rm spin}_y\propto-S^y$ in physical
axes, with
$[A^{\rm spin}_x,A^{\rm spin}_y]\neq0$. A position-dependent spin twist removes it
at linear order but not beyond, since the field strength
$[A^{\rm spin}_x,A^{\rm spin}_y]$ is of order $\lambda^2$ and gauge
covariant, its norm being invariant (Sec.~\ref{sec:crossover}). Even at
linear order the twist removes $\lambda$ exactly only from spin-singlet
observables, since it rotates the spin frame from site to site.

\subsection{Gradient order}
Genuine gradient couplings correspond to symmetric Majorana coefficient
matrices, since antisymmetric ones are total derivatives of the
zeroth-order bilinears. Beyond the two kinetic invariants the
classification gives two structures for $D'_\perp$,
\begin{align}
O_A&=\Tr[\Xb\,\sigma^z\,\ii \overset{\leftrightarrow}{D}_0\,\X],\nonumber\\
O_B&=\Tr[\Xb\,\mu^y\sigma^z(\gamma^y \ii \overset{\leftrightarrow}{D}_x
+\gamma^x \ii \overset{\leftrightarrow}{D}_y)\,\X],
\label{eq:OAB}
\end{align}
with $\overset{\leftrightarrow}{D}=\tfrac12(\overset{\rightarrow}{D}
-\overset{\leftarrow}{D})$ and $\sigma^z$ the spin matrix. In the $\psi$
form of Eq.~\eqref{eq:Ldirac} they read
$O_A\propto\bar\psi\,\ii\overset{\leftrightarrow}{D}_0\psi$ and
$O_B\propto\bar\psi\mu^y(\gamma^y\ii\overset{\leftrightarrow}{D}_x
+\gamma^x\ii\overset{\leftrightarrow}{D}_y)\psi$. They are
spin-dependent velocity and tilt deformations that move but cannot gap
the cones. For in-plane DM, the classification gives two further
dimension-3 structures subleading to Eq.~\eqref{eq:ODM}, six gradient
invariants in total. The operator $O_A$ is redundant. In the Euclidean
conventions of Appendix~\ref{app:integrals} the off-shell identity
$\bar\psi\,\ii\overset{\leftrightarrow}{D}^{\nu}\psi
=\tfrac12\varepsilon^{\nu\mu c}\partial_\mu(\bar\psi\gamma_c\psi)
+\text{EOM}$ holds, with the sign of the curl term reversed in the
Minkowski signature of this section. The identity exhibits $O_A$ as a total
derivative of the conserved $S^z$ spin current $\bar\psi\gamma_c\psi$ plus
equation-of-motion (EOM) terms, with no physical effect at linear order. The
single physical dimension-3 operator induced by $D'_\perp$ is therefore
$O_B$, the $(xy)$ component of the flavored spin-2 current
$\Tr[\Xb\,\Gamma_T\gamma^{(\mu}\ii\overset{\leftrightarrow}{D}{}^{\nu)}\X]$
with the flavor matrix $\Gamma_T=\mu^y\sigma^z$, a component of the SO(5)
vector. Its $\psi$ form is
$2\bar\psi\mu^y\gamma^{(\mu}\ii\overset{\leftrightarrow}{D}{}^{\nu)}\psi$.
It is a conformal primary, since the only descendants with its quantum
numbers, derivatives of the currents with the flavor $\mu^y\sigma^a$,
vanish identically in the Majorana form. Its fate is computed in
Sec.~\ref{sec:spin2}.

\subsection{Lattice derivation of the induced coupling}
\label{sec:lattice}
The operator selection can be derived directly on the lattice. We Wick-decouple the DM term $H_{\rm DM}$ of Eq.~\eqref{eq:H} in the
$\pi$-flux and Z3000 ground states, using the ground-state correlations of
one unit cell of a periodic lattice and summing the result over all
lattice translations, and project onto the Dirac subspace. The projection
yields zero for $D'_\perp$, whose contributions cancel
over the eight bonds of the cell, and zero for the intra-dimer $D$ on the
$\pi$-flux state, whose mean-field amplitudes connect only opposite
sublattices of the square lattice while the dimer bonds join sites of
the same sublattice. For the in-plane
inter-dimer components it yields pure $\mathcal O_{\rm DM}$. The decoupling can be done in closed form
(Appendix~\ref{app:wick}). Let
$\chi_0=|\langle f^\dagger_{i\sigma}f_{j\sigma}\rangle_{\rm NN}|=0.23952$ be
the nearest-neighbor (NN) singlet bond amplitude per spin (no sum over $\sigma$) of the
$\pi$-flux state in the thermodynamic limit ($0.23963$ on the $16\times16$
antiperiodic lattice of the finite-lattice calculations). Then
\begin{align}
\delta\mathcal L&=\lambda\,\mathcal O_{\rm DM},\qquad
\lambda=\frac{\chi_0\,d}{2\sqrt2},\nonumber\\
d&\equiv\frac{D'_\Sigma}{t},\qquad
D'_\Sigma\equiv D'_{\parallel,s}+D'_{\parallel,ns},
\label{eq:lambda}
\end{align}
in units of the spinon hopping $t$, in agreement with the numerical
projection. The staggered and non-staggered components contribute equally
at this order. At mean-field order on the $\pi$-flux state the intra-dimer $D$ enters
only through the Higgs sector (Sec.~\ref{sec:yukawa}). This follows from the
sublattice selection rule $\chi_{ij}=0$ on the dimer bonds. Symmetry
alone does not forbid it, since $D$ and $D'_\parallel$ transform
identically under the locked group. Beyond mean field $D$ can renormalize $\lambda$. In the dimer phase, by contrast, $D$ and
$D'_{\parallel,s}$ enter the triplon Hamiltonian of
Ref.~\cite{Romhanyi2015} at linear order only through the combination
$\tilde D_\parallel$, and $D'_{\parallel,ns}$ does not appear at that
order (Sec.~\ref{sec:crossover}).

\emph{Coupling conventions.} Throughout, $d$ denotes the lattice DM
amplitude in units of $t$, and $\lambda=\chi_0 d/(2\sqrt2)\simeq0.085\,d$ is
the continuum coefficient of $\mathcal O_{\rm DM}$. For the material,
$|D'_\Sigma|\simeq0.16$--$0.29$\,K (Sec.~\ref{sec:model}) and the spinon
hopping $t\simeq20$--$22$\,K of Sec.~\ref{sec:crossover} give
$|d|\simeq0.007$--$0.015$ and $\lambda\simeq10^{-3}\,t$.
The lattice calculations of Secs.~\ref{sec:chiral}--\ref{sec:field} use
$d=0.6$, about 40 to 80 times the material value, in order to resolve the
quadratic band touching of Sec.~\ref{sec:qbt} on finite grids. The sign of
the plaquette harmonic $b_2$ is the same for both signs of $d$ between
$0.01$ and $0.6$ and for both in-plane textures
(Appendix~\ref{app:plaquette}). The field-selection signs are the same at
$d=\pm0.6$, and the Chern numbers also at $d=0$, where the chirality
carries no moment (Appendix~\ref{app:chiral}). The continuum
Kubo responses below $T^*$ (Sec.~\ref{sec:qbt}) use $\lambda=0.075\,t$, and
the spectrum shown in Sec.~\ref{sec:qbt} uses $\lambda=0.12\,t$.

\section{Linear-order Higgs--Yukawa sector and Z3000 node stability}
\label{sec:yukawa}

The gauge-adjoint bilinear space consists of 28 multiplets: the 10
SO(5)-adjoint scalars $\Tr[\sigma^a\Xb\,T^j\X]$, the 15 SO(5)-vector
currents $\Tr[\sigma^a\Xb\,\Gamma^j\gamma^\mu\X]$, and the 3 gauge currents
$\Tr[\sigma^a\Xb\gamma^\mu\X]$. We classify the Yukawa couplings
$\Phi^a_s\,\mathcal B^a$ under the locked groups, with the Higgs
representations fixed by the invariance of the Lagrangian of
Sec.~\ref{sec:continuum}. Without DM the classification returns the two
zeroth-order Yukawa structures of Ref.~\cite{Feuerpfeil2026}. With DM,
$D'_\perp$ admits no new Yukawa coupling, and the in-plane components admit
three new channels,
\begin{align}
&\Phi_{1}^a\Tr[\sigma^a\Xb\mu^y\sigma^x\gamma^0\X]
-\Phi_{2}^a\Tr[\sigma^a\Xb\mu^y\sigma^y\gamma^0\X],\nonumber\\
&\Phi_1^a\Tr[\sigma^a\Xb\X\sigma^y]-\Phi_2^a\Tr[\sigma^a\Xb\X\sigma^x],
\label{eq:newyuk}\\
&\Phi_3^a\Tr[\sigma^a\Xb\mu^y(\sigma^x\gamma^y+\sigma^y\gamma^x)\X],
\nonumber
\end{align}
the last being the first zeroth-order $\Phi_3$ coupling of the theory.
At the multicritical point $\Phi_{1,2}$ are critical and $\Phi_3$ is not,
so the first two channels are relevant there, and their size decides
whether they matter. The $\gamma^0$ channel is an SO(5) and Lorentz
partner of the SO(5)-breaking Yukawa coupling $y$ of
Ref.~\cite{Feuerpfeil2026}, whose flavor matrices $\mu^z\gamma^x$,
$\mu^x\gamma^y$ lie in the same SO(5) vector and Lorentz vector, so at the
SO(5) point it has the same scaling dimension. The scalar channel differs
from $y$ only in the Dirac structure of the vertex; at $O(1/N)$ only the
gauge exchange between the two fermion legs distinguishes the two, and it
makes the scalar channel less relevant than $y$ by
$64/(3\pi^2N)\simeq0.18$ at $N=12$. Reference~\cite{Feuerpfeil2026} obtains
this difference for Yukawa vertices with and without Dirac structure [its
Eq.~(100)], in agreement with Appendix~\ref{app:yukawa}.
The bare sizes follow from the lattice.
Deforming the $\pi$-flux state along the Z3000 amplitudes and projecting
the hopping and the Wick-decoupled DM operator on the nodes yields the
Yukawa coupling $y\propto\Delta_d+\Delta_{d'}+2\Delta_{g,1}$ of
Ref.~\cite{Feuerpfeil2026} and shows that only the intra-dimer $D$
generates the first two channels; the in-plane inter-dimer couplings feed
only the $\Phi_3$ channel, and $D'_\perp$ none. Each channel starts at
$0.006$--$0.054$ of $y$ per unit $D/t$ for the individual Z3000
amplitudes $\Delta_d,\Delta_{d'},\Delta_{g,1}$ (Appendix~\ref{app:yukawa}) and at $0.033$ for representative ones, i.e.\
at a ratio $r\simeq6\times10^{-4}$--$2\times10^{-2}$ for $D/t\simeq0.1$--$0.3$. These
ratios refer to a fixed direction of the Higgs field in amplitude space
and do not bound the DM channels if $y$ is small through a cancellation
among the Z3000 amplitudes. Barring such a cancellation, the
$\gamma^0$ channel stays a factor $r$ below $y$ at every scale
and the scalar channel falls further behind, so the pseudocritical window
is terminated by $y$ itself and the DM-induced Yukawa couplings remain
subleading within it. The conclusion does not rest on the sign of the $O(1/N)$ difference
(Appendix~\ref{app:yukawa}). In the Z3000
phase the DM-induced Yukawa couplings become quadratic terms
$\propto D\,\langle\Phi\rangle$ in the spinon Hamiltonian, and the absence of gauge-singlet masses does not decide whether they gap the Z3000
nodes. Diagonalization of the Bloch-resolved Z3000 Bogoliubov
Hamiltonian together with the Wick-induced DM operator, with the nodes
located by continuous minimization, shows that they remain gapless
for each of the four DM components up to couplings far exceeding the
material values, and also with all four components applied simultaneously
in ranges of ratios of relevance to the material.

Analytically, the protection follows from a chiral symmetry. The
full spinon Bogoliubov Hamiltonian including every DM coupling anticommutes
with the unique staggered chiral operator
$\mathcal S=(-1)^{i_x+i_y}\,\mathcal C_M$, where $\mathcal C_M$ is the
on-site Majorana-component rotation. The staggered sign maps onto
$\rho^y\mu^y$ in the continuum, and $\mathcal S$ is the product of time
reversal and particle--hole conjugation, so the protection is a symmetry
statement. Each node therefore carries a chiral winding number
and can disappear only by pairwise annihilation. The intra-dimer $D$ shifts
the nodes along the zone diagonal. The in-plane inter-dimer components move
them off the diagonal. In momenta conjugate to the translations by $2a$ ($a$ the lattice spacing)
of the $2\times2$ cell, each node splits into a pair displaced by
$\pm(\delta k_x,\delta k_y)\simeq\pm(0.067,-0.070)\,D'_{\parallel,s}/t$
($|\delta k|\simeq0.097\,D'_{\parallel,s}/t$) for the staggered component
and by $\pm(0.059,-0.059)\,D'_{\parallel,ns}/t$ for the non-staggered one,
with slopes evaluated at the reference Z3000 ansatz.
Qualitatively this displacement is the momentum-space counterpart of the
$\pm4\lambda$ splitting of Sec.~\ref{sec:qbt}. $D'_\perp$ leaves the nodes
in place at linear order.

\section{\texorpdfstring{Fate of $D'_\perp$: flavored spin-2 currents at $O(1/N)$}{Fate of D'perp: flavored spin-2 currents at O(1/N)}}
\label{sec:spin2}

\begin{figure*}[t]
\centering
\includegraphics[width=0.96\textwidth]{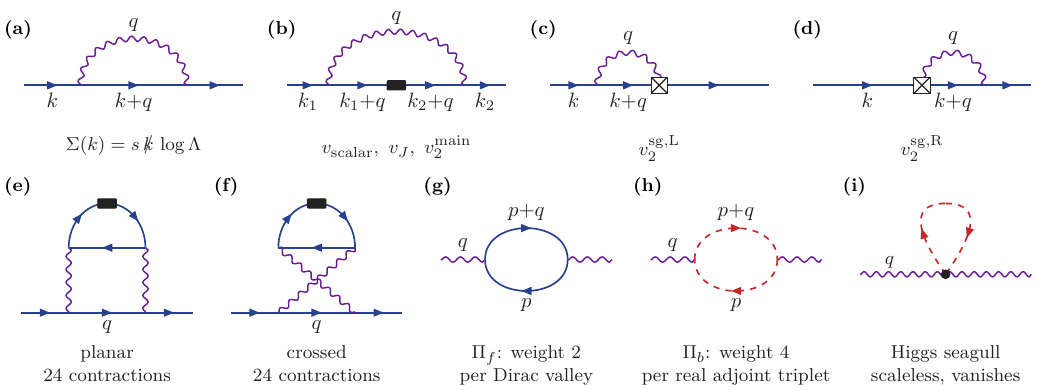}
\caption{Diagrams of the large-$N$ sector (Appendix~\ref{app:integrals}).
Solid indigo lines are spinons (Majorana-matrix fermions), wavy violet
lines the SU(2) gauge field with the induced propagator $D_{\mu\nu}$,
dashed red lines the adjoint Higgs fields, the black box a flavored
operator insertion and the crossed box its seagull vertex. Top row,
$O(1/N)$ renormalization of a flavored bilinear: (a) fermion self-energy,
giving $s$, with external momentum $k$ and internal momentum $k+q$. (b)
Vertex correction with the operator inserted at momentum transfer
$k_2-k_1$, giving $v_{\rm scalar}$, $v_J$ or $v_2^{\rm main}$ according to
the insertion. (c),(d) The two seagull graphs of the covariant-derivative
operator, $v_2^{\rm sg}$ (labeled L and R for the left and right
attachment of the gauge line), with the same momentum labels as (a). Bottom
row: (e),(f) the two annihilation topologies of the flavor-singlet mass
(planar and crossed, 24 labeled contractions each), which produce the
singlet--adjoint splitting of Eq.~\eqref{eq:lifting}. (g),(h) The fermion
and adjoint-Higgs polarization loops, whose sum $\Pi=N|q|/16$ with
$N=2N_f+4N_b$ defines $D_{\mu\nu}$; the arrows on the real adjoint Higgs
lines mark the orientation of the generator $t^b$ in the vertex. (i) The
Higgs seagull tadpole (dot: the seagull vertex), which is scaleless and
vanishes.}
\label{fig:diagrams}
\end{figure*}

The physical operator $O_B$ of Eq.~\eqref{eq:OAB} renormalizes as the
flavor non-singlet spin-2 current
$O_T^{\mu\nu}=\Tr[\Xb\,\Gamma_T\gamma^{(\mu}\ii\overset{\leftrightarrow}{D}{}^{\nu)}\X]$,
in $\psi$ form
$2\bar\psi\mu^y\gamma^{(\mu}\ii\overset{\leftrightarrow}{D}{}^{\nu)}\psi$ with
$D_\mu=\partial_\mu-\ii A^a_\mu\sigma^a$. Its flavor matrix
$\Gamma_T=\mu^y\sigma^z$ is a component of the SO(5) vector, and its
QED$_3$ analogue is the flavor-adjoint spin-2 current. The flavor matrix
acts on valley and spin indices and commutes with the gauge generators.
Being flavored, the operator cannot mix with gauge-field or Higgs
operators at $y=0$, and a single anomalous dimension governs it. We compute this dimension at leading order
in $1/N$, with $N=2N_f+4N_b=12$ the polarization count of
Sec.~\ref{sec:continuum} and Appendix~\ref{app:polarization}, using the
diagrams of Fig.~\ref{fig:diagrams}. The log-divergent integrals are
collected in Appendix~\ref{app:integrals}. The same rules give the
following results for related operators.
\begin{enumerate}
\item The flavored scalar bilinear, whose dimension $\Delta_V$ governs the
VBS channel of Secs.~\ref{sec:instab} and~\ref{sec:disorder}, acquires
$\Delta_V=2-64C_g/(3\pi^2N)=1.46$ at $C_g=\sum_a\sigma^a\sigma^a=3$, the
Casimir factor of the SU(2) gauge vertex, independent of the gauge
parameter. Reference~\cite{Feuerpfeil2026} obtains the same dimension
(its Eq.~(79)), with the count $N=2N_f+4N_b$ of
Appendix~\ref{app:polarization}, which agrees with the rule of the companion U(1)
analysis of Ref.~\cite{Feuerpfeil2026PRB} and is used throughout this paper.
\item The flavored conserved current has $\Delta=2$ exactly, in every gauge
and component.
\item With the gauge line replaced by a Gross--Neveu scalar (no seagull),
the same rules give the published non-singlet spin-2 value
$\gamma_2^{\rm GN}=16/(15\pi^2n)$ for $n$ two-component flavors, i.e.\
$32/(15\pi^2n_{\rm comp})$ with $n_{\rm comp}=2n$ components [Eq.~(3.24) of Ref.~\cite{GKS2017} at $d=3$, $s=2$; Refs.~\cite{MutaPopovic,ManashovSkvortsov2017}] (Appendix~\ref{app:integrals}).
\item The result is independent of the external-momentum routing for
the symmetric (two-sided) derivative vertex, with no admixture of other
tensor structures.
\end{enumerate}
The gauge-parameter dependence cancels between vertex and self-energy, and
the seagull graphs are gauge-parameter free. The relative sign of the
seagull and tree vertices is fixed by gauge covariance. With
$D_\mu=\partial_\mu-\ii A^a_\mu\sigma^a$ the operator $O_T^{\mu\nu}$ has
tree vertex $-\mu^y\gamma^{(\mu}k^{\nu)}$ and seagull vertex
$+\mu^y\gamma^{(\mu}\delta^{\nu)}_\alpha\sigma^a$ in $\psi$ form, and each seagull graph
then equals $-16C_g/(3\pi^2N)$ times the tree vertex
(Appendix~\ref{app:integrals}). The result contains no free parameter,
\begin{equation}
\gamma_2^{\rm adj}=\frac{64\,C_g}{5\pi^2N}
=\frac{16}{5\pi^2}\simeq+0.324,
\label{eq:gamma2}
\end{equation}
the last equality at $C_g=3$, $N=12$, so $\Delta_{O_B}=3.32$ and the
$D'_\perp$-induced coupling is irrelevant at leading order in $1/N$.
Independently of the expansion, a non-conserved spin-2 primary of a
unitary conformal field theory (CFT) has $\Delta>3$ strictly, equality
holding only for a conserved current~\cite{Minwalla1998,Evans1967}, and
$O_B$ is not conserved at $O(1/N)$. Within the assumed CFT
the operator is therefore irrelevant, and the value 3.32 is its
leading-order dimension. Non-conservation is
an $O(1/N)$ effect, since at $N=\infty$ the operator is a conserved
higher-spin current. In the walking scenario with complex fixed points
unitarity bounds do not apply strictly~\cite{WangNahum2017,Zhou2024fuzzy}. For $y\neq0$, SO(5) is broken and
$O_B$ may mix with Higgs spin-2 bilinears of the same lattice quantum
numbers. In this sense, within the assumed CFT at $y=0$, the largest
inter-dimer DM coupling of SrCu$_2$(BO$_3$)$_2$ is inert at the critical
point at linear order among the fermion bilinears. Beyond bilinears,
$D'_\perp$ enters four-fermion operators at linear order, an SO(5)
$\mathbf{35}$ Lorentz vector with $\Delta=4$ and a $\mathbf5$ spin-2 part
with $\Delta=3.91$ at $O(1/N)$ (Appendix~\ref{app:treeseeds}), both
irrelevant. At second order it seeds the rank-2 SO(5) tensor of
Sec.~\ref{sec:selection}.

At $C_g=1$, Eq.~\eqref{eq:gamma2} reproduces the QED$_3$ flavor-adjoint
spin-2 anomalous dimension of Zhou and He, Eq.~(79) of
Ref.~\cite{ZhouHe2023} at spin $\ell=2$, obtained with the same two-component counting and
photon normalization $\Pi=N|q|/16$, and the flow $d\delta/d\ell=-64\delta/(5\pi^2N)$
of the flavored velocity anisotropy $\delta$ of the QED$_3$ Dirac spin liquid
[Eq.~(17) of Ref.~\cite{Hermele2005}], a component of the same spin-2 current. Because the seagull sign is fixed a
priori, this agreement is an independent test of Eq.~\eqref{eq:gamma2}, and fixing the
seagull vertex by minimal coupling alone gives the same result. The non-abelian placement of $C_g$ follows Gracey's general-$d$,
general-gauge-group result, Eq.~(14) of Ref.~\cite{Gracey1994}, with Casimir
structure $C_2(R)/[T(R)N_f]$, whose master formula at $d=3$ (his $\mu=3/2$) gives
$\gamma_2=64/(5\pi^2N)$ in two-component abelian conventions. The
protected limits, the adjoint current at $\ell=1$ and the singlet spin-2
operator at $\ell=2$, have vanishing anomalous dimension, as they must~\cite{ZhouHe2023}. The Chern--Simons-matter results of
Ref.~\cite{CharanPrakash2018}, for U($M$) Chern--Simons theory with $N$ fermion flavors, give vanishing $O(M/N)$ anomalous dimensions for the spin-1 and spin-2 singlet currents at every 't~Hooft coupling [its Eqs.~(4.16), (4.17)]; at $M=1$ and $\lambda_M\to\infty$ (QED$_3$) its odd-spin result, Eq.~(4.17), coincides with the tower below at $C_g=1$. For reference, the full adjoint tower
of spin-$\ell$ currents in our theory reads
\begin{equation}
\gamma_\ell^{\rm adj}=\frac{16\,C_g}{N\pi^2}
\Bigl[\textstyle\sum_{i=1}^{\ell}\frac1{i-1/2}
-\frac23\,\frac{11\ell^2-2}{4\ell^2-1}\Bigr],
\end{equation}
adapted from Ref.~\cite{ZhouHe2023}. At $C_g=3$ and $N=12$ the prefactor is
$4/\pi^2$, and $\ell=1$, $2$ return $0$ and Eq.~\eqref{eq:gamma2}, respectively.
Continued to $\ell=0$, outside the range $\ell\ge1$ stated in
Ref.~\cite{ZhouHe2023}, the formula returns $\Delta_V-2$, as does
Eq.~(14) of Ref.~\cite{Gracey1994} at $n=0$.

\emph{Perspective at $O(1/N^2)$.} No controlled next-order result exists
for the mixed gauge--Higgs theory. The
combination $64/(\pi^2N)$ that sets the scale of every one-loop anomalous
dimension above equals 0.54 at $N=12$, so the $1/N$ results of this
section are qualitative and fix only signs and orderings. The known analogues indicate the size and direction of the next
order. The abelian QED$_3$ adjoint dimension is $2-2.162/N-0.470/N^2$
for $N$ two-component flavors [Ref.~\cite{Gracey1993mass}, Eq.~(27);
the leading term also in Ref.~\cite{BK2019}, Eq.~(3.5)]. Its next-to-leading/leading (NLO/LO) ratio is
$0.217/N\simeq1.8\%$ at $N=12$. The non-abelian QCD$_3$ analogue~\cite{CDGM2000} carries a larger, $C_A$-dominated
next-order term. For SU(2) its NLO/LO ratio is about $3.9/N_f$ for $N_f$
four-component flavors, so the abelian $1.8\%$ is not a size estimate for
the present theory. Neither theory
contains the adjoint-Higgs screening of the present one, whose $O(1/N^2)$
contribution is uncomputed. Both known analogues shift $\Delta_V$ further
below 2, the direction that strengthens the conclusions drawn from it. We are not aware of an $O(1/N^2)$ result for a flavored spin-2 current in any gauge theory. In the Gross--Neveu prototype the non-singlet
spin-2 current at next order has the same sign as its leading term, $\gamma_2^{\rm GN}\simeq0.216/n_{\rm comp}+0.645/n_{\rm comp}^2$ for $n_{\rm comp}$ fermion components [Ref.~\cite{ManashovSkvortsov2017}, Eqs.~(3.5a) and (B.1)]. The $O(1/N^2)$ computation for the
present theory, whose new ingredient is the adjoint-Higgs polarization at
criticality (Appendix~\ref{app:polarization}), is left to future work.
\section{DM-seeded SO(5) anisotropies}
\label{sec:selection}
This section treats the DM-seeded perturbations at second order, which
act above the crossover scale $T^*=4\lambda$ and below the length
$\xi_{\rm DM}=v/(4\lambda)$ of Sec.~\ref{sec:crossover}, with $v$ the Dirac velocity. The mass-channel
responses are computed in the $\lambda$-deformed spectrum, whose quadratic band
touching (QBT) below $T^*$ and its gap channels are described in
Sec.~\ref{sec:qbt}.
At leading order in $1/N$ the fermions are governed by the quadratic
theory, so the mass-channel susceptibilities of the $\lambda$-deformed
theory constitute the leading-order result, to all orders in $\lambda$.
We compute all six mass-channel responses with the full static Kubo kernel at temperature $T$,
\begin{equation}
\Pi_O=\tfrac12\sum_{\bm k,ij}\frac{f(E_i)-f(E_j)}{E_j-E_i}\,|O_{ij}|^2,
\end{equation}
with $f$ the Fermi function and $E_i$ the band energies. The degenerate limit $-f'(E)$ of the kernel supplies the intraband (Pauli)
response. The Pauli term carries the intraband response of the Fermi pockets and
is required for the pocket-forming N\'eel$_z$ channel. We
quote $\delta\Pi_O=\Pi_O(\lambda)-\Pi_O(0)$ in a continuum
normalization defined by a momentum sum over the grid without
the factor $(2\pi)^{-2}$ and with mass matrices of eigenvalues $\pm2$, so
only signs and ratios are meaningful. The factor $\tfrac12$ removes the
double counting of the Majorana basis. At $\lambda=0$ all six responses coincide. The locked symmetry allows
three rank-2 tensors on the SO(5) vector: the two diagonal ones below and
an off-diagonal one proportional to $V_xV_y$. The valley-blind identity of
the quadratic theory, in which the Hamiltonian $h(\bm k)$ of Sec.~\ref{sec:qbt} contains no $\mu$ matrices,
forces the third coefficient to vanish at this order, which symmetry alone
does not. Within the SO(5) vector the splitting decomposes into a singlet
shift and the two diagonal tensors,
\begin{align}
\delta\Pi\big|_{\text{rank-2}}
&=\tau_1\,\mathrm{diag}(3,3,-2,-2,-2)\nonumber\\
&\quad+\tau_2\,\mathrm{diag}(0,0,1,1,-2),
\label{eq:rank2}
\end{align}
in the channel order $(V_x,V_y,N_x,N_y,N_z)$ and in the normalization
above.

The regime relevant for the pseudocritical theory is $T\gg4\lambda$,
where the responses are of order $\lambda^2$. This regime is evaluated at
$\lambda=0.005\,t$ for $T/\lambda=5$--$100$, with circular and square
cutoffs between $1.2$ and $2.4$; the $\lambda^2$ scaling is confirmed at
$\lambda=0.0025\,t$. For $T/\lambda\le20$ the N\'eel responses are
$\delta\Pi_{N_{x,y}}=-\beta$ and $\delta\Pi_{N_z}\simeq-2\beta$, with
$T\beta/\lambda^2=184$--$196$ independent of the cutoff. The gap channels
show no universal term of this form, and their small remainder changes
by a factor of up to six with the cutoff. Hence, $\tau_1=4\beta/15>0$ and
$\tau_2=\beta/3>0$, with $\tau_1/\tau_2=0.80$, and the signs of both tensors are
the same as in the QBT regime below $T^*$ (Sec.~\ref{sec:qbt}). In the critical regime they come from a
suppression of the N\'eel channels, and the singlet channels carry no
universal term. The cutoff-independent content is therefore a
N\'eel$\to$VBS splitter and an easy-plane splitter of definite sign, both
of order $\lambda^2/T$. Both DM-seeded
tensors are traceless and lie in the rank-2 symmetric traceless
representation of SO(5), the $\mathbf{14}$. Whether they matter depends
on the dimension of that operator, on which the two descriptions of the
SO(5) point disagree. In the large-$N$ theory of this paper the
$\mathbf{14}$ has $\Delta=3.46$ at $O(1/N)$ (the rank-2 four-fermion
tensor of Appendix~\ref{app:treeseeds}) and is irrelevant, so the SO(5)-breaking four-fermion scalar generated
by DM at second order does not destabilize the fixed point above $T^*$ at
this order. Every other SO(5)-breaking eigenvalue of the $O(1/N)$
four-fermion spectrum also exceeds $3$, but in the representations
$\mathbf5$ and $\mathbf{10}$ mixing with gauge and Higgs composites is not
computed, so we make no statement at higher order in DM. The DM-induced Yukawa couplings of Sec.~\ref{sec:yukawa} stay
below the Yukawa coupling that bounds the pseudocritical window.
For the SO(5) nonlinear sigma model with a Wess--Zumino--Witten term, the
fuzzy-sphere study of Ref.~\cite{Zhou2024fuzzy} instead finds this
operator relevant, with $\Delta_{\mathbf{14}}=1.456(12)<3$, a value obtained in the
walking region off the critical point; its authors identify the
SO(3)$\times$SO(2)-invariant component, the $\tau_1$ tensor of
Eq.~\eqref{eq:rank2}, as the tuning operator of the N\'eel--VBS
transition. If that relevance carries over to the pseudocritical regime,
the $O(\lambda^2)$ tensors are not the leading DM-seeded ones. The
out-of-plane coupling, whose mean-field image is the irrelevant bilinear
$O_B$, generates the same tensor at second order, because $O_B$ carries
the N\'eel$_z$ flavor $\mu^y\sigma^z$ and the product $O_B\times O_B$
contains the $\mathbf{14}$. The same Kubo kernel evaluated with the
Wick-decoupled $D'_\perp$ operator gives, per $(D'_\perp/t)^2$ and in the
normalization of Eq.~\eqref{eq:rank2}, $\tau_1=-0.18$ and $\tau_2=-1.94$,
nearly independent of temperature, quadratic (to numerical precision), and ultraviolet dominated
(the continuum $O_B$ model gives $\tau_2=-(3/2\pi)\Lambda_{\rm eff}(D'_\perp/t)^2$
with an effective cutoff $\Lambda_{\rm eff}\simeq4$). The $D'^2_\perp$ tensor is
easy-axis, opposite to the easy-plane $\lambda^2$ tensor, and the ratio of
the two is $-0.030\,(T/t)(D'_\perp/\lambda)^2$. For $D'_\perp=0.10$--$0.18$\,meV
and the Fock values $\lambda=0.008$--$0.025$\,K the $D'^2_\perp$ tensor is therefore
comparable to the $\lambda^2$ one at $T^*$ and dominates above
$T_\times\simeq0.01$--$0.3$\,K. With $y_T=3-\Delta_{\mathbf{14}}\simeq1.54$ a
coupling $g_T$ of the $\mathbf{14}$ becomes of order one at
$E_T=\Lambda g_T^{1/y_T}$, with $\Lambda\simeq t$ unless stated. The
mean-field coefficient,
$g_T\simeq0.014(D'_\perp/t)^2$, gives $E_T\simeq0.03$--$0.07$\,K, of order
$T^*$; a bond-level XXZ estimate, $D'^2_\perp/2J'^2\simeq0.08(D'_\perp/t)^2$,
gives $0.1$--$0.2$\,K ($0.2$--$0.5$\,K with $\Lambda=J'$), and a coefficient of order one would give
$0.5$--$1$\,K. These estimates concern the magnitude only. The bond-level
XXZ anisotropy is easy-plane, whereas the easy-axis sign found above is a
free-spinon mean-field result dominated by lattice-scale momenta, so the
sign of the $D'^2_\perp$ tensor is not a universal statement. The
intra-dimer $D$ has no mean-field image on the $\pi$-flux state, since
its Wick amplitude vanishes on the dimer bonds, so its $O(D^2)$ tensor,
like the symmetric intra-dimer anisotropies $J_{zz}$ and $J_{xy}$ of the
neutron fit of Ref.~\cite{McClarty2017} and the dipolar coupling, which
are of the same order,
lies outside this theory. The symmetric anisotropy cannot be dropped
against the second-order DM effects: for a single bond, DM and the
symmetric anisotropy of the same order are equivalent to isotropic
exchange by a local spin rotation, and only their frustration across
bonds has observable consequences~\cite{Shekhtman1992}.

If the $\mathbf{14}$ is relevant, the DM-seeded SO(5) breaking above $T^*$
is a second-order anisotropy, easy-axis at mean field (a sign that is
not universal, see above) and, with $\tau_1<0$, tilted toward the
N\'eel side. It acts at a scale of order $T^*$ or above, up to about
$1$\,K, and the statement that DM cannot destabilize the pseudocritical
regime above $T^*$ then holds at linear order only. The same
second-order seed also carries an SO(5)-singlet component, $-2.1$ per
$(D'_\perp/t)^2$, which shifts the tuning parameter but breaks no
symmetry. Beyond
$\xi_{\rm DM}$ the competition among these channels is decided by the QBT
flow of Sec.~\ref{sec:qbtir}. Within the $\lambda^2$ tensors alone the
deconfined N\'eel--VBS competition is tilted toward the singlet side and
residual magnetism is easy-plane, statements that hold below $T_\times$.
The signs of $\tau_1$ and $\tau_2$ are leading-order statements
(Sec.~\ref{sec:spin2}).

\section{Crossover scale and the band touching}
\label{sec:rg}

\subsection{RG eigenvalue and scales}
\label{sec:crossover}
$\mathcal O_{\rm DM}$ is built from the Noether currents of the spin
SU(2) symmetry, which is conserved at the lattice level in the
unperturbed model. Its scaling dimension is therefore exactly 2 at any
genuine fixed point of the unperturbed theory, and $\lambda$ has RG
eigenvalue $+1$ with no anomalous correction at any order. In the
pseudocritical (walking) regime of Ref.~\cite{Feuerpfeil2026}, the flow
stays close to the SO(5)-symmetric fixed point, expected to be conformal
at vanishing Yukawa coupling, until the Yukawa coupling $y$ drives it away.
That coupling is relevant in the large-flavor expansion of
Ref.~\cite{Feuerpfeil2026}, with RG eigenvalue $\simeq1.38$ at
$N_f=N_b=2$, and its authors regard the fate of $y$ at these values as unclear.
A window reaching down to $T^*$ requires a correspondingly small bare
value of $y$, which grows by a factor of
$3\times10^3$--$3\times10^4$ between $J'$ and $T^*$.
In this regime the RG eigenvalue of $\lambda$ is 1 up
to corrections controlled by the flow of the Yukawa coupling. The same
holds, with corrections controlled by the slow drift of the couplings,
in the pseudocritical scenario of Ref.~\cite{WangNahum2017}, where the
flow passes near a pair of complex fixed points and no fixed point is
reached at real couplings. Hence
\begin{equation}
\xi_{\rm DM}\simeq\frac{v}{4\lambda},\qquad
T^*\simeq4\lambda=\sqrt2\,\chi_0\,|D'_\Sigma|
\simeq0.34\,|D'_\Sigma| .
\end{equation}
The eigenvalue is exact at a fixed point. The prefactor of $\lambda$ carries the $O(1)$
uncertainty of the mean-field matching of Sec.~\ref{sec:lattice}.

$D'_\Sigma$ is not the triplon combination $\tilde D_\parallel$
constrained by dimer-phase spectroscopy~\cite{Cheng2007,Romhanyi2015},
and its value rests on the DFT vectors of Ref.~\cite{Mazurenko2008} (Sec.~\ref{sec:model}): $D'_\Sigma\simeq-0.014$\,meV,
or $|D'_\Sigma|\simeq0.007$--$0.025$\,meV with the tabulated precision and
an ESR rescaling. At the
Fock level this gives $T^*\simeq0.03$--$0.1$\,K. The
value of
Sec.~\ref{sec:lattice} is the Fock one: at linear order in $D$ the DM term
is a background spin gauge field, a twist of the Fock hopping $\chi_0 J'$ by
the angle $D'_{\rm bond}/J'$, so the node splitting scales with the spinon
velocity, $4\lambda=\sqrt2\,(t/J')\,|D'_\Sigma|$ for a hopping $t$, and it
is the length $\xi_{\rm DM}=v/(4\lambda)=\sqrt2\,(J'/|D'_\Sigma|)\,a$ that
does not depend on $t$. We take the $\pi$-flux Dirac velocity $v=2ta$,
$a$ the square-lattice spacing, and assume $t\simeq0.4J'$, a value not
derived here and $1.7$ times the Fock value $\chi_0 J'\simeq0.24J'$. With
it the same twist gives $T^*\simeq0.57|D'_\Sigma|\simeq0.05$--$0.17$\,K. We therefore quote $T^*\simeq0.03$--$0.1$\,K at the
Fock level, up to $0.17$\,K with $t\simeq0.4J'$. The
range $T^*\simeq0.03$--$0.17$\,K used elsewhere in
this paper includes the tabulated precision, the ESR rescaling and this
prefactor. For $J'=51$--$54$\,K this gives $t\simeq20$--$22$\,K and
$\xi_{\rm DM}\simeq250$--$470$ lattice spacings ($940$ at $|D'_\Sigma|=0.007$\,meV).

The energy $4\lambda$ is the scale of the node reconstruction. The
thermodynamic crossover lies an order of magnitude lower, because the
reconstructed band $\sqrt{v^2k^2+4\lambda^2}-2\lambda$ is quadratic only for
energies well below $\lambda$. For free spinons the specific heat of
$h(\bm k)$ of Sec.~\ref{sec:qbt} stays within $5\%$ of its Dirac value down
to $T\simeq0.15\times4\lambda$. Its local exponent $d\ln C/d\ln T$ is $1.42$
at $0.1\times4\lambda$ and $1.12$ at $0.02\times4\lambda$. For the material
the thermodynamic crossover is therefore at about $3$--$25$\,mK, and
$C\propto T$ is approached only near $0.02\,T^*$, i.e.\ $0.6$--$3$\,mK, comparable to the instability scale $T^{**}$ of
Sec.~\ref{sec:qbtir}.
Above $T^*$ the Dirac spin liquid and the SO(5) pseudocritical window
are protected against the DM perturbation at linear order by the RG
eigenvalue 1 and the small prefactor; the second-order anisotropies were
treated in Sec.~\ref{sec:selection}. This range covers
nearly all temperatures explored experimentally in SrCu$_2$(BO$_3$)$_2$. The lowest NMR
temperature under pressure, $0.07$\,K~\cite{Cui2023}, lies near $T^*$.

The twist structure of $\mathcal O_{\rm DM}$ parallels, but does not reproduce, the
reducible/irreducible classification of Cheng
\emph{et al.}~\cite{Cheng2007}. There the non-staggered in-plane component, after a uniform
dimer rotation, sums to zero around closed loops and is removed to first
order, whereas $D'_\perp$ and $\tilde D_\parallel$ are not. In the Dirac theory, at strictly linear order, the
coupling $\lambda\,\mathcal O_{\rm DM}$, which contains
$D'_{\parallel,s}$ and $D'_{\parallel,ns}$ with equal weight, is
twist-equivalent to spin-dependent node displacements. No rotation
removes the $\pm4\lambda$ splitting into the
quadratic band touching or the $O(\lambda^2)$ anisotropies of Sec.~\ref{sec:selection}. The
obstruction to removing them is non-abelian, $[A^{\rm spin}_x,A^{\rm spin}_y]\neq0$, and
differs from the loop-sum criterion of Ref.~\cite{Cheng2007}.

\subsection{\texorpdfstring{Quadratic band touching below $T^*$}{Quadratic band touching below T*}}
\label{sec:qbt}
\begin{figure}[t]
\includegraphics[width=\linewidth]{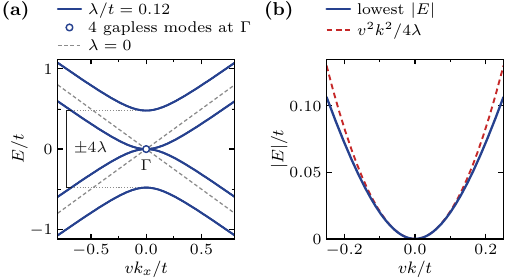}
\caption{Node reconstruction by $\lambda\,\mathcal O_{\rm DM}$ at the
$\pi$-flux point. (a) Spectrum of
$h(\bm k)=v(k_x\rho^x-k_y\rho^z)+\lambda h_{\rm DM}$, with $h_{\rm DM}$ the
node matrix of $\mathcal O_{\rm DM}$, along $k_y=0$ at
$\lambda/t=0.12$ (solid; every band is doubly degenerate, so the eight
bands appear as four curves) and the $\lambda=0$ Dirac cone (dashed):
two branches split to $\pm4\lambda$ (bracket); four modes stay gapless
at $\Gamma$ (open circle). (b) The residual dispersion (solid) is an
isotropic quadratic band touching, $|E|\simeq v^2k^2/4\lambda$ (dashed).
Energies and momenta are plotted in units of $t$ ($E/t$, $vk_x/t$,
$vk/t$).}
\label{fig:qbt}
\end{figure}
Diagonalizing $h(\bm k)=v(k_x\rho^x-k_y\rho^z)+\lambda h_{\rm DM}$ at
$\Gamma$ splits the eightfold multiplet into branches at $\pm4\lambda$
and four gapless modes (Fig.~\ref{fig:qbt}). Second-order
$k\!\cdot\!p$ theory gives the canonical $d$-wave quadratic band touching
\begin{equation}
H_{\rm QBT}(\bm k)=\frac{v^2}{4\lambda}
\bigl[(k_x^2-k_y^2)\,\mathcal A_1+2k_xk_y\,\mathcal A_2\bigr],\ \
\{\mathcal A_1,\mathcal A_2\}=0,
\label{eq:qbt}
\end{equation}
with no $k^2\,\one $ term. Such a term is forbidden by the chiral symmetry $\mathcal C=\rho^y$, which anticommutes with the full
$h(\bm k)$ including the DM term and acts within the band-touching
subspace. The Berry winding is 2, and the spectrum is isotropic at all
$k$, with no cubic warping. The spinon density of states is finite at
energies well below $\lambda$, giving $C\sim T$ in the free-spinon regime
for $T\ll0.1\times4\lambda$ (Sec.~\ref{sec:crossover}). Since
$T^{**}\simeq(0.008$--$0.025)\times4\lambda$ (Sec.~\ref{sec:qbtir}), this regime spans at most a
factor of a few in temperature before the instability. We do not infer
a constant spin susceptibility or a finite intrinsic $\kappa_0/T$ from
this. The projected uniform-$z$ spin equals minus the flux mass and
shares the gap-channel infrared logarithm, and a clean quadratic band
supplies no momentum-relaxation mechanism. Transport requires disorder
or gauge scattering and is not computed here.

\emph{Mass content of the QBT.} Projecting the six masses onto the QBT
subspace gives the following. N\'eel$_{x,y}$ vanish identically, since
they live entirely on the $\pm4\lambda$ branches. VBS$_x$, VBS$_y$, and
the flux (scalar-chirality) mass anticommute with both
$\mathcal A_{1,2}$ and open full gaps. N\'eel$_z$ survives with equal
norm but commutes with $\mathcal A_{1,2}$, so it splits the bands into
Fermi pockets without opening a gap, and a zero-energy ring persists
in its spectrum. The QBT therefore has three gauge-singlet gap channels, and N\'eel
(staggered-magnetic) gapping is excluded kinematically. On the QBT
subspace the projected uniform spin density obeys
$P(J^0\sigma^z)=-P(\text{flux})$ exactly, so the flux gap coincides there
with the uniform-$z$ magnetization channel. Chirality and uniform
$z$ magnetization are locked, so a chiral gap carries weak $z$
ferromagnetism, and a field $\bm B\parallel[001]$ couples linearly to
the flux channel. The $[001]$ entry of the field classification of
Sec.~\ref{sec:field} follows independently from this identity.

\emph{Mass-channel responses below $T^*$.} With the Kubo kernel of
Sec.~\ref{sec:selection}, Table~\ref{tab:kubo} lists the responses at
$\lambda=0.075t$ on the grid $|k_x|,|k_y|\le1.2$ with $360^2$ points.
With $4\lambda=0.3t$, all tabulated temperatures lie below $T^*$.
\begin{table}[t]
\caption{Static Kubo responses $\delta\Pi_O=\Pi_O(\lambda)-\Pi_O(0)$
of the mass channels at $\lambda=0.075t$ ($4\lambda=0.3t$), in the
continuum normalization of Sec.~\ref{sec:selection}, in which only signs and ratios are meaningful, on the grid $|k_x|,|k_y|\le1.2$
($360^2$ points). All temperatures lie below $T^*=4\lambda$; the quadratic
regime proper sets in below about $0.1\times4\lambda$.
The three gap channels VBS$_x$, VBS$_y$ and flux share one value at every
$T$, and N\'eel$_x$ and N\'eel$_y$ share another.}
\label{tab:kubo}
\begin{ruledtabular}
\begin{tabular}{lrrr}
$T/t$ & $\delta\Pi_{V,{\rm flux}}$ & $\delta\Pi_{N_{x,y}}$ &
$\delta\Pi_{N_z}$\\
\hline
$0.1$ & $-0.02$ & $-7.76$ & $-12.90$\\
$0.025$ & $+5.50$ & $-12.49$ & $-18.35$\\
$0.0016$ & $+24.78$ & $-14.11$ & $-19.97$\\
\end{tabular}
\end{ruledtabular}
\end{table}
The three gap channels are equal at every $T$. This equality is the
kinematic identity of the valley-blind quadratic theory and is lifted
only at $O(1/N)$. Their common value grows logarithmically as $T\to0$
(the gap-channel infrared logarithm, regulated here by $T$), and it
crosses zero near $T\sim0.1t$ at a cutoff-dependent
position. The individual $\delta\Pi_O$ are matching-dependent and carry
no invariant sign. The N\'eel channels saturate. At low $T$ the
N\'eel$_z$ response is the sum of an interband depression ($-27.4$) and
the pocket Pauli term ($+7.4$). In the decomposition of Eq.~\eqref{eq:rank2},
$\tau_1(T)>0$ grows logarithmically, to $8.17$ at $T=0.0016t$, and
$\tau_2=+1.953$ ($|k_i|\le1.2$) or $+2.092$ ($|k_i|\le1.6$) saturates below
$T\simeq0.02t$ and must be quoted with its ultraviolet prescription. The
logarithmic growth of $\tau_1$, the positive gap-channel response and its
zero crossing are properties of the QBT regime.

\section{Instabilities of the band touching}
\label{sec:instab}
The regime below $T^*$ treated in Sec.~\ref{sec:qbt} and in this section
rests on five conditions, each stated where it arises and collected here.
(i) A pseudocritical window that reaches $T^*$, which requires a small
bare Yukawa coupling (Sec.~\ref{sec:crossover}), and the second-order anisotropies must not act
first, which they may if the $\mathbf{14}$ is relevant
(Sec.~\ref{sec:selection}). (ii) Proximity to the critical pressure. The
band touching exists between the Higgs splitting $\delta'$ of Eq.~\eqref{eq:splitter} and $\lambda$,
a window open for Higgs-deviation scales $s_H\lesssim0.03$--$0.08$ at $d=0.6$ and for smaller
$s_H$ at the material $\lambda$ (Sec.~\ref{sec:whichvbs}), and no
measurement fixes $s_H$. (iii) A two-dimensional sample. The interlayer
coupling $J_c=0.053(3)$\,meV${}\simeq0.6$\,K, fitted at
$5.5$\,GPa in the monoclinic phase~\cite{Fogh2024PRL} and assumed here to carry over to the tetragonal one, exceeds $T^*$ by a
factor $4$--$20$ and $T^{**}$ by two to three orders of magnitude. If $J_c$
couples the N\'eel vectors of adjacent layers, through
$\bm N_1\cdot\bm N_2$ of dimension $2\Delta_N$, with $\Delta_N$ the scaling
dimension of the N\'eel order parameter, the dimensional crossover
lies near $J'(J_c/J')^{1/(3-2\Delta_N)}$, negligible at the leading-order
$\Delta_N=1.46$ and about $4$\,K at $\Delta_N\simeq0.63$. (iv) Clean
samples. The random valence-bond field of point defects becomes strong
near $0.1$--$0.3$\,K for free-spin fractions of $0.1$--$0.4\%$,
comparable to $T^*$ (Sec.~\ref{sec:disorder}).
(v) No order above $T^*$. The samples measured so far order above
$1$\,K. Within these conditions the instability scale is
$T^{**}\simeq(0.03$--$0.1)\lambda$, $0.2$--$2.5$\,mK at the Fock level (up to $4$\,mK with $t\simeq0.4J'$), an order-of-magnitude one-loop estimate, and the basin
fractions found below depend on the sampling measure of the initial
couplings.

\subsection{Infrared fate: interactions at the QBT}
\label{sec:qbtir}
In the Higgsed (Z3000) phase the SU(2) gauge field acquires a Higgs
mass $m_A$, proportional to the Higgs amplitude and vanishing at the
critical point, and below $m_A$ the gauge sector is gapped to
$\mathbb Z_2$. The band touching exists between the Higgs-induced
splitting $\delta'$ of Eq.~\eqref{eq:splitter} and the upper end of its
quadratic regime, an energy of order $\lambda$ (Sec.~\ref{sec:crossover}).
Near the critical point, where the window is open, $m_A$ lies below
$\lambda$ over most of this range, and between $m_A$ and $\lambda$ the QBT
couples to the unscreened SU(2) gauge field. The band touching screens
this field itself (Appendix~\ref{app:qbtgauge}). Its finite density of
states gives the color density of the eight QBT modes the static
polarization $\Pi_{00}=4\ln2\,\rho_0$, which removes the long-range
color-Coulomb field. The transverse field couples through the velocity
vertex, and its effective strength falls as $(E/\lambda)^{1/2}$. The Higgs
polarization, of relative size $1.1(E/\lambda)^{1/2}$ at the momenta of the
window, is smaller than $\Pi_{00}$ well below $\lambda$, and so is the
Higgs-mass term as long as $m_A\lesssim\lambda$. On both sides of $m_A$
the gauge field therefore leaves a short-range interaction
$\tfrac U2\sum_a\rho^a\rho^a$ between the color densities $\rho^a$, with
the strength $\rho_0U=1/(4\ln2)\simeq0.36$ at low energy, where
$\rho_0=\lambda/(\pi v^2)$ is the single-band density of states. The
problem then falls in the marginally relevant QBT class of
Refs.~\cite{Sun2009,VafekYang2010}. On a single spinless QBT a weak
repulsion $V$ grows as $dV/d\ell=\rho_0V^2$~\cite{Sun2009}, with $\ell$ the logarithm
of the momentum cutoff. With several channels a pure density coupling has no
one-loop term of its own but generates the others, which run away~\cite{VafekYang2010}.
Here, the color interaction, which couples the
fourteen channels, grows faster (Appendix~\ref{app:qbtgauge}). It drives
an instability into one of the gap channels analyzed below at
$T^{**}\sim E_2\,e^{-2\ell^*}$, with $E_2$ the upper end
of the short-range regime and $\ell^*$ the flow time at which the
couplings diverge. N\'eel gapping is excluded kinematically (Sec.~\ref{sec:qbt}). The three
channels are degenerate at quadratic level by valley-blindness, but
gauge fluctuations lift the degeneracy at $O(1/N)$. The flux mass is a
flavor singlet and receives the two-photon annihilation (closed-loop)
correction shown in Fig.~\ref{fig:diagrams}(e),(f), familiar from the
QED$_3$ singlet~\cite{HermeleErratum,ChesterPufu2016,BK2019},
whereas the VBS (valley-adjoint) and N\'eel loops are traceless. In
our theory this class is evaluated in Appendix~\ref{app:integrals}. The
topology weight $|C_{\rm topo}|=1/2$ follows from the 48 labeled
contractions of the annihilation topology and the routing sign from
momentum conservation on the fermion line, and together they give the
QED$_3$ value $\Delta_{\rm singlet}-\Delta_{\rm adj}=+64/(\pi^2N)$ for $N$
two-component fermions. In the
present theory the flavor loop carries ${\rm tr}(\sigma^a\sigma^b)
=2\delta^{ab}$ per valley, summed over the $N_f$ valleys, the fermion
line carries $\sum_a\sigma^a\sigma^a=C_g=3$, and the two gauge
propagators carry $(16/N)^2$ with $N=2N_f+4N_b=12$
(Sec.~\ref{sec:continuum}). The result is
\begin{equation}
\Delta_{\rm flux}-\Delta_{\rm VBS}
=\frac{64}{\pi^2}\,\frac{6N_f}{N^2}
=\frac{16}{3\pi^2}\simeq+0.54,
\label{eq:lifting}
\end{equation}
so $\Delta_{\rm VBS}\simeq1.46$ and $\Delta_{\rm flux}=2+O(1/N^2)$.
The $O(1/N)$ correction to the latter cancels accidentally, since
$6N_f\cdot64/(\pi^2N^2)=64C_g/(3\pi^2N)$ holds for $N_b=N_f$. The chirality
channel is suppressed relative to VBS as $T^*$ is approached from above.

The Fierz-complete one-loop flow in this symmetry class
(Appendix~\ref{app:qbtflow}) settles the competition. The locked group
admits 14 invariant quartic couplings of the eight QBT Majorana modes,
and with the analytic $z{=}2$ bubble the $\beta$ function closes on
this space. On a single spinless QBT it gives the marginal relevance of
repulsion and the dominance of the time-reversal-breaking gap channel
found in Ref.~\cite{Sun2009}. The pocket-splitting N\'eel$_z$ channel has identically vanishing
gap-channel rate in every flow, as the kinematic exclusion requires. The flow has two stable runaway rays: a VBS-symmetric ray
with equal VBS$_x$/VBS$_y$ vertex growth and suppressed chirality, and
a chirality ray with dominant flux vertex growth and suppressed VBS.
Attraction in either gap channel selects its own ray, repulsion in one
funnels into the other, and N\'eel$_z$-type seeds of either sign flow
to VBS. We scan random initial couplings, uniform on the Frobenius
sphere, as a proxy measure for the unknown microscopic couplings. This
measure is canonical for orthonormal Majorana modes but carries no
microscopic information. Labeled by the fastest-growing channel when the
couplings first reach $|g|=200$, $487$ of $600$ seeds ($81\%\pm1.6\%$) are
VBS, $111$ are flux, and two do not diverge up to the integration cutoff
$\ell=60$ (in the physical normalization of Appendix~\ref{app:qbtflow}). On the asymptotic ray $83$--$84\%$ are VBS, a value reproduced by a
second sample of 600 seeds (Appendix~\ref{app:qbtflow}). Neither fraction is universal, since the ultraviolet matching and the
sampling measure enter, but for this sampling the flux basin is clearly
the smaller one. At the critical point only the six gauge-singlet
quartics are legitimate (Appendix~\ref{app:qbtflow}). A census of 600
seeds drawn inside that subspace, which the flow preserves to $10^{-15}$,
gives valence-bond order for $73\%\pm2\%$, the flux mass for $25\%$ and no
instability for $3\%$, so the majority verdict holds there as well. On the
majority ray the enhanced source sector is fivefold: the VBS pair together with the three gauge-adjoint
$\mu^y$-flavored mass directions, with equal rates. The full local mass
space of the eight modes is six-dimensional. The adjoint
directions have no overlap with the six symmetry-allowed bilinears on the
QBT modes discussed below, so they are symmetry-breaking order parameters. At the critical point they are not
gauge-invariant, but in the Higgsed phase, where the flow applies, they
become gauge-invariant once dressed by the condensate. On the ray itself the one-loop
flow does not select among these five directions, so its majority
terminus is VBS or adjoint-mass order unless the approach to the ray
selects, as the gauge interaction below does in favor of VBS. The VBS$_x$/VBS$_y$ rates on the ray are equal, and
the selection within the doublet is discussed in
Sec.~\ref{sec:whichvbs}. The dimension splitting of
Eq.~\eqref{eq:lifting} enhances the magnitude of the VBS quartic relative
to the flux quartic at the matching scale but does not fix its sign. It
favors the VBS terminus only if that quartic is attractive. At tree level
the exchange of the material makes it attractive, $g=J'$, and the flux
channel repulsive (Sec.~\ref{sec:whichvbs}), and the seed flows to the
VBS ray. The $O(1/N)$ running between the lattice scale and $\lambda$
keeps the signs of the VBS and N\'eel$_z$ channels and turns the flux
channel attractive, and the running seed ends on the VBS ray for every
value of the constraint-field dimension $\Delta_\sigma$, which is not
controlled at $N=12$ (Sec.~\ref{sec:whichvbs}). In the static limit the screened
gauge interaction is attractive with equal Hartree--Fock weight, $+1.08$
in the units of Appendix~\ref{app:treeseeds}, in the flux, VBS and
N\'eel$_z$ channels, and repulsive in the gauge-adjoint masses, and it
reaches the VBS ray at $\ell^*\simeq0.8$. Its full one-loop propagator
contains the frequency dependence of the color polarization, the
transverse field, screened only by the orbital response of the band
touching and of the gapped Dirac background, and the polarization of the
critical Higgs fields (Appendix~\ref{app:qbtgauge} and
Fig.~\ref{fig:fourfermion}(e),(f)). With it the gauge
interaction projects onto a gauge-singlet seed whose overlap with the
static one exceeds $0.9995$ below $\lambda/4$ and whose norm is $1.3$--$1.4$
times the static one at $10^{-4}\lambda$ and $0.8$ times at $\lambda/4$.
Dynamical screening tilts it toward the chirality channel, and the
transverse field tilts it back. The seed lies in the VBS basin at every
scale of the window, in every scale-covariant gauge and momentum routing, and
reaches the VBS ray at $\ell^*\simeq0.6$--$1.1$ at fixed scale. The margin
stays small: across the window and the schemes a chirality-attractive
admixture of about $2$--$4\%$ of its norm sends it to the chirality ray,
against $2.6$--$3.1\%$ for the static seed. Higgs exchange through the
SO(5)-breaking Yukawa coupling reaches the band touching only through the
$\pm4\lambda$ states and is chirality-attractive. It moves the seed across
the separatrix only if it is at least as strong as the gauge exchange at
the scale $\lambda$, which requires the Yukawa coupling to have ended the
pseudocritical window, and the DM-induced Yukawa channels change the
margin by about $10^{-4}$. On the way the static seed favors the VBS pair
over the gauge-adjoint masses (rates $0.36$ and $0.28$ at $|g|=200$), and
the asymptotic ray restores their degeneracy. Added to the renormalized
lattice seeds with any relative weight between $0.03$ and $10^3$, it
leaves both the VBS terminus and the sign of the plaquette selection of
Sec.~\ref{sec:whichvbs} unchanged. Being valley-blind, it selects no
plaquette pattern itself. The VBS verdict for the material therefore rests
mainly on the VBS attraction of the exchange seed, which dominates after
the running. The gauge interaction alone reaches the VBS ray only by a
small margin. A
repulsive VBS seed funnels into the chirality ray. The gapped chiral
$\mathbb Z_2$ spin liquid, with spontaneous scalar chirality, weak $z$
ferromagnetism by the chirality--magnetization locking above, and a
quantized thermal Hall response (Sec.~\ref{sec:chiral}), survives as a
minority-basin alternative that requires chirality-attractive residual
interactions that overcome the VBS attraction of the running exchange seed and the small VBS preference
of the gauge interaction. Ordering through the N\'eel$_z$ pockets or
through high-energy modes is a separate question not addressed by the
mass projections. For a single spinless QBT, weak repulsion drives the time-reversal-breaking
quantum anomalous Hall state~\cite{Sun2009}; with the many channels of the
present band touching, the corresponding chiral outcome is instead the
minority one for uniformly sampled couplings.

\emph{Protection, and the Higgsed-phase window.} At the $\pi$-flux/critical point the
QBT is protected beyond the chiral argument: the unique locked-invariant
gauge-singlet bilinear is $\mathcal O_{\rm DM}$ itself, whose projection on the
QBT subspace vanishes because the QBT is its null space. No
symmetry-allowed bilinear exists there at all, so the vanishing
one-loop tadpole is symmetry-required. In the Higgsed Z3000 phase the
locked group admits six gauge-adjoint $k^0$ bilinears on the QBT modes,
two triplets such as the $\mu^y$-flavored analogue of $\mathcal O_{\rm DM}$.
The condensates dress them into allowed couplings by the same mechanism
as the zeroth-order $\Phi_3$ Yukawa of Sec.~\ref{sec:yukawa}. These
split the touching without gapping it, and the nodes remain gapless
throughout. On the lattice the splitting depends on the lattice DM
amplitude $d$ and on the Higgs-deviation scale $s_H$, which multiplies the deviations of the Z3000
ansatz from the $\pi$-flux state, $\delta\theta=0.30\,s_H$ and
$(\Delta_d,\Delta_{d'},\Delta_{g,1},\Delta_{g,2})=s_H(0.4,0.25,0.15,0.10)$
in units of $t$. The splitting $\delta'\equiv\min_{\bm k}E_{(3)}(\bm k)$,
with $E_{(3)}$ the third-lowest excitation energy, vanishes identically at
$s_H=0$ and obeys a second-order law,
\begin{equation}
\delta'=\frac{(c_\Phi\,s_H\,t)^2}{4\lambda},\qquad c_\Phi\simeq0.67,
\label{eq:splitter}
\end{equation}
with $\delta'\cdot4\lambda$ independent of $d$ to a few percent over a
factor $4$ in $d$. It arises from virtual excursions through the $\pm4\lambda$ bands,
quadratic in the Higgs deviation. The quartic flow thus runs over
the energy window $[\delta',\lambda]$, since the screened gauge interaction
is present on both sides of $m_A$ for $m_A\lesssim\lambda$. For the screened gauge interaction
alone $\ell^*\simeq0.8$ in the static limit and $0.6$--$1.1$ with its full
one-loop propagator at fixed scale (Appendix~\ref{app:qbtgauge}). Followed
scale by scale, with the Higgs polarization weakening the interaction
toward the upper end of the window, $\ell^*\simeq0.9$--$1.6$ for $E_2$
between $\lambda/4$ and $4\lambda$. Random seeds of strength $0.5$ in bubble units
diverge at $\ell^*=4.6$--$52$ (Appendix~\ref{app:qbtflow}), with median
$10.8$. At $z=2$ an energy window $[E_1,E_2]$ corresponds to the flow time
$\tfrac12\ln(E_2/E_1)$. The quadratic regime ends at an energy $E_2$ of
order $\lambda$, so the basin verdict requires
$\ln[2\lambda/(c_\Phi s_Ht)]\gtrsim\ell^*$, i.e.\ $c_\Phi s_Ht\lesssim
0.5\lambda$, and the instability sets in at $T^{**}\simeq E_2e^{-2\ell^*}$.
This is $(0.03$--$0.1)\lambda$ for $E_2$ between $\lambda/4$ and
$\lambda$, and up to $0.2\lambda$ if the flow is extended to
$E_2=4\lambda$. Gauge fluctuations also flatten the band touching, raising
the dynamical exponent to $z\simeq2.2$ at low energy, which lowers these
values by $5$--$15\%$. They are order-of-magnitude estimates, since the couplings reach order
one already at $\ell^*\lesssim1$, outside the range where a one-loop flow
is controlled. The window opens logarithmically on
approach to the critical point and closes deep in the Higgsed phase. At the
reference Z3000 parameters with all four DM couplings the node
disperses linearly down to the smallest resolved momenta. Where the window closes, the low-energy theory is a spin liquid with
displaced, chirally protected linear Dirac nodes. Short-range quartic
interactions are irrelevant there, and the stronger VBS tendency of
Eq.~\eqref{eq:lifting}, a property of the critical theory, does not by
itself drive the system into an ordered phase.
Possible strong-coupling instabilities of the deep-Z3000 Dirac phase
require a separate analysis that we do not perform. The window
criterion therefore bounds the regime of every $T^{**}$ statement in
this paper.

\subsection{The chiral endpoint: lattice ansatz, Chern number, and thermal Hall response}
\label{sec:chiral}
The chiral outcome of the flow, in which the flux mass grows, can be
realized as an explicit mean-field state on the lattice, whose Chern
number and thermal Hall response we compute. On the lattice the flux mass is found by projecting every
$\mathcal T$-odd spin-singlet amplitude in the Z3000 gauge onto the 36
continuum bilinears (Appendix~\ref{app:chiral}). The only gauge-singlet
ones are imaginary singlet hoppings $u^0$ on the diagonal bonds. Each of
them projects onto the flux mass ($\pm\tfrac14$) together with the
$\mathcal T$-odd valley density $\mathrm{Tr}[\bar{\mathcal
X}\gamma^0\mu^y\mathcal X]$ ($\pm\tfrac14$), and the latter cancels
between the two diagonals of every square when they carry opposite
signs. The pure flux mass is therefore the chiral pattern $\mathcal P_\chi$ of
Eq.~\eqref{eq:Fpattern}, with $\mp u^0$ on the dimer and orthogonal
diagonals of the crossed squares and $\pm u^0$ on the two diagonals of
the empty squares.It projects onto the flux mass alone, with no admixture of the valley
density. Under the Z3000 PSG it is translation-even, odd under $G_x$,
$\sigma_{xy}$ and $\mathcal T$, and even under $\mathcal TG_x$ and
$\mathcal T\sigma_{xy}$. Among the four $\mathcal T$-broken descendants
of Z3000 classified in Ref.~\cite{Maity2025} this signature selects
Z3000$_{11}$ uniquely, and the representative mean-field spectrum of
that descendant is fully gapped. Z3000$_{00}$ (Dirac) and
Z3000$_{01,10}$ carry different signatures and are unrelated to the
flux instability. The two runaway sectors thus admit explicit
descendants of the same parent. The empty-plaquette VBS route leads to
P5A, the plaquette descendant of Z3000 identified in Appendix~G of
Ref.~\cite{Maity2025}, and the chirality route leads to Z3000$_{11}$.

Adding $m\mathcal P_\chi$ to the full Z3000$(s_H)$+DM Bloch Hamiltonian opens a full gap
close to $4|m|$, for example $0.39$--$0.40$ at $|m|=0.1$. The BdG Chern number of
the occupied Majorana bands, evaluated over the full zone with the
lattice formula of Ref.~\cite{FHS2005} with determinant link variables [its Eq.~(16)], is
\begin{equation}
C_{\rm BdG}=-4\,\mathrm{sgn}(m)\qquad(s_H=0\text{--}0.2,\ d=0\text{ and }0.6).
\label{eq:chern}
\end{equation}
On a
cylinder the gap
hosts four chiral Majorana branches per edge with opposite chirality on
the two edges ($\pm4$ signed zero-energy crossings). For $s_H>0$ the
Higgs condensate leaves a gapped $\mathbb Z_2$ gauge field, and the chiral
central charge of the state is $c=C_{\rm BdG}/2=\mp2$~\cite{Kitaev2006,ReadGreen2000}, two valleys times two spins of the same
chirality. At $s_H=0$ the same bands carry the SU(2) invariant gauge
group, and projection then gives an SU(2)$_1$ chiral spin liquid with
$|c|=1$ instead. The quantized value below therefore requires the Higgsed
phase, and temperatures below both the Chern gap and the $\mathbb Z_2$
flux gap, which is small near the critical point. Then
\begin{equation}
\frac{\kappa_{xy}}{T}=c\,\frac{\pi^2k_B^2}{3h}=\mp2\,\frac{\pi^2k_B^2}{3h}
\label{eq:kxy}
\end{equation}
is the $T\to0$ thermal Hall coefficient of the minority phase,
$\mp1.9\times10^{-12}$\,W\,K$^{-2}$ per layer. Finite-$T$ bulk
corrections are exponentially small in the gap. The quantized value requires the $\mathbb Z_2$ gauge field to be
deconfined and gapped. The sign of $\kappa_{xy}$ follows that of the
scalar chirality and of the accompanying $z$ magnetization; at mean
field $M_z\kappa_{xy}>0$ (Appendix~\ref{app:chiral}).
The PSG class does not fix the Chern number. The Z3000$_{11}$ family of
the locked PSG is 42-dimensional and contains other topological sectors
separated by gap closings. Along 32 random $\mathcal T$-odd rays the
minimal state $m\mathcal P_\chi$ keeps $C_{\rm BdG}=-4\,\mathrm{sgn}(m)$ at half its
amplitude on all rays and at equal amplitude on 31, and on every ray
whose phase boundary is determined its gapped component extends to $\mathcal T$-odd
admixtures comparable to or larger than the flux mass
(Appendix~\ref{app:chiral}). The weak uniform $z$ moment is DM-specific:
it follows from the projected identity $P(J^0\sigma^z)=-P({\rm flux})$ of
the spin--orbit-locked QBT (Sec.~\ref{sec:qbt}), vanishes in the
spin-rotation-invariant limit, and is not a property of the
Z3000$_{11}$ PSG class as such.

The spin-singlet Z3000 and Z3000$_{11}$ ans\"atze are seeds of the
mean-field theory of the spin--orbit-locked problem and do not exhaust it. Besides the
seven singlet amplitudes, the locked Z3000 PSG admits fourteen
spin-triplet amplitudes in the buckled structure (two with $D'_\perp$
alone), and the Wick-induced DM operators lie inside this space and are
purely triplet. Restricted self-consistency on a fixed Z3000
background changes the in-plane DM operator by $0.06\%$ in norm and leaves
the nodes gapless (Appendix~\ref{app:chiral}).

\subsection{Which VBS: plaquette selection}
\label{sec:whichvbs}
The VBS$_x$/VBS$_y$ degeneracy has two layers. In the quadratic sector
it is exact. The valley charge $Q=\int\mathrm{Tr}[\bar{\mathcal
X}\gamma^0\mu^y\mathcal X]$ generates a $U(1)_{\rm valley}$ under which
$(V_x,V_y)$ rotates with charge 2 while the flux mass, $\mathcal O_{\rm DM}$ and
the kinetic term are singlets, as follows from exponentiating the
bilinears of Sec.~\ref{sec:inventory}. No zeroth-gradient bilinear, and no mean-field effect
of the linearized continuum theory, can therefore split the doublet. The
splitting enters through the irrelevant gradient terms of the lattice
theory, the $O(k^2)$ band curvature and the $O(k)$ form factors of the DM
and VBS vertices, which break $U(1)_{\rm valley}$ and are contained in the
lattice determinant computed below. The quartic sector is not fully valley-symmetric, since
$9$ of the $14$ invariant couplings preserve $U(1)_{\rm valley}$ and
$5$ break it (the charges appear as singular values $4,2,2,2,2$ of the
generator's action), the $V_xV_y$-type quartic among them. A $V_xV_y$ seed feeds the combinations
$V_x\pm V_y$ with opposite rates through the one-loop source mixing. On
the majority ray the tie is restored dynamically, so the degeneracy along the flow from random seeds is an
emergent property of that ray. Selection within
the doublet therefore has two contributions: flow amplification of
valley-breaking quartic seeds, which we estimate at tree level below, and
the lattice determinant. We compute the determinant contribution and the bare valley-breaking
couplings.

Which anisotropies are allowed is fixed by the locked group. On the
$(V_x,V_y)=|V|(\cos\theta,\sin\theta)$ circle the tables of
Sec.~\ref{sec:inventory} give $\hat S_4:(V_x,V_y)\to(V_y,V_x)$, a swap,
i.e.\ a mirror along the $(1,1)$ diagonal,
$\hat\sigma_{xy},\hat\sigma_{x\bar y}:(V_x,V_y)\to(-V_y,-V_x)$, and
$\hat G_x:(V_x,V_y)\to-(V_x,V_y)$. The first three are mirrors along the
diagonals, and $\hat G_x$ is the rotation by $\pi$ that they generate. Hence
\begin{equation}
F(\theta)=F_0+b_2\sin2\theta+a_4\cos4\theta+\cdots,
\label{eq:aniso}
\end{equation}
with $\cos2\theta$ and $\sin4\theta$ forbidden. Higher allowed
harmonics such as $\sin6\theta$ appear at higher order in $m$ and are
negligible at the parameters below. The four
high-symmetry directions are lattice states, identified by
projecting explicit bond patterns onto the continuum masses. The axes are the two columnar dimerizations,
$(-1)^{i_x}$ on $x$ bonds giving $-V_x$ and $(-1)^{i_y}$ on $y$ bonds
giving $+V_y$. The diagonal $\theta=\pi/4$ is the empty-plaquette
pattern $P_{\rm e}$, a bond strengthening alternating between the two
sublattices of void squares (the singlet-plaquette state on empty
squares), which as a lattice operator equals the columnar combination
$-{\rm col}_x+{\rm col}_y$. The diagonal $\theta=3\pi/4$ is the
full-plaquette pattern $P_{\rm f}$, the alternation of the two
dimer-carrying crossed-square sublattices, equal to $+{\rm col}_x+{\rm
col}_y$, and $\theta=-\pi/4$ is its partner $-P_{\rm f}$. Thus, $b_2$ splits empty from full plaquette, which on the
Shastry--Sutherland lattice are inequivalent phases, while $a_4$
controls diagonals versus axes. With both harmonics present the extrema
of Eq.~\eqref{eq:aniso} satisfy $\cos2\theta\,(b_2-4a_4\sin2\theta)=0$.
The diagonals are always stationary, while the columnar axes are
stationary only at $b_2=0$. For $a_4<0$ and $|b_2|<4|a_4|$ the minima lie
off the diagonals at $\sin2\theta=b_2/(4a_4)$, and they approach the
columnar axes as $|b_2|/|a_4|\to0$.

\begin{figure}[t]
\includegraphics[width=\columnwidth]{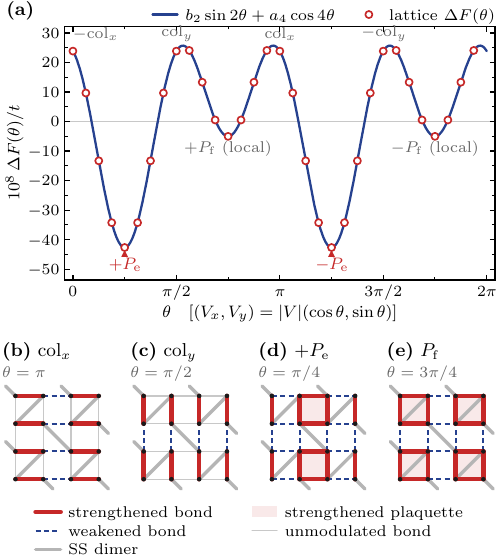}
\caption{Selection within the VBS doublet. (a) Lattice free energy
$\Delta F(\theta)$ on the $(V_x,V_y)=|V|(\cos\theta,\sin\theta)$ circle
at $d=0.6$, $m=0.05$ (circles), lying on the two allowed harmonics
$b_2\sin2\theta+a_4\cos4\theta$ (line; $b_2=-18.82\times10^{-8}\,t$,
$a_4=+23.84\times10^{-8}\,t$); the forbidden harmonics vanish. The global
minima of this determinant contribution (red triangles) are the two
empty-plaquette states $\pm P_{\rm e}$ ($\theta=\pi/4,5\pi/4$), one on each
void sublattice. The full-plaquette directions $\pm P_{\rm f}$
($\theta=3\pi/4,7\pi/4$) are only shallow local minima at $d=0.6$. Gray
labels mark the columnar directions. (b)--(e)
The four lattice states col$_x$, col$_y$, $+P_{\rm e}$, $P_{\rm f}$ with
their $\theta$ (thick red: strengthened bonds; dashed indigo: weakened;
gray: Shastry--Sutherland dimers; thin gray: unmodulated bonds; shading:
strengthened plaquettes).}
\label{fig:whichvbs}
\end{figure}

Both signs follow from the lattice determinant
(Fig.~\ref{fig:whichvbs}) of the Bloch-reduced $\pi$-flux
ansatz plus the Wick-induced DM operator at lattice amplitude $d$,
perturbed by $m[\cos\theta\,(-{\rm col}_x)+\sin\theta\,{\rm col}_y]$,
with $F(\theta)$ evaluated on momentum grids up to $200^2$. Only the harmonics allowed by the locked group, $\sin2\theta$ and
$\cos4\theta$, appear in $F(\theta)$. At $d=0$ the $\pi$-flux determinant is square-lattice
symmetric, so $\sin2\theta$ vanishes while the standard square-lattice
$\cos4\theta$ survives. The results are identical for both in-plane DM
textures and even under $d\to-d$. We find
\begin{equation}
a_4>0\ (\propto m^4),\qquad b_2<0\ (\propto d^4),
\label{eq:selection}
\end{equation}
for $m\gtrsim3\times10^{-3}$ (smaller $m$ is treated below), so $a_4>0$ puts minima on the diagonals, and $b_2<0$ selects
$\sin2\theta=+1$, the empty-square plaquette direction, which is the
plaquette channel of the pure Shastry--Sutherland model~\cite{CorbozMila2013}. The scaling $b_2\propto d^4$ holds for $m\gg\lambda$
(Appendix~\ref{app:plaquette}). For $m\ll\lambda$ the gap
removes the finite density of states of the band touching between the
gap scale $2mt$ and the reconstruction scale, and
\begin{equation}
b_2=-A\,m^2\ln(m_0/m),\quad A=8.28\times10^{-5}\,t,\ m_0=0.103,
\label{eq:b2asym}
\end{equation}
at $d=0.6$ (Appendix~\ref{app:plaquette}). To second order in $k$, the
eight zero modes at $\Gamma$ carry
\begin{align}
H(\bm k)={}&\frac{v^2}{4\lambda}\Bigl[\Bigl(1+\frac{2\lambda^2}{t^2}\Bigr)(k_x^2-k_y^2)\mathcal A_1
+2k_xk_y\mathcal A_2\Bigr]\nonumber\\
&+\frac{v^2}{4t}\,2k_xk_y(V_y-V_x),
\label{eq:qbtdxy}
\end{align}
where $V_{x,y}$ are the projected VBS vertices, which anticommute with
$\mathcal A_{1,2}$ and with each other. The last term is the remainder of
the $O(k)$ form factor of the DM vertex after the $\pi$-flux band
curvature cancels its $s$-wave and $d_{x^2-y^2}$ parts. It is a $d_{xy}$
form factor along the full-plaquette direction and breaks
$U(1)_{\rm valley}$ at relative order $\lambda/t$. With the probe the
spectrum is $\pm|k^2\bm n(\phi)+2mt\,\bm u(\theta)|$, each fourfold,
and the component of the mass along $\bm n$ shifts the bands instead of
gapping them. The $\sin2\theta$ part of the $m^2\ln(1/m)$ term is, per
magnetic unit cell,
\begin{equation}
A=\frac{4\lambda}{\pi^2}\,\frac{(1+\epsilon)E(\kappa)-K(\kappa)}{(1+\epsilon)^2}
=\frac{2\lambda^3}{\pi t^2}\Bigl[1-\frac{15}{4}\frac{\lambda^2}{t^2}
+O\Bigl(\frac{\lambda^4}{t^4}\Bigr)\Bigr],
\label{eq:Aclosed}
\end{equation}
with $\epsilon=2\lambda^2/t^2$, $\kappa^2=\epsilon/(1+\epsilon)$ and $K$,
$E$ the complete elliptic integrals, so that $b_2\propto|d|^3$ in this
regime. The constant $m_0$ is a zone integral of the second-order lattice
kernel. It lies close to $2\lambda$, so the
logarithm runs from the probe gap to the reconstruction scale. The
lattice $b_2$ approaches Eq.~\eqref{eq:b2asym} to $10^{-4}$ for
$m\le1.6\times10^{-3}$. The empty-plaquette mass $V_x+V_y$ anticommutes
with every term of $H(\bm k)$, whereas the full-plaquette mass $V_y-V_x$
commutes with the $d_{xy}$ term and gains less condensation energy, hence
$b_2<0$. A model linearized in $k$ keeps the DM form factor but not the
band curvature that cancels its $s$-wave and $d_{x^2-y^2}$ parts. Its
logarithmic coefficient is $2A$ for $\lambda\ll t$ ($1.99A$ at $d=0.6$),
and its larger subleading terms raise the ratio to $3$ over the window
$m\le0.0125$.

For $m\ll\lambda$ both harmonics have a controlled form. The touching also
generates an infrared $\cos4\theta$ term, fourth order in the $d_{xy}$
form factor, so that $a_4=-(3\lambda^5/16\pi t^4)\,m^2+a_4^{(4)}m^4$ to
leading order in $\lambda/t$, with $a_4^{(4)}=0.039\,t$ at $d=0.6$, and $a_4$
changes sign near $m\simeq7\times10^{-4}$; $a_4\propto m^4$ holds for
$m\gtrsim3\times10^{-3}$. The ratio $|b_2|/4|a_4|$ exceeds unity for
$m\lesssim0.03$ and approaches $(8/3)(t/\lambda)^2\ln(m_0/m)\simeq10^3\ln(m_0/m)$
for $m\ll7\times10^{-4}$. It therefore grows without bound as $m\to0$,
logarithmically, provided the Higgs splitting $\delta'$ and the
temperature lie below the gap $2mt$. Within the determinant contribution
the minima at $T^{**}$ are
therefore the empty-plaquette pair $\theta=\pi/4,\,5\pi/4$ (the two void
sublattices), a two-state Ising order parameter. The total selection also receives the flow-amplified
valley-breaking quartic seeds. Their tree-level estimate, given below,
is larger than the determinant contribution, but after the running to
$\lambda$ it need not be. If the
thermal transition is continuous, 2D Ising universality is expected, but the transition temperature and its order are not
computed here. Since the Wick-induced DM operator is held fixed while
the singlet links are modulated, the result is a determinant
contribution, and the self-consistent total energy is not computed.

In the full Z3000$(s_H)$+DM Bloch background,
with the Higgs-deviation scale $s_H$ of Eq.~\eqref{eq:splitter}, the
global minimum is located directly, because the sign of $b_2$ alone is not a
phase criterion once $a_4<0$. The forbidden harmonics again vanish, $b_2$ stays between $-1.88$ and
$-2.05\times10^{-7}\,t$ for $s_H\le0.1$, and the two empty-plaquette states remain
the global minima for $s_H\le0.055$ (at $m=0.05$, $d=0.6$). This range
covers most of the open-QBT window at $d=0.6$, which extends to
$s_H\simeq0.03$--$0.08$ for the gauge-driven flow ($\ell^*\simeq0.6$--$1.6$,
depending on the screening and on the upper energy of the quadratic
regime), and to smaller $s_H$ at the material $\lambda$. At larger $s_H$, $a_4$
changes sign near $s_H\simeq0.055$, and once $4|a_4|$ exceeds $|b_2|$, near
$s_H\simeq0.057$, the minima leave the diagonal. They lie $12^\circ$ from
the nearest columnar axis at $s_H=0.06$, within $3^\circ$ of it at
$s_H=0.07$, and within $0.5^\circ$ for $s_H\ge0.1$. Within the determinant, the empty-plaquette selection is therefore a
near-critical statement, and the near-critical regime is where the
phase diagram proposed in Ref.~\cite{Cui2024two} places the
empty-plaquette phase, next to the antiferromagnet (see below).
Deeper in the Higgsed phase, where no weak-coupling instability exists,
the same determinant favors columnar-like directions.

The determinant result is compared here with the selection already
present in the Heisenberg problem. Reference~\cite{LYSV2019} describes the
N\'eel--VBS transition on this lattice in monopole language, with the
VBS doublet $(v_x,v_y)$ combined into a complex monopole field $M$.
There the double-monopole term $\lambda_2\,{\rm Re}\,M^2$, the most
relevant monopole perturbation allowed on this lattice, breaks the
U(1) phase rotation of $M$ down to $\mathbb Z_2$. For $\lambda_2>0$
it selects the square (empty) plaquette state, with the
twofold-degenerate order parameter ${\rm Im}\,M$. At the transition
${\rm Im}\,M$ is argued to combine with the N\'eel vector into an
O(4) vector, with an emergent O(4) symmetry in place of the
SO(5) of the theory without that term. Tensor-network
calculations for the pure Shastry--Sutherland model find the same
empty-plaquette phase~\cite{CorbozMila2013}. The harmonic
$b_2\sin2\theta$ of Eq.~\eqref{eq:aniso} plays the role of that
selector in the fermionic language, reducing the $(V_x,V_y)$ circle to
one diagonal pair, $\pm P_{\rm e}$ or $\pm P_{\rm f}$. The $(V_x,V_y)$
doublet is the counterpart of $(v_x,v_y)$, and hence of $M$, in
Ref.~\cite{LYSV2019}. The empty plaquette lies at $\theta=\pi/4$ here
and at $v_x=-v_y$ there, where ${\rm Re}\,M^2=2v_xv_y<0$, so $b_2<0$
corresponds to $\lambda_2>0$. In the material the Heisenberg
couplings, with $J'/J\simeq0.6$, are the dominant selector, whereas the
DM couplings enter the determinant only through $d$, with
$|D'_\Sigma|\sim2\times10^{-3}J$. The empty-plaquette selection therefore comes from the Heisenberg
couplings, and the determinant adds only a small correction of the
same sign. Its own correction has the sign
of the Heisenberg selection. Because $b_2$ is even in $d$, this correction
is the same in the two inversion twins of the crystal~\cite{Sparta2001},
which reverse the sign of all in-plane DM components. The absence of a $d^2$ term in $b_2$, which scales
as $d^4$ down to $d=0.01$ for $m\gg\lambda$ and as $|d|^3$ in the logarithmic
regime (Appendix~\ref{app:plaquette}), together with the vanishing of the
dimer-bond amplitude $\langle
f^\dagger_if^{\vphantom\dagger}_j\rangle_{\rm dimer}=0$ on the
$\pi$-flux state (the bipartite chiral selection rule of Sec.~\ref{sec:lattice}), makes the DM
channel the only Shastry--Sutherland-aware term in the $\pi$-flux
determinant at this level. In the Z3000$(s_H)$ background the dimer and
orthogonal-dimer amplitudes add a DM-independent contribution,
$b_2\simeq-1.0\times10^{-4}s_H^4\,t$ at $m=0.05$ and small $s_H$. It has
the same sign and dominates the DM channel for $s_H\gtrsim0.4\,d$. Exchange fluctuations beyond the determinant renormalize $b_2$, and
a sign change would require them to overcome the computed channel.

\emph{Tree-level quartic seeds.} The bare quartic couplings follow at tree level from projecting the
exchange Hamiltonian of Eq.~\eqref{eq:H}, written in terms of the
partons, onto the eight QBT modes at zeroth order in gradients
(Appendix~\ref{app:treeseeds}). We denote by $g(\theta)$ the resulting
coupling in the VBS channel along $\theta$, with $g>0$ attractive. The
$J$--$J'$ exchange gives a valley-symmetric seed for every $J'/J$. Each
nearest-neighbor bond alone breaks $U(1)_{\rm valley}$, but the four
nearest-neighbor bond classes cancel in the symmetric sum. The dimer
bonds preserve it, and so do diagonal couplings of any range, which at
$\Gamma$ connect the same sublattice pairs. The exchange gives $g=J'$ at
every $\theta$, an attractive N\'eel$_z$ channel ($+1.51J$ at
$J'/J=0.63$) and a repulsive flux channel ($-1.01J$). The valley-breaking
seed comes from the DM exchange. The in-plane inter-dimer components
$D'_{\parallel,s}$ and $D'_{\parallel,ns}$ break $U(1)_{\rm valley}$ but
contribute nothing on the VBS circle. This is required by the inversion
twins, which reverse these components and map each plaquette pattern to
itself. The intra-dimer $D$ drops out of the QBT. The out-of-plane
component gives
\begin{equation}
g(\theta)=J'+2D'_\perp\sin2\theta ,
\label{eq:gtree}
\end{equation}
so that $g(P_{\rm e})-g(P_{\rm f})=4D'_\perp$, with $D'_\perp$ normalized
as in Table~\ref{tab:dm}. The term exists only on the
spin--orbit-locked QBT modes. On the sixteen Dirac modes the same operator
has no component on the VBS circle, in line with the inertness of
$D'_\perp$ above $T^*$ (Sec.~\ref{sec:spin2}).

The fractional anisotropy of Eq.~\eqref{eq:gtree} is
$2|D'_\perp|/J'\simeq0.04$--$0.08$ for $|D'_\perp|=0.10$--$0.18$\,meV.
The corresponding ratio for the determinant in the band-touching channel
is $A/C_{\rm iso}=(\lambda^2/2t^2)[1+O(\lambda^2/t^2)]=1.29\times10^{-3}$
at $d=0.6$, where $C_{\rm iso}\simeq4\lambda/\pi$ is the coefficient of the
isotropic $m^2\ln(1/m)$ term, and it falls as $d^2$ toward the material
value. It is a ratio of the singular parts; the full ratio
$b_2/\Delta F_{\rm iso}$, which includes the regular parts, is
$(1$--$2)\times10^{-4}$ at $d=0.6$. In the one-loop flow the tree-level seed lies in the VBS basin
(Sec.~\ref{sec:qbtir}). Without the running described below, the vertex of
the plaquette pattern favored by $D'_\perp$ outgrows the other by
$\Delta\ln\Gamma\simeq2.0$ for $|D'_\perp|=0.10$\,meV and $\simeq3.3$ for
$0.18$\,meV, independently of the seed scale (a directional measure, since
the tree seeds themselves diverge only after $\ell^*\simeq40$--$160$). At
tree level the DM-induced selection within the doublet is therefore set by
the sign of $D'_\perp$, with the determinant a subleading correction; the
running described below can reverse this ordering.
That sign is $D'_\perp<0$ by the NMR determination of Ref.~\cite{Kodama2005}
and by Table~II of Ref.~\cite{Mazurenko2008}
(Sec.~\ref{sec:model}); this sign favors the full-plaquette pair
$\pm P_{\rm f}$. Reference~\cite{Cheng2007} read the opposite sign, which
would favor $\pm P_{\rm e}$, from the far-infrared analysis of
Ref.~\cite{CepasZiman2004}, a discrepancy that Ref.~\cite{Mazurenko2008}
attributes to conventions. The seed exists
only on the band-touching modes, i.e.\ for $D'_\Sigma\neq0$ and below
$T^*$. Perturbatively, in the gapped plaquette phases, a global $\pi$ spin
rotation about an in-plane axis reverses $D'_\perp$ and leaves the exchange
invariant, so an energy difference odd in $D'_\perp$ is at least of order
$D'_\perp D_\parallel^2$ there. The selection that the pure
Shastry--Sutherland model makes beyond tree level is not computed.

\emph{Running to the band-touching scale.} The seeds are defined at the
lattice scale $\Lambda\sim J'$--$J$. Between $\Lambda$ and $\lambda$, over
$L=\ln(\Lambda/\lambda)\simeq7$--$9$, they run in the Dirac regime with
the scaling dimensions of four-fermion operators at the SO(5) fixed point.
These dimensions are obtained at $O(1/N)$ for all 336 gauge-invariant local
four-fermion operators, from the fermion self-energy, one-gauge-line
exchange between every pair of legs, and the two-gauge-line annihilation
of flux-singlet pairs and, through the induced three-gauge vertex, of
gauge-current pairs (Appendix~\ref{app:treeseeds}). The same rules give the bilinear
dimensions of Appendix~\ref{app:integrals} and, in their abelian form, agree with the
large-$N$ QED$_3$ four-fermion dimensions of Ref.~\cite{ChesterPufu2016},
including the flavor-singlet pair whose mixing with $F^2$, the square of the gauge field strength, enters through
the current square. The $D'_\perp$ seed acquires no anomalous dimension at this order: its plaquette selector lies entirely in
the SO(5) representation $\mathbf{35}$ as a Lorentz vector, an eigenspace with
$\Delta=4$ at $O(1/N)$, whose anomalous graphs cancel
(Appendix~\ref{app:treeseeds}) and which cannot mix with operators
containing gauge or Higgs fields. Its remaining half, a vector $\mathbf5$ with Lorentz spin 2 at
$\Delta=3.91$, carries no QBT channel. The isotropic VBS attraction it
competes with does run. Its rank-2-tensor part has $\Delta=3.46$ and grows
by a factor of $40$--$100$ over $L$ relative to the seed. Its SO(5)-singlet scalar part grows
fastest. Unlike in QED$_3$, in SU(2) the current square does not represent $F^aF^a$. For
the $\pi$-flux SU(2) state, which the fermions alone screen, the
four-fermion block yields the singlet instability found in
Ref.~\cite{Xu2008}, and the dimension of $F^aF^a$ removes it at $O(1/N)$
(Appendix~\ref{app:treeseeds}). At the multicritical point the $O(1/N)$
mixing problem of the channel is solved in Appendix~\ref{app:treeseeds}.
At $N=\infty$ it contains five operators of
dimension four that are SO(5) and O(2) singlets, Lorentz scalars and even
under $\mathcal P$ and $\mathcal T$: $(\mathrm{Tr}\,\Xb\X)^2$; the square
$R=J^a_\perp J^a_\perp$ of the relative current $J_\perp=x_bJ_\psi-x_fJ_\Phi$,
which the gauge-field equation of motion makes equal to $J_\psi^2$,
$J_\Phi^2$ and $-J_\psi\cdot J_\Phi$, with $x_b$ and $x_f=1-x_b$ the Higgs
and fermion shares of the screening; $F^aF^a$, whose $O(1/N)$ dimension with
fermion and Higgs loops is $\Gamma_{FF}=5.14$ [Eq.~\eqref{eq:gammaFF}];
$\sigma^2$, with $\sigma$ the constraint field of the Higgs sector; and the
square of the covariant gradient of the gauge-spin-2 Higgs bilinear. The
lowest mode dominated by the four-fermion operators has $\Delta_1=2.99$
($2.91$ in the physical-$N_f$ counting), independent of the gauge
parameter, so the lowest four-fermion singlet is relevant, or close to
marginal, at this order, as is the
lowest singlet of the SO(5) deconfined point on the fuzzy
sphere~\cite{Zhou2024fuzzy}, although the two values are not
quantitatively comparable (Appendix~\ref{app:treeseeds}). The channel also
contains a mode that is $92\%$ $\sigma^2$, at $\Delta=0.46$. The constraint
field has $\Delta_\sigma=2-72u=0.38$ at $N=12$, with $u=8/(3\pi^2N)$
[Eq.~\eqref{eq:Dsigma} of Sec.~\ref{sec:spin2}], below the unitarity bound
$1/2$~\cite{Minwalla1998,Evans1967}, because the adjoint Higgs field
carries the Casimir $8$ against $3$ for the fermions, so the $1/N$
expansion of the Higgs sector is not controlled at $N=12$. We therefore
scan $\Delta_\sigma$. For $\Delta_\sigma$ between $0.38$ and $1.5$,
$\Delta_1$ lies between $2.99$ and $3.24$ ($2.91$ and $3.18$), and for
$\Delta_\sigma=1.75$--$2$ the level ordering changes and the lowest
four-fermion-dominated mode lies at $2.47$--$2.65$. The two further
Higgs operators of the channel, $\sigma\,\mathrm{Tr}\,\tilde M^2$ and
$(\mathrm{Tr}\,\tilde M^2)^2$, decouple at this order. Their mixing with the
five operators is at most of order $N^{-3/2}$ in the normalized basis,
since the graphs of lower order are scaleless, vanish by the tracelessness
of $\tilde M_{ab}$ or are removed by the equation of motion, and it moves
$\Delta_1$ only at $O(1/N^2)$ (Appendix~\ref{app:treeseeds}).

The seeds at $\lambda$ follow from the full matrix. At the $O(1/N)$ value of
$\Delta_\sigma$ the $\sigma^2$ mode dominates the running, and the
fractional anisotropy at $\lambda$ is reduced from $0.04$--$0.08$ to
$10^{-13}$--$10^{-10}$; over the scan of $\Delta_\sigma$ up to $2$ it stays
below $10^{-5}$, and the vertex asymmetry of the QBT flow below
$\Delta\ln\Gamma\simeq10^{-4}$. If a single dimension $\Delta_1\simeq3$
is instead assigned to the four-fermion singlet, the fractional
anisotropy reaches a few times $10^{-5}$ and the asymmetry
$2\times10^{-4}$ (Appendices~\ref{app:treeseeds} and~\ref{app:qbtgauge}). In every variant the flow ends on the VBS ray
and the $D'_\perp$ seed keeps its sign. The VBS and N\'eel$_z$ couplings
keep their tree-level signs, whereas the flux coupling, repulsive at tree
level, becomes attractive at $\lambda$ in the leading counting for every
$\Delta_\sigma$ of the scan, and in the physical-$N_f$ counting for
$\Delta_\sigma\le1.75$, because the lowest four-fermion-dominated mode is
dominated by the current square, which carries the same sign in all
three channels. Below
$\lambda$ the screened gauge interaction of
Sec.~\ref{sec:qbtir} adds a valley-blind VBS attraction, which dilutes the
anisotropy further and keeps its sign. At the material couplings
($d\simeq0.007$--$0.015$) the fractional anisotropy of the
determinant in the band-touching channel is
$1.8\times10^{-7}$--$8.1\times10^{-7}$, and the full ratio
$b_2/\Delta F_{\rm iso}$ is smaller; the Wick
term of $D'_\perp$ in the determinant, which also favors $P_{\rm f}$ for
$D'_\perp<0$, contributes about $10^{-10}$--$10^{-9}$. If a single
dimension $\Delta_1$ is assigned to the four-fermion singlet, with the
rank-2 part of the attraction at $\Delta=3.46$, the $D'_\perp$ seed leads
the determinant only if $\Delta_1$ exceeds a crossover value
$\Delta_1^*=1.90$--$2.21$ at $L=7$ and $2.27$--$2.52$ at $L=8.5$ with
$b_2/\Delta F_{\rm iso}$, the consistent choice at the gap-equation saddle,
or $2.25$--$2.47$ and $2.56$--$2.74$ with $A/C_{\rm iso}$ (for
$|D'_\perp|=0.10$\,meV, about $0.08$ lower for $0.18$\,meV); if the rank-2
tensor is instead relevant, as on the fuzzy sphere, that part of the
attraction grows by a larger factor and the seed never leads. With the full
matrix, in the leading counting, the seed leads only for
$\Delta_\sigma\gtrsim1.5$ ($1.75$) at $L=7$ with the first (second)
ratio, and at $L=8.5$ only for $\Delta_\sigma\gtrsim1.75$ with the first
and for no $\Delta_\sigma\le2$ with the second; at the $O(1/N)$ value of
$\Delta_\sigma$ the determinant leads. The direction of the DM correction
is therefore not determined at $O(1/N)$: where the seed leads it favors,
for $D'_\perp<0$, the full plaquette, and where the determinant leads it
favors the empty one. After the running the DM correction is in every
case a small bias, the flow ends on the VBS ray, and the pattern that
forms is decided by the pure-model term unless that term is smaller still.

\section{Magnetic field: direction-selective mass activation}
\label{sec:field}
Classifying the 36 operators under the stabilizer of $(\text{DM},\bm
B)$, the locked elements $O$ with $O\bm B=\bm B$ and the products $\mathcal TO$ with $O\bm B=-\bm B$, gives
Table~\ref{tab:field}. There $N^a$ denotes the N\'eel mass along the
crystal axis $a$, the continuum image of the staggered magnetization
$\sum_i\epsilon_iS_i^a$ with $\epsilon_i=+1$ on the $[110]$ dimers and
$-1$ on the $[1\bar10]$ dimers. Field and moment directions refer to crystal
axes, related to the spin labels of Tables~\ref{tab:unlocked}--\ref{tab:hightemp}
as stated in Appendix~\ref{app:protocol}.
\begin{table*}[t]
\caption{Field-activated mass channels of the spin--orbit-locked $\pi$-flux state. For each
field direction (tetragonal axes; the dimers lie along $[110]$ and
$[1\bar10]$) the table lists the mass channels activated by the
stabilizer of $(\text{DM},\bm B)$, the lowest orders in $B$ and in the DM
couplings at which symmetry permits them, and the Gaussian (mean-field)
coefficient of Sec.~\ref{sec:field}. Coefficients are quoted per site in
units of $t$, in the normalization of $\lambda$ (the Wick decoupling acts
on the central $25$ cells of a $64$-cell torus, and the listed coefficients
include the factor $64/25$), with $B$ in units of $t$ and the DM operator in
the state at $d=0.6$; $\lambda$ is the continuum DM coefficient of
Eq.~\eqref{eq:lambda} and $D$ here the in-plane coupling of the decoupled DM source (taken as $D'_{\parallel,s}$), not the intra-dimer $D$. Orders
refer to the spin model. An in-plane $g$ tensor that depends on the dimer
orientation adds a staggered field of first order in spin--orbit coupling
(see text). The Gaussian coefficients of the in-plane rows are evaluated
at $T=0.5t>4\lambda$ on an $8^2$ torus and are not computed below $T^*$;
the $[001]$ entry is a statement on the QBT, below $T^*$. $^a$The coupling
requires spin--orbit locking but carries no factor of $\lambda$ on the QBT
modes.}
\label{tab:field}
\begin{ruledtabular}
\begin{tabular}{lp{3.6cm}ccp{6.4cm}}
$\bm B$ direction & activated mass & order in $B$ & order in DM & mean-field coefficient\\
\hline
$[001]$ & flux (scalar chirality) & $B$ & $0$ on the QBT$^{\,a}$ & Zeeman term $=$ flux mass on the QBT, $P(J^0\sigma^z)=-P({\rm flux})$; lowest spinon excitation $g\mu_BB/2$ for $g\mu_BB/2\ll4\lambda$, saturating near $0.4\times4\lambda$\\
$[100]$ & $N^y$ ($\perp\bm B$) & $B$ & 2 (in-plane) & $\propto B\lambda D$\\
$[110]$ & $(N^x+N^y)/\sqrt2$ ($\parallel\bm B$) & $B$ & 2 (in-plane) & $\propto B\lambda D$\\
 & $V_x-V_y$ ($=-P_{\rm f}$) & $B_xB_y$ & 2 (in-plane) & $-1.6\times10^{-5}$ per unit $D$ at $B=0.2t$ (DM channel); $-1.1\times10^{-6}$ per unit $J'$ (NN exchange)\\
\end{tabular}
\end{ruledtabular}
\end{table*}
The stabilizer analysis determines which masses are allowed. The leading
power follows separately.
The field induces an effective staggered field, a term linear in the
N\'eel order parameter. It is $\mathcal T$-odd and appears at linear order in
$B$. Their order in the DM couplings follows from
$\hat\sigma_h=\one \otimes C_{2z}$, which acts on the layer as the
global spin rotation $C_{2z}$. It maps the model with in-plane DM
components, the
out-of-plane components $D'_\perp$ and the field $\bm B$ exactly onto the
model with the in-plane components reversed, the same $D'_\perp$ and the
field $C_{2z}\bm B$, and it reverses the in-plane components of $\bm N$.
An in-plane staggered field linear in an in-plane field is therefore even
in the in-plane DM couplings, and the term of first order is forbidden.
The locked group of the high-temperature structure, which contains
$\hat\sigma_h$ and so classifies the terms of even order, keeps the tensor $N^x\propto B_y$,
$N^y\propto B_x$ of Table~\ref{tab:field}. The leading order is thus the
second, $\lambda D$ at the Gaussian level computed below, where here and in Table~\ref{tab:field} $D$ denotes the in-plane inter-dimer source $D'_{\parallel,s}$, not the intra-dimer $D$. With
$D'_\perp$ alone the model keeps the spin rotations about $c$, and no
staggered field appears at any order. Within the spin model the staggered
field is therefore of second order in spin--orbit coupling. A
dependence of the in-plane $g$ tensor on the dimer orientation, which
the Cu site symmetry allows and the model omits, enters at first order.
With in-plane components $g_1$ along the dimer and $g_2$ across it, the
Zeeman term $\mu_B\sum_i\bm B\cdot g_i\bm S_i$, $g_i$ being the $g$
tensor of site $i$, contains
$\frac12(g_1-g_2)\mu_B\sum_i\epsilon_i(B_yS_i^x+B_xS_i^y)$. This is a
staggered field with the same tensor, and we do not estimate $g_1-g_2$.
An alternating $g$ tensor and DM interactions together generate the
staggered field behind the field-induced gap of the spin chain Cu
benzoate~\cite{OshikawaAffleck1997}, the one-dimensional precedent of
this mechanism.
The field conjugate to the valence-bond imbalance is $\mathcal T$-even and must be even in
$\bm B$, and for $[110]$ its leading form is $B_xB_y$ times a DM-even
coefficient. Since $V_x-V_y=-P_{\rm f}$ (Sec.~\ref{sec:whichvbs}), it is a
full-plaquette field, of opposite sign for $[110]$ and $[1\bar10]$. At order $B^2$ and at the
mean-field (Gaussian) level that coefficient receives contributions from three channels. The single-particle
channel is the Zeeman term on the 16 zero modes with the DM-split QBT
multiplet as intermediate space (second-order Schrieffer--Wolff). The
second channel is the Hartree--Fock (Wick) response of $H_{\rm DM}$, and
the third is the Hartree--Fock response of the Heisenberg quartics $J'$
(NN) and $J$ (dimer). The last two are evaluated on the field-polarized
thermal Gaussian state of the $\pi$-flux ansatz including the
Wick-induced DM operator ($d=0.6$, $T=0.5t$ on an $8^2$ torus, comparable
to its finite-size single-particle gap $0.46t$, magnetization linear in
$B$). This temperature exceeds $4\lambda\simeq0.2t$ at $d=0.6$, so the
magnitudes and signs of the Gaussian coefficients below are those of a
high-temperature state and are not computed below $T^*$. The orders in
$B$ and in the DM couplings are analytic statements and do not depend on
this choice. The
single-particle channel gives no VBS-imbalance mass at $O(B^2)$, only
the spin-current anisotropies $J^x\sigma^x-J^y\sigma^y$ ($[100]$) and
$J^x\sigma^y+J^y\sigma^x$ ($[110]$), in the spin labels of the tables. On a DM-free Gaussian state the
Wick channel gives none either. With the DM operator in the state the
mean-field responses follow the symmetry table. For
$\bm B\parallel[100]$ the DM channel produces the N\'eel mass $N^y$
linear in $B$ and no VBS imbalance. For $\bm B\parallel[110]$ it produces $(N^x+N^y)/\sqrt2$,
the N\'eel component along $\bm B$, linear in $B$, and the VBS imbalance
$V_x-V_y$ at order $B^2$ (coefficient $-1.6\times10^{-5}$ per unit $D$ at
$B=0.2t$). Both
N\'eel coefficients are $\propto B\lambda D$, of second order in the
in-plane DM couplings as the $\hat\sigma_h$ argument requires. They change sign
with $\lambda$ at fixed $D$ and are linear in $B$ and in $\lambda$. The NN exchange
channel adds an $O(B^2)$ imbalance of $-1.1\times10^{-6}$ per unit $J'$,
and the dimer exchange adds none. Since the imbalance vanishes when the
DM operator is removed from the state, it is of second order in the DM
couplings. The DM-channel coefficient scales linearly with $d$ and the
exchange-channel one quadratically, so
$V_x-V_y\propto B^2\lambda D$ and $B^2\lambda^2J'$. At the mean-field
level the imbalance is smaller than the N\'eel activation by a factor
$\propto B$ (ratio $0.07$ at $B=0.2t$ for the unit-normalized combinations
$(V_x-V_y)/\sqrt2$ and $(N^x+N^y)/\sqrt2$). Fluctuation
corrections beyond the Gaussian decoupling are not computed. The symmetry statement, even in $\bm B$ and allowed as $B_xB_y$, holds
independently of the decoupling. For
$\bm B\parallel[001]$ the single-particle channel confirms the locking
of Sec.~\ref{sec:qbt}. The Zeeman term projects on the QBT onto the flux
mass alone, linear in $B$ with no $O(B^2)$ remainder and with
the same coefficient for both signs of $\lambda$, and it gaps the QBT
modes at $\Gamma$ to $g\mu_BB/2$, the lowest spinon excitation (a spin
flip costs $g\mu_BB$). The coefficient carries no factor of
$\lambda$, but the coupling requires spin--orbit locking. Without
spin--orbit coupling $M_z$ belongs to a spin triplet and cannot couple
linearly to the flux mass, a spin singlet. On the lattice ($d=0.3$ and
$0.6$) the field-induced gap is linear for $g\mu_BB/2\lesssim0.1\times4\lambda$
and saturates near $0.4\times4\lambda$, and $C_{\rm BdG}$ is unchanged
up to $g\mu_BB/2=2.5\times4\lambda$ (Appendix~\ref{app:chiral}).

The field-induced N\'eel moment for an in-plane field is to be
distinguished from the staggered magnetization that the intra-dimer DM
interaction and the staggered $g$ tensor induce in the dimer phase~\cite{Kodama2005,Miyahara2004,Miyahara2007}. The buckling tilts the principal $g$ axes alternately
toward the $\langle110\rangle$ directions and allows the intra-dimer
DM coupling. Both mix the dimer singlet with the lowest triplet at first
order, whereas an inter-dimer DM coupling, which acts on two dimers, has no matrix element
between the singlet product state and a single-triplet state~\cite{Miyahara2004,Takigawa2004}. The dimer-phase staggered moment therefore does not
measure, at leading order, the in-plane inter-dimer couplings that set
$\lambda$. In the two geometries of Ref.~\cite{Kodama2005}, the induced
intra-dimer staggered moments lie in-plane along each dimer's own axis
for $\bm B\parallel[001]$ and along $c$ on the dimers parallel to the
field for $\bm B\parallel[110]$.
The two responses differ in two ways. In both geometries one of field
and moment lies in the plane and the
other along $c$. Under $\hat\sigma_h$ those moments are therefore odd
in the in-plane couplings and appear at first order in $D$ and $g_s$,
whereas the N\'eel response to an in-plane field is even.
They are intra-dimer
staggered moments of the gapped dimer phase and differ from the
$(\pi,\pi)$ N\'eel masses of Table~\ref{tab:field}. For $\bm B\parallel[001]$ their
in-plane dimer-staggered moment is compatible with our classification
only in the weaker sense that the N\'eel mass channel is forbidden there
while staggered spatial-current channels remain allowed. A positive
identification of the observed response with a specific allowed channel
requires a matrix-element calculation we do not perform. The $[100]$
prediction has no measured counterpart.

The critical scaling of the $[100]$ response is $m_s\propto
B\,\mathcal D\,\xi^{\,3-2\Delta_N}$, with $\mathcal D$ the coupling
factor of the source, $\lambda D'_{\parallel,s}$ at mean field or $g_1-g_2$. That scaling is the singular part, and a regular
ultraviolet contribution coexists with it. With the leading-order value
$\Delta_N=\Delta_V=1.46$ of Sec.~\ref{sec:spin2} the exponent is
$3-2\Delta_N=+0.08$, so the $[100]$ N\'eel response is weakly enhanced at
criticality. An exponent this small leaves the regular contribution
able to dominate the total response over accessible ranges of $\xi$.
The leading-order value fixes only the sign of the exponent (Sec.~\ref{sec:spin2}). For the much smaller $\Delta_N$ found in
numerics on the bosonic deconfined quantum critical point (DQCP)~\cite{Nahum2015} the exponent would be of order one and the
response strongly enhanced, itself a diagnostic of the anomalous
dimension. The sharp prediction is
the orientation selectivity of the N\'eel response, which is
symmetry-activated for $\bm B\parallel[100]$ and symmetry-forbidden for
$\bm B\parallel[001]$. The test is $^{11}$B NMR at $2.4$--$2.8$\,GPa
comparing $\bm B\parallel[100]$ with $\bm B\parallel[001]$ at matched
Zeeman energy $g\mu_BB$, below the field-induced transition near
$5.5$--$5.7$\,T, provided the in-plane staggered moment transfers a nonzero
hyperfine field to the boron site; the existing data do not make this comparison
(Sec.~\ref{sec:byprobe}). A $c$-axis field also splits the
lines, through the intra-dimer staggered moments, so the test is the
pattern of the splitting, a N\'eel pattern, with the
two moments of each dimer parallel, for $[100]$ and none for $[001]$ at
the same $g\mu_BB$, measured above the plaquette ordering temperature.
Exact diagonalization of Eq.~\eqref{eq:H} with the $I\bar42m$ DM vectors on
16-, 20- and 8-site clusters, the last at finite temperature, confirms
the rule independently of the gauge theory: the linear response is $(N^x,N^y)\propto(B_y,B_x)$, even and quadratic in the
in-plane couplings, absent for $D'_\perp$ alone and for
$\bm B\parallel c$, with a sign that depends on $J'/J$ and $T$
(Appendix~\ref{app:ed}). The selection rule follows
from the lattice symmetry and holds in the dimer phase as well. Only the dependence of the $[100]$ response on the distance to the
critical pressure is specific to the spin-liquid regime. It is weak at
leading order in $1/N$ and strong if $\Delta_N$ is as small as in
bosonic-DQCP numerics~\cite{Nahum2015}. We quote no number
for the size of the staggered moment, because the Gaussian coefficient
is a mean-field estimate whose fluctuation corrections are not computed,
and we do not estimate $g_1-g_2$.

\emph{Tilted fields.} The same tensor bears on field-tuned tests of
deconfined criticality. Any in-plane component of the field, as in a
field tilted by $\theta$ from $c$, is linearly conjugate to the in-plane
N\'eel order parameter, with the same magnitude for every azimuth. For a
general azimuth it removes every symmetry that distinguishes the
field-induced antiferromagnet from the paramagnet; for a tilt within a
$\{100\}$--$c$ or $\{110\}$--$c$ plane a twofold symmetry survives. Only a
field strictly along $c$ keeps the full N\'eel symmetry. The conjugate
staggered field, $h_s\sim\tfrac12(g_1-g_2)\mu_BB\sin\theta$ plus a term of
second order in the in-plane DM couplings, is relevant with RG eigenvalue
$3-\Delta_N$ and rounds a critical point on the scale
$T_h\sim\Lambda(h_s/\Lambda)^{1/(3-\Delta_N)}$. For an illustrative
$|g_1-g_2|=10^{-2}$, the $8.6^\circ$ tilt of Ref.~\cite{Cui2023} at
$H_c\simeq5.7$\,T and a cutoff $\Lambda\simeq J'$--$J\simeq50$--$85$\,K give $T_h\sim0.1$\,K at the large-$N$
value $\Delta_N=1.46$ and $\sim1$\,K at $\Delta_N\simeq0.63$, the value
implied by the loop-model estimate $\eta_{\rm N\acute eel}\simeq0.26$~\cite{Nahum2015}. The sharp ordering signatures reported with that tilt
therefore bound $|g_1-g_2|$ from above if $\Delta_N$ is small.

\section{Quenched disorder}
\label{sec:disorder}
Pressure and strain inhomogeneity make quenched randomness integral to the
SrCu$_2$(BO$_3$)$_2$ phase diagram. Its effect on the theory above is
fixed by two inputs from the operator tables, the $\mathcal T$ parities of
the 36 operators and their dimensions. Non-magnetic disorder preserves
$\mathcal T$ in every realization while breaking lattice symmetries
locally, so it couples random fields to every $\mathcal T$-even local
operator and to no $\mathcal T$-odd one. Exchange randomness $\delta J$
(spin-singlet, $\mathcal T$-even) couples linearly to the VBS doublet
$(V_x,V_y)$ and to the valley currents. Random DM $\delta D$ couples to
the spin currents, as a random $\lambda(\bm r)$, and to the $\mathcal
T$-even staggered spin-current densities $\gamma^0\mu^{x,z}\sigma^a$.
Being $\mathcal T$-even, these densities are not magnetization-like. Their
natural microscopic content is staggered vector-chirality and spin-current
textures, the operators that DM itself conjugates. No randomness couples
linearly to the flux mass or to the N\'eel masses. Chirality and
magnetism are random-field protected by $\mathcal T$, and the VBS is not.
Beyond the 36 fermion bilinears, the critical Higgs sector adds
gauge-invariant $\mathcal T$-even bilinears at the multicritical point,
where $N=12$ and the eigenvalue $y_\Delta=1.08$ below is defined:
$\Phi_1\cdot\Phi_1+\Phi_2\cdot\Phi_2$, the tuning parameter, to which bond
randomness couples as a random $T_c$, and $\Phi_1\cdot\Phi_1-\Phi_2\cdot
\Phi_2$ and $\Phi_1\cdot\Phi_2$, which break the lattice symmetry and
receive random fields. Their dimensions, derived below, are not controlled at $N=12$, and they
are $\mathcal T$-even, so they do not affect the protection of chirality, but
random $T_c$ and random Higgs-anisotropy fields may well dominate at the
multicritical point.

\emph{Disorder in SrCu$_2$(BO$_3$)$_2$ and its strength.}
Four sources of quenched randomness are documented. (i) Point defects
that free a spin $1/2$: low-temperature Curie tails correspond to
$0.14\%$ of free spins per Cu in traveling-solvent floating-zone
crystals~\cite{KageyamaJPSJ1999} and to $0.7$--$1\%$ in powders~\cite{Kageyama1999,Sibav2024}. Substitution does not leave only spinless
vacancies: Mg on the Cu site and La or Na on the Sr site each prevent a dimer
singlet from forming and free a Cu spin~$1/2$~\cite{Aczel2007}, Mg at $2.5\%$ of the Cu sites leaves unpaired spins $1/2$
antiferromagnetically correlated with the surrounding singlets~\cite{Haravifard2006}, $^{11}$B NMR on Zn-doped crystals images an
extended spin polaron around each Zn~\cite{Yoshida2015}, and Mg
incorporation in solid-state synthesis levels off near
$x\approx0.05$--$0.07$ in SrCu$_{2-x}$Mg$_x$(BO$_3$)$_2$~\cite{Sibav2024}. (ii) Inversion twins, formed at $T_s$
as incoherent microtwins of equal volume fraction and, according to
Ref.~\cite{Sparta2001}, presumably with high wall concentrations; their
size has not been measured. They reverse every in-plane DM component~\cite{Suetsugu2022}, while $J$, $J'$ and $D'_\perp$, which are allowed
above $T_s$, are the same in both twins. (iii) Pressure gradients. All
pressure media are solid or glassy at the measurement temperatures~\cite{TateiwaHaga2009}:
Daphne 7373 is solid at room temperature above $2.2$\,GPa, with solidification setting in near $2.0$\,GPa~\cite{Yokogawa2007}, and in a diamond-anvil-cell survey of fourteen media non-hydrostatic effects (ruby-line broadening from pressure gradients and uniaxial stress) developed in all of them at low temperature, weakest for helium, nitrogen and argon~\cite{TateiwaHaga2009}. The narrow coexistence of the
plaquette and antiferromagnetic phases, which Ref.~\cite{Guo2025}
attributes to a tiny pressure inhomogeneity, suggests an in situ spread
of about $\delta P\approx0.02$--$0.05$\,GPa; Ref.~\cite{Cui2023}
attributes its own dimer--plaquette coexistence to the slope of the
first-order line. This spread corresponds to
$\delta(J'/J)\approx(0.6$--$1.6)\times10^{-3}$ with
$d(J'/J)/dP\simeq0.032$\,GPa$^{-1}$~\cite{Sakurai2018}. (iv) Shear strain
from non-hydrostatic stress and dislocations, not measured and plausibly
$10^{-5}$--$10^{-3}$. These sources enter the classification above
differently. A hydrostatic gradient is fully symmetric and acts as
random-$T_c$ disorder, not as a random VBS field. The two plaquette
states differ in symmetry: a full-plaquette order that is the same in all
layers transforms like $\epsilon_{xy}$ (irrep $B_2$ of $\bar42m$; $Fmm2$
is the subgroup Ref.~\cite{Waki2007} inferred from the $^{11}$B shifts at
$2.4$\,GPa, and for layer-alternating order no clear orthorhombic
distortion is expected~\cite{Boos2019}) and couples linearly to shear strain along
$[110]$, whereas the empty-plaquette order parameter ($A_2$) couples
linearly to no strain component. With the strain derivative of the spin
gap, $\partial E_{\rm gap}/\partial\epsilon\approx700$--$900$\,K from the
longitudinal $c_{11}$ mode~\cite{Zherlitsyn2000},
as the magnetoelastic scale, a shear strain above about $10^{-5}$ exceeds
the full--empty splitting of the pure model, which is below $10^{-4}J'$~\cite{Xi2023}, and near the critical pressure the strain is a uniform
valence-bond field that outweighs the DM seed of
Sec.~\ref{sec:whichvbs} by orders of magnitude at the band-touching
scale. Random DM from the same local distortions scales as
$\delta D\sim(D/J)\,\delta J$, so its covariance is smaller than that of
$\delta J$ by $(D/J)^2\sim10^{-3}$, and over the available RG time
$\ln(J'/T)\lesssim7$ its marginal flow changes it by less than a percent;
the sign of its one-loop coefficient below has no experimental
consequence. The random VBS field is different. With a lattice-scale
variance of order the defect density $x$, it becomes strong near
$T_{\rm dis}\sim J'x^{1/y_\Delta}$, which for $y_\Delta=1.08$ and $x\approx0.1$--$0.4\%$
is of order $0.1$--$0.3$\,K, comparable to $T^*$, and exceeds it by
more than an order of magnitude if the valence-bond dimension is as small
as in bosonic-DQCP numerics~\cite{Nahum2015}. The defects are magnetic. In an
applied field their polarized polarons, imaged at ambient pressure for a
$c$-axis field in Ref.~\cite{Yoshida2015}, are $\mathcal T$-odd random fields on the
N\'eel channels for every field direction, because a local defect removes
the lattice symmetry that forbids a uniform N\'eel moment for
$\bm B\parallel[001]$. The random-field protection of chirality and
magnetism stated above therefore holds for the zero-field Hamiltonian
and fails below any freezing of the orphan spins.

The static-randomness eigenvalues $y_\Delta=2+2z-2\Delta_O$ in two
spatial dimensions then read as follows. A random VBS field has
$y_\Delta=4-2\Delta_V=1.08$ at leading order in $1/N$ and is relevant.
Current couplings have $y_\Delta=0$ at tree level, protected by
conservation, and random velocities have $y_\Delta=-2$. For the
marginal set the one-loop variance flow of Appendix~\ref{app:disorderrg},
which contains the disordered-Dirac results of Ref.~\cite{LFSG1994}, gives
the coefficients $+2$ (potential), $-2$ (mass) and $0$ (abelian vector
potential) in units of $\Delta^2/(2\pi)$. For a smooth random modulation
of the DM scale, $\delta\lambda(\bm r)\,\mathcal O_{\rm DM}$, the static vertex
tensors vanish and the canonical normalization of the matrix-valued
frequency renormalization gives
\begin{equation}
\frac{d\Delta_\lambda}{d\ell}=-\frac{2}{\pi}\Delta_\lambda^2 ,
\label{eq:randomlambda}
\end{equation}
so this component is marginally irrelevant. DM randomness uncorrelated on
the bond scale behaves differently. Matched on the $\pi$-flux state, its
covariance lies half in the spin-current sextet and half in the
$\mathcal T$-even staggered densities, with no $\mathcal T$-odd or mass
component, and for independent Gaussian components on every bond the
projected one-loop flow is
\begin{equation}
\frac{d\Delta}{d\ell}=+\frac{2}{3}\,
\frac{\Delta^2}{2\pi}\quad(+\text{new directions}),
\label{eq:multivertex}
\end{equation}
with an ensemble-dependent coefficient. Every bond-uncorrelated ensemble
generates random $\mathcal T$-even VBS masses, however, and these carry
the relevant eigenvalue $1.08$ at the SO(5) point, so bond-scale DM
randomness adds to the random VBS field of $\delta J$ instead of being
inert. Its strength in the material is not derived.

\emph{Gauge fluctuations.} With one SU(2) gauge field and one pair of Higgs
triplets per replica, the covariance of the $\mathcal T$-even random fields
flows at $O(\Delta^2,\Delta/N)$ as
\begin{equation}
\frac{d\Delta_{ij}}{d\ell}=(4-\Delta_i-\Delta_j)\,\Delta_{ij}
+\frac{1}{2\pi}\,B_{ij}[\Delta],
\label{eq:disflowN}
\end{equation}
where $\Delta_i$ is the $O(1/N)$ scaling dimension of the operator that
carries the random field $i$ and $B$ is the $N$-independent one-loop kernel
of Appendix~\ref{app:disorderrg}. Gauge fluctuations enter only through
$\Delta_i$: the $O(\Delta^2)$ graphs with one gauge propagator carry
$\operatorname{tr}\sigma^c=0$ (Fig.~\ref{fig:disorder}(h)), and those with two need a nonzero flavor
trace, which no $\mathcal T$-even singlet has. The VBS components keep
$\Delta_V=2-64C_g/(3\pi^2N)$, so the random plaquette field has
$y_\Delta=128C_g/(3\pi^2N)=1.08$ at every value of the gauge parameter, and
the conserved currents keep $\Delta=2$, so current randomness is marginal
at linear order to all orders in $1/N$. For the plaquette pair, with
variances $\Delta_+$ of the empty ($A_2$, $\mu^x+\mu^z$) and $\Delta_-$ of
the full ($B_2$, $\mu^x-\mu^z$) component and $\Delta_J$ of the
VBS-rotation current $\gamma^{x,y}\mu^y$, the flow closes:
\begin{align}
\frac{d\Delta_\pm}{d\ell}&=y_\Delta\Delta_\pm+\frac{1}{2\pi}\bigl[-2\Delta_\pm^2
+2\Delta_+\Delta_-\nonumber\\
&\qquad\mp4\Delta_J(\Delta_+-\Delta_-)\bigr],\nonumber\\
\frac{d\Delta_J}{d\ell}&=\frac{\Delta_+\Delta_-}{\pi}.
\label{eq:plaqflow}
\end{align}
The one-loop terms cancel on $S=\Delta_++\Delta_-$, which grows with the
exponent $y_\Delta$, while the full--empty imbalance
$\delta\Delta=\Delta_+-\Delta_-$ obeys $d\,\delta\Delta/d\ell=[y_\Delta-(S+4\Delta_J)/\pi]\,\delta\Delta$ and
decays at strong disorder. The only fixed point with a single random
plaquette component, $\Delta^*=\pi y_\Delta$ with
$z^*=1+y_\Delta/2\simeq1.54$, lies outside the weak-disorder regime and is
unstable, with eigenvalue $(1+\sqrt3)\,y_\Delta=2.95$ in the
$(\Delta_-,\Delta_J)$ plane. Every $\mathcal T$-even ensemble with a
plaquette component therefore flows to strong disorder after a flow time
of order $y_\Delta^{-1}\ln(1/\Delta_0)$, which sets the scale $T_{\rm dis}$
used above. From $0.05$ times the normalized covariance this takes
$\ell^*=5.15$ for bond-uncorrelated interdimer exchange, $5.80$ for
vacancies and $9.6$--$10.6$ for bond-uncorrelated DM randomness, whose
trace without the gauge dressing grows only to $0.07$--$0.08$ by $\ell=40$
(Appendix~\ref{app:disorderrg}).

Matched on the $\pi$-flux state, bond-uncorrelated interdimer exchange
disorder has the weights $1/4$, $1/4$ and $1/2$ on $\Delta_+$, $\Delta_-$
and $\Delta_J$. Dimer-bond randomness has no zero-mode projection at
mean-field order, but its site group allows it to feed $\Delta_-$ beyond
it. The two opposite interdimer bonds of a site project with opposite
signs, so a spinless vacancy couples at mean-field order only through the
imbalance $\delta J'_{\rm near}-\delta J'_{\rm far}$ between the
modifications of the interdimer bonds on the side of its dimer partner and
on the far side, with a variance of order
$x(\delta J'_{\rm near}-\delta J'_{\rm far})^2/J'^2$ rather than $x$. The
mirror through the Cu site forbids the empty-plaquette component to all
orders, so vacancies feed only the full plaquette and the VBS-rotation
current, and the empty plaquette acquires its random field only through
Eq.~\eqref{eq:plaqflow}. The mean-field lattice with one site removed
agrees (Appendix~\ref{app:disorderrg}). In the $\pi$-flux parent the
vacancy binds two quasi-localized zero modes that carry the freed spin
$1/2$ as a Kramers doublet, with neither magnetization nor chirality in
its $\mathcal T$-invariant average, and the induced bond-energy pattern is
$C_{4v}$ symmetric. In the Z3000 state the empty-plaquette harmonic
vanishes after averaging over boundary twists, and the twist-averaged full plaquette and VBS-rotation current are
$(3$--$10)\times10^{-3}$ of the bond energy per vacancy on clusters of
$12$--$24$ sites per side, the full plaquette extrapolating to
$(2$--$6)\times10^{-3}$,, an order of magnitude below the isotropic harmonic that feeds
random $T_c$. The zero modes correspond to the bound state of the earlier
mean-field theories discussed in Ref.~\cite{Kolezhuk2006} for the
staggered-flux state, where gauge fluctuations replace the free-spin Curie
constant by a universal one and remove the staggered induced moment; that
U(1) analysis is not carried over here to the SU(2) and Z3000 cases.

In large-$N$ noncompact QED$_3$, time-reversal-invariant disorder that
keeps a U(1)$\times$SU($N$) flavor symmetry, on the kagome lattice a random
DM coupling along one spin axis, flows to a fixed line at finite disorder
with continuously varying $z>1$, whereas disorder that breaks the flavor
symmetry further, including random-bond disorder through the valence-bond
masses, flows to strong disorder, and random gauge flux becomes relevant
beyond leading order~\cite{ThomsonSachdev2017}. Gauge screening stabilizes
that theory against a weak random chemical potential, which couples to the
U(1) gauge charge, while random Dirac masses drive it to a fixed point at
finite disorder~\cite{GoswamiGoldmanRaghu2017}. In that abelian limit the
one-gauge-line graphs survive as parity-odd triangles, and
Eq.~\eqref{eq:disflowN} with them reproduces both
results (Appendix~\ref{app:disorderrg}). In the SU(2) theory the triangles
vanish, the chemical potential of the gauge charge is not a gauge-invariant
coupling, and no conserved flux current exists at the multicritical point,
so neither the fixed line nor the screening of a random chemical potential
carries over. In compact QED$_3$ with four flavors,
time-reversal-invariant random spin currents, a non-abelian vector
potential, lower the dimensions of monopoles of every
charge~\cite{Dey2020}. This bears on the single allowed monopole of U800,
already relevant in the large-$N$ extrapolation.

\emph{Chirality filter beyond $N=\infty$.} Time reversal holds in every
realization of non-magnetic disorder, so the replicated action is invariant
under $\mathcal T$ acting on a single replica while the others are left
unchanged. A replica-bilocal term $\int O_i^aO_j^b$ with a $\mathcal T$-odd
factor changes sign under this operation and is absent to all orders in
$\Delta$ and $1/N$ and for every cumulant of the disorder. The selection
rule is stronger than invariance under a common $\mathcal T$ of all
replicas, which would allow a variance $\Delta\,\Tr[\Xb\X]_a\Tr[\Xb\X]_b$.
The random-field protection of chirality and magnetism is therefore exact
within the zero-field Hamiltonian, and gauge fluctuations change only the
dimensions of the $\mathcal T$-even channels. $\mathcal T$-odd operators
enter the averaged theory only through $\mathcal T$-even composites such as
$(\Tr[\Xb\X])^2$ in the SO(5)-singlet quartic channel, as random-$T_c$
couplings of the chiral transition. Once $\mathcal T$ is broken, in a field
or below the freezing of orphan spins, the N\'eel masses carry random fields
with $y_\Delta=1.08$ and the flux mass a random field that is marginal at
linear order to $O(1/N)$: its dimension $2+256(N_f-N_b)/(\pi^2N^2)$ equals
$2$ at $N_f=N_b=2$, an accidental cancellation of the field content that no symmetry
enforces.

\emph{Random $T_c$.} The tuning parameter is represented by the constraint
field $\sigma$ of the Higgs sector. At $O(1/N)$
\begin{equation}
\Delta_\sigma=2-\frac{32}{3\pi^2M}+\frac{64C_s}{3\pi^2N}-\frac{64\,WC_s}{\pi^2N^2},
\label{eq:Dsigma}
\end{equation}
with $M=6$ real Higgs components, the adjoint Casimir $C_s=8$ in the
normalization $C_g=3$, and the Higgs loop weight $W=8$. The three terms are
the constraint-field exchange, one gauge line with its seagulls, and two
gauge lines attached to a closed Higgs loop, the graphs of
Fig.~\ref{fig:higgs}(a)--(c) (Appendix~\ref{app:integrals}); the
constraint-field triangle gives no logarithm in $d=3$, and the mixing with the flux singlet $\Tr[\Xb\X]$,
degenerate with $\sigma$ at $N=\infty$, vanishes because the Higgs loop is
parity even and the fermion loop parity odd, as time reversal requires.
Every term is independent of the gauge parameter, and in the abelian limit
Eq.~\eqref{eq:Dsigma} reproduces the correlation-length exponent of
Ref.~\cite{KaulSachdev2008} [its Eq.~(34)] for all $N_b$ and $N_f$,
including the CP$^{N-1}$ value
$\nu=1-48/(\pi^2N)$~\cite{IrkhinKataninKatsnelson1996,VasilevNalimov1983}
at $N_f=0$. At $N=12$, $\Delta_\sigma=0.38$ lies below the unitarity bound
$1/2$~\cite{Minwalla1998,Evans1967}, and the random-$T_c$ eigenvalue is
$y_{T_c}=4-2\Delta_\sigma=3.24$. Random $T_c$ is therefore relevant at
$O(1/N)$, and the Harris criterion fails at this order, but a correction of
this size is not controlled at $N=12$, and the $O(1/N)$ value of $\nu$
carries no quantitative information there.

\emph{Higgs sector.} The anisotropy bilinears $\Phi_1\cdot\Phi_1-\Phi_2\cdot
\Phi_2$ and $\Phi_1\cdot\Phi_2$ have
$\Delta=1-64C_s/(3\pi^2N)+32/(3\pi^2M)$, whose gauge term, $-1.44$ at
$N=12$, is $C_s/C_g=8/3$ times the fermion value and takes $\Delta$ below
the unitarity bound. The expansion thus fixes only the direction: random
Higgs anisotropies, with $y_\Delta=2$ at $N=\infty$, become more relevant. The
gauge-invariant composites $\Phi^a\Tr[\sigma^a\Xb M\X]$, through which
gauge-adjoint couplings can be random, have $\Delta=1.80$ (mass type) and
$1.62$ (current type) at $N=12$, against $5/2$ at $N=\infty$
(Appendix~\ref{app:disorderrg}), shifts that leave their relevance
undecided. All of these operators are $\mathcal T$-even where they carry
random fields, so none of them affects the protection of chirality.

The consequences favor, but do not by themselves establish, an inversion
of the clean-limit competition. The VBS channel condenses a discrete
order parameter, the Ising plaquette pair of
Sec.~\ref{sec:whichvbs}, exposed to a relevant random field. Under
strictly two-dimensional short-range random fields, the domain argument
of Imry and Ma~\cite{ImryMa1975}, marginal in two dimensions for discrete order and
made rigorous by Aizenman and Wehr~\cite{AizenmanWehr1989}, at $T>0$ for any field distribution in the ferromagnetic random-field Ising model and otherwise for continuously distributed disorder, with its
quantum extension, including ground states, for continuously distributed disorder~\cite{GreenblattAizenmanLebowitz2009},
shows that these fields destroy long-range VBS order for arbitrarily weak $\delta J$, leaving a
plaquette domain glass with $\xi_{\rm IM}\sim a\,\exp[c_{\rm IM}(J_I/h)^2]$, where
$h$ is the rms random field per cell, $J_I$ the effective Ising coupling
of the plaquette variables; the $(J_I/h)^2$ form is the $T=0$ domain-breakup estimate of Ref.~\cite{Binder1983}, with a constant of order unity, and exact ground states of the square-lattice random-field Ising model give $c_{\rm IM}\simeq2.1\pm0.2$ for Gaussian fields ($1.9\pm0.2$ for bimodal ones)~\cite{Seppala1998}. If each free spin pins its cell,
$h\simeq4J_I\sqrt{4x}$ for a free-spin fraction $x$ per Cu, and $x=0.14\%$,
$0.4\%$ and $2.5\%$ give $\xi_{\rm IM}$ of order $e^{23}a$, micrometers
and a few cells; these estimates are exponentially sensitive to the
assumed pinning strength. In three dimensions weak random fields do not
destroy order~\cite{Imbrie1984,BricmontKupiainen1987}, and a layered
crystal crosses over to that behavior once the interlayer coupling of the
plaquette variables exceeds the random-field energy of a two-dimensional
domain. For the full plaquette, whose order parameter couples linearly to
strain, this coupling is elastic and three-dimensional, and smooth strain
templates the domains instead of softening their formation. A plaquette
domain glass is therefore expected in substituted crystals rather than in
the nominally pure crystals studied so far, and a calorimetric anomaly
does not decide between the two, since a 2D Ising peak that is sharp on
the scale $|T-T_P|/T_P\sim0.1$ probes correlation lengths of about ten
cells. Random-singlet-like behavior does not follow from this: in
Ref.~\cite{KimchiNahumSenthil2018} it requires spin-$1/2$ VBS defects,
which a Lieb--Schultz--Mattis-type constraint enforces for half-odd-integer
spin per unit cell, whereas the Shastry--Sutherland cell holds four spins
$1/2$ and the plaquette order is Ising-like, as for the two-leg ladder,
where the unpaired spins at a domain wall form a singlet.
Whether plaquette domain walls bind low-energy spins is not determined
here. The chiral $\mathbb Z_2$ channel
sees at its thermal transition only random-$T_c$-type disorder, which is
Harris-marginal for the 2D thermal Ising class~\cite{Harris1974,Vojta2006}. Weak bond disorder is in
fact marginally irrelevant there, with logarithmic corrections~\cite{DotsenkoDotsenko1983,Shankar1987,LFSG1994}: the $\ln\ln$ specific heat of Ref.~\cite{DotsenkoDotsenko1983} is confirmed, while its spin-sector result, correlations decaying more slowly than any power, is contradicted by the exact bound of Ref.~\cite{Shankar1987}. The ordered phase
survives. The quantum transition with columnar-in-time disorder is a
separate problem. Non-magnetic, $\mathcal T$-invariant quenched disorder
in zero field therefore supplies a chirality-filter mechanism, degrading
the majority VBS channel while sparing the minority chiral one, in
strictly two dimensions or at disorder strong enough that $\xi_{\rm IM}$
falls below the interlayer length. Whether the system then orders
chirally or remains a nonchiral glass depends on the sign and size of the
residual chiral couplings. The random field destroys long-range VBS order
but leaves local VBS gaps inside the domains, so the chiral channel gains
no energetic advantage from it. Whether the filter makes thermal-Hall and
weak-ferromagnetic signatures more likely in inhomogeneous samples is
therefore a scenario that this analysis does not prove, and a sharp calorimetric anomaly
does not mark a clean sample. Below $T^*$ the counting must be applied to projected vertices.
Disorder with a nonzero zero-derivative projection on the QBT, the VBS
random field foremost, has $y_\Delta=2+2z-2\Delta_O=+2$ and is strongly relevant,
and a diffusive spinon regime preempting $T^{**}$ becomes possible. Its
stability against localization and gauge constraints is not settled
here. Random $\lambda$ itself projects to zero on the QBT, which is the
null space of $h_{\rm DM}$ (Sec.~\ref{sec:qbtir}), so its leading
projected coupling carries two derivatives and has $y_\Delta=-2$. The
scale-forming coupling is inert to its own randomness. At the
pressure-tuned plaquette--antiferromagnet point, if the clean correlation-length
exponent satisfies $\nu<1$, so that the two-dimensional Harris criterion $2\nu>2$ fails, random-$T_c$ disorder is relevant~\cite{Harris1974,Chayes1986}. If the transition is instead weakly first
order, quenched disorder rounds it in two dimensions~\cite{AizenmanWehr1989,HuiBerker1989,GreenblattAizenmanLebowitz2009}. The clean DQCP scaling
should be rounded, while a distinct disordered critical point is not
excluded, and rare-region (Griffiths) effects may accompany either outcome~\cite{Vojta2006}. The finite-size-scaling correlation-length exponent of such a disordered critical point must obey
$\nu_{\rm FS}\ge1$ in two dimensions~\cite{Chayes1986}, a theorem for independent site or bond disorder given a suitable finite-size-scaling event. Existing high-pressure NMR does
not report such rounding. Reference~\cite{Cui2023} attributes the satellite
line broadening it observes at $2.1$\,GPa and $0.2$\,T to the lattice
distortion of the full-plaquette state. It reports quantum-critical
scaling of $1/T_1$ between $0.2$ and $2$\,K at $2.4$\,GPa, in near-$c$
fields just below the plaquette--antiferromagnet transition field.
Reference~\cite{Cui2024two}, with $\bm B\parallel c$ at $2.4$\,GPa and
$4$\,T, assigns its broad low-frequency $^{11}$B line to an
empty-plaquette phase coexisting with the full-plaquette phase at a
first-order, pressure-driven transition.

\section{Experimental signatures}
\label{sec:experiment}
Table~\ref{tab:predictions} collects the predictions by probe.

\begin{table*}[t!]
\caption{Experimental signatures by probe. ``Order'' as in
Sec.~\ref{sec:intro}: ``exact'' (symmetry or current conservation),
``leading order'' ($O(1/N)$ or one loop), ``estimate'' (mean-field
prefactors); ``mean field'' marks Gaussian-level coefficients.
$\kappa_{xy}$ is per layer unless stated.}
\label{tab:predictions}
\begin{ruledtabular}
\begin{tabular}{p{3.0cm}p{7.9cm}p{3.5cm}p{1.6cm}}
Probe & Prediction & Order & Section\\
\colrule
Any probe, $T\gg T^*$ &
Dirac spin liquid or pseudocritical regime with DM corrections
controlled by $T^*/T$. No symmetry-allowed local gauge-singlet fermion mass at any
order in the DM couplings. $D'_\perp$ irrelevant at linear order,
$\Delta_{O_B}=3.32$; its second-order tensor acts above $T^*$ only
if the $\mathbf{14}$ is relevant. &
no-mass exact. $D'_\perp$: bound $\Delta>3$ at a unitary critical point,
value leading order &
\ref{sec:inventory}, \ref{sec:spin2}\\
Specific heat, zero field, $2.6$--$2.9$\,GPa &
Node reconstruction at $T^*\simeq0.03$--$0.17$\,K, and a thermodynamic
crossover near $(0.1$--$0.15)\,T^*$. Below it, near criticality, a finite
spinon density of states, with $C\propto T$ at most a few times
$T^{**}$; $T^{**}\simeq(0.03$--$0.1)\lambda\simeq0.2$--$2.5$\,mK at the Fock level. &
eigenvalue exact. Prefactor estimate. $C\propto T$ at mean field &
\ref{sec:crossover}, \ref{sec:qbt}\\
$^{11}$B NMR, $\bm B\parallel[100]$ vs $[001]$ at matched $g\mu_BB$,
$2.4$--$2.8$\,GPa &
$[100]$: N\'eel moment $\perp\bm B$, linear in $B$, second order in
the in-plane DM couplings ($\lambda D'_{\parallel,s}$ at mean field) or first order in
an in-plane $g$ anisotropy, singular part weakly enhanced (exponent
$+0.08$). $[001]$: no N\'eel moment, and the flux channel is the only activated
mass. Intra-dimer staggered moments, allowed for this field direction by
the crystal symmetry used in Ref.~\cite{Kodama2005}, are not excluded. &
selectivity exact. Coefficient estimate. Exponent leading order &
\ref{sec:field}\\
NMR, $\bm B\parallel[110]$ &
N\'eel component along $\bm B$, $(N^x+N^y)/\sqrt2\propto B$, plus
valence-bond imbalance $V_x-V_y\propto B_xB_y$, both second order in the
DM couplings, the imbalance smaller by a factor $\propto B$. The N\'eel component also arises at first order in an in-plane $g$ anisotropy. &
symmetry exact. Coefficients estimate &
\ref{sec:field}\\
Thermal Hall, zero field below $T^{**}$ after cooling in $\bm B\parallel[001]$ &
Chiral minority phase: $\kappa_{xy}/T\to\mp2\pi^2k_B^2/3h\simeq
\mp1.9\times10^{-12}$\,W\,K$^{-2}$ per layer, bulk $\mp5.7\times10^{-3}$\,W\,K$^{-2}$\,m$^{-1}$ for two layers per $c$. Remanent after field
cooling, with the sign of $\kappa_{xy}$ in
the cooling field ($\kappa_{xy}B_c>0$ at mean field), reversed with that
field, hysteretic in $B$ below $T^{**}$ only; in-field $\kappa_{xy}$ is
no test. &
mean field. Conditional on a gapped deconfined $\mathbb Z_2$ sector and
on the minority basin &
\ref{sec:chiral}, \ref{sec:field}, \ref{sec:byprobe}\\
Magnetization, zero field &
Weak uniform $M_z$ in the chiral phase, with $M_z\kappa_{xy}>0$ at mean
field. Remanent after cooling in $\bm B\parallel[001]$, parallel to the
cooling field, and hysteretic below $T^{**}$. &
mean field. Direction of the remanent $M_z$ set by the cooling field &
\ref{sec:chiral}, \ref{sec:byprobe}\\
Neutron, Raman, ESR &
Below $T^*$ the gauge-singlet gap channels are the two VBS components and the scalar
chirality only. Reconstruction at $4\lambda\simeq0.003$--$0.015$ ($0.03$--$0.10$)\,meV.
Spectra not computed. &
gap channels exact in the quadratic theory. Energy scale estimate &
\ref{sec:qbt}\\
Plaquette type near $P_c$ (NMR, calorimetry, neutrons) &
Two-state Ising plaquette pair. The DM correction is a bias of at most
of order $10^{-4}$ after the running: the seed linear in
$D'_\perp$ favors $P_{\rm f}$ for $D'_\perp<0$ (NMR, DFT), the
determinant favors $P_{\rm e}$, and which one leads is not fixed at
$O(1/N)$. Twin-insensitive. Transition temperature and order not
computed. &
$O(1/N)$ running; direction not established &
\ref{sec:whichvbs}\\
Disorder (calorimetry vs NMR linewidths) &
Random VBS field relevant, $y_\Delta=1.08$: under strictly 2D random fields a
plaquette domain glass at any disorder strength, with exponentially large
domains at weak disorder. Interlayer and elastic coupling restore order at
weak disorder, and calorimetry does not discriminate. Shear strain along
$[110]$ selects full plaquettes. Chirality random-field protected for
$\mathcal T$-invariant disorder in zero field. Random DM negligible next
to $\delta J$. Clean critical point Harris-unstable at $O(1/N)$. &
$O(1/N)$ flows, uncontrolled at $N=12$; protection exact, filter a scenario &
\ref{sec:disorder}\\
\end{tabular}
\end{ruledtabular}
\end{table*}

\subsection{Units, couplings, and scales}
\label{sec:units}
We use $J=85$\,K and $J'=54$\,K~\cite{MiyaharaUeda2003,Cepas2001}; the
other parameter sets are listed in Sec.~\ref{sec:model}. The spinon
hopping $t\simeq0.4J'\simeq20$--$22$\,K for $J'=51$--$54$\,K
($J'/J=0.60$--$0.635$ at $J=85$\,K) is a mean-field estimate fixed by no measurement.
It enters $T^*$ through the node splitting and the second-order scales
through $D'_\perp/t$, whereas the length $\xi_{\rm DM}$ does not depend on
it (Sec.~\ref{sec:crossover}).

The DM couplings and their extraction are given in
Sec.~\ref{sec:model}: $D'_\perp\simeq0.09$--$0.18$\,meV, the only
intralayer DM coupling above $T_s$; the intra-dimer
$D\simeq0.22$--$0.35$\,meV, the largest; the spectroscopic combination
$\tilde D_\parallel\simeq0.07$--$0.08$\,meV; and
$D'_\Sigma\simeq-0.014$\,meV from Table~II of Ref.~\cite{Mazurenko2008}. The in-plane
components require the buckling, which persists down to $2$\,K at
ambient pressure~\cite{Vecchini2009}; under pressure the cell stays
tetragonal up to the monoclinic onset at $4.0$--$5.0$\,GPa~\cite{Loa2005,Zayed2014}, near $4$\,GPa at low temperature~\cite{Wu2026}, but no refinement determines the buckling there, and
Ref.~\cite{Loa2005} suggests that it decreases. The crystals are inversion twins, which carry
opposite in-plane DM; the magnetization, $\kappa_{xy}$, the NMR spectra
and the plaquette selector $b_2$, which is even in $\lambda$, are
twin-insensitive. All lattice energies quoted in units of $t$ or in kelvin
are physical excitation energies, in the convention of
Appendix~\ref{app:protocol}.

The scales of the theory are ranges. The crossover energy is
$T^*\simeq0.03$--$0.17$\,K (Sec.~\ref{sec:crossover}).
The thermodynamic crossover of the free-spinon specific heat lies near
$(0.1$--$0.15)\,T^*$, i.e.\ at $3$--$25$\,mK. The
associated length is $\xi_{\rm DM}=\sqrt2\,(J'/|D'_\Sigma|)\,a\simeq250$--$940$
lattice spacings. The lower edge of the
quadratic-band-touching window is the Higgs-induced splitting
$\delta'=(c_\Phi s_Ht)^2/4\lambda$ with $c_\Phi\simeq0.67$,
Eq.~\eqref{eq:splitter}, where the Higgs deviation $s_H$ is fixed by no
measurement. The flow applies on $[\delta',\lambda]$, where the band
touching screens the gauge field into a color-density
interaction (Sec.~\ref{sec:qbtir}). That interaction diverges at
$\ell^*\simeq0.6$--$1.6$, so the window requires
$\ln[2\lambda/(c_\Phi s_Ht)]\gtrsim\ell^*$, i.e.\ $c_\Phi s_Ht\lesssim0.5\lambda$. It
is open only near the critical pressure and closed deep in the Higgsed
phase. Within it the instability temperature is
$T^{**}\simeq(0.03$--$0.1)\lambda$, or at the Fock level $0.2$--$2.5$\,mK
(up to $4$\,mK with $t\simeq0.4J'$), an
order-of-magnitude one-loop estimate at order-one coupling.

\subsection{Signatures by probe}
\label{sec:byprobe}

\subsubsection{Specific heat}
Above $T^*$ the Dirac spin liquid and the pseudocritical regime survive
with DM corrections controlled by $T^*/T$, and at linear order in the DM
couplings the specific heat carries no DM-induced feature; the
second-order anisotropy of Sec.~\ref{sec:selection}, if the rank-2 SO(5)
tensor is relevant, would instead produce a symmetry-breaking crossover
at a scale of order $T^*$ or above. Below $T^*$ and near the critical pressure the
Dirac cones reconstruct into the quadratic band touching of
Sec.~\ref{sec:qbt}, which has a finite spinon density of states at mean
field. The resulting $C\propto T$ appears only well below the
thermodynamic crossover near $(0.1$--$0.15)\,T^*$, i.e.\ in the
millikelvin range, and only if $T^{**}\lesssim0.1\lambda$
(Sec.~\ref{sec:crossover}). We quote no coefficient
for this term, which requires the QBT curvature beyond mean field, and we
do not compute the thermodynamic signature of the crossover itself. All
existing zero-field data near the plaquette--antiferromagnet boundary lie above
$T^*$. The a.c.\ calorimetry of Ref.~\cite{Guo2025} reaches $0.4$\,K and
locates a first-order boundary at $2.78$\,GPa ending in a bicritical
point. That work fits $C/T=a_0+a_1T^2+(a_2/T^3)\,e^{-\Delta/T}$ and concludes
that a clean system is gapless above $2.7$\,GPa. Its lowest-temperature $C/T$, roughly $a_0$,
is enhanced around the critical pressure of $2.78$\,GPa, which the
authors tentatively attribute to enhanced fluctuations near the
bicritical point. These data lie at least a factor of two above $T^*$
(four at the Fock level). They lie inside the plaquette and N\'eel phases, whose ordering temperatures
exceed $1$\,K, so the enhancement does not test the spinon density of
states of Sec.~\ref{sec:qbt}. The calorimetry of
Ref.~\cite{LarreaJimenez2021} reaches $0.1$\,K, and its zero-field data
cover $0$--$26.5$\,kbar. That work expects the peak near $2$\,K, seen for
$P\geq18$\,kbar up to its highest pressure, to mark the thermal
transition out of the gapped plaquette phase. On its own pressure scale
its data below that peak therefore lie in the plaquette phase. Its
lowest temperatures lie within the range of $T^*$, still inside that phase.

In applied fields Ref.~\cite{Guo2026} finds, at fields where the
zero-field order (plaquette at $2.68$ and $2.75$\,GPa, antiferromagnetic
at $2.9$\,GPa) is suppressed and the field-induced antiferromagnet has
not set in, a specific heat linear in $T$ from about $1$--$1.7$\,K down
to $0.4$\,K, over $1.5$--$4$\,T at
$2.68$\,GPa, $3$--$7$\,T at $2.75$\,GPa and $2.5$--$4$\,T at $2.9$\,GPa,
with intercepts quoted as $C/T\simeq1.1$--$1.4$, $1.1$--$1.6$ and
$1.2$\,J\,K$^{-2}$\,mol$^{-1}$ at the three pressures, and reproduced on
a second sample at $2.92$\,GPa. It regards fractionalized spin excitations
as one possible origin and leaves the mechanism open. A thermal
tensor-network study of the SS model in a Zeeman field, on
$3\times12\times4$-site cylinders, finds a $T$-linear specific heat at
intermediate fields inside the plaquette phase and near its boundary with
the antiferromagnet, within the reach of its system size and bond
dimension~\cite{Song2026}; that model has full spin-rotation symmetry, so
the field direction does not enter there, whereas with DM it does. In Ref.~\cite{Guo2026} the field was applied essentially along the
$c$ axis, and $C/T$ is quoted per mole of formula
units~\cite{GuoPrivComm}. At mean field a Zeeman field splits the Dirac cones into
spinon Fermi pockets, which for $\bm B\parallel c$ the DM couplings gap at
about $0.4\times4\lambda$ (Sec.~\ref{sec:field}). For
$0.4\times4\lambda\ll k_BT\ll g\mu_BB/2$ their density of states, and hence
$C/T$, grows linearly with $B$. With $t\simeq20$\,K its value at $4$\,T is
about $0.07$\,J\,K$^{-2}$ per mole of formula units ($0.16$ with the Fock
hopping), seven to twenty-three times below the reported intercepts, so
the free-spinon pockets do not account for them. The field also activates
different masses for different orientations (Sec.~\ref{sec:field}): the
chirality mass alone for $\bm B\parallel c$, and the N\'eel components
for an in-plane field, each requiring at least first order in
spin--orbit coupling. The $T$-linear data, taken with $\bm B\parallel c$,
probe the orientation that activates only the chirality mass; a
measurement with $\bm B\parallel[100]$ in the same regime would test the
selection rule where the spinon density of states is largest. These data lie above
$T^*$.

\subsubsection{NMR}
The classification of Sec.~\ref{sec:field} gives an orientation-selective
N\'eel response. For $\bm B\parallel[100]$ a staggered (N\'eel) moment
perpendicular to $\bm B$ is activated at linear order in $B$. It is of
second order in the in-plane DM couplings, $\lambda D'_{\parallel,s}$ at mean field, and
of first order in a dimer-dependent in-plane $g$ anisotropy. Its singular part scales as
$\xi^{3-2\Delta_N}$ with $3-2\Delta_N=+0.08$ at leading order in $1/N$, a
weak enhancement that becomes strong for the smaller $\Delta_N$ of
bosonic-DQCP numerics~\cite{Nahum2015}. For $\bm B\parallel[001]$ no N\'eel moment is
allowed and the flux channel is the only activated mass. Intra-dimer
staggered moments of the type of Ref.~\cite{Kodama2005} are not excluded. For $\bm B\parallel[110]$ the
N\'eel component along $\bm B$, $(N^x+N^y)/\sqrt2$, appears linearly
in $B$ and a
valence-bond imbalance $V_x-V_y$ appears at order $B_xB_y$, second order
in the DM couplings.
The magnitude of the moment in $\mu_B$, which requires a matrix-element
calculation, and the spin-lattice rate $1/T_1$ are not computed. The
proposed test is $^{11}$B NMR at $2.4$--$2.8$\,GPa comparing $\bm
B\parallel[100]$ with $\bm B\parallel[001]$ at matched $g\mu_BB$. The
matching uses $g_a=2.05$ and $g_c=2.28$, which Ref.~\cite{Nojiri1999}
obtained by ESR at ambient pressure from the paramagnetic resonance at
$25$\,K, and it assumes that pressure changes these values little. The
comparison should be made below the field-induced transition, which lies
at $5.5$--$5.7$\,T at $2.4$\,GPa for a field along or near $c$~\cite{Cui2023,Cui2024two}. It should also be made above the plaquette
ordering temperature, $T_P=1.6$\,K at $2.4$\,GPa in a $c$-axis field of
$4$\,T~\cite{Cui2024two}. If the plaquette phase lowers the tetragonal symmetry, it
allows additional DM components that are not classified here
(Sec.~\ref{sec:scope}). The induced staggered field is even under the sign reversal of the in-plane DM
components between inversion twins, so the induced moment has the same
sign in both twin domains. No existing experiment makes this comparison.
Reference~\cite{Waki2007} compared $\bm B\parallel c$ with
$\bm B\parallel[110]$ at $2.4$\,GPa and $7$\,T down to $2.1$\,K and found,
for $\bm B\parallel c$ below $30$\,K, a loss of fourfold symmetry in the
magnetic shifts; no $[100]$ data exist.
Reference~\cite{Cui2023} measured at pressures up to $2.4$\,GPa, in
$0.2$--$15$\,T and down to $0.07$\,K, with the field $8.6^\circ$ from
$c$. Its in-plane component, about $0.15B$, points along an in-plane
direction that was not characterized. Reference~\cite{Cui2024two} used $\bm B\parallel c$ to $2.65$\,GPa
between $0.3$ and $5$\,K. Reference~\cite{Wu2026} applied all fields in the $ab$ plane and made no
orientation comparison. At $2.4$\,GPa it has spectra for $\bm
B\parallel a$ at $8$\,T down to $1.8$\,K, and they show no line splitting
on cooling. These temperatures lie above the N\'eel temperature found at
$2.4$\,GPa for $\bm B\parallel c$, which increases monotonically with
field to $1.5$\,K at $10$\,T~\cite{Cui2024two}. The $\bm B\parallel
a{+}b$ spectra of Ref.~\cite{Wu2026} are at $4$--$14$\,GPa, and its other in-plane data, at
$1$--$8$\,T, span $2.65$--$14$\,GPa. Above the monoclinic onset at
$4.0$--$5.0$\,GPa~\cite{Zayed2014} these data probe a lower lattice
symmetry than the one classified here. The antiferromagnet above $H_c\simeq5.7$--$6.2$\,T with
moments along $a$ or $b$~\cite{Cui2023} is the thermodynamic
field-induced phase and is distinct from the DM-activated mass.

\subsubsection{Thermal Hall effect and magnetization}
The chiral minority phase of Sec.~\ref{sec:chiral} has $C_{\rm
BdG}=\mp4$ and $c=\mp2$, hence a quantized $\kappa_{xy}/T\to\mp2\pi^2
k_B^2/3h\simeq\mp1.9\times10^{-12}$\,W\,K$^{-2}$ per layer as $T\to0$,
conditional on a gapped deconfined $\mathbb Z_2$ sector. With two
CuBO$_3$ layers per $c$-axis lattice constant of $6.649$\,\AA{}, from single-crystal
x-ray diffraction at ambient conditions~\cite{SmithKeszler1991,KageyamaJCG1999}
($6.641$\,\AA{} at $2$\,K~\cite{Vecchini2009}),
the bulk value is
$\kappa_{xy}/T\simeq\mp5.7\times10^{-3}$\,W\,K$^{-2}$\,m$^{-1}$, that is,
$|\kappa_{xy}|\simeq6$\,$\mu$W\,K$^{-1}$\,m$^{-1}$ at $1$\,mK, the scale
of $T^{**}$, if the plateau is reached there. The $c$
axis shortens under pressure, by about $6\%$ at $5.5$\,GPa~\cite{Fogh2024PRL}, so the bulk value in the pressurized chiral liquid is
slightly larger. A weak
uniform magnetization $M_z$ accompanies the chirality. Its magnitude is
not computed, and at mean field its sign satisfies $M_z\kappa_{xy}>0$ for
the magnetic moment $M_z$ and $\bm J^Q=-\kappa\nabla T$
(Appendix~\ref{app:chiral}). A $c$-axis field couples linearly to the
flux channel, because $M_z$ and the flux mass carry the same
one-dimensional representation of the buckled and of the
high-temperature spin--orbit-locked groups. Without spin--orbit coupling
they do not. The twin operation leaves both invariant, so the two twin
domains respond alike. Cooling through $T^{**}$ in $\bm B\parallel[001]$
therefore selects the chirality domain whose moment is parallel to the
field. For the Zeeman coupling $g_c\mu_B\bm B\cdot\bm S$ of Cu$^{2+}$
that domain has $\kappa_{xy}B_c>0$ at mean field, with $B_c$ the field
component along $+c$. On the QBT the Zeeman term is itself the flux mass
(Sec.~\ref{sec:field}). Below $T^*$ the field therefore also opens the
flux gap without spontaneous order, with lowest excitation energy
$g_c\mu_BB/2$ at weak field, and at mean field this field-only state has
the Chern number of the domain the field selects. A thermal Hall signal
measured in the field, including a plateau, therefore does not
distinguish the chiral phase from the field-gapped QBT. The test of
spontaneous chirality is the remanent $\kappa_{xy}$ at zero field after
field cooling. It has the sign of $\kappa_{xy}$ in the cooling field,
reverses with that field, and comes with hysteresis of $\kappa_{xy}(B)$
and $M_z(B)$ below $T^{**}$ that disappears above $T^{**}$. The cooling
field must keep the field-induced gap $g_c\mu_BB/2$ below
$k_BT^{**}$, so that it selects a domain and does not decide the phase.
For $T^{**}\simeq0.2$--$2.5$\,mK this requires $B\lesssim0.07$--$0.8$\,mT
($g_c\mu_BB/2k_B\simeq0.05$--$0.6$\,mK). The field needed to select a
single domain depends on the magnitude of $M_z$ and is not computed. The ambient-pressure thermal Hall
measurements of Refs.~\cite{PritchardCairns2020,Suetsugu2022}, both with
$\bm B\parallel c$ and the heat current along $a$, resolved no
$\kappa_{xy}$. Reference~\cite{PritchardCairns2020} reports no signal of
the predicted triplon magnitude at $7.5$ and $10$\,K in fields up to
$3$\,T. Reference~\cite{Suetsugu2022} bounds $|\kappa_{xy}|$ at $7$,
$8.5$, and $10$\,K by $20$--$30\%$ of its own non-interacting-triplon
calculation. That calculation is opposite in sign to, and smaller by a
factor $2\pi$ than, those of Refs.~\cite{McClarty2017,Romhanyi2015}. These
nulls constrain the triplon mechanism of the dimer phase and leave the
pressurized chiral liquid untested, since no thermal Hall measurement
under pressure exists. Reference~\cite{Cui2023} infers the uniform magnetization from the
$^{11}$B Knight shift at $2.1$\,GPa and $0.07$\,K in a near-$c$ field.
The magnetization stays below $2\%$ of the saturated moment from $5$ to $15$\,T, and at
$15$\,T the system is inside the field-induced antiferromagnet. The
zero-field $M_z$ below $T^{**}$ is therefore untested.

\subsubsection{Neutron, Raman, and ESR spectroscopy}
Below $T^*$ the only gauge-singlet gap channels of the reconstructed spectrum are the
two valence-bond components and the scalar chirality, and no N\'eel gap
channel exists, a kinematic statement. The node reconstruction takes
place at the energy $4\lambda\simeq T^*$, that is, $0.003$--$0.015$\,meV.
We do not compute the dynamic structure factor, the Raman response, or
the ESR spectrum. The $D'_\perp$-dominated one-triplon splitting of
$85$\,GHz~\cite{Nojiri2003,Cepas2001,Gaulin2004} and the triplon Chern
bands~\cite{Romhanyi2015,McClarty2017} are quantities of the dimer
phase and do not test the present theory.

\subsubsection{Plaquette phase type versus pressure}
Within the VBS doublet, and near the critical pressure, the order
parameter is a two-state Ising pair (Sec.~\ref{sec:whichvbs}). Below $T^*$ the
DM-induced correction to the selection that the pure Shastry--Sutherland
model makes is a bias of at most of order $10^{-4}$ after the running to the
band-touching scale. Which of its two contributions leads, a seed linear
in $D'_\perp$, which favors the full-plaquette pair for $D'_\perp<0$, or
the determinant term, which favors the empty one, depends on the Higgs
sector of the SO(5)-singlet channel, which is not controlled at $N=12$, so
the direction is not established (Sec.~\ref{sec:whichvbs}). The pure-model
selection itself is not computed here. The iPEPS study of
Ref.~\cite{CorbozMila2013} and the infinite density-matrix renormalization-group study of Ref.~\cite{LYSV2019}
find the empty (square) plaquette state, and Ref.~\cite{LYSV2019}
chooses the sign $\lambda_2>0$ of its monopole anisotropy to match it. The high-pressure $^{11}$B NMR of
Ref.~\cite{Cui2024two}, a preprint, resolves two plaquette phases that
coexist across a first-order pressure-driven transition. The
empty-plaquette fraction rises to $70\%$ at $2.65$\,GPa, and the
empty-plaquette phase borders the antiferromagnet. The same work
reports, at $2.4$\,GPa with $\bm B\parallel c$, a field-induced transition
from the empty-plaquette phase to the antiferromagnet near $5.5$\,T, with
$1/T_1\propto T^{0.6}$ between $0.3$ and $1.5$\,K at $5.5$\,T. It also
reports a plaquette ordering at $T_P=1.6$\,K at $2.4$\,GPa and $4$\,T,
which the authors assign to the $\mathbb Z_2$ (Ising) breaking of the
twofold degeneracy shared by both plaquette phases. The Ising thermal
transition near $2$\,K in zero field is seen in calorimetry~\cite{LarreaJimenez2021,Guo2020}. The empty and full labels of
Ref.~\cite{Cui2024two} rest on an interpretation of NMR lineshapes and
relaxation. Other work favors the full-plaquette phase at similar pressures. The
neutron intensities of Ref.~\cite{Zayed2017} at $2.15$\,GPa tentatively favor it,
and Ref.~\cite{Cui2023} attributes an NMR satellite broadening at
$2.1$\,GPa and $0.2$\,T to a full-plaquette lattice distortion.
Reference~\cite{Shi2022} interprets two high-field phases, measured with
$\bm B$ in the $ab$ plane, as full-plaquette descendants, although its iPEPS
calculations also find both phases in the undeformed model, whose
low-field phase is the empty plaquette. The deformed-coupling calculation of
Ref.~\cite{Boos2019} reaches the same conclusion by comparison with the
neutron data of Ref.~\cite{Zayed2017} and with NMR, ESR and
specific-heat data, so it is not independent of them.
Reference~\cite{LarreaJimenez2021} also expects the state it observes above
$1.8$\,GPa and below $2$\,K to be the full-plaquette phase. In the model
itself the empty plaquette remains the ground state, but near the
upper edge of the plaquette phase ($J'/J\approx0.79$) the full
plaquette lies less than $10^{-4}J'$ above it~\cite{Xi2023}. In the
gapped plaquette phases a correction odd in $D'_\perp$ is at least of
order $D'_\perp D_\parallel^2$ (Sec.~\ref{sec:whichvbs}), and the
band-touching seed linear in $D'_\perp$ exists only below $T^*$, far below
the plaquette ordering temperature, so DM does not decide between the two
plaquette phases at $T_P$. Shear strain, to which the full-plaquette order
parameter couples linearly and the empty one does not
(Sec.~\ref{sec:disorder}), is the more likely selector at that scale. The present theory speaks only to the phase adjacent to
the critical point, and it does not compute the plaquette transition
temperature or its order.

\subsubsection{Disorder}
Section~\ref{sec:disorder} predicts, for strictly two-dimensional random
fields or for substituted crystals, a plaquette domain glass, with
chirality random-field protected for $\mathcal T$-invariant disorder in
zero field to all orders in the disorder and in $1/N$, random $T_c$
relevant at $O(1/N)$, a full-plaquette order parameter that couples
linearly to shear strain along $[110]$ while the empty one does not, and
vacancies that feed only the full-plaquette component of the random
field.
Several high-pressure studies invoke pressure inhomogeneity.
Reference~\cite{Cui2023} suggests that it may cause the $^{11}$B NMR line
splittings reported at $2.4$\,GPa and $7$\,T in Ref.~\cite{Waki2007},
whose authors ascribed them instead to an intrinsic magnetic origin and a
novel ordered phase. Reference~\cite{Cui2024two} notes that neutron
scattering just above the dimer--plaquette boundary favored the
full-plaquette phase~\cite{Zayed2017}, while zero-field specific-heat
data~\cite{LarreaJimenez2021,Guo2025} align with its own NMR data for the
plaquette-liquid regime above the empty-plaquette phase. It names the
degree of hydrostaticity achieved in each experiment as one possible
source of this discrepancy. It also offers a more homogeneous
high-pressure phase as one possible interpretation of its N\'eel
temperatures of the field-induced antiferromagnet at $2.4$\,GPa, which
are over $50\%$ higher than those of Ref.~\cite{Cui2023}.
Reference~\cite{Guo2025} attributes the narrow overlap of its plaquette and
antiferromagnetic phases to a tiny inhomogeneity of the pressure.
Zero-field calorimetry in pressure cells loaded with liquid media resolves a specific-heat
peak at the plaquette ordering temperature, at $T\lesssim 2$\,K~\cite{LarreaJimenez2021,Guo2020,Guo2025}.
References~\cite{LarreaJimenez2021,Guo2020} interpret it as a thermal Ising
($\mathbb{Z}_2$) transition out of the plaquette phase, and
Ref.~\cite{Guo2025} does not discuss its universality class. These peaks
are consistent with the clean-sample expectation but do not discriminate
it from weak disorder. The one controlled-disorder experiment under
pressure, magnetometry on SrCu$_{2-x}$Mg$_x$(BO$_3$)$_2$ with $x\le0.05$ at
$2.1$--$2.4$\,GPa, suggests that the dopants destabilize the plaquette phase
in favor of the antiferromagnet~\cite{Shi2022}. Neither it nor any other
experiment tests the domain-glass or chirality-filter statements, and the
strength and correlation structure of the material's DM randomness are
not known. The symmetry analysis of Sec.~\ref{sec:disorder} suggests four
tests. (1) Because shear strain along $[110]$ couples linearly to the
full-plaquette order parameter and not to the empty one, the
full-plaquette fraction resolved by NMR~\cite{Cui2024two} should grow
with the non-hydrostatic stress of the pressure environment and depend on
the loading history; a uniaxial stress of $0.01$--$0.1$\,GPa along
$[110]$ should convert empty into full plaquettes and select one
full-plaquette domain, while stress along $[100]$ couples linearly to
neither and serves as a control. (2) A full-plaquette order that is the
same in all layers carries an orthorhombic strain, estimated at
$10^{-4}$--$10^{-3}$, that splits the $(hh0)$ reflections in
high-resolution diffraction under helium pressure below $T_P$; the peak
widths then measure the domain size. (3) Substituted crystals across the
critical pressure should show a broadened or absent $2$\,K plaquette
anomaly and, if the freed spins form a random-singlet-like network, the
scaling collapse $C/T\sim H^{-\gamma}F_q(T/H)$ of
Ref.~\cite{KimchiSheckelton2018}, with $\gamma$ the exponent of the zero-field
$C\sim T^{1-\gamma}$ and $\chi\sim T^{-\gamma}$ and an integer $q\in\{0,1,2\}$ set by
how spatial symmetries constrain the DM couplings; random-singlet states also
arise from bond disorder in a $\mathbb Z_4$ columnar valence-bond solid, with spinons where four domain walls meet, whereas site dilution of the same model gives antiferromagnetic order~\cite{LiuShaoLinGuoSandvik2018}. (4) Tests of the physics below $T^*$
require free-spin fractions below about $10^{-3}$, residual shear strain
below about $10^{-5}$, and a subtraction of the orphan-spin contribution,
whose entropy $xR\ln2$ is released near their mutual coupling scale. The
comparison of $[100]$ and $[001]$ fields in NMR must separate a uniform
line splitting from the polaron satellites that defects produce~\cite{Yoshida2015}, and muon-spin relaxation
is not a clean probe of $\mathcal T$ breaking here, since muons on the Cu--O--Cu
path appear to break the dimer singlet locally~\cite{Aczel2007}.

\subsection{Status of the predictions and decisive tests}
\label{sec:convergences}

\emph{Crossover scale.} The prediction is a node reconstruction at
$T^*\simeq0.03$--$0.17$\,K near the critical pressure, with a thermodynamic crossover to the QBT
regime near $(0.1$--$0.15)\,T^*$. All existing
low-temperature data lie above or near $T^*$. The NMR data of Refs.~\cite{Cui2023,Cui2024two} that come closest are taken in near-$c$ fields of several tesla. Near the field-induced
transition at $2.4$\,GPa, Ref.~\cite{Cui2024two} reports
$1/T_1\propto T^{0.6}$ from $0.3$ to $1.5$\,K at $5.5$\,T along $c$. At the same pressure, with the field $8.6^\circ$ from
$c$, Ref.~\cite{Cui2023} fits the shifted rate $1/T_1+b_H\propto
T^{0.2}$, with a field-dependent offset $b_H$, for several fields close
to $H_c\simeq5.72$\,T. Each fit holds only above a field-dependent onset
$T_P$, below which the rate is thermally activated, and the fits
together span $0.2$ to $2$\,K. At $5.4$--$5.6$\,T this onset, which
Ref.~\cite{Cui2023} identifies with the plaquette transition, lies
between about $0.4$ and $0.7$\,K. A $c$-axis field couples only to the
flux channel, and the in-plane component of about $0.15B$ in
Ref.~\cite{Cui2023} also activates a N\'eel channel
(Table~\ref{tab:field}). The theory makes no statement about $1/T_1$ in
either case. The crossover is untested. A test
requires measurements below about $0.1$\,K near the critical pressure in
samples that remain disordered there, and the free-spinon $C\propto T$
requires millikelvin temperatures.

\emph{Stability of criticality above $T^*$.} The absence of gauge-singlet masses and the
irrelevance of $D'_\perp$ imply that DM cannot cause a departure from
criticality above $T^*$ at linear order. The second-order anisotropies of
Sec.~\ref{sec:selection} act there only if the rank-2 SO(5) tensor is
relevant; their SO(3)$\times$SO(2)-invariant part has the form of the
N\'eel--VBS tuning operator~\cite{ThomsonSachdev,WangNahum2017,Zhou2024fuzzy}
and only shifts the critical pressure, so the candidate is the easy-axis or
easy-plane spin anisotropy, at a scale of order $T^*$ or above, up to
about $1$\,K. The DM-induced Yukawa couplings stay below the Yukawa
coupling that bounds the pseudocritical window (Sec.~\ref{sec:yukawa}).
Reference~\cite{Guo2025} reports zero-field transitions into the plaquette and
antiferromagnetic phases at about $1.2$--$1.6$\,K at $2.6$--$2.9$\,GPa,
extrapolated to a bicritical point at $2.78$\,GPa near $1.3$\,K, and
infers a first-order transition from the abrupt change of phase and from
the drop of the fitted activation gap.
These transitions and the gap fits, down to $0.4$\,K, lie above $T^*$. In the field-tuned
NMR of Ref.~\cite{Cui2023} the plaquette--antiferromagnet transition is
first order at $2.1$ and $2.4$\,GPa, with a discontinuity that weakens
with pressure, and $1/T_1$ shows critical scaling only at $2.4$\,GPa;
Ref.~\cite{Cui2024two} finds the field-induced transition out of the
phase it assigns as empty plaquette continuous or nearly so. Apart from a
relevant rank-2 tensor, the present theory attributes departures from
continuous criticality to the interlayer coupling $J_c\simeq0.6$\,K, to
disorder, to the multicritical character of the SO(5) point once a
singlet is relevant, or to an intrinsically weakly first-order
transition, as iPEPS finds at the plaquette boundary of the pure model~\cite{Corboz2026}. The finite-temperature iPEPS calculations of
Ref.~\cite{LarreaJimenez2021}, with an intra-dimer DM coupling of the size
known for the material ($D/J\simeq0.03$), show no qualitative effect on
the dimer--plaquette critical point, which the same work locates
experimentally at $P_c\approx19$\,kbar and $T_c\approx3.3$\,K. The spin-wave analysis of Ref.~\cite{Fogh2024PRL}
attributes the $1.90(8)$\,meV excitation at $(0,1,0)$ of the
antiferromagnet at $5.5$\,GPa to the interlayer coupling
$J_c=0.053(3)$\,meV, since a two-dimensional model would need an
intra-dimer DM coupling of about $0.7J'$, close to $3$\,meV for its
fitted $J'=4.28$\,meV, and $D'_\perp$ opens no gap
there; inelastic neutron data at $4.2$\,GPa favor the same interlayer
model~\cite{Cederholm2026}. These data lie at or above the monoclinic
onset~\cite{Zayed2014}, and neither they nor the iPEPS calculation, which
omits $D'_\perp$, tests the inertness of $D'_\perp$, which dominates only
the triplon spectrum of the dimer phase.

\emph{Plaquette type.} The DM correction is a small bias below $T^*$
whose direction the theory does not fix: the seed linear in $D'_\perp$
favors, for the NMR sign $D'_\perp<0$, the full plaquette, the pattern of
the interpretations of Refs.~\cite{Zayed2017,Shi2022,Boos2019}
summarized in Sec.~\ref{sec:byprobe}, and the lattice determinant favors
the empty plaquette of Ref.~\cite{Cui2024two} adjacent to the
antiferromagnet; the pure-model selection outweighs both at $T_P$. In
Ref.~\cite{Boos2019} the full-plaquette phase is an instability of the
coupled spin--lattice system, stabilized by inequivalent intra- or
inter-dimer couplings, and a direct transition into it from the dimer
phase would have to occur below $J'/J=0.675$, where the
Shastry--Sutherland model enters the empty-plaquette phase; the plaquette
type thus indicates which parent state controls each pressure window. The full-plaquette evidence of
Refs.~\cite{Zayed2017,Shi2022} comes from pressures of at most
$2.4$\,GPa. Reference~\cite{Cui2024two} finds the full-plaquette phase
coexisting with the empty-plaquette phase from $1.85$\,GPa up to its
highest pressure of $2.65$\,GPa. The decisive tests are an independent
determination of the sign of $D'_\perp$, a bound on the SO(5)-singlet spectrum, and a zero-field
structural or spectroscopic identification of the plaquette pattern
adjacent to the antiferromagnet. The Ising order-parameter symmetry does not
discriminate, since both plaquette phases are twofold degenerate~\cite{LarreaJimenez2021,Cui2024two}.

\emph{Chiral minority phase.} The chiral phase is untested. The decisive test is a
thermal Hall measurement under pressure in zero field below $T^{**}$,
which lies below $T^*\simeq0.03$--$0.17$\,K, after cooling in a weak
$c$-axis field of $0.07$--$0.8$\,mT, chosen so that $g_c\mu_BB/2\lesssim k_BT^{**}$
(Sec.~\ref{sec:byprobe}). The remanent $\kappa_{xy}/T$ is
compared with the plateau value given there. Its sign must follow the
sign of the cooling field, and $\kappa_{xy}(B)$ must be hysteretic below
$T^{**}$ only. A weak remanent $M_z$ parallel to the cooling field
completes the test.

\emph{Field orientation.} The $[100]$ versus $[001]$ N\'eel response and
the $[110]$ valence-bond imbalance are untested; the test is the
matched-field NMR comparison of Sec.~\ref{sec:byprobe}.

\emph{Disorder.} The resolved calorimetric plaquette peaks are consistent
with the clean-sample prediction, but the chirality filter is untested as such.
The decisive test is a controlled-disorder study under pressure, for
example calorimetry and NMR on the same crystal in media of different
hydrostaticity, or Mg-substituted crystals~\cite{Shi2022,Dabkowska2007} across the
critical pressure. At ambient pressure Mg substitution keeps the spin gap
open, although the effective gap of powders falls from $26$ to $20.5$\,K
at nominal $x=0.05$~\cite{Sibav2024}, and muon-spin relaxation on powders with up to
$6\%$ nominal Mg finds no doping-induced static order, only a
quasi-static component below $3.75$\,K that is present already at $x=0$~\cite{Lappas2003}. Each substituted Cu frees a spin $1/2$ on its dimer
partner~\cite{Haravifard2006,Shi2019}, so the substitution adds magnetic
polarons rather than spinless vacancies, and it tests the random-field
statements only in zero field and above the freezing of those moments.

\section{Scope and limitations}
\label{sec:scope}
The following items are not computed in this paper or are computed only
at the stated level.
\begin{itemize}
\item The $1/N$ expansion parameter $64/(\pi^2N)$ is $0.54$ at $N=12$,
so the leading-order numbers $\Delta_V=1.46$, $\gamma_2^{\rm
adj}=0.324$, the lifting $\Delta_{\rm flux}-\Delta_{\rm VBS}=0.54$, and
$y_\Delta=1.08$ are qualitative and fix only signs and orderings. The abelian next-to-leading-order analogue,
$0.217/N=1.8\%$, applies to the abelian problem only, and no
controlled next-order result exists for the mixed gauge--Higgs theory.
\item The prefactor of $T^*$ is mean field, uncertain by a factor of
order one (the Fock value and the twist of the assumed hopping
$t\simeq0.4J'$ differ by a factor $1.7$). $D'_\Sigma$ rests on one DFT calculation~\cite{Mazurenko2008}, whose tabulated vectors are covariant under the
crystal symmetry only pair by pair; the
bond-axis component ($0.01$\,meV) is the same on all four bonds, at the limit of its precision, and
vanishes in the arXiv version. Spectroscopy does not constrain $D'_{\parallel,ns}$. All
couplings are ambient-pressure values; at $2.4$--$2.9$\,GPa the
buckling that allows the in-plane components is expected to decrease~\cite{Loa2005}, which would lower $T^*$. The
interlayer coupling $J_c=0.053(3)$\,meV${}\simeq0.6$\,K, fitted at
$5.5$\,GPa in the monoclinic antiferromagnet~\cite{Fogh2024PRL} and
consistent with $J_c=0.049(6)$\,meV from inelastic neutron data at
$4.2$\,GPa~\cite{Cederholm2026}, exceeds $T^*$ and
is not included.
\item The window $[\delta',\lambda]$ requires proximity to the critical
pressure through the Higgs deviation $s_H$, which no measurement fixes.
$T^{**}\simeq(0.03$--$0.1)\lambda$ is an order-of-magnitude one-loop
estimate at order-one coupling. With dynamical screening, the transverse
gauge field and the Higgs polarization included, the gauge interaction
alone lies $2$--$4\%$ of its norm from the boundary between the VBS and
chirality basins, and Higgs exchange through the Yukawa coupling crosses
it only outside the pseudocritical window. Deep in the Higgsed phase no weak-coupling instability
follows, and strong-coupling instabilities of the displaced Dirac phase
are not analyzed.
\item The basin fractions of the one-loop flow are scheme-dependent, and
the majority verdict is a one-loop statement.
\item The quartic seeds are computed at tree level, of order $J'$, and
run to $\lambda$ with the $O(1/N)$ four-fermion dimensions of
Appendix~\ref{app:treeseeds}, a linear running that is not controlled for
seeds of that size. The SO(5)-singlet scalar channel is solved at
$O(1/N)$ with five operators, $\Delta_1=2.99$ ($2.91$ in the physical-$N_f$
counting), but its Higgs sector is not controlled at $N=12$, where the
constraint field has $\Delta_\sigma=0.38$, below the unitarity bound; the
two further Higgs operators of the channel decouple at this order, and the
Yukawa coupling is not included.
Whether the seed linear in $D'_\perp$ or the lattice-determinant term leads
the DM correction to the plaquette pattern depends on $\Delta_\sigma$ and
on the determinant ratio, and is not established.
The sign of $D'_\perp$ is that of the NMR determination of
Ref.~\cite{Kodama2005}; the opposite sign read from the far-infrared
analysis of Ref.~\cite{CepasZiman2004} rests on an assumed relative sign
of the phonon-induced DM vectors, and Ref.~\cite{Mazurenko2008}
attributes the discrepancy to conventions. The selection that the pure
Shastry--Sutherland model makes beyond tree level is not computed. The
coefficient $A$ of Eq.~\eqref{eq:b2asym} is derived only for the
band-touching channel of the determinant,
exchange fluctuations beyond the determinant renormalize $b_2$, and the Ising
transition temperature and order are not computed.
\item The Chern number $C_{\rm BdG}=\mp4$ holds locally for the
Z3000$_{11}$ state, and the PSG class does not fix it. The magnitude of $M_z$ is
not computed, and the thermal Hall quantization assumes a gapped
deconfined $\mathbb Z_2$ sector.
\item The field-induced coefficients are Gaussian, fluctuation
corrections are not computed, and no moment is quoted in $\mu_B$. The
dimer-phase moments of Ref.~\cite{Kodama2005} are not identified with a
specific channel, which would require a matrix-element calculation.
\item The U800 monopole classification (Appendix~\ref{app:monopole})
is kinematic. The scaling dimension of the allowed SO(5)-singlet monopole
in the present theory is not computed; the QED$_3$ value with four two-component flavors is
about $1$~\cite{BKW2002,Pufu2014}, and the monopole's topological phases are taken from the mean-field
spinon state.
\item The disorder flows are evaluated at $O(\Delta^2,\Delta/N)$, with
$\Delta$ the disorder strength. The dimensions of the $\mathcal T$-even
Higgs bilinears and of the constraint field, which carry random fields and
random $T_c$ at the multicritical point, receive corrections of order one
at $N=12$ and are not controlled; the vacancy results are mean field. The
strengths of the material's disorder are estimates
from Curie tails, pressure inhomogeneity and strain, the stability of a
diffusive spinon regime is not settled, and the domain-glass conclusion
holds for strictly two-dimensional random fields or for substituted
crystals.
\item If the plaquette phase lowers the tetragonal symmetry, additional
DM components become symmetry-allowed and are not classified here; the
symmetry of that phase is not settled (Sec.~\ref{sec:model}). The same
holds above the monoclinic onset at $4.0$--$5.0$\,GPa~\cite{Loa2005,Zayed2014}.
\end{itemize}

\section{Conclusions}
\label{sec:conclusions}
We obtain the following results for the Shastry--Sutherland Dirac
spin liquids with the DM couplings of SrCu$_2$(BO$_3$)$_2$.
\begin{enumerate}
\item \emph{DM interactions cannot gap the spinons or destabilize the
SO(5) point above $T^*$ at linear order.} No symmetry-allowed local
gauge-singlet fermion mass is
generated at any order. The largest inter-dimer coupling, $D'_\perp$,
induces at linear order only a flavored spin-2 current, irrelevant at any
unitary critical point, with $\Delta_{O_B}=3.32$ at leading order. At second
order the DM couplings seed the rank-2 SO(5) tensor, easy-axis from
$D'_\perp$ and easy-plane from $\lambda$ at mean field; the tensor is
irrelevant at leading order in $1/N$ and, if the fuzzy-sphere dimension
applies, acts at a scale of order $T^*$ or above, up to about $1$\,K. The in-plane couplings generate one conserved-current
operator of dimension exactly 2, with RG eigenvalue 1 and scale
$T^*\simeq0.03$--$0.17$\,K for the DFT couplings. The intra-dimer $D$ also induces Yukawa couplings to the
critical Higgs fields, which start at $6\times10^{-4}$--$2\times10^{-2}$
of the Yukawa coupling that bounds the pseudocritical window and,
barring a cancellation among the Z3000 amplitudes, remain subleading
within it.
\item \emph{Below $T^*$ the spinons form a quadratic band touching with
two weak-coupling instability sectors at one loop.} Near the critical pressure the Dirac cones
reconstruct into a chirally protected $d$-wave band touching whose only
gauge-singlet gap channels are the valence-bond doublet and the scalar chirality, with
N\'eel gapping excluded kinematically. The one-loop flow makes a fivefold
sector containing valence-bond order the majority outcome, also for seeds
restricted to the gauge-singlet couplings, and the seed of
the material exchange lies in that basin, also after the $O(1/N)$
running. The gauge interaction, including its dynamical screening by the band
touching, the transverse field and the Higgs polarization, lies in the
same one-loop basin, but only $2$--$4\%$ of its norm from its boundary, so
gauge exchange alone does not decide the outcome; it sets the
order-of-magnitude one-loop estimate $T^{**}\simeq(0.03$--$0.1)\lambda$. A gapped chiral $\mathbb Z_2$ liquid with
$C_{\rm BdG}=\mp4$ and quantized thermal Hall conductance is the minority
outcome.
\item \emph{The DM correction to the plaquette pattern is a small bias
whose direction is not fixed.} A seed linear in $D'_\perp$,
$g(P_{\rm e})-g(P_{\rm f})=4D'_\perp$ at tree level, and the lattice
determinant, of opposite preference, compete within the Ising plaquette
pair. The seed's selector does not run at $O(1/N)$, while the isotropic
valence-bond attraction grows through its SO(5)-singlet part, whose
running at the $O(1/N)$ value of $\Delta_\sigma$ is dominated by the
constraint-field mode, so at the band-touching scale the DM correction is
a bias of at most of order $10^{-4}$ and the flow ends on the VBS ray.
Which of the two leads depends on the Higgs sector of the singlet channel,
which is not controlled at $N=12$, so the direction is not determined at
$O(1/N)$. For $D'_\perp<0$, the sign found by NMR, the seed favors the
full plaquette.
\item \emph{Magnetic fields and disorder act selectively.} A field along
$[100]$ activates a N\'eel mass at linear order and a field along $[001]$
does not, a selection rule of the lattice symmetry that holds in the
dimer phase as well. Non-magnetic, time-reversal-invariant quenched
disorder in zero field couples a relevant random field to the plaquette
doublet and, to all orders in the disorder and in $1/N$, none to the
chirality; at the multicritical point random $T_c$ is relevant at
$O(1/N)$ and the $\mathcal T$-even Higgs bilinears also receive random
fields, but the dimensions behind both statements are not controlled at
$N=12$. 
\end{enumerate}

The limitations are those listed in Sec.~\ref{sec:scope}. The large-$N$
numbers are qualitative at $N=12$, and the crossover prefactor and the
field-induced coefficients are mean field. The QBT window is conditional
on proximity to the critical pressure and on a strictly two-dimensional
sample, whereas the interlayer coupling of the material exceeds $T^*$ and
the samples measured so far order above $1$\,K. The running of the
plaquette correction is $O(1/N)$, with an uncontrolled Higgs sector in the
SO(5)-singlet channel, and the sign of $D'_\perp$ rests on one NMR determination,
with an apparent far-infrared discrepancy that has been attributed to
conventions. The disorder verdict depends on
the correlation structure and the magnetic character of the material's
randomness.

Four measurements bear on the scenario. A $^{11}$B NMR comparison of
$\bm B\parallel[100]$ and $\bm B\parallel[001]$ at matched $g\mu_BB$
between $2.4$ and $2.8$\,GPa, below the field-induced transition, tests
the orientation-selective N\'eel response. Zero-field calorimetry near the critical
pressure would test the finite spinon density of states in samples in
which the pseudocritical regime survives down to $T^*$. It requires
temperatures below about $0.03$--$0.17$\,K. The linear term is
approached only near $0.02\,T^*$, comparable to
$T^{**}\simeq(0.008$--$0.025)\,T^*$, so the instability may preempt it. In the existing samples, order sets in above $1$\,K
(Sec.~\ref{sec:byprobe}). A zero-field thermal Hall measurement under pressure
below $T^{**}$, after cooling in a weak $c$-axis field of either sign,
would detect the
chiral minority phase if it forms. A null result is the outcome expected
for the majority basin. An independent determination of the sign of $D'_\perp$ fixes the
direction of the DM correction to the plaquette pattern only once the
constraint-field dimension $\Delta_\sigma$ and the determinant ratio are
known, since they decide whether the seed or the determinant leads. The absence of gauge-singlet masses requires only the time-reversal
oddness of the N\'eel and chirality masses, the lattice parities of the
spin-singlet VBS masses, which locking cannot change, and the absence of
quantum-spin-Hall masses in the SU(2) theory. It therefore carries over to any
lattice realization of this gauge theory whose SO(5) vector consists of
N\'eel and VBS order parameters. The locked-symmetry classification
applies to other Dirac-spin-liquid candidates with spin--orbit coupling, and on the Shastry--Sutherland lattice it also delivers the
complete gauge-singlet bilinear tables without derivatives
(Tables~\ref{tab:unlocked} and~\ref{tab:locked}). These complement the
published gauge-adjoint classifications of
Refs.~\cite{ThomsonSachdev,Feuerpfeil2026}.

\section*{Acknowledgments}
The work of Y.I. was performed, in part, at the Aspen Center for Physics, which is supported by a grant from the Simons Foundation (1161654,
Troyer). This research was also supported in part by grant NSF PHY-2309135 to the Kavli Institute for
Theoretical Physics (KITP). Y.I. acknowledges support from the Abdus Salam International Centre for Theoretical Physics (ICTP) through the Associates Programme, from the Simons Foundation through Grant No. 284558FY19, from IIT Madras through the Institute
of Eminence (IoE) program for establishing QuCenDiEM (Project No. SP22231244CPETWOQCDHOC), and the International Centre for Theoretical Sciences (ICTS), Bengaluru for participation in the Discussion Meeting ``Fractionalized Quantum Matter'' (code: ICTS/DMFQM2025/07).
The author acknowledges the use of the AI tools Claude Opus 5.5 and
Fable 5.1 (Anthropic) to assist in developing code, cross-checking some
of the calculations, and preparing parts of the manuscript. The author
reviewed all results and changes and takes full responsibility for the
content.

\section*{Data availability}
The data that support the findings of this article are available from the
author upon reasonable request.

\appendix

\section{Conventions}
\label{app:protocol}
This appendix collects the conventions used in the analytic and numerical
calculations.

\emph{Spin axes.} A uniform lattice Zeeman field along a physical axis
projects onto the single bilinear $\gamma^0\sigma^a$, and a staggered one
with the N\'eel pattern onto the single bilinear $\mu^y\sigma^a$. The
component along $x$ enters with the sign opposite to those along $y$ and
$z$, so the spin label of the bilinears in
Tables~\ref{tab:unlocked}--\ref{tab:hightemp} and in Eq.~\eqref{eq:ODM} is
related to the physical axes by $x\to-x$. The field and moment directions
of Sec.~\ref{sec:field} and Table~\ref{tab:predictions} are physical and
include this sign.

\emph{Energies.} The lattice Majorana Bloch kernels are built by applying
a local antisymmetric-bilinear map to the kinetic terms of each ansatz.
They are used for number-conserving quantities only, namely spectra,
susceptibilities, and band-touching matrices, for which they agree with
the complex-fermion construction of the same states. The Majorana kernel
$H=\ii\sum_{KL}A_{KL}M_KM_L$ with $\{M_K,M_L\}=\delta_{KL}$ has
excitation energies $2\,{\rm eig}(\ii A)$. All lattice energies quoted in
this paper, in units of $t$ or in kelvin, are physical. Kelvin values
carry the mean-field prefactor uncertainty stated in
Sec.~\ref{sec:crossover}.

\emph{Published tables.} Figure~4 of Ref.~\cite{ThomsonSachdev} contains three rows
$\mu^y\sigma^b\gamma^\mu$ with $\mu=0,x,y$ (printed there as $\mu^y\sigma^a\gamma^\mu$), where $b$ is the spin index.
The caption states that the entries give the transformation of the
trace $\Tr[\sigma^a\Xb\,\Gamma^j\gamma^\mu\X]$. The printed $\mathcal T$
signs of these rows are $(+,-,-)$. The corresponding Hermitian operators
transform under $\mathcal T$ with the opposite signs $(-,+,+)$, for all
nine combinations of the gauge index $a$ and the spin index $b$, in
agreement with Table~3 of Ref.~\cite{Feuerpfeil2026}, whose $G_x$ signs
$(+,+,-)$ also agree with ours.
A phase convention for $\mathcal T$ cannot produce this flip, because it
is a relative sign within one gauge-adjoint SO(5) multiplet. The
conserved charge of the generator $\mu^z\sigma^b$ is $\mathcal T$-even, and
it rotates $\mu^x\gamma^\mu$ into $\mu^y\sigma^b\gamma^\mu$ through
$[\mu^z\sigma^b,\mu^x]=2\ii\mu^y\sigma^b$. Because $\mathcal T$ is
antiunitary, the factor $\ii$ requires the two rows to carry opposite
$\mathcal T$ signs, whereas Fig.~4 of Ref.~\cite{ThomsonSachdev} prints equal signs. We therefore regard the printed signs of these three rows there as
misprints. All Lagrangian-relevant rows and
all gauge-singlet statements are unaffected.

\section{Vanishing of the SO(5)-adjoint scalars}
\label{app:proof}
Write the sixteen Majorana fields as $\eta_K$ and any bilinear as
$\mathcal O=\ii\,\eta^T A\,\eta$ with $A$ real antisymmetric. In the
Majorana basis the matrices representing $\gamma^0 M$ for $M\in\{T^j\}$,
the would-be adjoint scalars, are symmetric. Their antisymmetric part,
the only part that survives between Grassmann fields, therefore vanishes
identically. For $M\in\{\Gamma^j,\one \}$ the matrices are
antisymmetric and survive. The counting $6+30=36$ then follows from
$\dim[\mathrm{antisym}(16)]=120$ reduced by gauge invariance. Physically, for $N_f=2$ valleys with an SU(2) gauge
redundancy, the SO(5)-adjoint (quantum-spin-Hall-type) masses exist only
as gauge adjoints, and the reality structure removes their gauge-singlet
projections. The spin-carrying N\'eel masses belong to the SO(5) vector
and survive.

\section{DM textures}
\label{app:dm}
The allowed DM components are the null space of the invariance condition
$\bm D_{O(i)O(j)}=\det(O)\,O\,\bm D_{ij}$ over one magnetic cell, for each bond class.
In the high-temperature structure the intra-dimer bonds carry 0
components, the inter-dimer bonds carry 1 ($D'_\perp$, alternating as in
Table~\ref{tab:dm}), and both the orthogonal diagonals of the crossed
squares and the diagonals of the empty squares carry 0. In the
buckled structure the intra-dimer bonds carry 1 component, in-plane,
perpendicular to the dimer, and staggered between orientations. This pattern matches Supplementary
Eq.~(4) of Ref.~\cite{Romhanyi2015} in
the relative sign of the two dimer orientations. The
inter-dimer bonds carry 3 components. The orthogonal diagonals of the
crossed squares carry 1, in-plane and perpendicular to the bond, and the
diagonals of the empty squares carry 2, both in-plane. These
next-nearest-neighbor couplings are of negligible range and are not
considered further. The Romh\'anyi frame maps to ours by
$\hat x_R=(\hat x{+}\hat y)/\sqrt2$, $\hat y_R=(-\hat x{+}\hat y)/\sqrt2$,
with the sublattice dictionary $A_1{=}(A,A)$, $A_2{=}(B,B)$,
$B_1{=}(B,A)$, $B_2{=}(A,B)$. With this mapping our reference vector
$\bm D'$ on $A_1\to B_1$ (Table~\ref{tab:dm}) coincides with that of
Ref.~\cite{Romhanyi2015}, while our $D$ equals $-D$ of
Ref.~\cite{Romhanyi2015}. The intra-dimer vectors of Fig.~\ref{fig:dm}(a)
therefore agree with Supplementary Fig.~2 of Ref.~\cite{Romhanyi2015}
only up to this overall sign. The combination
$\tilde D_\parallel=D'_{\parallel,s}-(J'/2J)D$ of Ref.~\cite{Romhanyi2015}
[Supplementary Eq.~(12)] reads $\tilde D_\parallel=D'_{\parallel,s}+(J'/2J)D$
in our convention.

\section{Mean-field decoupling of the DM interaction}
\label{app:wick}
For a bond $\langle ij\rangle$ with DM vector $\bm D$, using
$\sum_{ab}\varepsilon^{cab}\sigma^a\otimes\sigma^b$ Fierz identities and
the singlet contractions
$\langle f^\dagger_{i\alpha}f_{j\beta}\rangle
=\chi_{ij}\delta_{\alpha\beta}$,
$\langle f_{i\alpha}f_{j\beta}\rangle=\Delta_{ij}\varepsilon_{\alpha\beta}$,
the Wick reduction of $\bm D\cdot(\bm S_i\times\bm S_j)$ gives the triplet
hopping
\begin{equation}
\delta H_{ij}=\frac{\ii}{2}\,\bm D\cdot
\bigl(\chi_{ij}\,f^\dagger_j\bm\sigma f_i
-\chi_{ji}\,f^\dagger_i\bm\sigma f_j\bigr)+(\text{pairing channel}),
\end{equation}
with $\chi_{ij}\equiv\langle f^\dagger_{i\sigma}f_{j\sigma}\rangle$
(no sum over $\sigma$), plus terms proportional to
$\langle\bm S\rangle=0$. Here the lattice spin operators are
$\bm S_i=\tfrac12f^\dagger_i\bm\sigma f_i$, written in the Majorana basis
of the Bloch kernels of Appendix~\ref{app:protocol}. The zero-mode matrix
$W$ used throughout the appendices collects the zero-energy eigenvectors of
the $\pi$-flux Bloch kernel at the Dirac point. It maps a lattice
quadratic form onto its continuum bilinear in the 36-operator basis of
Sec.~\ref{sec:inventory}. Summing the bond-resolved
patterns of Table~\ref{tab:dm} over one cell with the $\pi$-flux amplitudes
$\chi_{ij}=\pm\ii\chi_0$ and projecting onto the Dirac subspace yields
Eq.~\eqref{eq:lambda}, and a numerical Wick decoupling on a bulk cell
confirms this coefficient and the vanishing contributions.
For $D'_\perp$ the contributions
cancel over the eight bonds. For the intra-dimer $\bm D$ the chiral
selection rule $\chi^{\rm same\ sublattice}=0$ on the $\pi$-flux state
removes the contribution, and on Z3000 the intra-dimer term appears only
in the gauge-adjoint (Yukawa) sector.

\section{Quadratic-band-touching flow}
\label{app:qbtflow}
\begin{figure}[t]
\centering
\includegraphics[width=0.46\columnwidth]{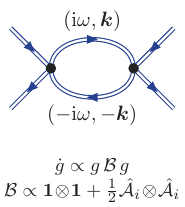}
\caption{The $z{=}2$ bubble of the quadratic-band-touching theory joining
two quartic couplings (dots). Double lines are the eight QBT Majorana
modes, and the bubble tensor $\mathcal B$ of Eq.~\eqref{eq:bubble} sets
the one-loop $\beta$ function of Appendix~\ref{app:qbtflow}. The
disorder graphs are shown in Fig.~\ref{fig:disorder}.}
\label{fig:loops}
\end{figure}
The 14 invariant quartic couplings span the invariant subspace of
$\Lambda^4(\mathbb R^8)$ under the locked group and $\mathcal T$,
projected to the QBT modes. The gap-channel candidates flux$^2$,
VBS$_x^2{+}$VBS$_y^2$, and N$_z^2$ lie in this subspace, and
flux$\cdot$N$_z$ does not. The one-loop bubble at the $z{=}2$ QBT, Fig.~\ref{fig:loops}, follows
from
$G(\ii\omega,\bm k)=(\ii\omega-H_{\rm QBT})^{-1}
=-(\ii\omega+H_{\rm QBT})/(\omega^2+E_{\bm k}^2)$ with
$H_{\rm QBT}=E_{\bm k}(\hat d_1\hat{\mathcal A}_1+\hat d_2\hat{\mathcal A}_2)$,
$\hat d=(\cos2\phi,\sin2\phi)$, $E_{\bm k}=v^2k^2/4\lambda$, and
$\hat{\mathcal A}_i^2=\one $. Since $H_{\rm QBT}(-\bm k)=H_{\rm QBT}(\bm k)$,
the product $G(\ii\omega,\bm k)\otimes G(-\ii\omega,-\bm k)$ has the
even-in-$\omega$ part
$[\omega^2\,\one \otimes\one +H\otimes H]/(\omega^2+E^2)^2$. The
frequency integrals
$\int\!\frac{d\omega}{2\pi}\frac{\omega^2}{(\omega^2+E^2)^2}=\frac1{4E}$ and
$\int\!\frac{d\omega}{2\pi}\frac{1}{(\omega^2+E^2)^2}=\frac1{4E^3}$ give
$\frac{1}{4E_{\bm k}}[\one \otimes\one +\hat d_i\hat d_j\,
\hat{\mathcal A}_i\otimes\hat{\mathcal A}_j]$. The angular average
$\langle\hat d_i\hat d_j\rangle=\tfrac12\delta_{ij}$ and the radial
measure $\int\!\frac{d^2k}{(2\pi)^2}\frac{1}{4E_{\bm k}}
=\frac{\lambda}{2\pi v^2}\!\int\!\frac{dk}{k}$ then yield, per
logarithmic interval $d\ell$,
\begin{equation}
\mathcal B=\frac{\lambda}{2\pi v^2}\Bigl[\one \otimes\one 
+\tfrac12\bigl(\hat{\mathcal A}_1\otimes\hat{\mathcal A}_1
+\hat{\mathcal A}_2\otimes\hat{\mathcal A}_2\bigr)\Bigr]d\ell ,
\label{eq:bubble}
\end{equation}
whose prefactor only sets the unit of RG time. The $\beta$ function is
the antisymmetrized double contraction of two coupling tensors through
$\mathcal B$, and it closes on the invariant space.

The Majorana propagator poles sit at the physical energies. On a single
spinless QBT the same construction gives the marginally relevant
repulsion and the dominant time-reversal-breaking channel of
Ref.~\cite{Sun2009}, with channel rates $2.0$ and $1.0$, and the
single-channel repulsion grows as $dV/d\ell=\rho_0V^2$ as in that work. The $2{:}1$ ratio follows from the kinematic identity
that also annihilates the N\'eel$_z$ rate: a channel matrix that commutes
with $H_{\rm QBT}(\bm k)$ for all $\bm k$ has a vertex rate proportional
to $\int d\omega\,(\omega^2-E^2)/(\omega^2+E^2)^2=0$. Vertex growth rates
use the same bubble.

Basin statistics in Sec.~\ref{sec:qbtir} use 600 random unit seeds of
strength $0.5$ in bubble-log units. Two seeds do not diverge up to
$\ell=60$ and are counted in neither basin. The label is the
fastest-growing gap channel when $|g|$ first exceeds $200$, and the
asymptotic-ray labels follow the flow to $|g|=2\times10^4$. Across the
divergent seeds $\ell^*$ ranges over $4.6$--$52$, with median $10.8$.
The same $\beta$ function follows from a Grassmann enumeration of the $72$
contraction patterns and from a second-order Schrieffer--Wolff
calculation on eight Majoranas, which fix its normalization: for
$H_4=\frac1{4!}g_{ijkl}m_im_jm_km_l$ with $\{m_i,m_j\}=\delta_{ij}$ the
one-loop correction is $\frac32$ times the antisymmetrized double
contraction through $\mathcal B$, and a bilinear source renormalizes with
the factor $\frac12$. Both factors are kept explicitly, so that $\ell^*$
and the vertex exponents are physical for seeds of known normalization
such as the screened gauge interaction of Appendix~\ref{app:qbtgauge}.
Rays, basins and the sign of every asymmetry do not depend on them.

The lattice QBT curvature is isotropic between the two $d$-wave channels,
with the computed ratio $r_2\equiv c_2/c_1=0.9994$, as the continuum
derivation assumes. A hypothetical anisotropy $r_2$ up to $3$, implemented by replacing the bracket in
Eq.~\eqref{eq:bubble} by
$\one \otimes\one +w_1\hat{\mathcal A}_1\otimes\hat{\mathcal A}_1
+w_2\hat{\mathcal A}_2\otimes\hat{\mathcal A}_2$ with angular weights
$w_i$, raises the VBS fraction for a sample of $300$ seeds from
$84\%$ (isotropic) to $87$--$88\%$, and the N\'eel$_z$ rate remains zero.

Only 6 of the 14 invariant quartics are gauge singlets. The 14-coupling
flow describes the Higgsed (Z3000) phase, where all 14 are legitimate
short-range couplings. At the $\pi$-flux critical point only the
gauge-singlet subset survives, and the screened color interaction of
Appendix~\ref{app:qbtgauge} lies inside it. The flow preserves this
subset to $10^{-15}$, and 600 seeds drawn inside it (same measure and
criterion as above) end on valence-bond order for $435$ ($73\%\pm2\%$), on
the flux mass for $149$, and do not diverge for $16$.

\section{Lattice anisotropy computation}
\label{app:plaquette}
The four candidate VBS bond patterns are defined on the nearest-neighbor
square bonds: ${\rm col}_x$, with weight $(-1)^{\min(i_x)}$ on $x$ bonds,
${\rm col}_y$, and the two plaquette alternations $P_{\rm e/f}$. The
latter carry $\pm1$ according to the sublattice of the unique void
(respectively crossed) square adjacent to each bond, the $\alpha/\beta$
voids for $P_{\rm e}$ and the $\gamma/\delta$ crossed squares for
$P_{\rm f}$. Each pattern modulates the ansatz links,
$u_{ij}\to(1+m\,w_{ij})u_{ij}$, and is projected onto the continuum
36-operator basis through the zero-mode matrix $W$ of
Appendix~\ref{app:wick}. The projections give the dictionary quoted in
Sec.~\ref{sec:whichvbs}, including the lattice operator identities
$P_{\rm e}=-{\rm col}_x+{\rm col}_y$ and $P_{\rm f}={\rm col}_x+{\rm
col}_y$. The free energy $F(m,\theta;\lambda)= -\tfrac12\sum_{E>0}E$,
with $E$ the physical excitation energies (twice the eigenvalues of
$\ii A$), is evaluated from the $16\times16$ Bloch reduction of the
$2\times2$-cell magnetic unit cell of $\pi$-flux plus the Wick-induced DM
operator, on momentum grids up to $200^2$. Harmonics are extracted by
projection on $\{1,\sin2\theta,\cos4\theta\}$, and the forbidden
$\cos2\theta$ and $\sin4\theta$ vanish. The results are the same for
both in-plane DM textures and both signs of $d$. The harmonic $b_2$ scales
as $d^4$ down to $d=0.01$, with $b_2/d^4\simeq-1.91\times10^{-6}\,t$ at
$m=0.05$ for $d\le0.05$, so there is no $d^2$ term. A positive $d^6$
correction shifts the ratio to $-1.88$, $-1.73$ and $-1.45\times10^{-6}\,t$
at $d=0.1$, $0.3$ and $0.6$, the last giving the value
$b_2=-1.88\times10^{-7}\,t$ of Fig.~\ref{fig:whichvbs}; material couplings
lie in the quartic regime for $m\gg\lambda$, whereas for $m\ll\lambda$
$b_2\propto|d|^3$, Eq.~\eqref{eq:Aclosed}. At $\lambda=0$ the $\sin2\theta$
harmonic vanishes. The pure $\pi$-flux determinant is therefore
square-lattice symmetric, and the Shastry--Sutherland structure enters
solely through the DM channel. The dimer-bond amplitude $\langle
f^\dagger_i f^{\vphantom\dagger}_j\rangle$ vanishes identically on the
$\pi$-flux state by the same sublattice rule that annihilates the
intra-dimer DM contribution.

\emph{Small-$m$ asymptotics of $b_2$.}
Regular grids cannot resolve the QBT crossover $k_m\sim\sqrt{4\lambda m}$
for $m\ll10^{-2}$. The zone is therefore split into a disc $|\bm k|<0.6$
around $\Gamma$, integrated on a polar grid with log-spaced radii from
$10^{-6}$ to $0.6$, and the remainder, integrated on a regular grid. With
$m$ halved from $0.05$ to $3.9\times10^{-4}$, followed by
$m=2.4\times10^{-4}$, the effective exponent of $b_2$ is
$1.29,1.48,1.60,1.68,1.74,1.78,1.81,1.83$. The exponent of $a_4$ is
$4$ for $m\ge3\times10^{-3}$; below $m\simeq10^{-3}$ the infrared term
$-(3\lambda^5/16\pi t^4)m^2$ dominates, and $a_4$ changes sign near
$m\simeq7\times10^{-4}$. A two-parameter fit
$b_2=A'm^2\ln(1/m)+B'm^2$ over the seven masses $m\le0.0125$ gives
$A'=-8.1\times10^{-5}$ and $B'=+1.8\times10^{-4}$ with a maximal relative
deviation of $1\%$, and a power law $m^p$ gives $p=1.74$ and deviations of
$8\%$. These fit values are the finite-window counterparts of the
constants $A=8.28\times10^{-5}\,t$ and $m_0=0.103$ of
Eqs.~\eqref{eq:b2asym} and \eqref{eq:Aclosed}: the fit over the window,
whose upper end $m=0.0125$ is $\lambda/4$, lies $2\%$ below $A$, and on a disc
quadrature down to $m=3\times10^{-6}$ the local logarithmic slope
reproduces $A$ to $10^{-4}$. The constant $m_0$ follows from the analytic
band-touching constant inside a small disc and the zone integral of the
second-order lattice kernel outside it, and lies within $1.5\%$ of
$2\lambda$ at $d=0.6$. The isotropic coefficient is $C_{\rm iso}=0.0644\,t$
at $d=0.6$, within $0.5\%$ of $4\lambda/\pi$, and $A/C_{\rm iso}$ follows
$(\lambda/t)^2/2$ to $0.5\%$ from $d=0.007$ to $d=0.6$.

The $16$-band $k{\cdot}p$ Hamiltonian at $\Gamma$ expanded to second
order in $k$ reduces on the eight zero modes to
Eq.~\eqref{eq:qbtdxy}. With the same disc quadrature it gives the lattice disc value of
$b_2$ to $0.6\%$ at $m=0.0125$ and the same $A'$ to $0.4\%$. A model linear
in $k$, which does not contain the QBT curvature, has the asymptotic
logarithmic coefficient $2A$ ($1.99A$ at $d=0.6$), and its subleading terms
raise the fitted $A'$ to three times the lattice value over the window; it
overestimates $|b_2|$ by a factor $10$--$26$ across the window.

\emph{Derivation of Eq.~\eqref{eq:Aclosed}.} With the probe, Eq.~\eqref{eq:qbtdxy}
reads $H=k^2\,\bm n(\phi)\cdot\bm\Gamma+M\,\bm u(\theta)\cdot\bm\Gamma$ on
$\bm\Gamma=(\mathcal A_1,\mathcal A_2,V_x,V_y)$, with $M=2mt$,
$\bm u=(0,0,\cos\theta,\sin\theta)$ and
\begin{align}
\bm n&=(a_1\cos2\phi,\,a_2\sin2\phi,\,-c\sin2\phi,\,c\sin2\phi),\nonumber\\
a_2&=\frac{v^2}{4\lambda},\quad a_1=a_2(1+\epsilon),\quad c=\frac{v^2}{4t},
\label{eq:a1a2c}
\end{align}
in magnetic-cell units ($v=t$). The four positive bands
$E=|k^2\bm n+M\bm u|$ enter $F=-\tfrac12\sum_{E>0}E$ per magnetic cell, with the
physical energies (twice the eigenvalues of $\ii A$). The radial integral
$\int k\,dk\,|k^2\bm n+M\bm u|$ contains
$\tfrac{M^2}4(\mathcal N^2-s^2)\mathcal N^{-3}\log(1/m)$, with
$\mathcal N=|\bm n|$ and $s=\bm n\cdot\bm u$, so that with $\psi=2\phi$
\begin{align}
F&\supset-\frac{M^2}{8\pi^2}\log\frac1m\int_0^{2\pi}\!d\psi\,
\frac{\mathcal N^2-s^2}{\mathcal N^3},\nonumber\\
\mathcal N^2&=a_1^2\cos^2\psi+b^2\sin^2\psi,\quad b^2=a_2^2+2c^2,\nonumber\\
s^2&=c^2\sin^2\psi\,(1-\sin2\theta).
\end{align}
The $\sin2\theta$ part is $b_2=A'M^2\log(1/m)$, i.e.\ $A=-4t^2A'$, with
\begin{align}
A'&=-\frac{c^2}{8\pi^2}J,\qquad
J=\int_0^{2\pi}\!\frac{d\psi\,\sin^2\psi}{(a_1^2\cos^2\psi+b^2\sin^2\psi)^{3/2}},
\nonumber\\
J&=\frac{4}{a_1^3\kappa^2}\Bigl[\frac{E(\kappa)}{1-\kappa^2}-K(\kappa)\Bigr],
\qquad\kappa^2=1-\frac{b^2}{a_1^2},
\label{eq:Jell}
\end{align}
where $K(\kappa)=\int_0^{\pi/2}(1-\kappa^2\sin^2\vartheta)^{-1/2}d\vartheta$ and
$E(\kappa)$ are the complete elliptic integrals with parameter $\kappa^2$. Since
$b^2=a_2^2(1+\epsilon)$, $\kappa^2=\epsilon/(1+\epsilon)$, and
$A=t^2c^2J/(2\pi^2)$ becomes Eq.~\eqref{eq:Aclosed}; its expansion in $\epsilon$
gives the series there, and $A=8.282\times10^{-5}\,t$ at $d=0.6$. The
$\theta$-independent part gives the matching integral
\begin{align}
C_{\rm iso}&=\frac{t^2}{2\pi^2}\int_0^{2\pi}\!d\psi\,
\frac{\mathcal N^2-c^2\sin^2\psi}{\mathcal N^3}\nonumber\\
&=\frac{t^2}{2\pi^2}\Bigl[\frac{4K(\kappa)}{a_1}-c^2J\Bigr]\simeq\frac{4\lambda}{\pi},
\end{align}
$0.06439\,t$ at $d=0.6$. The nonlogarithmic $m^2$ term depends on $s$ through
$\tfrac12(\mathcal N^2-s^2)\mathcal N^{-3}\log(1+s/\mathcal N)$, whose $s^4$ term
gives
\begin{equation}
a_4\simeq-\frac{M^2c^4}{64\pi^2}\int_0^{2\pi}\!d\psi\,\frac{\sin^4\psi}{\mathcal N^5}
\simeq-\frac{3\lambda^5}{16\pi t^4}\,m^2
\end{equation}
to leading order in $\lambda/t$; the full band-touching value is
$-2.000\times10^{-8}\,t\,m^2$ at $d=0.6$.

\section{One-loop disorder RG}
\label{app:disorderrg}
\begin{figure*}[t]
\centering
\includegraphics[width=0.96\textwidth]{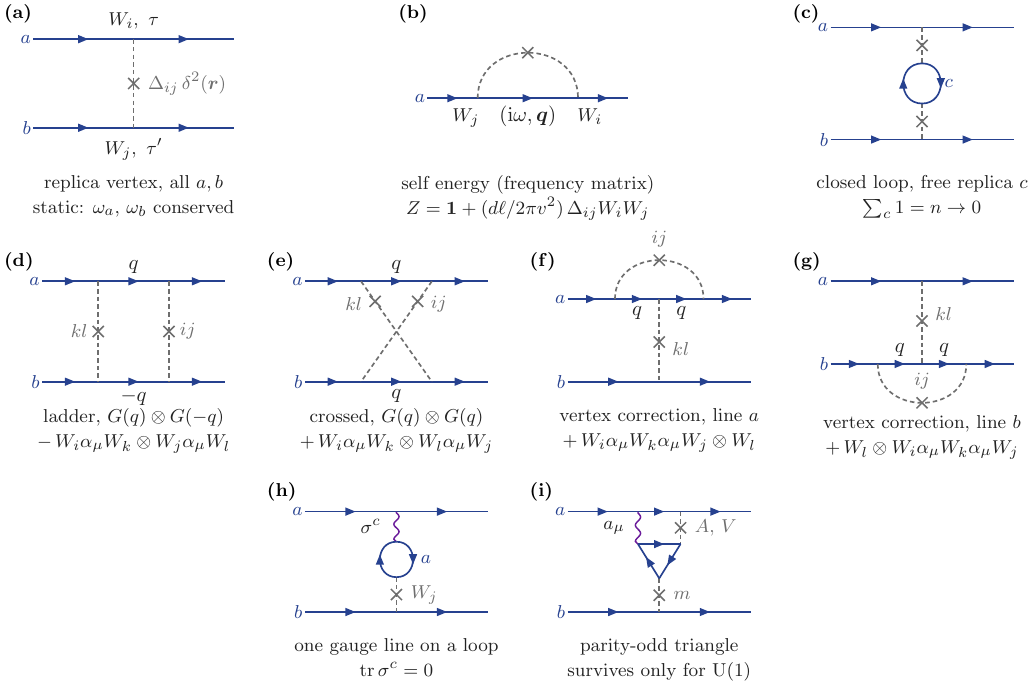}
\caption{One-loop diagrams of the replicated static-disorder problem, in
the spinon basis $\psi=(\text{Dirac})\otimes(\text{valley})\otimes(\text{spin})$
in which all 36 singlets are number conserving. Solid indigo lines:
$\langle\psi\psi^\dagger\rangle=(\ii\omega+H)/(\omega^2+v^2k^2)$; $a,b,c$
are replica indices. Gray dashed lines with $\times$: the static
correlator $\Delta_{ij}\delta^2(\bm r)$, which transfers momentum but no
frequency. Matrix products are read against the arrow, and $ij$, $kl$
mark the covariance entries. (a) Replica vertex
$-\tfrac12\sum_{ab}\Delta_{ij}(\psi^\dagger_aW_i\psi_a)_\tau(\psi^\dagger_bW_j\psi_b)_{\tau'}$.
(b) Self-energy, the only $O(\Delta)$ term. It is momentum independent
and renormalizes the frequency term into the matrix $Z$. For the
random-$\lambda$ vertex it is the entire one-loop flow,
Eq.~\eqref{eq:randomlambda}. (c) Closed loops carry a free replica sum and
vanish as $n\to0$. (d)--(g) Renormalization of the replica-off-diagonal
vertex. Each entry is multiplied by $\Delta_{ij}\Delta_{kl}\,d\ell/(4\pi
v^2)$ and summed over $\mu=x,y$ and all indices: (d) ladder,
$G(q)\otimes G(-q)$; (e) crossed, $G(q)\otimes G(q)$; (f), (g) vertex
corrections on either replica line, the multiplicity two of the
same-line dressing. The gauge dressing of each random-field vertex is the
vertex correction of Fig.~\ref{fig:diagrams}(b), which enters through
$\Delta_i$ in Eq.~\eqref{eq:disflowN}. (h) A random-field vertex that ends
on a fermion loop tied to the fermion line by one gauge propagator
(wavy). For the $\mathcal T$-even gauge singlets the color trace of the
loop is $\operatorname{tr}\sigma^c=0$, also with further random-field
vertices on the loop. (i) The parity-odd triangle of a zero-momentum mass
$m$, a random potential or vector potential and one gauge line, which
survives only for a U(1) gauge field $a_\mu$; in SU(2) it carries a single
generator and vanishes.}
\label{fig:disorder}
\end{figure*}

Static disorder couples as $\sum_j\int d\tau\,d^2r\,V_j(\bm
r)\,\psi^\dagger W_j\psi$ with $\overline{V_iV_j}=\Delta_i
\delta_{ij}\delta^2(\bm r)$. The replica quartic is treated at second
order in the cumulant expansion with the fast-shell propagator
$G(k)=-(\ii\omega+H)/(\omega^2+k^2)$, $H=k_xa_x+k_ya_y$,
$a_i=\gamma^0\gamma^i$ (written $\alpha_\mu$ in Fig.~\ref{fig:disorder}
and below). The three contraction topologies, ladder,
crossed, and same-line vertex dressing with multiplicity two,
Fig.~\ref{fig:disorder}(d)--(g), are enumerated
directly. Angular
averages use $\langle k_mk_n\rangle=\tfrac12\delta_{mn}k^2$ and the
shell measure $\int'\!d^2k/(2\pi)^2k^{-2}=d\ell/2\pi$. The frequency
renormalization must be kept as a matrix,
$Z=\one +\kappa\sum_jW_jW_j$ with $\kappa=\Delta\,d\ell/2\pi$, and
canonically normalized with $R=Z^{-1/2}$ acting on the velocity
($R\,a_iR$) and on the vertex ($R\,W R$) alike. Throughout this appendix
the vertices are normalized to $\operatorname{tr}W^2=\dim W$, so that
the potential vertex is $\one $, and flow coefficients are quoted in
units of $\Delta^2/(2\pi)$. For scalar $Z$ the velocity and the vertex
rescale identically and the vertex-tensor values stand: potential
$+2\Delta^2/(2\pi)$ (marginally relevant), mass $-2\Delta^2/(2\pi)$
(marginally irrelevant), abelian gauge pair $\equiv0$ (exact line, since here vertex and velocity
are the same matrices). Reference~\cite{LFSG1994} obtained these three
results for a single Dirac fermion: the mass flow as its Eq.~(75), the
potential by mapping onto an imaginary mass (its Sec.~VII), and the exact
vector-potential fixed line (its Sec.~VI).

The one-loop content is collected in Fig.~\ref{fig:disorder}. Because the
disorder is static and $\delta$-correlated, the self-energy (b) is
independent of momentum and gives, for a general covariance,
$Z=\one+(d\ell/2\pi v^2)\sum_{ij}\Delta_{ij}W_iW_j$ with no velocity
term. Closed loops (c) vanish in the replica limit. The four diagrams
(d)--(g) are the complete set of second-order terms with four slow legs on
two replica lines; the remaining contractions are self-energies on
external legs, already contained in $Z$, or disconnected. Writing
$W_i\alpha_\mu W_k=S^\mu_{ik}+A^\mu_{ik}$ with
$A^\mu_{ik}=\tfrac12(W_i\alpha_\mu W_k-W_k\alpha_\mu W_i)$, the sum of (d)
and (e) is $-(d\ell/2\pi v^2)\sum\Delta_{ij}\Delta_{kl}A^\mu_{ik}\otimes
A^\mu_{jl}$. The gauge-charged parts $S\otimes S$ cancel between ladder
and crossed because $G(-q)=-G(q)$ at zero frequency, and (f) and (g) are
gauge singlets by construction, so a gauge-singlet covariance generates
only gauge singlets. For the density vertex, (f) and (g) each equal
$Z-\one$ (Ward identity), and the $+2$ of the potential class is the
reduction of the velocity. For a single mass the same diagrams give $-2$.
For the abelian pair, (d) and (e) cancel and (f) and (g) vanish. For the
random-$\lambda$ vertex every diagram (d)--(g) contains
$W_0\alpha_iW_0=0$, and the flow comes from (b) alone. With
$W_0=a_x\sigma^x+a_y\sigma^y$, $\beta=-\ii a_xa_y$ and
$P_\pm=(\one\pm\beta\sigma^z)/2$ one has $W_0^2=4P_-$, and we write the
frequency matrix by sector, $Z=z_+P_++z_-P_-$. Every self-energy
diagram begins and ends on $W_0=P_-W_0P_-$, so $\Sigma=P_-\Sigma P_-$ to
all orders, $z_+$ and $v$ do not run, and $g=\Delta z_+/(v^2z_-)$ is
proportional to the Born scattering rate in units of the energy. Its
decrease, Eq.~\eqref{eq:randomlambda}, is therefore a decrease of the
scattering rate, independent of the normalization convention.

The random-DM vertex is valley-blind, so the per-valley $4\times4$
space closes. Written without normalization, $W_0=a_x\sigma^x+a_y\sigma^y$
has $W_0^2=2(\one -\beta\sigma^z)$ and $W_0a_iW_0=0$ identically,
and it is the one case with non-scalar
$Z=\one +2\kappa(\one -\beta\sigma^z)$. With
$R=P_++P_-/\sqrt{1+4\kappa}$ one finds $Ra_iR=a_i/\sqrt{1+4\kappa}$ but
$RW_0R=W_0/(1+4\kappa)$, hence $\Delta_{\rm eff}=\Delta/(1+4\kappa)$ and
$d\Delta_\lambda/d\ell=-4\Delta_\lambda^2/(2\pi)=-2\Delta_\lambda^2/\pi$,
which is Eq.~\eqref{eq:randomlambda} for the variance of the
unnormalized vertex $W_0$. The trace-normalized vertex is $W_0/\sqrt2$,
whose variance is $\Delta'_\lambda=2\Delta_\lambda$ and obeys
$d\Delta'_\lambda/d\ell=-2\Delta'^2_\lambda/(2\pi)$, the value $-2$ of
the ensemble list below. The two statements describe the same flow in
the two normalizations, a marginal irrelevance without a fixed line.
Retaining the generated $\beta\sigma^z$ frequency tensor explicitly as a
coupling closes the same flow: only $z_-$ runs,
$\dot z_-=2\Delta z_+/(\pi v^2)$ from $\sum_iW_0\alpha_iZ\alpha_iW_0
=8z_+P_-$, and $g(\ell)=g_0/(1+2g_0\ell/\pi)$. The full $16\times16$ tensor
projection of every one-loop topology vanishes because each factor
contains $W_0\alpha_iW_0=0$.

\emph{Multi-vertex flow of the matched DM covariance.}
The 36 gauge singlets are represented as $8\times8$ Hermitian matrices
in the (Dirac, valley, spin) basis, a complex structure in which all 36
are number-conserving, and completed to the 64-dimensional
$\rho^a\mu^b\sigma^c$ basis of $u(8)$, on which the three topologies
close. The 28 gauge-charged elements are carried along. In units of
$\Delta^2/(2\pi)$ for trace-normalized vertices the multi-vertex flow gives
potential $+2$, mass $-2$, abelian gauge $0$, and random-$\lambda$ vertex
tensors $0$ with normalization $-2$, in agreement with the values above. Matching proceeds bond by
bond. For each of the 8 NN and 2 dimer bonds of the reference cell and
each $\delta D_\alpha$, the Wick-decoupled quadratic operator is
projected with the zero-mode matrix $W$ on the 36 singlets. On the NN
bonds the projection has the same norm, $2.65\times10^{-3}$ per unit $D$,
for every bond and component, and the dimer-bond operators have no
projection. The covariance $\Delta_0=\sum c\,c^{\sf T}$ of the projected vectors $c$ has
rank 12 with a twelvefold degenerate spectrum. Its class weights are
$1/2$ ($J^{x,y}\sigma^a$), $1/2$ ($\gamma^0\mu^{x,z}\sigma^a$), and
$0$ (masses, magnetization, $\mu^y$). The flow gives
$\operatorname{tr}(\dot\Delta\Delta_0)/\operatorname{tr}\Delta_0^2=+2/3$,
all from the vertex tensors, the normalization contributing $0$ because
$Z-\one \propto\one $ for $\Delta_0$. The largest
eigendirections of $\dot\Delta$ are pure singlets, and the generated
mass diagonal is $\mu^x{:}\,\mu^z=1{:}1$ with all other masses $0$.
Integrating from $0.05\,\Delta_0$ to $\ell=80$ gives
$\operatorname{tr}\Delta=0.050\to0.130$, with weights random-$\lambda$
$0.50\to0.51$, staggered density $0.50\to0.33$, masses $0\to0.13$, the
remaining $0.04$ in other $\mathcal T$-even singlets, and gauge-charged $0$
throughout. For this ensemble and normalization the
linear extrapolation of $1/\operatorname{tr}\Delta$ reaches zero at
$\ell\simeq130$, a property of the model covariance that sets no material
scale.

\emph{Ensemble dependence.} For other covariances built from the
single-bond vectors, with the same matching and flow, we list
the rank, the projected coefficient
$\operatorname{tr}(\dot\Delta\hat\Delta)/\operatorname{tr}\hat\Delta^2$,
and $\operatorname{tr}\Delta$ after integrating from $0.05$ to
$\ell=30$:
\begin{itemize}
\item Independent components with equal variance (the isotropic case
above): rank 12, $+2/3$, $0.064$.
\item Independent in-plane components: rank 8, $0$, $0.065$.
\item Independent out-of-plane components: rank 4, $+1/2$, $0.071$.
\item Independent bond amplitudes of the allowed staggered
(non-staggered) in-plane pattern: rank 8, $0$, $0.065$ ($0.065$).
\item Independent bond amplitudes of the out-of-plane pattern: rank 4,
$+1/2$, $0.071$.
\item One cell-correlated amplitude of the staggered or non-staggered
in-plane pattern: rank 1, $-2$, $0.034$, which is the random-$\lambda$
vertex of Eq.~\eqref{eq:randomlambda}.
\item The cell-correlated out-of-plane pattern: zero projection.
\end{itemize}
The growth of the zero-coefficient ensembles comes entirely from the
generated VBS-mass directions. The DM-generated covariance is thus
marginally relevant at one loop for every bond-uncorrelated ensemble
and marginally irrelevant for the smooth (cell-correlated) one,
Eq.~\eqref{eq:multivertex} being the isotropic bond-uncorrelated case.

\emph{Gauge fluctuations at $O(1/N)$.} The replicated critical theory is
$S_n=\sum_aS_*[\X_a,A_a,\Phi_a]-\tfrac12\sum_{ab}\Delta_{ij}\int d^2r\,d\tau\,
d\tau'\,O_i(\bm r,\tau)_aO_j(\bm r,\tau')_b$, with one gauge field and one
pair of Higgs triplets per replica and $O_i$ running over the 16
$\mathcal T$-even gauge singlets: the two VBS masses, the six spin currents
$\gamma^{x,y}\sigma^a$, the six staggered densities
$\gamma^0\mu^{x,z}\sigma^a$ and the two VBS-rotation currents
$\gamma^{x,y}\mu^y$. In the $8\times8$ basis time reversal is
$\Theta=\rho^y\mu^y\sigma^y$ followed by complex conjugation, which
reproduces the parities of Table~\ref{tab:unlocked} and
$H(\bm k)\to H(-\bm k)$. For exchange disorder the covariance lives on the
four spin singlets and has three invariants under the layer group,
$\Delta_+$, $\Delta_-$ and $\Delta_J$, the plaquette components forming two
one-dimensional representations; for DM disorder the locked group leaves 24
invariants on the 16 channels.

The gauge line is replica diagonal, so at $O(\Delta/N)$ it dresses one end
of a disorder line at a time: vertex correction plus self-energy, which for
a static insertion have the same logarithm as at zero momentum. With the
rules of Appendix~\ref{app:integrals} a mass vertex has
$v=8C_g(3-\zeta)/(\pi^2N)$ and a density or spatial-current vertex
$v=8C_g(1-3\zeta)/(3\pi^2N)=-s$, so the dimensions entering
Eq.~\eqref{eq:disflowN} are $2-64C_g/(3\pi^2N)$ for the VBS and N\'eel
masses and $2$ for all 30 current components, independent of $\zeta$. At
$O(\Delta^2)$ the only graphs beyond the free kernel are fermion loops tied
to the external line by gauge propagators (Fig.~\ref{fig:disorder}(h),(i)), since loops tied by disorder
lines alone carry a free replica sum. Charge conjugation removes the loops
with vertex sets $\{m,m,A\}$, $\{j,j,A\}$ and $\{j,A,A\}$, and the
zero-momentum current insertion on $\{j,m,A\}$ vanishes by the Ward
identity. The remaining triangle, a zero-momentum mass with a random vector
potential or potential and one gauge line, is parity odd and transverse,
hence independent of the gauge parameter, and changes the mass vertex by
$\kappa g\,d\ell$ per unit variance with $\kappa_A=-8F/(2\pi N)$ and
$\kappa_V=+8F/(2\pi N)$, where $F$ is the flavor trace of the loop. In
QED$_3$, $F=N$ for $T_X=T_Y=\sigma^a$, so that the triangles add
$8g_tg_A-8g_tg_v$ to $\pi\beta_t$ in the notation of
Ref.~\cite{ThomsonSachdev2017}, and Eq.~\eqref{eq:disflowN} with these
terms reproduces its Eqs.~(50) and~(51) term by term. The fixed line with
U(1)$\times$SU($N$) flavor symmetry that follows from its Eq.~(51) is
$g_t=c-6g_A$ with $z=1+(c-4g_A)/(2\pi)$. A single random Dirac mass has the
stable fixed point $z^*=1+64/(3\pi^2N)=1+2.16/N$ for $N$ two-component
flavors; the random mass of Ref.~\cite{GoswamiGoldmanRaghu2017} is a flavor
non-singlet in two-component language, and its $\epsilon$-expansion value
$z^*=1+9\epsilon/(8N_4)$ at $\epsilon=1$, with $N_4=N/2$ four-component
flavors, equals $1+2.25/N$. In the SU(2) theory the triangle carries a
single gauge generator and vanishes, and the two-gauge-line class requires
$\operatorname{tr}T_X\neq0$, which holds only for the $\mathcal T$-odd flux
mass. The same normalization reproduces the annihilation value
$-64/(\pi^2N)$ of the flavor-singlet mass. A random potential on the U(1)
gauge charge is screened at both replica ends, its correlator vanishing as
$k^2$. Reference~\cite{GoswamiGoldmanRaghu2017} dresses a single end and finds a correlator $\propto|k|$ [its Eq.~(C5)]; in both cases the random potential is irrelevant at the QED$_3$ fixed point.

The kernel $B$ is the one of this appendix, and Eq.~\eqref{eq:plaqflow}
follows from it on the invariant subspace
$\{\Delta_+,\Delta_-,\Delta_J\}$, with no weight generated outside it. A
random $\mathcal T$-even initial covariance generates no $\mathcal T$-odd
and no gauge-charged entries, in accord with the per-replica selection
rule. The times to strong disorder quoted in Sec.~\ref{sec:disorder} are
those at which the largest class variance reaches $\pi$; for vacancies the
initial covariance is that of the near--far imbalance. The
cell-correlated random $\lambda$ generates no mass and stays marginally
irrelevant, Eq.~\eqref{eq:randomlambda}. At $\ell^*$ the covariance is
dominated by the two plaquette components with equal variance and the
VBS-rotation current at half of it.

The Higgs dimensions follow from the same logarithms with the scalar
propagator $1/p^2$, the gauge vertex $(p+p')_\mu t^b$, the seagull, and the
constraint field with $\langle\sigma\sigma\rangle=16|q|/M$. The flavor
current of the scalars keeps $\Delta=2$ at every $\zeta$ once both seagull
graphs are included; the gauge part of a flavor-traceless gauge-singlet
bilinear is $-64C_s/(3\pi^2N)$, independent of $\zeta$, and the constraint
part $+32/(3\pi^2M)$. For $N$ complex unit-charge scalars this gives
$\eta_N=1-32/(\pi^2N)$, the CP$^{N-1}$ value of
Ref.~\cite{IrkhinKataninKatsnelson1996} (its Table~I at $d=3$). For the
composites $\Phi^a\Tr[\sigma^a\Xb M\X]$ the fermion--Higgs exchange graph
is proportional to $(1-\zeta)$; independence of the gauge parameter fixes
its coefficient to the Casimir value $8|T_\psi\cdot T_\Phi|/(\pi^2N)$ with
$T_\psi\cdot T_\Phi=-C_s/2$, and the neutral limit $C_s\to0$ returns
$\Delta_\Phi+\Delta_{\bar\psi M\psi}$. This gives
$\Delta=5/2-(8C_s+64C_g)/(3\pi^2N)+4/(3\pi^2M)=1.80$ for a mass-type and
$5/2-40C_s/(3\pi^2N)+4/(3\pi^2M)=1.62$ for a current-type $M$ at $N=12$.

\emph{Vacancy at mean-field level.} The Nambu Hamiltonian
$H=-\sum\psi_i^\dagger u_{ij}\psi_j+{\rm h.c.}$ of the $\pi$-flux and
reference Z3000 ans\"atze is number conserving in
$\psi=(f_\uparrow,f_\downarrow^\dagger)$, so each state is a Slater
determinant of $\psi$, with $S^z=(\psi^\dagger\psi-1)/2$ and
$\langle\bm S_i\cdot\bm S_j\rangle=\tfrac14[(n_i-1)(n_j-1)-\operatorname{tr}
G_{ij}G_{ji}]+\operatorname{Re}\det G_{ij}$, $G_{ij}=\langle\psi_i^\dagger
\psi_j\rangle$. A vacancy removes one site and its bonds on clusters of
$(2L)^2$ sites, $L=6$--$16$, with antiperiodic and open boundaries. The
zero-field state is the average of the two members of the Kramers doublet,
in which the pair of $\psi$ modes closest to zero energy is filled or left
empty, $S^z=\pm1/2$. In the $\pi$-flux state $(-1)^{x+y}$ anticommutes
with $H$ and the removed site leaves an index of two, matched by two
zero modes at every size and for both boundaries, with weight below
$10^{-28}$ on the vacancy sublattice. The weight inside $r\le2$ and $r\le4$
follows $1/W=a\ln L+b$ with residuals below $10^{-3}$ (antiperiodic) and
$10^{-2}$ (open). The bond-energy response $\delta e$ on nearest-neighbor
bonds is projected on the lattice harmonics of the zero-gradient channels,
$V_x=\sum_x(-1)^x\delta e$ and $V_y=\sum_y(-1)^y\delta e$, with
$P_\mp=V_x\pm V_y$ for the full and empty plaquette, and the staggered
harmonics $(-1)^{x+y}$ for the VBS-rotation current, whose signs follow
from the zero-mode projection of single bonds. With antiperiodic boundaries
every harmonic is below $10^{-14}$ and the pattern is invariant under $C_4$
and the site mirror to $10^{-15}$; on open clusters, where the vacancy sits
half a site from the center, the local harmonics decrease as $1/L$
together with the $C_4$ residual. Relaxing the amplitudes of all
interdimer and dimer bonds within $r\le3$ self-consistently, with the bulk
fixed at the $\pi$-flux saddle, keeps $C_{4v}$ to $10^{-14}$ and the dimer
amplitudes at zero. In the Z3000 state the chiral operator is
$(-1)^{x+y}\tau^z$, with $\tau^z$ acting on the Nambu index, whose trace
over a site vanishes, and the pair of $\psi$
modes closest to zero energy splits by $1.8\times10^{-3}$ to
$1.2\times10^{-1}$, depending on size and boundary twist, while the Kramers
doublet stays degenerate. For individual twists that break the site mirror
the empty-plaquette harmonic reaches $1.3\times10^{-3}$; for diagonal twists
and after averaging over $4\times4$ twists it is below $10^{-15}$, as the
site mirror requires. Within $r\le3$ the twist-averaged full plaquette is
$1.0\times10^{-2}$, $6.8\times10^{-3}$, $7.1\times10^{-3}$ and
$6.1\times10^{-3}$ and the current $-4.9\times10^{-3}$ to
$-3.3\times10^{-3}$ for $L=6$--$12$; the extrapolation of the full
plaquette in $1/L$ stays finite, $(2$--$6)\times10^{-3}$, and both
harmonics grow faster than linearly with the Z3000 amplitudes. The
$\mathcal T$-odd observables, the magnetization and the scalar chirality on
the triangles of the sixteen plaquettes around the vacancy, are below
$10^{-14}$ in the $\mathcal T$-invariant state in every case.

\section{Induced gauge propagator: the adjoint-Higgs polarization}
\label{app:polarization}

The critical theory carries no Maxwell term. The gauge propagator is
generated entirely by matter loops, so the count $N=2N_f+4N_b$ of
Sec.~\ref{sec:continuum} and the propagator $D_{\mu\nu}$ that
Appendix~\ref{app:integrals} takes as input are themselves computed
quantities. The fermions of each valley form one
complex two-component Dirac gauge doublet with covariant derivative
$D_\mu=\partial_\mu-\ii A^a_\mu\sigma^a$ and bare vertex
$\gamma_\mu\sigma^a$, so that $\operatorname{tr}\sigma^a\sigma^b
=2\delta^{ab}$. The Higgs fields are real adjoint triplets with
$(D_\mu\Phi)_a=\partial_\mu\Phi_a-\ii A^b_\mu(t^b)_{ac}\Phi_c$ and
$(t^b)_{ac}=-2\ii\epsilon_{bac}$, so that
$\operatorname{tr}(t^at^b)=8\delta^{ab}$. The one-loop polarization is
transverse and reads
\begin{equation}
\Pi^{ab}_{\mu\nu}(p)=\delta^{ab}
\Bigl(\delta_{\mu\nu}-\frac{p_\mu p_\nu}{p^2}\Bigr)
\frac{(2N_f+4N_b)\,|p|}{16}.
\label{eq:PiLO}
\end{equation}
The two weights arise as follows. A two-component Dirac fermion of unit
charge screens with $|p|/16$ (Appendix~\ref{app:integrals}), and
the doublet loop with two Pauli vertices multiplies this by
$\operatorname{tr}\sigma^a\sigma^b=2\delta^{ab}$, giving weight 2 per
valley. Writing a valley as eight real Majorana components
(Appendix~\ref{app:proof}) does not change this weight: realification
doubles the trace to $4$ and the Majorana loop halves it back to $2$. A
complex scalar of unit charge also screens with $|p|/16$
(Appendix~\ref{app:integrals}). A real adjoint triplet integrates to
$\tfrac12\operatorname{Tr}\log$, so its weight is
$\tfrac12\operatorname{tr}(t^at^b)=4$ per triplet. The critical sector
contains the two real triplets $\Phi_1,\Phi_2$, hence $N_b=2$ real
triplets and $N=2N_f+4N_b=4+8=12$.

Two limits fix the fermion weight independently. In the U(1)
sublimit in which only $A^3_\mu$ is kept, $\sigma^3=\pm1$ on the two
doublet components, so each valley contributes two unit-charge
two-component fermions and $\Pi^{33}=2N_f|p|/16=4|p|/16$, the $N=4$
QED$_3$ that is the standard count for the $\pi$-flux Dirac spin liquid.
The companion U(1) analysis of Ref.~\cite{Feuerpfeil2026PRB} uses the
same rule, $\Pi_a=(N_f+4N_b)|p|/16$ in its notation. There each
two-component fermion carries weight 1, with $N_f=4$ for its triangular
and kagome theories and $N_f=12$ for its maple-leaf theory, and each
complex charge-2 scalar carries weight 4. Reference~\cite{Feuerpfeil2026} uses the same count, $\Pi_A=(2N_f+4N_b)|p|/16$ in its Eq.~(64). Inverting Eq.~\eqref{eq:PiLO} gives
$D_{\mu\nu}=\frac{16}{N|q|}(\delta_{\mu\nu}-\zeta\hat q_\mu\hat q_\nu)$
and hence $\dim[\bar\psi\mu^i\psi]=2-64/(\pi^2N)=1.46$ at $N=12$
(Sec.~\ref{sec:spin2}). The $O(1/N)$ correction to
Eq.~\eqref{eq:PiLO}, comprising two-loop matter blocks, a ghost loop, and the
induced three-gauge-boson vertex class generated by matter triangles, is
what an $O(1/N^2)$ determination of $\Delta_V$ would require. It is not
evaluated here.

\section{DM-induced Yukawa couplings}
\label{app:yukawa}
\emph{Scaling dimensions.} At the SO(5)- and Lorentz-invariant point the
$\gamma^0$ channel of Eq.~\eqref{eq:newyuk},
$\Phi_{1,2}^a\Tr[\sigma^a\Xb\mu^y\sigma^b\gamma^0\X]$, and the Yukawa
coupling $y$ of Ref.~\cite{Feuerpfeil2026},
$\Phi_{1,2}^a\Tr[\sigma^a\Xb\mu^{z,x}\gamma^{x,y}\X]$, are components of the
same multiplet, SO(5) vector times Lorentz vector times gauge adjoint, and
have the same dimension to all orders. The scalar channel
$\Phi_{1,2}^a\Tr[\sigma^a\Xb\X\sigma^b]$ has the Dirac structure $\one$.
Reference~\cite{Feuerpfeil2026} evaluates the two vertex graphs for a Yukawa
vertex with and without Dirac structure [its Eqs.~(97)--(99)], and its
Eq.~(100) gives at $N_f=N_b=2$ the RG eigenvalues $1.378$ for
$\Gamma\propto\gamma^\mu$, which includes $y$, and $1.198$ for
$\Gamma\propto\one$; the scalar DM channel is of the second type. The
following evaluation corroborates their difference.
The fermion and Higgs self-energies do not depend on the
Dirac structure $\Gamma$ of the vertex. In the graph with the gauge line
on the Higgs leg a single fermion propagator separates the two vertices,
and the Higgs--gauge vertex $q_\nu$ contracts with the adjacent
$\gamma_\mu$, $\gamma_\mu\slashed q\,\Gamma\,q_\nu(\delta_{\mu\nu}-\zeta
q_\mu q_\nu/q^2)=(1-\zeta)q^2\Gamma$, so this graph is the same for every
$\Gamma$. In the graph joining the two fermion legs, in $d=3$ with
$2\times2$ Dirac matrices, $\gamma_\mu\slashed q\Gamma\slashed q\gamma_\mu$
averages to $q^2\Gamma/3$ for $\Gamma=\gamma^\nu$ and to $3q^2\Gamma$ for
$\Gamma=\one$, and the gauge-adjoint color factor is
$\sigma^b\sigma^a\sigma^b=-\sigma^a$. In units $u=8/(3\pi^2N)$, this graph gives
$(1-3\zeta)u$ for $\gamma^\nu$ and $(9-3\zeta)u$ for $\one$, so
\begin{equation}
\Delta_{\rm scalar}-\Delta_y=8u=\frac{64}{3N\pi^2},
\end{equation}
gauge independent. The same graph with the singlet factor
$\sigma^b\sigma^b=3$ gives $-24u=-64/(N\pi^2)$, the shift $\Delta_V-2$
of Eq.~(79) of Ref.~\cite{Feuerpfeil2026}, and for U(1) the QED$_3$
flavor-adjoint mass, of dimension $2-64/(3\pi^2N)$~\cite{RantnerWen2002,Hermele2005} (Ref.~\cite{RantnerWen2002} counts four-component flavors, $N/2$). In the gauge-adjoint channel the
exchange is repulsive: at $N=12$ the scalar channel is less relevant
than $y$ by $0.18$, with eigenvalues $\omega_y\simeq1.38$ and
$\omega_s\simeq1.20$.
These values are qualitative, since the Higgs sector that enters them
is not controlled at $N=12$: its constraint field has $\Delta_\sigma=0.38$,
Eq.~\eqref{eq:Dsigma}, below the unitarity bound.

\emph{Bare sizes.} The Z3000 amplitudes of Ref.~\cite{Feuerpfeil2026}
condense $\Phi_{1,2}$ in the gauge $x$--$y$ plane ($\Delta_d$,
$\Delta_{d'}$, $\Delta_{g,1}$ at zeroth order, $\Delta_{g,2}$ with a
gradient) and $\Phi_3^z$ ($\delta\theta$). The couplings follow from
deforming the $\pi$-flux ansatz along each amplitude $s$ separately, with
the ground-state covariance on $8\times8$ cells, the DM bonds of one
central cell Wick-decoupled on that covariance as in Appendix~\ref{app:wick}, and both
the derivative of the hopping and that of the DM operator with respect to
$s$ projected on the 16 node modes. The kinetic image coincides with the coupling of
Ref.~\cite{Feuerpfeil2026}, $\Tr[\sigma^x\Xb(\gamma^x\mu^z-\gamma^y\mu^x)\X]
+\Tr[\sigma^y\Xb(\gamma^x\mu^z+\gamma^y\mu^x)\X]$ with weight
$\Delta_d+\Delta_{d'}+2\Delta_{g,1}$, and vanishes for $\Delta_{g,2}$ and
$\delta\theta$ at zeroth order. The nonvanishing DM images decompose with
residuals below $10^{-14}$. The intra-dimer $D$ gives the scalar and $\gamma^0$
channels with equal norms, and relative to the kinetic image per unit
$D/t$ they are $0.054$ for $\Delta_d$, $0.0064$ for $\Delta_{d'}$ and
$0.012$ for $\Delta_{g,1}$; $\Delta_{g,2}$ adds $0.048\,\delta$, with
$\delta=\Delta_{g,2}/(\Delta_d+\Delta_{d'}+2\Delta_{g,1})$ the gradient
ratio of Ref.~\cite{Feuerpfeil2026}. For the representative amplitudes
$(\delta\theta,\Delta_d,\Delta_{d'},\Delta_{g,1},\Delta_{g,2})
\propto(0.30,0.4,0.25,0.15,0.10)$ the ratio is $0.033$. The in-plane
inter-dimer couplings give only the $\Phi_3$ channel, through
$\delta\theta$, and $D'_\perp$ gives none. The ratios are stable to
$2\times10^{-3}$ under a change of the finite-difference step, and at $s=0$
the same projection gives $\lambda=0.0847\,D'_\Sigma$, the coefficient of
Eq.~\eqref{eq:lambda}.

\emph{Consequence.} Both channels are generated at the lattice scale with
the ratio $r$ to $y$. Under the flow $y(\ell)=y_0e^{\omega_y\ell}$ the
$\gamma^0$ channel stays at $ry(\ell)$, and the scalar channel grows as
$ry_0e^{\omega_s\ell}$ with $\omega_s-\omega_y=-0.18$, so its
ratio to $y$ decreases. For $r\simeq6\times10^{-4}$--$2\times10^{-2}$
both stay at or below $ry$ throughout a pseudocritical window bounded by
$y$. This does not rest on the sign of the $O(1/N)$ difference: were the
scalar channel more relevant than $y$ by as much as the singlet shift,
$64/(N\pi^2)\simeq0.54$, it would reach $y$ only at
$\ell_\times=\ln(1/r)/(\omega_s-\omega_y)\simeq7$--$14$, a scale factor of
$10^3$--$10^6$. In the Z3000 phase both channels become the
$D\langle\Phi\rangle$ terms treated in
Sec.~\ref{sec:yukawa}.

\section{The chiral endpoint on the lattice: flux mass, Chern number, triplet amplitudes}
\label{app:chiral}

\emph{$\mathcal T$-odd singlet amplitudes and the flux mass.} We work in
the Z3000 gauge of Ref.~\cite{Feuerpfeil2026}, with
$\psi_i=(f_{i\uparrow},f^\dagger_{i\downarrow})$,
$H=-\sum\psi^\dagger_iu_{ij}\psi_j$ and
$u=u^0\tau^0+\Delta^1\tau^x+\Delta^2\tau^y+\chi^3\tau^z$, where $u^0$ is
the imaginary singlet hopping. The PSG time-reversal element is
$W_{\mathcal T}(i)=\pm\ii\tau^z$ with the sign $+$ on the $AA$, $BB$ and
$-$ on the $AB$, $BA$ sublattices, and the rule is
$u_{ij}\to W^\dagger_{\mathcal T}(i)(-u_{ij})W_{\mathcal T}(j)$. Bonds
joining sublattices of equal sign (all diagonals and on-site) therefore have
$\mathcal T$-odd $\tau^0,\tau^z$ and $\mathcal T$-even $\tau^{x,y}$
components, and NN bonds the reverse. Each $\mathcal T$-odd component,
placed on one bond of the $2\times2$ cell and translated, is projected
onto the 36 gauge-singlet continuum bilinears through the zero-mode
matrix $W$ of Appendix~\ref{app:wick}. The diagonal $\tau^z$, NN
$\tau^{x,y}$ and on-site $\tau^z$ components have zero gauge-singlet
projection (they are Higgs-sector, gauge-adjoint operators). Each
diagonal $\tau^0$ component projects onto $\{$flux,
$\mathrm{Tr}[\bar{\mathcal X}\gamma^0\mu^y\mathcal X]\}$. For the bond
orientations of Eq.~\eqref{eq:Fpattern} the coefficients are
$(-\tfrac14,-\tfrac14)$ on $(0,0)(1,1)$ and $(-\tfrac14,+\tfrac14)$ on
$(2,1)(1,2)$ (dimers), $(+\tfrac14,+\tfrac14)$ on $(1,1)(2,2)$ and
$(+\tfrac14,-\tfrac14)$ on $(1,0)(0,1)$ (orthogonal diagonals), and
$(+\tfrac14,-\tfrac14)$ on $(1,0)(2,1)$, $(+\tfrac14,+\tfrac14)$ on
$(1,1)(0,2)$, $(-\tfrac14,+\tfrac14)$ on $(0,1)(1,2)$ and
$(-\tfrac14,-\tfrac14)$ on $(2,0)(1,1)$ (empty-square diagonals). The
valley density cancels between the two diagonals of each square when they
enter with opposite signs, as in $\mathcal P_\chi$. With cell
sites $AA=(0,0)$, $BA=(1,0)$, $AB=(0,1)$, $BB=(1,1)$ (offsets from
$2\bm r$), the pure flux mass is
\begin{align}
\mathcal P_\chi&=\sum_{\bm r}u^0\Bigl[-u_{(0,0)(1,1)}-u_{(2,1)(1,2)}\nonumber\\
&\qquad+u_{(1,1)(2,2)}+u_{(1,0)(0,1)}+u_{(1,0)(2,1)}\nonumber\\
&\qquad+u_{(1,1)(0,2)}-u_{(0,1)(1,2)}-u_{(2,0)(1,1)}\Bigr]
\label{eq:Fpattern}
\end{align}
with $u\to\tau^0$ on the listed bond. Its projection is $-2\ii u^0$
times the flux bilinear and nothing else.
Its PSG signature is $(T_{2x},T_{2y},G_x,\sigma_{xy},\mathcal T,\mathcal
TG_x,\mathcal T\sigma_{xy})=(+,+,-,-,-,+,+)$. The last five entries are
the Z3000$_{11}$ row of Table~XVI (Appendix~I) of Ref.~\cite{Maity2025}.
Both translations are retained there with trivial PSG [its Eqs.~(I1) and
(I3)].

\emph{Chern number.} Let $H(\bm k)=\ii A(\bm k)$ be the $16\times16$
Majorana Bloch Hamiltonian of Z3000$(s_H)$ + $d\,h_{\rm DM}$ + $m\mathcal P_\chi$ in
the conventions of Appendix~\ref{app:protocol}. The Chern number of the
eight negative bands is evaluated with the gauge-invariant lattice
formula~\cite{FHS2005}, $C=\frac1{2\pi}\sum_{\square}\arg\prod
\det[\Psi^\dagger_{\bm k}\Psi_{\bm k'}]$, with $\Psi_{\bm k}$ the matrix of
occupied eigenvectors and the product of the four link determinants taken
counterclockwise, over the plaquettes of a $36\times36$ grid of the cell
zone. For
$(s_H,d)=(0,0)$, $(0.2,0)$, $(0,0.6)$, $(0.05,0.6)$, $(0.1,0.6)$,
$(0.2,0.6)$ at $|m|=0.1$ the minimum excitation energies on the $36^2$ grid are
$0.400$, $0.399$, $0.400$, $0.400$, $0.401$, $0.394$, and
$C_{\rm BdG}=-4\,\mathrm{sgn}(m)$ in every case. At $d=0$ the
complex-fermion $(f_\uparrow,f^\dagger_\downarrow)$ Bloch Hamiltonian
built from the same bonds has eigenvalues equal to twice those of
$H(\bm k)$, as the Majorana convention of Appendix~\ref{app:protocol}
requires, and Chern number
$-2\,\mathrm{sgn}(m)=C_{\rm BdG}/2$. On a cylinder of $24$ cells open in
$x$, branches localized on the outer four cells cross $E=0$ with net
signed count $+4$ on the left and $-4$ on the right edge. For a $\mathbb Z_2$ spin liquid with gapped gauge field the
chiral central charge is $c=C_{\rm BdG}/2$~\cite{Kitaev2006,ReadGreen2000}, giving
Eqs.~\eqref{eq:chern}--\eqref{eq:kxy} for $s_H>0$. At $s_H=0$ every
kernel of $\pi$-flux$+d\,h_{\rm DM}+m\mathcal P_\chi$ commutes with the on-site SU(2)
gauge generators, whereas the Higgs amplitudes at $s_H>0$ do not, so the
$s_H=0$ state has an SU(2) invariant gauge group.

\emph{Field selection of the chirality.} The Zeeman operator
$+b\sum_iS^z_i$, with $b=g_c\mu_BB_c$ and the lattice spin operators of
Appendix~\ref{app:wick}, is added to the Bloch Hamiltonian of $\pi$-flux
$+\,d\,h_{\rm DM}+m\mathcal P_\chi$ at $d=\pm0.6$. At $b=0$ the ground-state
$\langle S^z\rangle$ per cell has the sign of $m$ for
$0.002\le|m|\le0.2$, takes the same value for $d=0.6$ and $d=-0.6$, and
vanishes at $d=0$. The mean-field energies of the two domains differ by
$2b\langle S^z\rangle_m$ per cell, so $B_c>0$ favors $m<0$. The
field-only state ($m=0$) is gapped with $C_{\rm BdG}=+4\,{\rm sgn}(B_c)$
for $0.02\le|b|/t\le1.0$. Its lowest excitation energy is $|b|/2$ at
small $|b|$ and saturates near $0.4\times4\lambda$, and the path in
$(m,b)$ from the favored domain at $b=0$ to the field-only state stays
gapped. On an open flake of $24\times24$ sites the in-gap edge states of
$m=+0.1$ circulate clockwise seen from $+c$, as on the cylinder above. A
field along $+c$ therefore selects the domain with $C_{\rm BdG}=+4$,
counterclockwise Majorana edge modes and $\kappa_{xy}>0$ for
$\bm J^Q=-\kappa\nabla T$, the same as the field-only state. The
magnetic moment of that domain, $-g_c\mu_B\langle S^z\rangle$, is
parallel to the field, so $M_z\kappa_{xy}>0$. The same signs hold on
Z3000$(s_H=0.05)$.

\emph{Family scan.} The Z3000$_{11}$ family is the fixed space of
$\{T_{2x},T_{2y},\mathcal TG_x\sigma_h\otimes C_{2x},\mathcal
T\sigma_{xy}\otimes C_2^{(1,-1,0)}\}$ in the 280-parameter bond space
defined below, of dimension $42=21+21$, and its $\mathcal T$-odd part
contains $\mathcal P_\chi$. The composite gauge is
$W_{\mathcal T}\ii\tau^y\overline{W}_{O}$, fixed by requiring Z3000, the
DM operator and $\mathcal P_\chi$ to be invariant. Random $\mathcal T$-odd members $Y$
are drawn basis-invariantly, as random vectors of the ambient
parameter space projected with the unique projector onto the
$\mathcal T$-odd part, and scaled so that their largest bond block
equals that of $m\mathcal P_\chi$. The rays Z3000$(s_H)$+DM+$m\mathcal P_\chi+aY$ are followed at
$s_H=0.05$ and $0.2$ (16 rays each). At fixed amplitudes on a $40^2$
grid, $C_{\rm BdG}=-4$ in $32/32$ rays at $a=0.5$, in $31/32$ at $a=1$
(one $-2$), and in $26/32$ at $a=2$ (four $-2$, two $0$). For gaps below
about $0.03$ the Berry curvature is concentrated on the momentum scale of
the gap, and grids of at least $40^2$ are needed to resolve the integer.

Phase boundaries are located along the first six rays of each $s_H$. The
continuous gap $G(a)=\min_{\bm k}{\rm gap}(a,\bm k)$, with ${\rm
gap}(a,\bm k)$ the minimum excitation energy at $\bm k$, and the Chern
number are evaluated on
the ladder $a=0.1,0.2,\dots,3$. Every Chern change and every local
minimum of $G$ below $0.02$ is followed by a minimization of $G$ in $a$,
and a boundary is a zero of $G$ across which the Chern number changes.
The first zeros are as follows.
\begin{itemize}
\item $s_H=0.05$: $0.9705$ ($-4\to-2$, with a second zero at $1.236$,
$-2\to0$), $1.512$ ($-4\to-2$, then $2.701$, $-2\to+2$), and $2.551$
($-4\to-2$). Three rays are gapped to $a=3$, with minimum $G$ of
$0.41$, $0.020$ and $0.015$.
\item $s_H=0.2$: $1.391$ ($-4\to-2$, then $2.342$, $-2\to0$), $1.473$
($-4\to0$, then $2.978$, $0\to+4$), and $1.727$ ($-4\to-2$, re-entrant:
$C=-4$ again from $a=1.80$ across a nearly closed stretch
$G<3\times10^{-3}$ up to $1.85$). Three rays are gapped to $a=3$, with
minimum $G$ of $0.28$, $0.23$ and $0.41$.
\end{itemize}
``Gapped to $a=3$'' means that the gap does not close on the ladder or
on a denser ladder with amplitude step $0.02$. A closing narrower than the
ladder spacing with sampled gaps above $0.02$ is not excluded. The PSG class thus
contains several topological sectors, and the Chern number of Eq.~\eqref{eq:chern} is a local statement
about the minimal state.

\emph{Triplet amplitudes allowed by the locked PSG.} We parametrize
every translation-invariant real antisymmetric Majorana form $A$, with
$H=\ii\sum_{KL}A_{KL}M_KM_L$, on the NN (8 per cell), dimer (2),
orthogonal-dimer (2), empty-square-diagonal (4) and on-site (4) classes
by $4\times4$ bond blocks. This gives $16\times16+4\times6=280$ real
parameters covering both spins and both Majorana components, hence all
hoppings and pairings, singlet and triplet. A locked generator acts on
the four Majoranas of a site through the real orthogonal matrix induced
by the PSG gauge $W_O$ on the Nambu doublets and by the axial spin
rotation $\Omega_O$ on $(f_\uparrow,f_\downarrow)$, generated by the
lattice spin operators of Appendix~\ref{app:wick}. The construction follows the spin--orbit-locked PSG of
Ref.~\cite{Dodds2013}, which kept only $z$-component triplets on the
kagome lattice. Here it is extended to all three triplet components and
to the spin rotations that accompany the operations of the SS space
group, and it is written in Majorana form. As in Ref.~\cite{Dodds2013},
the $\mathcal T$ generator is the antiunitary spin flip combined with an
additional gauge rotation, here $W_{\mathcal T}\ii\tau^y$.
Reference~\cite{Feuerpfeil2026} writes its ans\"atze in Wen's Nambu
formalism~\cite{Wen2002} and gives its time-reversal PSG element in its
Eq.~(A.7). In that formalism $\mathcal T$ acts as $u_{ij}\to u_{ij}^*$,
which with Wen's fixed gauge rotation $-\ii\tau^2$ becomes
$u_{ij}\to-u_{ij}$ [his Eq.~(19)] before the PSG element is applied. In Wen's variables this
is the operation above. The generator $W_{\mathcal T}\ii\tau^y$ leaves both the Z3000 ansatz and the
Wick-induced DM operators invariant, whereas
$W_{\mathcal T}$ without the $\ii\tau^y$ does not act correctly on
triplets. The allowed ansatz
is the common fixed space of $\{T_{2x}, T_{2y},G_x\sigma_h\otimes
C_{2x},\sigma_{xy}\otimes C_2^{(1,-1,0)}, \mathcal T\}$ (buckled) or of
$\{T_{2x},T_{2y},G_x\otimes C_{2y},\sigma_{xy} \otimes
C_2^{(1,-1,0)},\sigma_h\otimes C_{2z},\mathcal T,{\rm U(1)}_z\}$
(high-temperature), split into singlets and triplets by the global spin
rotations. With trivial spin factors and full spin SU(2) the fixed space consists of seven singlet amplitudes
(two NN, one dimer, one orthogonal, two empty-square, one on-site), and
it contains no triplet. The Z3000 ansatz of Ref.~\cite{Feuerpfeil2026},
Eqs.~(8)--(17) there, lies inside it. That ansatz has six
real parameters and no on-site term, so it accounts for six of the seven
amplitudes. With the locked generators the singlet space is
unchanged, and the Wick-induced DM operators (in-plane inter-dimer,
intra-dimer and $D'_\perp$) lie inside the fixed space and are purely
triplet. The triplet complement is the orthogonal complement of the
singlet space within the allowed space. The buckled structure admits 14
triplet amplitudes, block-diagonal by bond class with 6 (NN), 2 (dimer),
2 (orthogonal), 4 (empty-square) and 0 (on-site). The high-temperature
structure admits 2 triplet amplitudes, both on NN bonds.

\emph{Fixed-background stability of the DM mean fields.} The analysis covers
the in-plane staggered inter-dimer channel only, with the
Z3000$(s_H)$ singlet amplitudes held fixed, at
$s_H=0.05$ and $0.1$ with $d=0.6$. The ansatz form is that of
Ref.~\cite{Feuerpfeil2026}, Eqs.~(8)--(17) there. The amplitudes form a one-parameter family fixed by hand and are not
obtained variationally. In the notation of
Ref.~\cite{Feuerpfeil2026} they are $t=1$, $\theta=\pi/2+0.30\,s_H$,
$\Delta_d=0.40\,s_H$, $\Delta_{d'}=0.25\,s_H$, $\Delta_{g,1}=0.15\,s_H$
and $\Delta_{g,2}=0.10\,s_H$. The same family underlies every
Z3000$(s_H)$ result in this paper. The lattice
is the $8\times8$-cell torus with periodic boundary conditions, on which
the
Z3000$(s_H)$ state has a finite-size gap (minimum excitation energy $0.047$ at $s_H=0.05$, $0.095$ at $0.1$). The Wick
decoupling of $H_{\rm DM}$ on the ground state of
Z3000$(s_H)$+$d\,h^{(n)}_{\rm DM}$ defines $h^{(n+1)}_{\rm DM}$. Starting
from the Wick operator $h^{(0)}$ of Appendix~\ref{app:wick}, the
iteration converges geometrically in a few steps. Relative to $h^{(0)}$
the converged operator has its norm changed by $0.06\%$ ($0.04\%$) and
its direction rotated by $4.4^\circ$ ($4.3^\circ$) for $s_H=0.05$
($0.1$). It lies inside the locked-PSG-allowed space, with $99.4\%$ of
its squared norm triplet and $0.6\%$ singlet. The NN singlet amplitude
moves from $0.2391$ to $0.2386$, $\delta'$ from $0.00548$ to $0.00549$
($0.02120\to0.02126$), and the nodes of the Bloch spectrum stay gapless.
This establishes the stability of one channel on a fixed background; it
is not a self-consistent solution of the full DM-coupled ansatz.

\section{Tree-level quartic seeds}
\label{app:treeseeds}
At zeroth order in gradients the parton fields of the $2\times2$ cell
reduce to the $\Gamma$-point modes of the $\pi$-flux state,
$f_{s\sigma}=\sum_{d,v}b_{s,(d,v)}c_{dv\sigma}$, with the orthogonal
basis map $b$ of Eq.~(26) of Ref.~\cite{Feuerpfeil2026} (Dirac index $d$,
valley index $v$, sublattice $s$). The eight complex modes $c_{dv\sigma}$
are represented on their full $2^8$-dimensional Fock space. The spin
operators $\bm S_s=\frac12 f_s^\dagger\bm\sigma f_s$ are built from them,
and the exchange operators of Eq.~\eqref{eq:H} per cell are formed with
two dimer bonds and eight nearest-neighbor bonds. The DM textures are
the symmetry-allowed patterns of Table~\ref{tab:dm}, each bond class summed
over its sublattice pair. The fully antisymmetric quartic part of each
operator is extracted in the Majorana basis by trace orthogonality,
projected onto the eight QBT Majoranas, and decomposed on the fourteen
invariant quartics of Appendix~\ref{app:qbtflow}. The residual outside
that space is below $10^{-14}$.

The VBS directions are taken from the lattice bond patterns themselves,
$V_x$ from $-{\rm col}_x$ and $V_y$ from ${\rm col}_y$ as in
Appendix~\ref{app:plaquette}. $P_{\rm e}$ and $P_{\rm f}$ are built
separately from their void and crossed squares and project onto
$\theta=45^\circ$ and $135^\circ$. In a gapped state with
$\langle\gamma_i\gamma_j\rangle=\delta_{ij}+i\varepsilon\hat B_{ij}$ the
Hartree--Fock energy of a quartic $\frac1{4!}T_{ijkl}\gamma_i\gamma_j
\gamma_k\gamma_l$ is $-\frac{\varepsilon^2}{8}T_{ijkl}\hat B_{ij}\hat
B_{kl}$. We quote the couplings in the normalization
$g(\hat B)=\frac12T_{ijkl}\hat B_{ij}\hat B_{kl}$, four times this
coefficient, with $g>0$ attractive; only signs and ratios of $g$ are used,
and the gauge seed of Appendix~\ref{app:qbtgauge} is quoted in the same
normalization. The results quoted in Sec.~\ref{sec:whichvbs} hold to
$10^{-14}$. The Heisenberg seed has zero valley charge for
$J'/J=0.4$, $0.63$ and $0.9$, with $g=J'$ on the whole VBS circle. For
unit $D'_\perp$ on the reference bond, $g(\theta)=2\sin2\theta$, while
$D'_{\parallel,s}$, $D'_{\parallel,ns}$ and $D$ give $g=0$ there.
Antiferromagnetic exchange makes the N\'eel$_z$ and VBS channels
attractive, as expected.

The flow uses the $\beta$ function of Appendix~\ref{app:qbtflow} from the
seed $J$--$J'$ exchange plus the DM terms at the DFT values, with
$J=7.3$\,meV. The vertex growth $\ln\Gamma_P=\int d\ell\,r_P(g(\ell))$,
evaluated with the running couplings up to $|g|=200$, is independent of
the overall seed scale (rescaling the seed by four changes it by less
than $2\%$). Without DM it is equal for $P_{\rm e}$ and $P_{\rm f}$. With
$D'_\perp=-0.10$ ($-0.18$)\,meV,
$\ln\Gamma_{P_{\rm f}}-\ln\Gamma_{P_{\rm e}}=+2.0$ ($+3.3$) with the
physical vertex normalization of Appendix~\ref{app:qbtflow}, and reversing
the sign of $D'_\perp$ reverses it. The in-plane components alone give no
splitting. The determinant ratio $A/C_{\rm iso}$ of the band-touching
channel, Eq.~\eqref{eq:Aclosed}, is $1.29\times10^{-3}$, $3.2\times10^{-4}$
and $3.6\times10^{-5}$ at $d=0.6$, $0.3$ and $0.1$.

\begin{figure*}[t]
\centering
\includegraphics[width=0.96\textwidth]{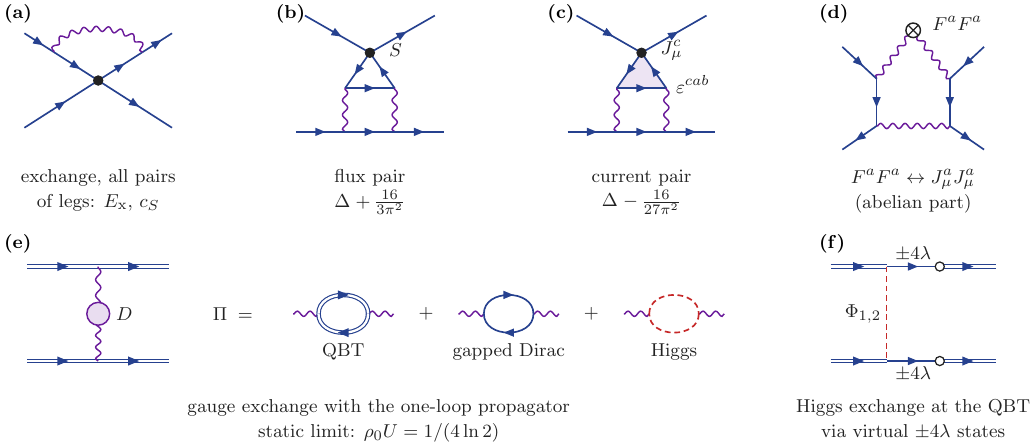}
\caption{Diagrams of the four-fermion running
(Appendix~\ref{app:treeseeds}) and of the gauge and Higgs exchange on the band
touching (Appendix~\ref{app:qbtgauge}). Line styles are those of
Fig.~\ref{fig:diagrams}. The black dot is a four-fermion operator, the
crossed circle $F^aF^a$, and double lines the QBT modes. (a) One-gauge-line
exchange, summed over all pairs of legs, with the self-energy. (b)
Two-gauge-line annihilation of a flux-singlet pair $S$, which lifts its
dimension by $16/(3\pi^2)$ at $N=12$. (c) Annihilation of an SO(5)-singlet
gauge-current pair through the induced three-gauge vertex (shaded,
$\propto\varepsilon^{cab}$), which lowers it by $16/(27\pi^2)$ at $N=12$ after the
gauge-field equation of motion. (d) $F^aF^a$ acting on fermions through
two gauge lines, the current square it corresponds to, and the exchange
that mixes it with the four-fermion operators. In QED$_3$ this is the
whole $F^2$ mixing, while in SU(2) the non-abelian dimension of $F^aF^a$
has to be added. (e) Gauge exchange between QBT modes with the one-loop
propagator $D$ (shaded blob), built from the polarization $\Pi$ of the QBT
modes (double lines), the gapped Dirac background (single lines) and the
critical Higgs fields (dashed), Eq.~\eqref{eq:pidyn}. In the static limit,
$\Pi_{00}=4\ln2\,\rho_0$ leaves a contact interaction of strength
$\rho_0U=1/(4\ln2)$. (f) Higgs exchange at the band touching. The Yukawa
vertices of $\Phi_{1,2}$ vanish on the QBT at first order, so each QBT line
couples to the Higgs line through a virtual $\pm4\lambda$ state (single
line), which returns to the band touching through the $k$-linear term of
the Hamiltonian (open circle); both orderings along each line contribute.}
\label{fig:fourfermion}
\end{figure*}
\emph{Running between the lattice scale and $\lambda$.} The
Dirac-regime dimensions are computed in the Majorana basis of the sixteen
$\Gamma$-point modes, with the kinetic matrices of the continuum limit,
the gauge generators in the same basis, and the propagator and angular
averages of Eqs.~\eqref{eq:rules} and \eqref{eq:angavg}. A local
operator with $n$ legs and coefficient tensor $T$ has
$\Delta=\Delta_{\rm tree}-E_{\rm x}-\tfrac n2 c_S+A_{\rm ann}$, where $c_S$ is the
self-energy coefficient, $E_{\rm x}$ the sum over all pairs of legs of the
one-gauge-line exchange, and $A_{\rm ann}$ the two-gauge-line annihilation of a pair
of legs, Fig.~\ref{fig:fourfermion}(a)--(c). The annihilation of an
arbitrary pair, obtained from the fermion triangle at zero pair momentum,
acts in two channels only. The triangle gives the flux-singlet pair the lifting
of Eq.~\eqref{eq:lifting}. The SO(5)-singlet gauge-current pairs $J^c_\mu$ annihilate
through the induced three-gauge vertex. At zero pair momentum the Ward
identity fixes this vertex to $4N_f\varepsilon^{cab}\,\partial_\lambda
\Pi_1^{\mu\nu}(q)$, with $\Pi_1$ the polarization of one two-component
Majorana fermion. Onto a fermion line it gives $-(3-2\zeta)/9$ of the lifting in
gauge $\zeta$, and onto a Higgs line a longitudinal part equal to the
gauge-dependent part of the fermion line. By the gauge-field equation of
motion $J_\psi+J_\Phi=0$ the two combine into the gauge-independent shift
$-\tfrac19\cdot16/(3\pi^2)=-128N_f/(3\pi^2N^2)$ per current pair. For
$n=2$ these rules give the self-energy coefficient of Eq.~\eqref{eq:s}
for every gauge parameter, $\Delta=2$ for all thirty current components,
$\Delta=2-64/(\pi^2N)=1.4596$ for the five SO(5)-vector masses and
$\Delta=2$ for the flux singlet, Eq.~\eqref{eq:lifting}. The fermion loop
they imply carries the factor $2N_f$ of the polarization. Two countings of
the annihilation are used. In the leading large-$N_f$ counting only pairs
of legs whose loop carries a free valley trace annihilate. In the
physical-$N_f$ counting, used for the spectrum and for
Table~\ref{tab:treerun}, every pair of legs annihilates with its actual
loop factor at $N_f=2$, which includes Fierz-crossed pairs that are
formally $O(1/N^2)$. The leading counting gives, in the abelian
version with four flavors, $4-64/(\pi^2N)$ and $4+64/(3\pi^2N)$ for the
two Lorentz-scalar four-fermion operators with one field of each flavor,
and $4+64(2\pm\sqrt7)/(3\pi^2N)$ for the flavor-singlet pair
$(\bar\psi\psi)^2$, $(\bar\psi\gamma_\mu\psi)^2$, the values of
Refs.~\cite{Xu2008,ChesterPufu2016} [Eqs.~(1.3) and (1.7) of
Ref.~\cite{ChesterPufu2016}, which obtains the singlet pair from the
mixing of $(\bar\psi\psi)^2$ with $F_{\mu\nu}F^{\mu\nu}$ and quotes
$4+64/(3\pi^2N)$ from Sec.~II.C of Ref.~\cite{Xu2008}], as well as the
normalization-independent entries of the U(1) flow of
Ref.~\cite{Xu2008}, its Eq.~(6); the physical-$N_f$
counting gives $4+(-25.3,+94.6)/(\pi^2N)$ for that pair instead of
$4+(-13.8,+99.1)/(\pi^2N)$. In QED$_3$ the current square stands in for
$F^2$: the gauge-field equation of motion makes it redundant, and its
Wilsonian coupling carries the $O(1/N)$ mixing of $F^2$
(Fig.~\ref{fig:fourfermion}(d)), so that mixing is included. The two
countings differ only in the SO(5) representations $1$, $5$ and $10$, by
$0.12$ in the lowest singlet of the four-fermion block ($2.477$ against
$2.592$) and by less than
$0.03$ elsewhere; we quote the physical-$N_f$ values and regard the
difference as part of the $O(1/N^2)$ uncertainty. On the 336
gauge-invariant four-fermion operators the dimension operator is
independent of the gauge parameter to $10^{-14}$, self-adjoint in the
free-field metric and SO(5) covariant. Its eigenvalues are $2.477$,
$3.219$, $3.405$, $3.460$,
$3.820$, $3.850$, $3.910$, $4$, $4.202$, $4.205$ and $4.540$. Mixing with
operators that contain gauge or Higgs fields, which have the same
dimension at $N=\infty$, can only occur in the SO(5) representations $1$,
$5$ and $10$. Among the channel-carrying eigenspaces of
Table~\ref{tab:treerun} this affects only the SO(5) singlets.
The eigenvalue $4$ of the $35$ Lorentz vector, which carries the
$D'_\perp$ plaquette selector, has no anomalous term for the following
reasons. The $35$ Lorentz vector occurs once among the 336 operators, so
it cannot mix. It contains no flux-singlet and no gauge-current pair, so
no annihilation graph contributes (below $10^{-17}$). It lies in the span
of products of two SO(5) currents (to $10^{-15}$), and on it the gauge
exchanges between legs of different currents sum to zero. The self-energy
and the exchanges within each current then cancel by the Ward identity,
for every gauge parameter: the self-energy is $+0.135$ at $\zeta=0$ and
$-0.270$ at $\zeta=1$, and the exchanges are the negatives. The $14$ spin-2 eigenspace, also a pure current product,
behaves in the same way. No symmetry argument for the cancellation between different currents
is known, and it does not hold in general: the $35'$ scalar, also
a pure current product, receives $+64/(\pi^2N)=0.54$. The value $4$ is
therefore an $O(1/N)$ statement.

\emph{The SO(5)-singlet scalar channel.} In the four-fermion basis this
channel is spanned by $(\mathrm{Tr}\,\Xb\X)^2$ and the gauge-current square
$J^a_\mu J^a_\mu$. Unlike in QED$_3$, in SU(2) the current square does not stand in for
$F^aF^a$. In the theory screened by the fermions alone, the $\pi$-flux
SU(2) state, the four-fermion block agrees in Landau gauge with the SU(2)
flow of Ref.~\cite{Xu2008}, its Eq.~(35), in both diagonal entries, in the
product of the off-diagonal ones and in the relevant singlet with
$\Delta=4-36.5/N_{\rm M}$ for $N_{\rm M}=2N$ Majorana flavors. The
current-square diagonal entry, $4-256(3-\zeta)/(3\pi^2N)$, is gauge
dependent, $4-512/(3\pi^2N)$ in Landau gauge, which by itself shows that
the current square is not a physical operator in SU(2). The $O(1/N)$
dimension of $F^aF^a$ in pure-fermion screening, Eq.~(11) of
Ref.~\cite{Gracey1996} with $\Delta=d-2\omega$, four-component flavor
number $N/4$, $T(R)=1/2$ and $C_2(G)=2$, is $4+320/(3\pi^2N)$. With that
dimension in place of the current-square diagonal, and the off-diagonal
entries of the stand-in, the two singlets have
$\Delta=4+32(17\pm\sqrt{85})/(3\pi^2N)$, both irrelevant; the lower one stays above $4$
unless the off-diagonal product exceeds $20/3$ times the stand-in value,
and above $3$ at $N=12$ unless it exceeds about $20$ times.
The instability of the $\pi$-flux SU(2) state that Ref.~\cite{Xu2008}
infers from this channel below $N_{\rm M}\simeq36.5$ Majorana flavors therefore
comes from the current-square stand-in, and it is absent at $O(1/N)$ once
$F^aF^a$ carries its own dimension, the comparison proposed in footnote~5 of Ref.~\cite{ChesterPufu2016}. At the physical $N_{\rm M}=8$ the expansion
is not controlled.

\emph{The dimension of $F^aF^a$ with Higgs loops.} At the multicritical
point the gauge field is screened by the fermions and by the adjoint Higgs
triplets, and the dimension of $F^aF^a$ receives Higgs-loop contributions
absent from Ref.~\cite{Gracey1996}. It follows from adding a Maxwell
term $(\epsilon/4)F^aF^a$ to the fixed point and extracting the logarithms
of the $O(1/N)$ gluon two-point function,
$\Gamma_2=\Pi_0(1+\beta_bL)+\epsilon q^2P_T(1+\beta_cL)$ with
$L=\ln(\Lambda/|q|)$, so that $\Delta(F^aF^a)=4-\beta_c+\beta_b$. The gluon
field renormalization $\beta_b$ follows from the fermion self-energy and
the vertex correction through the induced three-gluon vertex, fixed at
zero momentum by the Ward identity. The coefficient $\beta_c$ collects the
box, the two-triangle and the cubic-Maxwell diagrams built from the
matter-induced vertices and, for critical Higgs fields, the exchange of
the constraint field between two induced $AA\sigma$ vertices. Every
induced vertex is expanded by regions in the soft external momentum. Only
the hard regions, in which all loop momenta are of the order of the
internal gluon momentum, renormalize $F^aF^a$ into itself; the soft
regions, in which a matter loop carries the external momentum, are the
mixing of $F^aF^a$ into matter operators, above all the redundant square
of the total current, and drop out of the diagonal entry. The
prescription agrees with the three published values: $4+320/(3\pi^2N)$
for fermionic SU(2) screening~\cite{Gracey1996} (computed there in Landau gauge) in every covariant gauge,
the vanishing diagonal of fermionic QED$_3$~\cite{ChesterPufu2016}, and
$4+32/(3\pi^2N)$ for the critical CP$^{N-1}$ model [Eq.~(27) of Ref.~\cite{VasilevNalimov1983} at $d=3$], which the hard and soft regions together
would miss by a factor of two. With $x_b$ the Higgs share of the
screening and $x_f=1-x_b$,
\begin{equation}
\Delta(F^aF^a)=4+\frac{8}{3\pi^2N}\Bigl[40+16x_fx_b+\tfrac{32}{3}x_b\Bigr]
\label{eq:gammaFF}
\end{equation}
for critical Higgs fields, and without the last term for Gaussian ones,
i.e.\ $\Gamma_{FF}\equiv\Delta(F^aF^a)=5.14$ ($4.98$) at $N=12$, independent of the gauge
parameter. Two pieces have no published benchmark: the fermion--Higgs
cross term, corroborated by a direct extraction from the full
triangle, and the non-abelian part of the critical-scalar sector, which
the abelian CP$^{N-1}$ total cannot separate. Replacing $5.14$ by the
Gaussian-Higgs value $4.98$, or by the value $4.90$ that the hard and soft
regions together give in Feynman gauge ($\zeta=0$), moves the lowest four-fermion
singlet below by less than $10^{-3}$. $\Gamma_{FF}$ is a diagonal entry of a degenerate
block, not an eigenvalue; in CP$^{N-1}$ the corresponding diagonal
entries ($+32$ and $-288$ in units of $1/(3\pi^2N)$) differ strongly from
the eigenvalues ($-323$ and $+67$).

\begin{figure*}[t]
\centering
\includegraphics[width=0.96\textwidth]{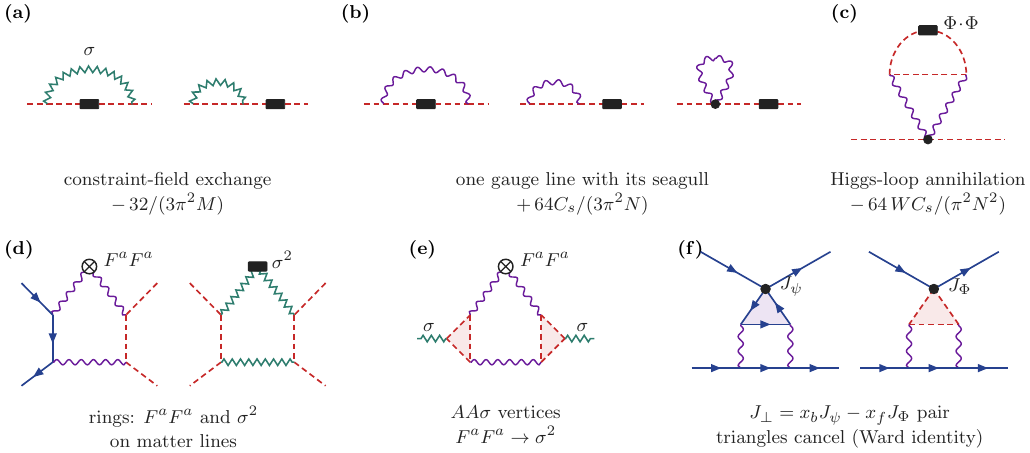}
\caption{Diagrams of the Higgs sector and of the SO(5)-singlet channel
(Sec.~\ref{sec:disorder} and Appendix~\ref{app:treeseeds}). Line styles
are those of Fig.~\ref{fig:diagrams}, without orientation arrows on the
Higgs lines, and teal zigzag lines are the constraint field $\sigma$. The
black box is an operator insertion, $\Phi\cdot\Phi$ in (a)--(c), the small
dot the Higgs seagull vertex, the crossed circle $F^aF^a$, and shaded
triangles are induced vertices. Top row, the dimension $\Delta_\sigma$ of
Eq.~\eqref{eq:Dsigma}: (a) constraint-field exchange between the Higgs legs
of the insertion and on either leg, which gives $-32/(3\pi^2M)$. (b) One
gauge line exchanged between the legs and on either leg, which give
$+64C_s/(3\pi^2N)$ for every gauge parameter, and the seagull tadpole,
which is scaleless and vanishes. (c) Two gauge lines attached to a closed
Higgs loop that carries the insertion and ending on the seagull vertex of
the external line, the annihilation topology of
Fig.~\ref{fig:diagrams}(e),(f). It gives $-64WC_s/(\pi^2N^2)$, the dominant
term, which takes $\Delta_\sigma$ below the unitarity bound. Bottom row,
graph classes of the singlet matrix of Eq.~\eqref{eq:singletmatrix}: (d)
ring graphs, in which $F^aF^a$ or $\sigma^2$ acts on two matter lines,
fermion or Higgs, through two gauge or two constraint-field lines, and the
two lines exchange a gauge line or, between Higgs lines, a constraint-field
line. (e) A gauge line exchanged between two induced $AA\sigma$ vertices
(Higgs triangles), which maps $F^aF^a$ onto $\sigma^2$. (f) Annihilation of
a pair of legs of a current square (black dot) through a fermion or Higgs
triangle into two gauge lines, as in Fig.~\ref{fig:fourfermion}(c). For the
relative current $J_\perp=x_bJ_\psi-x_fJ_\Phi$ the two triangles cancel by
the Ward identity, so the pairs of $J_\perp$ do not annihilate.}
\label{fig:higgs}
\end{figure*}

\emph{The singlet channel at the multicritical point.} The critical Higgs
current makes the relative current $J_\perp$ of Sec.~\ref{sec:whichvbs} an
independent operator. The
dilatation operator on $\{(\mathrm{Tr}\,\Xb\X)^2,R,F^aF^a,\sigma^2,M\}$, with
$M=(D_\mu\tilde M_{ab})^2$ and $\tilde M_{ab}$ the traceless gauge-tensor
bilinear of the Higgs fields, follows from the self-energies, the gauge and
constraint-field exchanges between all pairs of legs
(Fig.~\ref{fig:higgs}(a),(b)), the covariant
completions of the currents and of $M$, the annihilation of pairs of legs
through matter triangles (Fig.~\ref{fig:higgs}(c),(f)), the ring graphs
through which $F^aF^a$ and $\sigma^2$ act on matter lines
(Fig.~\ref{fig:higgs}(d)), and the exchange of a gauge line between
two $AA\sigma$ vertices (Fig.~\ref{fig:higgs}(e)). The entries not computed directly follow from
self-adjointness in the metric of the primaries. In the leading counting
$\hat\Delta=4+u\,\mathcal M$, with $u=8/(3\pi^2N)$ and
\begin{equation}
\mathcal M=\begin{pmatrix}
0&-\tfrac12&-\tfrac1{64}&0&0\\
-512&-\tfrac{128}3&0&8&-\tfrac{32}3\\
-4096&0&\tfrac{152}3&256&-\tfrac{1024}3\\
0&\tfrac{512}9&\tfrac{64}9&-144&0\\
0&-\tfrac{40}3&-\tfrac53&0&96
\end{pmatrix},
\label{eq:singletmatrix}
\end{equation}
where row $i$ is the image of the $i$th operator. The signs of the
off-diagonal entries depend on the relative orientation of the operators:
reversing $(\mathrm{Tr}\,\Xb\X)^2$ and $R$ together conjugates $\mathcal M$
by ${\rm diag}(-1,-1,1,1,1)$ and leaves the spectrum unchanged;
Table~\ref{tab:singletentries} gives the content of the entries. The
eigenvalues $0.46$, $2.99$, $4.09$, $5.11$ and $6.45$ are independent of the
gauge parameter. The lowest mode is $92\%$ $\sigma^2$, and the mode at
$2.99$ is $86\%$ $R$ and $10\%$ $(\mathrm{Tr}\,\Xb\X)^2$. In the
physical-$N_f$ counting of the annihilations, with the current-pair shift
of Fig.~\ref{fig:fourfermion}(c), the four-fermion-dominated mode lies at
$2.91$. The image of $F^aF^a$ on the current squares is a combination of
$J_\psi\cdot J_{\rm tot}$ and $J_\Phi\cdot J_{\rm tot}$ alone, so $F^aF^a$ and $R$
do not mix. The pairs of $J_\perp$ do not annihilate, since their hard
triangle vanishes by the Ward identity (Fig.~\ref{fig:higgs}(f)). The same rules reproduce the
singlet mixing of QED$_3$~\cite{ChesterPufu2016}, the $\sigma^2$--$F^2$
mixing of the CP$^{N-1}$ model~\cite{VasilevNalimov1983}, the exponents of
Refs.~\cite{BK2019,IrkhinKataninKatsnelson1996} and those of the U(1)
theory with fermions and bosons~\cite{KaulSachdev2008}, and conserved
currents come out at $\Delta=2$ for every gauge parameter. The Higgs
entries, $4-144u=2\Delta_\sigma$ and $4+96u$, are not controlled at $N=12$,
and the running uses the scan of Sec.~\ref{sec:whichvbs}, in which
$\Delta_\sigma$ is varied between $0.38$ and $2$ with the other entries
fixed. The two further Higgs operators of the channel,
$\sigma\,\mathrm{Tr}\,\tilde M^2$ and $(\mathrm{Tr}\,\tilde M^2)^2$, decouple
at this order and leave $\Delta_1$ unchanged (below). The lowest four-fermion singlet is therefore
relevant, or close to marginal, at this order, like the lowest
singlet of the SO(5) deconfined point on the fuzzy sphere,
$\Delta_S\simeq2.8$--$3.0$ at the largest size in its conformal region ($2.90(6)$ from its pseudocritical fit), which drifts toward relevance with system size~\cite{Zhou2024fuzzy}. The comparison is not clean, because the two
theories differ in $N$: for the flux singlet the same rules give
$2+128/(\pi^2N)=5.24$ at $N=4$, against $5.37$ on the fuzzy sphere, but
$\Delta=2$ at $N=12$, and the pure-fermion hybrid singlet at $N=4$ is
$6.1$ against the fuzzy-sphere $\simeq2.9$; whichever $N$ describes the
fuzzy-sphere point, one of the two comparisons misses by about $3$. The
seed is mapped at the lattice scale onto $(\mathrm{Tr}\,\Xb\X)^2$ and
$R=J_\psi^2$ and run with the full matrix. At $\lambda$ it is
$\mathsf s(\lambda)=\sum_b e^{(4-\Delta_b)L}P_b\mathsf s(\Lambda)$ over the
eigenspaces $b$, up to the common factor $e^{-L}$, and its $F^aF^a$,
$\sigma^2$ and $M$ components, which carry no four-fermion content at tree
level, are dropped. It is then projected onto the QBT and run in the
one-loop flow. The projection stays inside the
fourteen invariant quartics to $10^{-14}$.

\begin{table}[tb]
\caption{Content of entries of Eq.~\eqref{eq:singletmatrix}, in units of $u$,
with the graph classes of Fig.~\ref{fig:higgs} and $F\equiv F^aF^a$. The first
rows are the images of $F$ on the current squares, for
$J_\psi=\bar\psi\gamma_\mu\sigma^a\psi$ and $J_\Phi=(D_\mu\Phi)t^a\Phi$;
$F\to R$ is their $J_\perp^2$ component $a-b+c$. The pair diagonals contain the
exchanges, the covariant completions and the self-energies; $J_\Phi^2\to M=96$
and $M\to J_\Phi^2=40/3$. Entries are in the orientation of Eq.~\eqref{eq:singletmatrix}, in which
$(\mathrm{Tr}\,\Xb\X)^2$ and $R$ enter with the sign opposite to
$(\bar\psi\psi/4)^2$ and $J_\perp\cdot J_\perp$ in these conventions.
Entries marked $g$ follow from $g_im_{ij}=g_jm_{ji}$.}
\label{tab:singletentries}
\begin{ruledtabular}
\begin{tabular}{llc}
entry & graphs & value\\
\colrule
$F\to J_\psi^2$ ($a$) & (d) & $(1024-512\zeta)/9$\\
$F\to J_\psi\!\cdot\!J_\Phi$ ($b$) & (d) & $1024(1-\zeta)/9$\\
$F\to J_\Phi^2$ ($c$) & (d); $\sigma$ line & $512(1-\zeta)/9$; $-512/9$\\
$F\to R$ & $a-b+c$ & $0$\\
$F\to M$ & (d); $\sigma$ line & $-1024/3$; $0$\\
$F\to(\mathrm{Tr}\,\Xb\X)^2$ & (d) & $-4096$\\
$F\to\sigma^2$ & (e) & $256$\\
$\sigma^2\to R$ & (d) & $512/9$\\
$R\to R$ & $J_\psi^2$, $J_\psi\!\cdot\!J_\Phi$, $J_\Phi^2$ & $-48$, $-40$, $-32$\\
 & $x_b^2$, $2x_bx_f$, $x_f^2$ & $-128/3$\\
$R\to M$ & $x_f^2\times96$ & $-32/3$\\
$M\to R$ & (a), (b) & $-40/3$\\
$M\to M$ & (a), (b); (c) & $32$; $2\times32$\\
$R\to\sigma^2$, $M\to F$ & $g$ & $8$, $-5/3$\\
$(\mathrm{Tr}\,\Xb\X)^2\to F$ & $g$ & $-1/64$\\
\end{tabular}
\end{ruledtabular}
\end{table}
\emph{Entries of Eq.~\eqref{eq:singletmatrix}.} The ring graphs of
Fig.~\ref{fig:higgs}(d) are single integrals over the
momentum $q$ of the operator's boson pair,
$\int_q2K_0(q)D(q)D(q)\,L_1L_2\,D(q+P_1)$ for $F^aF^a$, with
$K_0=2(q^2\delta_{\mu\nu}-q_\mu q_\nu)$ its two-gluon vertex, $L_{1,2}$ the two
matter lines, $P_1$ the external momentum of the first, and
$D\to\langle\sigma\sigma\rangle$ for a constraint-field line. $K_0$ is
transverse, so $\zeta$ enters only through the exchanged line, and the
$\zeta$-dependent image, $-(512\zeta/9)(J_\psi+J_\Phi)^2$, vanishes by the
equation of motion. The two-triangle graph of Fig.~\ref{fig:higgs}(e) has the
induced $AA\sigma$ vertex $-\ii W\Pi^{(1)}_{\mu\nu}$ of
Eq.~\eqref{eq:higgsmassder} at zero $\sigma$ momentum and gives the entry
$3\int_q\mathrm{tr}[K_0DW\Pi^{(1)}DW\Pi^{(1)}D]=9216N_b^2u/N^2=256u$, the $3$
being the color trace; the graph with a single vertex contains
$\partial^2_{m^2}$ of the Higgs loop, whose hard part vanishes. The norms
$\pi^4x^8\langle O(x)O(0)\rangle$ are $128/9$ ($\sigma^2$), $512$ ($F^aF^a$),
$2$ ($R$), $5/2$ ($M$) and $1/512$ [$(\mathrm{Tr}\,\Xb\X)^2$, normalized as
$(\bar\psi\psi/4)^2$]. For $M$ the norm
is that of the primary: for a field $\varphi$ of dimension $1$,
$x^8\langle(\partial\varphi)^2(\partial\varphi)^2\rangle=88$, but modulo total
derivatives $(\partial\varphi)^2=-\tfrac23P$ with the primary
$P=\varphi\partial^2\varphi-\tfrac12(\partial\varphi)^2$, whose norm gives $8$;
the naive norm would turn $-5/3$ into $-55/3$. In $M\to M$ the graph of
Fig.~\ref{fig:higgs}(c) acts on each $\tilde M_{ab}$ pair with $1/4$ of its
$\Phi\cdot\Phi$ value $64WC_s/(\pi^2N^2)=128u$, the ratio of the color
factors $\sum_{cd}\mathrm{tr}(Xt^ct^d)\{t^c,t^d\}=128X$ for $X=\one$ and
$32X$ for $X=\tilde M_{ab}$.

\emph{Decoupling of $\sigma\,\mathrm{Tr}\,\tilde M^2$ and
$(\mathrm{Tr}\,\tilde M^2)^2$.} Written in flavor-singlet constituents (the
Higgs bilinear $\tilde M_{ab}$, the fermion bilinears and currents, $\sigma$
and $F^a$), the five operators of Eq.~\eqref{eq:singletmatrix} have two
constituents, $\sigma\,\mathrm{Tr}\,\tilde M^2$ has three and
$(\mathrm{Tr}\,\tilde M^2)^2$ four. In the basis normalized at $N=\infty$, a
graph with $I$ closed matter loops and $P$ boson lines between operators
$A$ and $B$ is of order $N^e$, $e=I-P-(n_A+n_B)/2$, with $n$ the number of
bilinears minus the number of boson fields; the entries of
Eq.~\eqref{eq:singletmatrix} have $e=-1$. Between three and two constituents
$e$ is half-integer, and its $e=-1/2$ graphs give no mixing: the $\sigma$ line of
$\sigma\,\mathrm{Tr}\,\tilde M^2$ absorbed on a Higgs leg gives a logarithm
proportional to the squared momentum of the leg, which the equation of
motion $D^2\Phi=\ii\sigma\Phi$ returns to $\sigma\,\mathrm{Tr}\,\tilde M^2$
itself; legs of the two $\tilde M_{ab}$ joined at the operator give
scaleless graphs; and a $\tilde M_{ab}$ pair closed on a single gauge or
$\sigma$ line vanishes, since $\mathrm{tr}\,\tilde M=\mathrm{tr}\,\tilde
Mt^c=0$. Between four and two constituents the $e=-1$ graphs fuse three
$\tilde M_{ab}$ into one, or two into one twice, and vanish in the same way.
The mixing with the channel is therefore at most of order $N^{-3/2}$ and
$N^{-2}$, respectively, and moves $\Delta_1$ only at $O(1/N^2)$. The dimensions are
$4+48u$ for $(\mathrm{Tr}\,\tilde M^2)^2$, from the exchanges and
self-energies, $-80u$, and the annihilation of each $\tilde M_{ab}$ pair,
$+32u$, and $\Delta_\sigma+\Delta(\mathrm{Tr}\,\tilde M^2)+32u=4-64u$ for
$\sigma\,\mathrm{Tr}\,\tilde M^2$, with $\Delta(\mathrm{Tr}\,\tilde
M^2)=2-24u$ and $32/(3\pi^2M)=8u$ per Higgs leg from the $\sigma$ line joined
to it. This is the value for which $\sigma\phi=-\ii\partial^2\phi$ has
dimension $\Delta_\phi+2$ at $C_s=0$. Like the other Higgs entries, these
dimensions are not controlled at $N=12$, where $\sigma\,\mathrm{Tr}\,\tilde M^2$
lies at $2.56$, close to $\Delta_1$; a mixing of normalized strength $20u$
between the two, formally of higher order, would move $\Delta_1$ by $0.03$.

\begin{table}[tb]
\caption{Eigenspaces of the $O(1/N)$ dimension operator that carry QBT
channels of the seeds, with their contributions to the VBS coupling $g_0$,
the plaquette anisotropy $g_P=[g(P_{\rm e})-g(P_{\rm f})]/2$, and the flux and
N\'eel$_z$ channels (Heisenberg in units of $J$ at $J'/J=0.63$;
$D'_\perp$ per unit amplitude on the reference bond). SO(5) irreps are
labeled by dimension, with $35$ the mixed-symmetry and $35'$ the
$[2,2]$ representation. The dimensions are those of the four-fermion
dimension operator in the physical-$N_f$ counting. The lowest SO(5)
singlet mixes with operators containing gauge and Higgs fields; its running
uses the five-operator $O(1/N)$ matrix of the text, Eq.~\eqref{eq:singletmatrix},
with $\Delta_\sigma$ scanned (the corresponding mode lies at $2.91$ at
$O(1/N)$), and the singlet spin-2 entry, which can mix
with the stress tensor, is the four-fermion value.}
\label{tab:treerun}
\begin{ruledtabular}
\begin{tabular}{lllrrrr}
seed & $\Delta$ & SO(5), Lorentz & $g_0$ & $g_P$ & flux & N\'eel$_z$\\
\colrule
$J$--$J'$ & $2.477$ & $1$, scalar & $+0.753$ & $0$ & $-0.467$ & $+0.753$\\
 & $3.460$ & $14$, scalar & $+0.123$ & $0$ & $-0.049$ & $0$\\
 & $3.850$ & $1$, spin 2 & $-0.378$ & $0$ & $-0.378$ & $-0.378$\\
 & $4$ & $14$, spin 2 & $+0.251$ & $0$ & $-1.005$ & $+1.508$\\
 & $4.202$ & $1$, scalar & $-0.183$ & $0$ & $+0.953$ & $-0.183$\\
 & $4.540$ & $35'$, scalar & $+0.063$ & $0$ & $-0.063$ & $-0.190$\\
$D'_\perp$ & $4$ & $35$, vector & $0$ & $+2$ & $0$ & $0$\\
\end{tabular}
\end{ruledtabular}
\end{table}

For $D'_\perp=-0.10$\,meV the fractional anisotropy $g_P/g_0$ at
$\lambda$, relative to its tree value, is $5\times10^{-10}$
($2.6\times10^{-12}$) at the $O(1/N)$ value of $\Delta_\sigma$,
$3.8\times10^{-7}$ ($9.7\times10^{-9}$) at $\Delta_\sigma=1$,
$1.7\times10^{-5}$ ($1.3\times10^{-6}$) at $\Delta_\sigma=1.5$ and
$1.0\times10^{-4}$ ($1.4\times10^{-5}$) at $\Delta_\sigma=2$, at $L=7$
($8.5$) in the leading counting; the physical-$N_f$ counting gives values
smaller by up to a factor $1.7$. The vertex asymmetry of the QBT flow is
$\Delta\ln\Gamma=2.8\times10^{-10}$ ($1.4\times10^{-12}$) at the $O(1/N)$
value of $\Delta_\sigma$ and $4.8\times10^{-5}$ ($6.3\times10^{-6}$) at
$\Delta_\sigma=2$, always in favor of $P_{\rm f}$ for $D'_\perp<0$, and
the flow ends on the VBS ray in every case. For $|D'_\perp|=0.18$\,meV the
ratios are the same and the asymmetries larger by a factor of up to $1.8$.
If instead a single dimension $\Delta_1$ is assigned to the eigenvector of
the four-fermion block, as in the crossover values $\Delta_1^*$ of
Sec.~\ref{sec:whichvbs}, the relative anisotropy is $4.6\times10^{-5}$
($5.7\times10^{-6}$), $1.9\times10^{-4}$ ($3.1\times10^{-5}$) and
$7.6\times10^{-4}$ ($1.7\times10^{-4}$) for $\Delta_1=2.6$, $2.8$ and $3$. On the same quadrature the Wick term of $D'_\perp$ in
the determinant has the linear coefficient $\partial b_2/\partial p\simeq-(1$--$2)\times10^{-5}
m^2\ln(1/m)$ at $d=0.6$ and $-(1$--$2)\times10^{-6}m^2\ln(1/m)$ at $d=0.3$
in physical units, where $p$ is the $D'_\perp$ amplitude in units of $t$;
its ratio to $m^2\ln(1/m)$ drifts by $40\%$ over the window. It vanishes
at $d=0$ and scales roughly as $d^{3}$--$d^{3.4}$. The gauge-mediated interaction of the QBT window
(Appendix~\ref{app:qbtgauge}) dilutes these asymmetries further and keeps
their sign.

\section{Gauge field on the band touching}
\label{app:qbtgauge}
Between $m_A$ and $\lambda$ the QBT modes couple to the SU(2) gauge field.
The color density of the eight modes is $\rho^a=\tfrac{\ii}{2}m^T\hat
r^am$, with $\hat r^a=P^Tr^aP$ the projected gauge generators. The generators $r^a$ commute
with $\mathcal O_{\rm DM}$, the QBT subspace is invariant under them, and $[\hat
r^a,\hat{\mathcal A}_i]=0$, so the color density commutes with $H_{\rm
QBT}$. Its static polarization, the Lindhard sum on a
scale-free logarithmic grid, is independent of $q$ and equals
$\Pi_{00}=2.7730\,\rho_0$, against $4\ln2=2.7726$. For a single complex
QBT the same evaluation gives the known value $2\ln2\,\rho_0$, both in the
complex and in the Majorana form. The polarization $\tfrac12|q|$ of the critical
adjoint Higgs triplets is $1.1(E/\lambda)^{1/2}\,\Pi_{00}$ at the momenta
$q\simeq2\sqrt{\lambda E}/v$ of the window for equal velocities, smaller
than $\Pi_{00}$ for $E\lesssim0.8\lambda$, and a Higgs mass adds a term
of order $m_A$, smaller than $\Pi_{00}$ for $m_A\lesssim\lambda$. The
color-Coulomb field is therefore screened on both sides of $m_A$. Its
exchange is the contact interaction $\tfrac U2\sum_a\rho^a\rho^a$ with
$U=1/\Pi_{00}$, i.e.\ $\rho_0U=1/(4\ln2)$ and the coupling $1/(8\ln2)$ in
the units of Eq.~\eqref{eq:bubble}. The transverse field couples through
the velocity vertex $\partial H_{\rm QBT}/\partial\bm k\propto v^2k/\lambda$
and is screened only by the orbital response,
$K_T(0,q)=(v^2q^2/4\pi\lambda)[\ln(\lambda/vq)+1.746]$ [Eq.~\eqref{eq:KTsmallq}] from the band touching
and the gapped Dirac background, which crosses over to $|q|/4$ above
$vq\simeq10\lambda$. At the momenta of the window the Higgs polarization
exceeds it, and the dimensionless transverse strength is of order
$(E/\lambda)^{1/2}$.

The interaction lies inside the fourteen invariant quartics and inside
their six-dimensional gauge-singlet subspace, both to $10^{-15}$. It is
valley-blind and attractive, with equal Hartree--Fock coupling $+1.08$ in
the flux, VBS$_x$, VBS$_y$ and N\'eel$_z$ channels. In the flow of
Appendix~\ref{app:qbtflow}, with the physical normalization of the
one-loop $\beta$ function stated there, it diverges at $\ell^*=0.83$ on
the VBS ray, with rates $0.36$ (VBS), $0.28$ (the three gauge-adjoint
masses, degenerate), $-0.07$ (flux) and $0$ (N\'eel$_z$) at $|g|=200$. At
$|g|=2\times10^4$ it reaches the majority ray of the random scan (overlap
$0.9999$), where the VBS and adjoint rates become equal. Rescaling it by
$0.5$ or $2$ changes $\ell^*$ in inverse proportion and not the ray. It
lies only $3\%$ of its norm from the separatrix between the VBS and
chirality basins: a chirality-attractive (flux$^2$) admixture of that
size, or one of $2.6\%$ along the least favorable gauge-singlet
direction, sends it to the chirality ray, and random admixtures of $5$--$10\%$ within the six
gauge-singlet couplings do so in about one case in ten. The nematic
(touching-splitting) bilinears, the leading instability of the screened
Coulomb QBT of bilayer graphene~\cite{VafekYang2010}, grow on the gauge
ray at $0.47$ times the VBS rate and never lead. With the static kernel
and the Higgs polarization included scale by scale, $U(E)=1/[\Pi_{00}+\Pi_H(E)]$ with
$\Pi_H/\Pi_{00}=1.13(E/\lambda)^{1/2}$ fed as a running source, the seed
diverges at $\ell^*=1.16$, $1.42$ and $1.82$ for the upper energies
$E_2=\lambda/4$, $\lambda$ and $4\lambda$, i.e.\ $T^{**}=E_2e^{-2\ell^*}
\simeq0.025\lambda$, $0.059\lambda$ and $0.105\lambda$, on the VBS ray in
every case (screening frozen at $E_2$ gives $0.017$--$0.029\lambda$).
With the renormalized lattice seed of Appendix~\ref{app:treeseeds},
whose absolute normalization in bubble units is not fixed, added at relative
weights from $0.03$ to $10^3$, every sum ends on the VBS ray, and the plaquette
asymmetry has the sign of the lattice seed. With $\Delta_1$ assigned to the
four-fermion singlet eigenvector it lies over the weights between
$5\times10^{-6}$ and $5\times10^{-5}$ at $\Delta_1=2.8$ and $L=7$, and
between $2\times10^{-5}$ and $2\times10^{-4}$ at $\Delta_1=3$. The
interaction is a gauge singlet, so the flow generates none of the
gauge-adjoint couplings of the fourteen, which are legitimate only below
$m_A$.

\emph{Dynamical and transverse screening.} Per color sector the eight QBT
modes form two chiral QBTs, whose color polarization is, with
$E_q=v^2q^2/4\lambda$ and $x=|\nu|/E_q$,
\begin{equation}
\frac{\Pi_{00}(\ii\nu,q)}{\rho_0}=2\ln\frac{1+x^2}{x^2+\frac14}
+\frac2x\Bigl[\frac\pi2-2\arctan\frac1x+\arctan\frac1{2x}\Bigr].
\label{eq:pidyn}
\end{equation}
It equals $4\ln2\,\rho_0$ in the static limit and falls as $\pi\rho_0/x$
at high frequency. It is the Kramers--Kronig transform of the spectral
function, which with $x=\omega/E_q$ is
\begin{equation}
{\rm Im}\,\Pi^R_{00}(\omega,q)=\begin{cases}0,&x<\tfrac12,\\[0.3ex]
\pi\rho_0\,(2x-1)/x,&\tfrac12<x<1,\\[0.3ex]
\pi\rho_0/x,&x>1,\end{cases}
\label{eq:impi}
\end{equation}
the threshold being the minimum $E_q/2$ of $E_{\bm k-\bm q/2}+E_{\bm k+\bm q/2}$.
The static moment $\tfrac2\pi\int_0^\infty d\omega\,{\rm Im}\,\Pi^R_{00}/\omega$
gives $4\ln2\,\rho_0$. The true compressibility, the limit $q\to0$ at
$T\gg E_q$, is $2\rho_0$ per sector; at the band touching the static and
uniform limits do not commute. The high-frequency tail
$\omega\,{\rm Im}\,\Pi^R_{00}\to\pi\rho_0E_q=c_0q^2$ gives $c_0=\tfrac14$.
The longitudinal current
kernel follows from the Ward identity, $K_L=(\nu/q)^2\Pi_{00}$, and the
density--transverse mixing vanishes. The transverse kernel of the band
touching and the gapped Dirac background is
\begin{align}
K_T(\ii\nu,q)&=\chi_T(0,0)-\chi_T(\ii\nu,q),\nonumber\\
\chi_T(\ii\nu,q)&=2\sum_{n,m}\int_{\bm k}
\bigl[f_n(\bm k_-)-f_m(\bm k_+)\bigr]\nonumber\\
&\quad\times\frac{|\langle n\bm k_-|\partial_{k_T}h|m\bm k_+\rangle|^2}{E_m(\bm k_+)-E_n(\bm k_-)-\ii\nu},
\label{eq:KT}
\end{align}
with $\bm k_\pm=\bm k\pm\bm q/2$, $h(\bm k)$ the four-band Hamiltonian of one
copy, the two band-touching modes and their $\pm4\lambda$ partners with
energies $\pm(\sqrt{4\lambda^2+v^2k^2}-2\lambda)$ and
$\pm(\sqrt{4\lambda^2+v^2k^2}+2\lambda)$, $f_n(\bm k)=f(E_n(\bm k))$, $\partial_{k_T}h$ the
current vertex transverse to $\bm q$, the factor $2$ for the two copies per
sector, and $\chi_T(0,0)$ the diamagnetic term. For $vq\ll\lambda$
\begin{equation}
K_T(\ii\nu,q)=\frac{v^2q^2}{4\pi\lambda}\Bigl[\ln\frac{\lambda}{vq}+c_T
+\Phi_T\Bigl(\frac{|\nu|}{E_q}\Bigr)\Bigr],
\label{eq:KTsmallq}
\end{equation}
where $c_T=\lim_{q\to0}[4\pi\lambda K_T(0,q)/v^2q^2-\ln(\lambda/vq)]=1.746$,
the momentum cutoff of Eq.~\eqref{eq:KT} being extrapolated to infinity
(the approach is $\propto1/\Lambda$). The universal function
$\Phi_T$, from the band touching alone, is $\Phi_T(x)\simeq0.307\,x^2$ for
$x\ll1$ and $\Phi_T\to\pi x/4$ for $x\gg1$, so that $K_T(\ii\nu,0)=|\nu|/4=c_0|\nu|$. The
critical Higgs fields add $\tfrac12|Q|$ times the relativistic transverse
projector. The gauge propagator is built in scale-covariant ($z=2$)
gauges (Fig.~\ref{fig:fourfermion}(e)). One exchange is inserted in the one-loop shell of the flow at all
frequencies, with the gauge boson on the shell, a routing in which the
pure-gauge exchange cancels against the fermion self-energy in every gauge,
so that the construction is gauge invariant; the routing with the fermion
lines on the shell is used for comparison. The projected seed is the contact
coupling $\theta_{\rm eff}=\sum_jc_js_j$ on the six gauge-singlet couplings $s_j$
that solves, in the least-squares sense,
\begin{equation}
P_6\bigl[2C(\theta_{\rm eff},s_j)\bigr]=P_6\,\mathcal L(s_j),\qquad j=1,\dots,6,
\label{eq:contactproj}
\end{equation}
where $\beta(g)=C(g,g)$ is the one-loop $\beta$ function of
Appendix~\ref{app:qbtflow}, $\mathcal L(s)$ the one-loop linear map of the
inserted exchange together with the fermion self-energy, both with the boson on
the shell, and $P_6$ the projection onto the singlet couplings. For the static
exchange $\mathcal L$ is itself of the form $2C(\theta,\cdot)$, and the
projection returns the static seed. The margins below refer to this projection.

At fixed scale $E$ the projected seed lies in the VBS basin from
$10^{-4}\lambda$ to $\lambda$, in every scale-covariant gauge and for both
routings of the loop momentum, with $\ell^*=0.6$ at $10^{-3}\lambda$ and
$1.0$--$1.1$ at $\lambda/4$. Along the flux$^2$ direction its margin is
$2.2$--$2.5\%$ of its norm at $10^{-4}\lambda$, $2.6\%$ at $10^{-2}\lambda$
and $2.9$--$3.4\%$ at $\lambda/4$, and along the least favorable
gauge-singlet direction $1.8$--$2.1\%$, $2.2\%$ and $2.4$--$2.9\%$, against
$3.1\%$ and $2.6\%$ for the static seed. Followed scale by scale from
$E_2$, it diverges at $\ell^*=0.96$, $1.07$ and $1.21$ for $E_2=\lambda/4$,
$\lambda/2$ and $\lambda$ ($0.92$, $1.02$ and $1.14$ with the fermion lines
on the shell),
so that $T^{**}=E_2e^{-2\ell^*}=(0.037$--$0.040)\lambda$,
$(0.059$--$0.065)\lambda$ and $(0.089$--$0.10)\lambda$, and
$(0.18$--$0.22)\lambda$ at $E_2=4\lambda$, on the VBS ray in every case
(screening frozen at $E_2$ gives $0.029$--$0.068\lambda$ for $E_2$ between
$\lambda/4$ and $\lambda$). The fermion self-energy renormalizes the band
curvature and raises the dynamical exponent to $z\simeq2.2$ at low energy,
which lowers $T^{**}$ by $5\%$ with the boson on the shell and by up to
$15\%$ with the fermion lines on the shell. Higgs exchange through the Yukawa couplings
vanishes on the QBT at first order and couples through the $\pm4\lambda$
states with vertices linear in $k$ (Fig.~\ref{fig:fourfermion}(f)). It is chirality-attractive and moves the
gauge seed across the separatrix only for $y^2\simeq1.6$--$1.8$ in units
in which the gauge exchange $16/(N|Q|)$ corresponds to $y^2\simeq1.3$ at
$|Q|=\lambda$. The DM-induced Yukawa channels change the margin by about
$10^{-4}$. The one-loop flow is used at order-one coupling from its first
e-fold, so $\ell^*$ and $T^{**}$ are order-of-magnitude estimates.

\section{Exact diagonalization of the field rule}
\label{app:ed}
Exact diagonalization of Eq.~\eqref{eq:H}, which does not use the gauge
theory, confirms the field rule of Sec.~\ref{sec:field}. The DM vectors are those of
Table~\ref{tab:dm}, the Zeeman term is $\bm B\cdot\sum_i\bm S_i$ with
isotropic $g$ unless stated, $\bm N=\sum_i\epsilon_i\bm S_i$, and
energies are in units of $J$, with $J'/J=0.63$ unless stated. Three
clusters are used: a 16-site torus spanned by $(4,0)$ and $(0,4)$, on
which all eight elements of the buckled locked group act exactly; a
20-site torus spanned by $(4,2)$ and $(-2,4)$, which keeps only $\hat S_4$
and its powers; and an 8-site cluster at $T=0$ and $T/J=0.1$, $0.3$ and $1$. The
DM couplings are either an enlarged reference set,
$(D,D'_\perp,D'_{\parallel,s},D'_{\parallel,ns})=(0.1,-0.05,0.05,-0.05)J$,
or a material set ($D$ from NMR, $D'_\perp$ from ESR, the in-plane
inter-dimer couplings from DFT),
$(-0.034,-0.025,0.0058,-0.0078)J$.

On 16 sites at $T=0$ the zero-field response
$\chi^{ab}=\partial\langle N^a\rangle/\partial B_b$ is the pure
off-diagonal tensor $\chi^{xy}=\chi^{yx}$, with every other entry below
$10^{-15}$. It is $-0.219$ for the reference set, $-5.7\times10^{-3}$ for
the material set, and $+0.611$ for the reference set at
$J'/J=0.70$, so its sign is not universal. Reversing every in-plane DM
component changes $\chi$ by less than $2\times10^{-14}$. With the in-plane
couplings scaled by $s$, $\chi^{xy}/s^2=-0.217$, $-0.220$ and $-0.219$ at
$s=0.25$, $0.5$ and $1$. $D'_\perp$ alone gives less than
$2\times10^{-15}$, the intra-dimer $D$ alone $-0.076$, and
$D'_{\parallel,s}$ alone $-0.036$. At finite field, $\bm B\parallel[100]$
induces $\bm N\parallel[010]$ and $\bm B\parallel[110]$ induces $\bm N$
along the field axis, with $|\bm N|/B=0.220$--$0.237$ for
$B=0.02$--$0.1J$; $\bm B\parallel[001]$ gives $|\bm N|/B<2\times10^{-13}$
up to $B=0.3J$. In a $c$-axis field of $0.1J$ the in-plane site moments,
$0.024$ per site for the reference set, form the intra-dimer staggered
pattern of Ref.~\cite{Kodama2005}, antiparallel within each dimer, with no
N\'eel component. For a field of $0.1J$ tilted by $8.6^\circ$ from $c$
at azimuth $\phi$, the in-plane N\'eel moment lies along
$(\sin\phi,\cos\phi)$ with $|\bm N_\parallel|/B_\parallel=0.2251$--$0.2253$
for every $\phi$.

On 20 sites the off-diagonal response is $-0.209$. The reduced symmetry of
this torus also allows a diagonal part, $\chi^{xx}=-\chi^{yy}=2.8\times
10^{-3}$, which the full group forbids. $D'_\perp$ alone gives less than
$10^{-16}$, and $\bm B\parallel[001]$ induces no N\'eel moment. On 8 sites
$\chi^{xy}=-0.0077$ at $T=0$ and $+0.034$, $+0.032$ and $+0.0019$ at $T/J=0.1$, $0.3$ and $1$,
unchanged under reversal of the in-plane DM, and zero for $D'_\perp$ alone
and for $\bm B\parallel[001]$ at $B=0.3J$.

With a dimer-dependent in-plane $g$ tensor, $g_1-g_2=2\delta$, and no DM,
the response equals $\delta\,\partial\langle\bm N\rangle/\partial h_s$ to
seven digits. On 16 sites at $J'/J=0.63$ the ground state is the exact
dimer singlet and the staggered susceptibility vanishes; it is finite for
$J'/J\ge0.8$ and at $T>0$ on 8 sites.

\section{Monopole operators of the U800 state}
\label{app:monopole}
At the SU(2) point the gauge flux is not conserved and there is no
monopole operator. The SO(5) vector of N\'eel and VBS order parameters
consists of fermion bilinears~\cite{WangNahum2017}, and these are
classified in Sec.~\ref{sec:inventory}. In Z3000 the gauge group is
$\mathbb Z_2$. In U800, $\langle\Phi_3\rangle\neq0$ breaks SU(2) to U(1),
and the low-energy theory is compact QED$_3$ with four two-component Dirac
fermions. A monopole operator inserts $2\pi$ flux. Each Dirac cone then
contributes one zero mode, and gauge neutrality fills two of the four,
which gives six monopoles. They decompose as $6=1\oplus5$ under the SO(5)
of the parent theory~\cite{SongPRX2020,WangNahum2017}: an SO(5) singlet and a vector
made of two spin-singlet (VBS-like) and three spin-triplet (N\'eel-like)
operators. The action of the lattice symmetries on them includes a phase in the
flux (topological) U(1). This phase is not contained in the PSG and has to
be computed~\cite{SongPRX2020,SongNC2019}.

In the U800 PSG [Eq.~(A.4) of Ref.~\cite{Feuerpfeil2026}] the
translations have gauge part only in the U(1) invariant gauge group, while $G_x$ and $\sigma_{xy}$ carry
a $\tau^1$-type gauge part that reverses the U(1) charge. Both operations
also reverse orientation, so they preserve the sign of the U(1) flux. So
does $\mathcal T$, which maps $u_{ij}e^{iA_{ij}\tau^3}$ to
$-u_{ij}e^{iA_{ij}\tau^3}$ before its gauge part, which commutes with
$\tau^3$, is applied. Every element therefore maps monopoles to
monopoles. An antiunitary element that sends a monopole $M$ to
$e^{i\varphi}M$ only fixes the phase of $g_M$ in $g_MM+{\rm
h.c.}$, so a monopole is symmetry-allowed if and only if it is invariant
under the unitary elements.

The monopole representation follows from inserting $2\pi$ flux into the
U800 mean-field state, $u_{ij}\to u_{ij}e^{iA_{ij}\tau^3}$, on an
$L_t\times L_t$ torus ($L_t=2L$) with flux $2\pi/L_t^2$ per plaquette. A symmetry
$g$ acts on Nambu modes through its PSG. In the flux sector it is combined
with the gauge transformation that restores $A$, lifted to Fock space as
$\exp(i\sum_i\vartheta_iQ_i)$ with $Q_i=\psi_i^\dagger\tau^3\psi_i$, which
acts trivially on Gutzwiller-projected states. Each eigenvalue is taken
relative to that of the zero-flux ground state with the same lift. A
single flux quantum on a torus is invariant only under point operations
that share a fixed point. We therefore use two geometries. The first is
$C_4$ about the center of an empty square, with holonomies
$(\pi+\pi/L_t,\pi-3\pi/L_t)$. The second is $\{\sigma_{xy},\sigma_{x\bar
y},C_2\}$ about a dimer center, with holonomies
$(\pi+\pi/L_t,\pi-\pi/L_t)$. The zero-flux reference has holonomies
$(\pi,\pi)$ and is gapped, whereas periodic boundary conditions keep Dirac
zero modes. The characters of $G_x$ and of the translations follow from
the relations $\sigma_{xy}C_4=T_{(-2,2)}G_x$ and $G_x^2=T_{2x}$.

For $L=3$ to $6$ and flux angles $\theta$ between $0.8$ and $2.4$ (the
$\pi$-flux point is $\theta=\pi/2$) the results do not change. The flux
sector has two zero modes per spin, separated from the rest of the
spectrum. The spin-triplet monopoles transform with a single character
$\eta$. The three spin singlets have eigenvalues $\{+1,+1,-1\}$ under
$C_4$ and under each mirror, which fixes the SO(5)-singlet character
$\chi_1$ as the repeated value. The two VBS monopoles $\Phi_{V\pm}$ are
identified directly. A hopping modulation with the symmetry of the VBS
mass $V_\mp$ splits the zero modes and leaves exactly one spin-singlet
monopole at zero energy, $\Phi_{V\pm}$, and its eigenvalue under the
preserved element is then read off. Table~\ref{tab:monopole} lists the
results.

\begin{table}[t!]
\caption{Characters of the six U800 monopoles under the unlocked SS
generators, obtained by $2\pi$-flux insertion. $\Phi_{V+}$
($\Phi_{V-}$) is the VBS monopole even (odd) under $C_4$. The spin-triplet
monopoles transform as a spin vector times the listed character.
$\mathcal T$ fixes the phase of the allowed coupling and does not enter
the selection.}
\label{tab:monopole}
\begin{ruledtabular}
\begin{tabular}{lccccc}
 & $T_{2x},T_{2y}$ & $G_x$ & $\sigma_{xy}$ & $\sigma_{x\bar y}$ & $C_4$ \\
\colrule
SO(5) singlet & $+$ & $+$ & $+$ & $+$ & $+$ \\
$\Phi_{V+}$ & $+$ & $-$ & $-$ & $-$ & $+$ \\
$\Phi_{V-}$ & $+$ & $-$ & $+$ & $+$ & $-$ \\
spin triplet & $+$ & $-$ & $+$ & $+$ & $-$ \\
\end{tabular}
\end{ruledtabular}
\end{table}

The SO(5)-singlet monopole is invariant
under the full unlocked group. For symmetries under which the Higgs field
is even, this is the statement of Ref.~\cite{SongPRX2020} that a U(1)
state with an SU(2) parent has a symmetric monopole. Here it also holds
for $G_x$ and $\sigma_{xy}$, under which $\Phi_3$ is odd and the U(1)
charge is conjugated. The VBS and triplet monopoles carry the same signs as the VBS and N\'eel mass rows of Table~\ref{tab:unlocked}.
Neither is invariant, and spin rotations exclude the triplet in any case.
The locked group adds an axial spin rotation to each element
[Eq.~\eqref{eq:locked}]. The spin singlets do not feel it, so the SO(5)
singlet remains allowed and the VBS monopoles remain forbidden. On the
triplet, $\hat S_4$ acts as $\eta(C_4)R_z(-\pi/2)=-R_z(-\pi/2)$, with
eigenvalues $-1,\pm i$, so no spin direction is invariant. The same holds
for the high-temperature locked group and for $D'_\perp$ alone. DM
therefore neither adds nor removes a symmetry-allowed monopole.

U800 thus carries one symmetry-allowed monopole, with or without DM. The symmetry argument excludes local masses in U800 but not confinement by monopole proliferation. Compact QED$_3$ with the SO(5)-singlet monopole
is a Higgs descendant of the SU(2) theory and shares its anomaly and its
possible infrared behaviors~\cite{WangNahum2017}. The leading monopole of
QED$_3$ with four two-component flavors has scaling dimension $0.265\times4\simeq1.06$
at leading large-$N_f$ order~\cite{BKW2002} and $1.02$ with the $O(N_f^0)$ correction~\cite{Pufu2014}, far below $3$, so
the allowed SO(5)-singlet monopole is expected to be strongly relevant and
U800 to confine; its dimension in the present theory, with the DM
couplings, is not computed here. The classification assumes that
the leading monopoles form the six-dimensional vector and that the
mean-field spinon state fixes their topological phases~\cite{SongPRX2020}.

\section{\texorpdfstring{Feynman rules and one-loop integrals of the large-$N$ sector}{Feynman rules and one-loop integrals of the large-N sector}}
\label{app:integrals}

This appendix gives the rules and the integrals that underlie every
large-$N$ number in the paper. The diagrams are those of
Fig.~\ref{fig:diagrams}, evaluated analytically with the gauge parameter
$\zeta$ and the gauge factor $C_g$ kept general.

\subsection{Rules}
\begin{table}[tb]
\caption{Euclidean $d=3$ integrals used in this appendix, per two-component flavor
where a loop is involved. Here $\int_p\equiv\int d^3p/(2\pi)^3$, and
$\perp$ denotes the transverse part. The third row is the raw
fermion-loop integral, which with the Euclidean rules of
Eq.~\eqref{eq:rules} enters the polarization with a positive sign. In the
fourth row the explicit minus sign comes from the expansion of the
bosonic $\operatorname{Tr}\log$, the raw integral's transverse part being
$-|q|(\delta_{\mu\nu}-\hat q_\mu\hat q_\nu)/16$, and the seagull graph is
scaleless. The fifth row is the mass derivative of the fourth, per complex
scalar of unit charge, with
$D_0=p^2+m^2$, $D_q=(p+q)^2+m^2$ and the infrared term $\propto1/m$ removed,
and the sixth is the scalar triangle; both enter the Higgs sector.}
\label{tab:masters}
\begin{ruledtabular}
\begin{tabular}{ll}
$\displaystyle\int\!\frac{d^3p}{(2\pi)^3}\frac{1}{p^2(p+q)^2}$ & $\dfrac{1}{8|q|}$\\[2ex]
$\displaystyle\int\!\frac{d^3p}{(2\pi)^3}\frac{(p+q)_\alpha}{p^2(p+q)^2}$ & $\dfrac{q_\alpha}{16|q|}$\\[2ex]
$\displaystyle\int_p\frac{\mathrm{tr}[\gamma_\mu\slashed p\gamma_\nu(\slashed{p}+\slashed{q})]}{p^2(p+q)^2}$ & $\dfrac{|q|}{16}\Bigl(\delta_{\mu\nu}-\dfrac{q_\mu q_\nu}{q^2}\Bigr)$\\[2ex]
$\displaystyle-\!\int_p\frac{(2p+q)_\mu(2p+q)_\nu}{p^2(p+q)^2}\Big|_{\perp}$ & $\dfrac{|q|}{16}\Bigl(\delta_{\mu\nu}-\dfrac{q_\mu q_\nu}{q^2}\Bigr)$\\[2ex]
$\displaystyle\partial_{m^2}\!\int_p\Bigl[\frac{(2p+q)_\mu(2p+q)_\nu}{D_0D_q}-\frac{2\delta_{\mu\nu}}{D_0}\Bigr]$ & $-\dfrac{1}{4|q|}\Bigl(\delta_{\mu\nu}-\dfrac{q_\mu q_\nu}{q^2}\Bigr)$\\[2ex]
$\displaystyle\int_p\frac{1}{p^2(p+p_1)^2(p-p_2)^2}$ & $\dfrac{1}{8|p_1||p_2||p_1+p_2|}$\\[2ex]
$\displaystyle\int_{k<|q|<\Lambda}\frac{d^3q}{(2\pi)^3}\,\frac{f(\hat q)}{|q|^3}$ & $\dfrac{\langle f\rangle}{2\pi^2}\log\dfrac\Lambda k$\\[1.5ex]
$\langle\hat q_\mu\hat q_\nu\rangle,\ \langle\hat q_\mu\hat q_\nu\hat q_\rho\hat q_\sigma\rangle$ & Eq.~\eqref{eq:angavg}
\end{tabular}
\end{ruledtabular}
\end{table}
We work in Euclidean $d=3$ with Hermitian $\gamma_\mu=(\rho^y,\rho^z,\rho^x)$,
$\{\gamma_\mu,\gamma_\nu\}=2\delta_{\mu\nu}$,
$\gamma_\mu\gamma_\nu=\delta_{\mu\nu}+\ii\varepsilon_{\mu\nu\rho}\gamma_\rho$,
so that $\gamma_\mu\gamma_\alpha\gamma_\mu=-\gamma_\alpha$ and
$\mathrm{tr}[\gamma_\mu\gamma_\alpha\gamma_\nu]=2\ii\varepsilon_{\mu\alpha\nu}$.
The critical theory carries no Maxwell term. The gauge propagator is the
inverse of the one-loop polarization of Fig.~\ref{fig:diagrams}(g),(h)
(Appendix~\ref{app:polarization}),
\begin{align}
G(p)&=\frac{\slashed p}{p^2},\qquad N=2N_f+4N_b,\nonumber\\
D_{\mu\nu}(q)&=\frac{16}{N|q|}\Bigl(\delta_{\mu\nu}-\zeta\,\hat q_\mu\hat q_\nu\Bigr),
\label{eq:rules}
\end{align}
with $\zeta$ the gauge parameter ($\zeta=0$ Feynman, $\zeta=1$ Landau).
The gauge vertex is $\gamma_\mu\sigma^a$ on the gauge doublet. For an
insertion that is a gauge singlet the two vertices of a one-loop graph
contract to $\sum_a\sigma^a\sigma^a=C_g=3$, kept general so
that $C_g=1$ gives two-component QED$_3$. A closed fermion loop
carries $-1$, and each valley is one complex Dirac gauge doublet, so a
flavor-blind loop carries $\mathrm{tr}(\sigma^a\sigma^b)=2\delta^{ab}$
per valley (Appendix~\ref{app:polarization}). The flavored operators
carry a traceless flavor matrix $T$ acting on valley and spin indices,
for example an SO(5) vector component, or $\mu^y$, the $\psi$ form of
$\Gamma_T$ for the operator of Sec.~\ref{sec:spin2}. It commutes with the
gauge generators, so the two gauge vertices of every one-loop graph with a
flavored insertion contract to $C_g$. The operators are inserted at zero
momentum transfer unless stated. The spin-2 operator is built from
the covariant derivative, and its two vertices follow from
$\ii D^\nu=\ii\partial^\nu+A^{a\nu}\sigma^a$ acting on a plane wave of
momentum $k$, which gives $-k^\nu$ from the derivative and
$+\sigma^a$ from the gauge field. Thus
\begin{align}
&\text{scalar: } T,\qquad
\text{current: } T\gamma^\lambda,\nonumber\\
&\text{spin-2: } T\gamma^{(\lambda}\ii D^{\nu)}\ \to\
-\tfrac12T\bigl[\gamma^\lambda(k{+}q)^\nu+\gamma^\nu(k{+}q)^\lambda\bigr]
\label{eq:insertions}
\end{align}
for the ordinary vertex, where $k{+}q$ is the loop momentum flowing
through the operator, Fig.~\ref{fig:diagrams}(b), together with the
seagull vertex
$S^\alpha=+\tfrac12T(\gamma^\lambda\delta^{\nu\alpha}+\gamma^\nu\delta^{\lambda\alpha})\sigma^a$
at which one gauge line ends on the operator,
Fig.~\ref{fig:diagrams}(c),(d). The overall sign of the operator is
immaterial. The relative sign of the two vertices is fixed by gauge
covariance, the opposite signs of $-k$ and $+A$ in $\ii D$, and it
carries no free parameter.

\emph{Log extraction.} Every one-loop graph is a single integral over
the gauge momentum $q$. With external momenta $k\ll q\ll\Lambda$ we
expand $G(k+q)$ in $k/q$,
\begin{align}
G(k+q)&=\frac{\slashed k+|q|\hat{\slashed q}}{q^2}\nonumber\\
&\times\Bigl[1-\frac{2k\!\cdot\!\hat q}{|q|}+\frac{4(k\!\cdot\!\hat q)^2-k^2}{q^2}
+\dots\Bigr],
\end{align}
keep the terms of the full integrand (including $D\propto1/|q|$) that
scale as $|q|^{-3}$, average over the direction $\hat q$ with
\begin{align}
\langle\hat q_\mu\hat q_\nu\rangle&=\tfrac13\delta_{\mu\nu},\nonumber\\
\langle\hat q_\mu\hat q_\nu\hat q_\rho\hat q_\sigma\rangle&=
\tfrac1{15}(\delta_{\mu\nu}\delta_{\rho\sigma}+\delta_{\mu\rho}\delta_{\nu\sigma}
+\delta_{\mu\sigma}\delta_{\nu\rho}),
\label{eq:angavg}
\end{align}
and use $\int_k^\Lambda\!\frac{d^3q}{(2\pi)^3}\,|q|^{-3}=\frac{1}{2\pi^2}\log\frac\Lambda k$.
A vertex graph then equals $v\times(\text{tree structure})\times\log\Lambda$
plus possible other tensor structures, and the self-energy equals
$s\,\slashed k\log\Lambda$. With these definitions the scaling dimension
of the inserted bilinear is
\begin{equation}
\dim[\bar\psi\,\mathcal M\,\psi]=\dim_{\rm tree}-(v+s),
\label{eq:dimrule}
\end{equation}
which follows from the scaling of the full three-point function, with
inverse propagator $\propto\slashed k(1-s\log\Lambda/k)$ and amputated
vertex $\propto\mathcal M(1+v\log\Lambda/k)$.

\subsection{Self-energy}
Figure~\ref{fig:diagrams}(a) is
$\Sigma(k)=C_g\!\int\!\frac{d^3q}{(2\pi)^3}\gamma_\mu G(k+q)\gamma_\nu D_{\mu\nu}(q)$.
The $k$-linear $|q|^{-3}$ terms are
$\gamma_\mu[\slashed k-2\hat{\slashed q}(k\!\cdot\!\hat q)]\gamma_\nu
(\delta_{\mu\nu}-\zeta\hat q_\mu\hat q_\nu)/q^2\times16/(N|q|)$. Using
$\gamma_\mu\slashed k\gamma_\mu=-\slashed k$,
$\langle\hat{\slashed q}(k\!\cdot\!\hat q)\rangle=\slashed k/3$ and
$\langle\hat{\slashed q}\slashed k\hat{\slashed q}\rangle=-\slashed k/3$
one finds the angular average
$\slashed k\,[-1+\tfrac23+\tfrac\zeta3+\tfrac{2\zeta}3]=\slashed k(3\zeta-1)/3$,
hence
\begin{equation}
s=\frac{8C_g(3\zeta-1)}{3\pi^2N}.
\label{eq:s}
\end{equation}

\subsection{Vertex corrections}
Figure~\ref{fig:diagrams}(b) with insertion $\mathcal M$ is
$C_g\!\int\!\frac{d^3q}{(2\pi)^3}\gamma_\mu G(k_1{+}q)\,\mathcal M\,G(k_2{+}q)\gamma_\nu D_{\mu\nu}(q)$.

\emph{Scalar} ($\mathcal M=T$). At leading order
$G(q)G(q)=\hat{\slashed q}\hat{\slashed q}/q^2=1/q^2$, so the integrand
is $(\gamma_\mu\gamma_\mu-\zeta\hat{\slashed q}\hat{\slashed q})/q^2=(3-\zeta)/q^2$
and
\begin{align}
v_{\rm scalar}&=\frac{8C_g(3-\zeta)}{\pi^2N},\nonumber\\
\dim[\bar\psi T\psi]&=2-\frac{8C_g(3-\zeta)+\tfrac83C_g(3\zeta-1)}{\pi^2N}\nonumber\\
&=2-\frac{64C_g}{3\pi^2N},
\end{align}
gauge-parameter independent and equal to $2-64/(\pi^2N)=1.46$ at
$C_g=3$, $N=12$. The formula has the form of Eq.~(79) of
Ref.~\cite{Feuerpfeil2026}, evaluated with the count of
Appendix~\ref{app:polarization}.

\emph{Conserved current} ($\mathcal M=T\gamma^\lambda$).
$\gamma_\mu\hat{\slashed q}\gamma^\lambda\hat{\slashed q}\gamma_\mu
=-2\hat q^\lambda\hat{\slashed q}+\gamma^\lambda\to\gamma^\lambda/3$ and
the $\zeta$ term gives $-\zeta\gamma^\lambda$, so
\begin{equation}
v_J=\frac{8C_g(1-3\zeta)}{3\pi^2N}=-s
\ \Rightarrow\ \dim=2\ \text{for every }\zeta,
\end{equation}
as required by the Ward identity. It
holds for each component $\lambda$ separately, with no admixture of
other structures.

\emph{Spin-2} ($\mathcal M$ of Eq.~\eqref{eq:insertions}). The
$k$-linear $|q|^{-3}$ terms now come both from the expansion of the two
propagators and from the explicit $(k{+}q)^\nu$ of the insertion, and
the rank-4 average of Eq.~\eqref{eq:angavg} enters. Decomposing the
result on the basis $\{\gamma^{(\lambda}k^{\nu)},\
\varepsilon^{\lambda\nu\rho}k_\rho\one ,\ \gamma^{[\lambda}k^{\nu]},\
k_\rho\gamma^\rho\delta^{\lambda\nu}\}$ gives a pure tree structure (all
other coefficients vanish identically), with
\begin{equation}
v_2^{\rm main}=\frac{8C_g(1-15\zeta)}{15\pi^2N}.
\end{equation}
The two seagull graphs of Fig.~\ref{fig:diagrams}(c),(d),
$C_g\!\int\!\gamma_\mu G(k{+}q)S^\alpha D_{\alpha\mu}(q)$ and its mirror,
are gauge-parameter free (their $\zeta$ parts average to zero) and
equal. With the relative sign of Eq.~\eqref{eq:insertions} each
contributes $-16C_g/(3\pi^2N)$ times the tree structure, so
\begin{align}
v_2^{\rm sg}&=-\frac{32C_g}{3\pi^2N},\nonumber\\
\gamma_2&=-(v_2^{\rm main}+v_2^{\rm sg}+s)
=\frac{64C_g}{5\pi^2N},
\label{eq:gamma2xi}
\end{align}
which is Eq.~\eqref{eq:gamma2}. The gauge parameter cancels between
Fig.~\ref{fig:diagrams}(a) and (b) alone, so $\zeta$-independence
places no constraint on the seagull contribution. Its sign and
magnitude follow from gauge covariance as stated above. At $C_g=1$,
Eq.~\eqref{eq:gamma2xi} gives $64/(5\pi^2N)$, the published abelian
value [Ref.~\cite{ZhouHe2023}, Eq.~(79); Eq.~(14) of Ref.~\cite{Gracey1994} at $\mu=3/2$, $C_2(R)=T(R)=1$, $N=2N_f$], with which the
result agrees without using it as input. Routing the external momentum
entirely through one propagator ($k_1=k$, $k_2=0$) or symmetrically
($k_1=k_2=k$) gives the same $v_2^{\rm main}$ with no admixture.

\subsection{Gross--Neveu comparison}
Replace the gauge line by a critical scalar with the Euclidean action
$\int\bar\psi(\slashed\partial+\sigma)\psi$, kernel
$D_\sigma(q)=8/(n|q|)$ for $n$ two-component flavors, Yukawa vertex $-1$
and no seagull. The same integrals then give $s=-\tfrac{4}{3\pi^2n}$
[from $\slashed k-2\langle\hat{\slashed q}(k\!\cdot\!\hat q)\rangle=\slashed k/3$],
$v_J=+\tfrac{4}{3\pi^2n}=-s$ (current protected) and
$v_2=+\tfrac{4}{15\pi^2n}$. Hence $\gamma_2^{\rm GN}=+16/(15\pi^2n)$ in
our two-component-flavor convention, i.e.\ $32/(15\pi^2n_{\rm comp})$ with
$n_{\rm comp}=2n$ components, the value of
Refs.~\cite{GKS2017,MutaPopovic} [Eq.~(3.20) of Ref.~\cite{MutaPopovic} at $D=3$ with its Eq.~(2.27), where $N\,{\rm Tr}[1]$ plays the role of $n_{\rm comp}$]. The fermion anomalous dimension
$-s/2=2/(3\pi^2n)$ is positive, as it must be. Dropping the factors of $\ii$ in propagators and vertices, which is
harmless for the gauge graphs above, would reverse the signs of $s$,
$v_J$ and $v_2$ here, because the Yukawa vertex carries $-1$ instead of
$\ii$.

\subsection{Annihilation (singlet-only) class}
For the flavor-singlet mass the operator can also sit on a closed
fermion loop connected to the external line by two gauge propagators,
Fig.~\ref{fig:diagrams}(e),(f). The loop with the zero-momentum mass
insertion is exact: $G(p)\,\one \,G(p)=1/p^2$, so per two-component
flavor of unit charge
\begin{equation}
L_{\mu\nu}(q)=\int\!\frac{d^3p}{(2\pi)^3}
\frac{\mathrm{tr}[\gamma_\mu(\slashed p+\slashed q)\gamma_\nu]}{p^2(p+q)^2}
=\frac{\ii}{8}\varepsilon_{\mu\alpha\nu}\hat q_\alpha,
\label{eq:Tloop}
\end{equation}
using the master integral of Table~\ref{tab:masters}. It is parity-odd
and transverse ($\hat q_\mu L_{\mu\nu}=0$), so the $\zeta$ terms of both
gauge propagators drop identically, and the two loop orientations
double it. The outer integral has the line propagator $G(q)$ between
the two attachments and the two orderings of the attachments, the
crossed one seeing $L(-q)=-L(q)$:
\begin{align}
\varepsilon_{\mu\alpha\nu}\hat q_\alpha
\bigl[\gamma_\mu\hat{\slashed q}\gamma_\nu-\gamma_\nu\hat{\slashed q}\gamma_\mu\bigr]
&=2\ii\,\varepsilon_{\mu\alpha\nu}\varepsilon_{\mu\rho\nu}\hat q_\alpha\hat q_\rho\nonumber\\
&=4\ii\,\one  ,
\end{align}
proportional to the tree mass structure with no admixture, so the log
coefficient of the outer structure is $2\cdot\frac{\ii}8\cdot4\ii/(2\pi^2)=-1/(2\pi^2)$.
Assembling with the fermion-loop sign, the flavor loop factor, the two
propagators $(16/N)^2$ and the topology constant $C_{\rm topo}$ that
converts the four labeled terms of the assembly (two orientations
$\times$ two orderings) into the two distinct diagrams,
\begin{align}
v_{\rm ann}^{\rm QED_3}&=(-1)\,N\Bigl(\frac{16}{N}\Bigr)^{2}C_{\rm topo}
\Bigl(-\frac1{2\pi^2}\Bigr)\nonumber\\
&=-\frac{64}{\pi^2N}\qquad(C_{\rm topo}=-\tfrac12),
\end{align}
i.e.\ $\Delta_{\rm singlet}-\Delta_{\rm adj}=-v_{\rm ann}=+64/(\pi^2N)$,
which is the QED$_3$ value~\cite{HermeleErratum,ChesterPufu2016,BK2019} [Eqs.~(3.5), (3.6) of Ref.~\cite{BK2019}].
The constant $C_{\rm topo}$ follows from the count of distinct
diagrams. A direct Wick expansion of the second-order term in the gauge
coupling with the mass insertion on the loop produces exactly two
distinct diagrams, the planar and the crossed attachment of the two gauge
lines to the external fermion line, each with unit weight (the $48$
labeled contractions of this topology, $24$ planar and $24$ crossed,
divided by the $4!$ of the vertex ordering). The four terms of the
assembly above, two loop orientations times two orderings, therefore
count each diagram twice, which is the magnitude $1/2$. The sign follows
from momentum conservation on the fermion line: with the gauge momentum
$\ell$ routed through the line, its propagator between the two
attachments is $G(k-\ell)\to-\hat{\slashed\ell}/|\ell|$ at the
logarithmic scale, opposite to the $G(+q)$ written above. Both the
magnitude and the sign follow from the Euclidean Feynman rules
alone, and the result agrees with the abelian literature value. For the
present theory the group factors are as follows: the loop carries
$\mathrm{tr}(\sigma^a\sigma^b)=2\delta^{ab}$ per valley, summed over the
$N_f$ valleys, the line carries $\sum_a\sigma^a\sigma^a=3$, and the
propagators carry $(16/N)^2$, so that $N$ in the QED$_3$ loop factor is
replaced by $6N_f$. The loop trace is symmetric in $a,b$ for both
orientations, so no $\varepsilon_{abc}$ part contributes. This gives
$\Delta_{\rm flux}-\Delta_{\rm VBS}=\frac{64}{\pi^2}\frac{6N_f}{N^2}
=\frac{16}{3\pi^2}\simeq0.54$ at $N_f=2$, $N=12$,
Eq.~\eqref{eq:lifting}. The VBS (valley-adjoint) and N\'eel loops are
traceless and receive no such term. Two points of the literature bear
on this class. To our knowledge the singlet value
$2+128/(3\pi^2N)$ first appears in the 2007 Erratum~\cite{HermeleErratum} to Ref.~\cite{Hermele2005}, whose original
appendix asserted equal singlet and adjoint dimensions. It is sometimes
attributed~\cite{ChesterPufu2016} to Ref.~\cite{RantnerWen2002}, which
computed only flavor-nonsinglet bilinears (the adjoint mass and conserved currents). The four-component large-$N_f$
values of $\gamma_m$~\cite{Gracey1993mass,CDGM2000} refer to a
parity-even bilinear that is an SU($2N_f$) adjoint of two-component
fermions in $d=3$. They receive no annihilation contribution and are not
singlet values.

\subsection{Higgs sector and the constraint field}
The adjoint Higgs fields are $M=3N_b$ real components $\Phi_A$ with
$S=\int[\tfrac12(D_\mu\Phi)^2+\tfrac{\ii}2\sigma\,\Phi\cdot\Phi]$,
$D_\mu\Phi=\partial_\mu\Phi-\ii A^b_\mu t^b\Phi$, and
$(t^b)_{ac}=-2\ii\varepsilon_{bac}$ on each triplet, the adjoint in the
normalization of $\sigma^b$. The rules are
\begin{align}
&G_\Phi(p)=\frac{\delta_{AC}}{p^2},\qquad
\Phi\Phi A^b_\mu:\ (p+p')_\mu\,t^b,\nonumber\\
&\Phi\Phi A^b_\mu A^c_\nu:\ -\delta_{\mu\nu}\{t^b,t^c\},\qquad
\Phi\Phi\sigma:\ -\ii,\nonumber\\
&\langle\sigma\sigma\rangle(q)=\frac{16|q|}{M},
\label{eq:higgsrules}
\end{align}
with $p$, $p'$ the momenta of the two Higgs lines along the line, and $D_{\mu\nu}$
of Eq.~\eqref{eq:rules}. Integrating out $\Phi$ gives
$\tfrac M2\operatorname{Tr}\log(-\partial^2+\ii\sigma)$, the factor $\tfrac12$ being
that of a real field, whose quadratic term is $\tfrac12\sigma\Pi_\sigma\sigma$ with
$\Pi_\sigma(q)=\tfrac M2\int_p[p^2(p+q)^2]^{-1}=M/(16|q|)$; its inverse is
$\langle\sigma\sigma\rangle$. In the same way a closed Higgs loop with gauge
vertices $a$ and $b$ carries $\tfrac12\sum\mathrm{tr}(t^at^b)=W\delta^{ab}$, summed
over the triplets, with $W=4N_b$ the Higgs part of $N=2N_f+4N_b$, and a Higgs line
carries $C_s=t^bt^b=8$ against $C_g=3$.

With the full propagator $G^{-1}=p^2-\Sigma\propto p^2(1+\eta L)$ and the
amputated $\Phi\cdot\Phi$ vertex $1+vL$, $L=\log(\Lambda/k)$, the insertion
$\Phi\cdot\Phi$ has dimension
$1+\eta-v$. Since $\langle\sigma\sigma\rangle=\Pi_\sigma^{-1}$ and $\Pi_\sigma$
is the two-point function of $\Phi\cdot\Phi$,
\begin{equation}
\Delta_\sigma=3-\Delta(\Phi\cdot\Phi)=2+v-\eta .
\label{eq:dsigmarule}
\end{equation}
The three classes of Fig.~\ref{fig:higgs}(a)--(c) give the following.

(a) Constraint-field exchange:
\begin{align}
\Sigma&=-\int_q\langle\sigma\sigma\rangle(q)\,G_\Phi(k{+}q),\nonumber\\
vL&=-\int_q\langle\sigma\sigma\rangle(q)\,G_\Phi(k{+}q)^2,\nonumber\\
\eta_\sigma&=\frac{8}{3\pi^2M},\qquad v_\sigma=-\frac{8}{\pi^2M}.
\end{align}

(b) One gauge line, with $V_\mu=(2k+q)_\mu$ (the seagull tadpole is
scaleless):
\begin{align}
\Sigma&=C_s\int_qV_\mu V_\nu D_{\mu\nu}(q)\,G_\Phi(k{+}q),\nonumber\\
vL&=C_s\int_qV_\mu V_\nu D_{\mu\nu}(q)\,G_\Phi(k{+}q)^2,\nonumber\\
\eta_A&=-\frac{8C_s(5+3\zeta)}{3\pi^2N},\qquad
v_A=-\frac{8C_s(\zeta-1)}{\pi^2N},
\end{align}
so that $v_A-\eta_A=64C_s/(3\pi^2N)$ for every $\zeta$.

(c) Higgs-loop annihilation. On a closed Higgs loop the insertion acts as a mass
term, and the loop with two gauge vertices and its seagull enters through
the mass derivative of Table~\ref{tab:masters},
\begin{align}
\Pi^{(1)}_{\mu\nu}(q)&=\partial_{m^2}\!\int_p\Bigl[
\frac{(2p+q)_\mu(2p+q)_\nu}{(p^2+m^2)((p+q)^2+m^2)}-\frac{2\delta_{\mu\nu}}{p^2+m^2}\Bigr]
\nonumber\\
&=-\frac{\delta_{\mu\nu}-\hat q_\mu\hat q_\nu}{4|q|},
\label{eq:higgsmassder}
\end{align}
after the infrared term $(\delta_{\mu\nu}-\hat q_\mu\hat q_\nu)/(4\pi m)$ is
removed; the longitudinal part vanishes identically. The two gauge lines end on the seagull
of the external line, so that
\begin{equation}
v_{\rm ann}=C_sW\!\int_q\mathrm{tr}\bigl[D(q)\Pi^{(1)}(q)D(q)\bigr]
=-\frac{64\,WC_s}{\pi^2N^2}.
\end{equation}
$\Pi^{(1)}$ is transverse, so the $\zeta$ terms of both propagators drop.

The constraint field generates no further logarithm. The insertion on a Higgs
loop coupled through two $\sigma$ lines contains $\partial_{m^2}$ of the
bubble $\arctan(|q|/2m)/(4\pi|q|)$, which has no hard part. The induced
$\sigma^3$ vertex is the triangle
\begin{equation}
\int_l\frac{1}{l^2(l+p)^2(l-q)^2}=\frac{1}{8|p||q||p+q|},
\end{equation}
and a graph with two of them falls as $1/(|q||p+q|)$ at large $|q|$, which has
no $|q|^{-3}$ term after the angular average. Adding the three classes in
Eq.~\eqref{eq:dsigmarule} gives Eq.~\eqref{eq:Dsigma}, $\Delta_\sigma=2-72u=0.379$
at $M=6$, $C_s=W=8$, $N=12$. With $\nu\simeq1+\Delta_\sigma-2$ the same rules
give $\nu=1-32/(3\pi^2M)$ for $C_s=0$ and $1-48/(\pi^2N)$ for CP$^{N-1}$
($M=2N$, $C_s=1$, $W=N$ for $N$ complex components).

\subsection{Master integrals}
The integrals are collected in Table~\ref{tab:masters}. Its third and
fourth rows are the polarization loops of
Fig.~\ref{fig:diagrams}(g),(h): a two-component Dirac fermion and a
complex scalar of unit charge screen identically, $|q|/16$, which with
the color weights of Appendix~\ref{app:polarization} (2 per Dirac
valley doublet, 4 per real adjoint triplet) gives Eq.~\eqref{eq:PiLO}
and hence the $16/(N|q|)$ of Eq.~\eqref{eq:rules}.

\bibliography{arXiv_v1_PRB}

\end{document}